\documentclass[oneside, twocolumn, conference]{IEEEtran}
\IEEEoverridecommandlockouts

\usepackage[utf8]{inputenc}
\usepackage[T1]{fontenc}

\usepackage{amsmath}
\usepackage{mathtools}
\usepackage{url}

\usepackage{balance}

\usepackage{caption}
\usepackage{subcaption}

\usepackage{amsthm}
\theoremstyle{definition}

\usepackage[noadjust]{cite}

\newcommand{\chart}[2]{%
\subfloat[#1]{%
    {\includegraphics[width=0.33\textwidth,clip]{#2}}%
}}

\title{Call Scheduling to Reduce Response Time of a~FaaS System
\thanks{Preprint of the paper accepted at IEEE Cluster 2022, Germany, 2022.}}

\author{
    \IEEEauthorblockN{%
        Paweł Żuk\IEEEauthorrefmark{1}, Bartłomiej Przybylski\IEEEauthorrefmark{2}, Krzysztof Rzadca\IEEEauthorrefmark{3}}
    \IEEEauthorblockA{%
        \emph{Institute of Informatics, University of Warsaw}\\
        Warsaw, Poland\\
        E-mail: \IEEEauthorrefmark{1}p.zuk@mimuw.edu.pl, \IEEEauthorrefmark{2}bap@mimuw.edu.pl, \IEEEauthorrefmark{3}krzadca@mimuw.edu.pl}
}

\usepackage{amsmath}
\usepackage{amssymb}
\usepackage{booktabs}
\usepackage{tabularx}
\usepackage{xcolor}

\begin{document}

\maketitle

\begin{abstract}
In an overloaded FaaS cluster, individual worker nodes strain under lengthening queues of requests. 
Although the cluster might be eventually horizontally-scaled, adding a new node takes dozens of seconds.
As serving applications are tuned for tail serving latencies, and these greatly increase under heavier loads, the current workaround is resource over-provisioning.
In fact, even though a service can withstand a steady load of, e.g., 70\% CPU utilization, the autoscaler is triggered at, e.g., 30-40\%  (thus the service uses twice as many nodes as it would be needed).
We propose an alternative: a worker-level method handling heavy load without increasing the number of nodes.

FaaS executions are not interactive, compared to, e.g., text editors: 
end-users do not benefit from the CPU allocated to processes often, yet for short periods.
Inspired by scheduling methods for High Performance Computing, we take a radical step of replacing the classic OS preemption by (1) queuing requests based on their historical characteristics; (2) once a request is being processed, setting its CPU limit to exactly one core (with no CPU oversubscription).

We extend OpenWhisk and measure the efficiency of the proposed solutions using the SeBS benchmark. In a loaded system, our method decreases the average response time by a factor of 4. The improvement is even higher for shorter requests, as the average stretch is decreased by a factor of 18. This leads us to show that we can provide better response-time statistics with 3 machines compared to a 4-machine baseline.
\end{abstract}


%%%%%%%%%%%%%%%%%%%%%%%%%%%%%%%%%%%%%%%%%%%%%%%%%%%%%%%%%%%%%%%%%%%%%%%%%%%%%%%%
\section{Introduction}
%%%%%%%%%%%%%%%%%%%%%%%%%%%%%%%%%%%%%%%%%%%%%%%%%%%%%%%%%%%%%%%%%%%%%%%%%%%%%%%%

With serverless computing \cite{Castro2019}, cloud customers execute their code without configuring or maintaining the production environment and the software stack.
One of the most popular classes of serverless products is called \emph{Function-as-a-Service (FaaS)} \cite{Fox2017}.
In FaaS, code snippets prepared by the customer define stateless functions.
Major cloud providers offer FaaS, e.g., Google Cloud Functions, AWS Lambda or Microsoft Azure Serverless.
Open-source software (e.g. Apache OpenWhisk) allows hosting FaaS infrastructure on-premises.

A recent study of the Azure Functions trace~\cite{Shahrad2020}, a major public provider of FaaS, shows that the rate of requests is uneven, with peaks of short duration. 
Even though these peak loads are short (minutes rather than hours), the performance of the service undergoing the peak load drives resource allocation in the steady state, because end-user serving workloads are often optimized for the tail (95th or 99th percentile) response latency \cite{Ding2019}.
Adding a new worker node to a FaaS cluster (horizontal autoscaling) does not address this problem, as it takes at least dozens of seconds \cite{Mao2012}.
Thus, currently the only way to handle peak loads without compromising the tail latency is to heavily over-provision: to run the services at low average CPU utilization (20\%-50\%), so that any peak can be handled gracefully on nodes. 
As the uneven load is common in end-user-serving systems (rather than specific to FaaS), studies of cloud (IaaS/PaaS) traces confirm low average CPU utilization: \cite{cortez2017resource} shows that more than half of Azure VMs have average CPU utilization below 20\%; 
in the Google's internal cloud, \cite{tirmazi2020borg} shows under-40\% CPU utilization for serving workloads; while \cite{bashir_take_2021} shows the CPU utilization of the median machine between 40\% and 50\%.   

Our approach reduces the response latency in loaded FaaS systems by optimizing the ordering of requests on a single worker node.
A worker node simultaneously serves multiple functions. 
In FaaS workloads, functions are short-lived (seconds), but called repeatedly.
Thus, scheduling strategies can rely on estimates of frequency and execution time of a call --- estimates based on local, node-level historical data.
In a loaded system, our node scheduler uses this information to prioritize short and infrequent calls.
We are also confident enough not to rely on OS-level preemption: 
instead, we set each executing container's CPU limit to a single core (and we don't oversubscribe the CPU).
As our experiments show, this approach improves response-related metrics from the perspectives of both a single node, as well as of a multi-node infrastructure.

Our approach has two main advantages.
First, with our node-level scheduling strategies, each worker node handles higher loads with significantly lower response time than the baseline method. 
Thus, our approach does not require so much CPU buffer to handle peak loads, so the amount of resources in the steady state can be reduced, resulting in lower infrastructure costs without compromising the latency.

Second, 
we do not modify the other two resource managers of a typical FaaS infrastructure: the controller (managing workers) and the load balancer (allocating individual invocations).
Our method is thus orthogonal to --- and can be applied in addition to --- the recent optimization efforts that concentrated on these elements of FaaS infrastructure:
function placement~\cite{Kaffes2019}, load balancing~\cite{Suresh2020}, autoscaling~\cite{Perez2019}, or choosing the repertoire of warm containers 
by predicting future calls and setting up containers in advance~\cite{Shahrad2020}.

The contributions of this paper are as follows:
\begin{itemize}
    \item We propose a new method of node-level container management that benefits from patience and (theoretically more optimal) non-preemptive strategies. This method reduces the number of preemptions (compared to interactive systems) and cold starts (compared to the baseline OpenWhisk).
    \item Within this general method, we introduce a number of node-level scheduling policies that are based on locally-gathered historical data on function calls (Sec.~\ref{sec:nlsp}).
    \item We implement our policies in OpenWhisk. We conduct experiments on a FaaS benchmark SeBS~\cite{copik2021sebs}, that we extend to handle OpenWhisk (Sec.~\ref{sec:experimental-setup}). 
    \item We show experimentally that our policies improve response time metrics on the node (Sec.~\ref{sec:experiments}) and infrastructure (Sec.~\ref{sec:multiple-workers}) levels. In particular, compared to the baseline using 4 machines, our solution produces shorter response latencies running on just 3 machines.
\end{itemize}

%%%%%%%%%%%%%%%%%%%%%%%%%%%%%%%%%%%%%%%%%%%%%%%%%%%%%%%%%%%%%%%%%%%%%%%%%%%%%%%%
\section{Problem definition}
\label{sec:pd}
%%%%%%%%%%%%%%%%%%%%%%%%%%%%%%%%%%%%%%%%%%%%%%%%%%%%%%%%%%%%%%%%%%%%%%%%%%%%%%%%

Our node-level scheduling problem can be formalized as follows.
A node (e.g. an OpenWhisk invoker) is capable of executing $n_f$ different functions (we use the terms \emph{function} and \emph{action} interchangeably), $f_1, f_2, \dots, f_{n_f}$. Each of these functions can be called multiple times in response to various events. 
In this paper, we focus on medium to heavy loads, i.e. we assume that in a short time-window, the total number of function calls, $|I|$, is close to or exceeds the steady-state throughput of the node.

The $i$-th action call (we use terms \emph{action call} and \emph{request} interchangeably) is described by the moment the request was generated by the end-user, $r(i)$, the requested function index, $f(i)$, the actual time spent by the node on executing the call, $p(i)$, and the moment when the results were obtained by the end-client, $c(i)$.
The system is on-line and non-clairvoyant: 
a call is unknown until $r(i)$ when it is generated; and its processing time $p(i)$ is known only when the execution ends. 

Our goal is to minimize performance metrics related to response time in such an uncertain environment.
A single call's \emph{response time} is $R(i) = c(i) - r(i)$. This takes into account that both the request and the response are transferred through a potentially latent network (as opposed to measuring the flow time just at the node level).
We additionally measure the \emph{stretch} defined as $S(i) = R(i)/p(i)$, i.e., response time expressed in units of the processing time. The stretch captures the intuitive expectation that a one-second delay of a $50$-second call is less noticeable than that of a $50$-millisecond call. Measuring both the response time and the stretch is especially important for workloads with highly diverse functions. 

To measure the system performance, we aggregate $R(i)$ across all the calls. We report the standard statistics: the \emph{average response time} $\sum_i R(i)/|I|$, a metric widely used in operational research and systems~\cite{Baker1974}; the \emph{average stretch} $\sum_i S(i)/|I|$; as well as order statistics: medians and quartiles.

Additionally, as the processing time $p(i)$ depends on (although it is not fully determined by) the function $f(i)$ being called, we will show aggregations of response time across all calls of the function $f(i)$. We do so to make sure that our methods do not discriminate against a certain class of function --- short, long, often- or rarely-called.

%%%%%%%%%%%%%%%%%%%%%%%%%%%%%%%%%%%%%%%%%%%%%%%%%%%%%%%%%%%%%%%%%%%%%%%%%%%%%%%%
\section{Resource management in OpenWhisk}
%%%%%%%%%%%%%%%%%%%%%%%%%%%%%%%%%%%%%%%%%%%%%%%%%%%%%%%%%%%%%%%%%%%%%%%%%%%%%%%%

Apache OpenWhisk \cite{OpenWhisk} is an open-source serverless platform.
OpenWhisk executes stateless functions (also called \emph{actions}) in response to events triggered by HTTP requests, fixed alarms or other functions.
The functions can be developed using numerous programming languages.
OpenWhisk isolates functions by invoking a call in a Docker container initialized with a runtime environment specific to a programming language in which the function was defined (a function can also request a customized Docker \cite{merkel2014docker} image).
All the processing done in OpenWhisk is asynchronous (following standard Scala~\cite{Scala} practices). 
One of the key roles of OpenWhisk is to manage the calls and the containers as the load of the deployed functions dynamically changes.

OpenWhisk operates on three types of components: \emph{controllers}, \emph{invokers} and \emph{action containers}. 
A controller manages other entities and routes actions invocations to invokers, acting as a load balancer if multiple invokers are available. 
With multiple invokers, the communication requires an intermediate event streaming system, such as Apache Kafka. 
A single invoker manages a single \emph{worker node} acting as its node-level resource manager. 
A node hosts multiple action containers, which in turn execute function calls and return the results.
The invoker is also responsible for assigning a function call, routed by the controller, to a specific action container. 
In a standard configuration, a single container does not execute multiple function calls in parallel. 
We now focus on the current algorithm used by the invoker to choose an action container, as this is what our method modifies.

When an invoker receives a new request and there are pending requests, the request is added to the queue; otherwise, the invoker tries to arrange a container on which the request can be executed immediately.
First, the invoker tries to locate a \emph{free pool container} matching the request. 
A free pool container is a container that is initialized with the execution environment and the requested function. 
Such containers can be usually found if a particular function is executed repeatedly. 
If a free pool container is not available, the invoker tries to find a \emph{prewarm pool container}. 
A prewarm pool container is initialized with the execution environment suitable for a group of functions (e.g. all the Python functions), but the specific function has not yet been initialized. 
If no free or prewarm containers match the request, 
the invoker tries to create a new container. 
This may be impossible if there is not enough free memory. 
In such a case, some non-matching free pool containers may be removed (evicted). 
If there are not enough free pool containers to be evicted, 
the action call is queued.

In the standard OpenWhisk, the controller assigns a request to an invoker (that then executes the above algorithm).
Although simple to implement, balancing the load in this method is more static, as the decision to execute a request at a certain invoker cannot be easily reversed (and, if the invoker fails, the assigned requests are lost).
To counteract that, OpenWhisk is currently gradually switching to a new action assignment model \cite{OWSP} based on global queues. In this new model, the controller does not decide which invokers will be responsible for executing actions. Instead, each invoker pulls requests from common Kafka queues: each function has its own global queue, and the invoker pulls requests from queues matching its free pool containers. However, the controller still decides what kinds of containers will be created by specific invokers.
Thus, this new action assignment model is orthogonal to the node-level scheduling policies we present later in our paper --- our policies can be still used once this new model is fully implemented.

%%%%%%%%%%%%%%%%%%%%%%%%%%%%%%%%%%%%%%%%%%%%%%%%%%%%%%%%%%%%%%%%%%%%%%%%%%%%%%%%
\section{Node-level scheduling policies}
\label{sec:nlsp}
%%%%%%%%%%%%%%%%%%%%%%%%%%%%%%%%%%%%%%%%%%%%%%%%%%%%%%%%%%%%%%%%%%%%%%%%%%%%%%%%

Our main goal is to improve the performance of a single FaaS node, and thus --- by the rule of scaling --- of a whole FaaS cluster.
We replace the current approach (that uses free pool containers and FIFO queues) with more sophisticated --- yet still greedy --- scheduling policies, which are implemented using priority queues. The priority of an incoming action call depends on factors such as the actual processing times of similar actions performed in the past. In fact, we estimate the expected processing time of an action by the average processing time of at most $10$ recent executions of the same action. It has been proven empirically that such a number is sufficient \cite{Bap2021}. Moreover, the processing time is estimated on the node-level and thus is not affected by network latency. To simplify implementation, once a priority of a particular action call is computed, it does not change.

Using the expected processing time as the action priority can starve some actions (if there is always a longer/shorter request waiting to be processed).
It is not crucial for our strategies to be starvation-free, as we consider only a short interval in which the system is overloaded. However, some of our strategies explicitly prevent starvation, because we determine the priorities based on the expected \emph{completion} time or the total processing time from the past. In particular, we introduce the five following strategies:
\begin{itemize}
    \item \emph{First-In, First-Out} (FIFO), the baseline, in which the action priority is the time $r'(i) \geq r(i)$ the action call is received by the invoker (pulled from a queue);
    \item \emph{Shortest Expected Processing Time} (SEPT), in which the action priority is the expected processing time of the action, $\mathbb{E}(p(i))$. As in~\cite{Bap2021}, we estimate $\mathbb{E}(p(i))$ as the average processing time $\bar{p}(j)$ over last $10$ finished calls of the same function $f(i)$;
    \item \emph{Earliest Expected Completion Time} (EECT), in which the action priority is $r'(i) + \mathbb{E}(p(i))$; this corresponds to the expected completion time of an action if a processor is immediately available;
    \item \emph{Recent Expected Completion Time} (RECT), in which the priority is $\bar{r}(i) + \mathbb{E}(p(i))$, where $\bar{r}(i)$ is the moment when the previous call of function $f(i)$ was received;
    \item \emph{Fair-Choice} (FC), in which we prioritize actions based on the estimation of the total processing time of the recently concluded calls of the same function. Namely, we define the priority of the $i$-th action as
    $\#(f(i), -T)\cdot \mathbb{E}(p(i))$, where $\#(f(i), -T)$ is the number of calls of function $f(i)$ during last $T$ seconds, for $T$ being a long time interval, e.g. 60 seconds. FC is related to FCP~\cite{Bap2021}.
\end{itemize}

EECT and RECT strategies prevent starvation. For EECT, consider two actions, $i$-th and $j$-th. If $r'(j) > r'(i) + \mathbb{E}(p(i))$, then the $j$-th call will be executed after the $i$-th one. For this reason, it is impossible for the $i$-th call to wait infinitely long for being executed. The same reasoning can be applied to RECT, as the value of $\bar{r}(i)$ is increasing in time.

We stress that our scheduler dynamically estimates the expected processing time $\mathbb{E}(p(i))$ based on at most $10$ most recent processing times of the same function $f(i)$ on the worker node. To present the results on stretch, we also use the run-time estimates based on off-line benchmarking of $f(i)$ (which we present in Table~\ref{tab:sebs-functions}), but these off-line results are never used to make any scheduling decisions. This way, our approach remains valid for other functions, with processing times known a posteriori.

\subsection{Replacing memory-based scheduling with CPU-based scheduling}
\label{subsec:mbcpu}

OpenWhisk limits the number of busy containers (action containers that execute actions at any given moment), run by a single node, by the amount of available operational memory. One large container can be exchanged for a few smaller ones. By default, OpenWhisk assigns to each container a share of CPU that is nearly linearly proportional to its memory requirement. Thus, some containers are assumed to use less than 1 CPU core, while others can be assigned more (however, these are soft, rather than hard limits).

This policy leads to OS-level preemption. If the number of concurrently executed actions is greater than the number of CPU cores, then multiple context switches might be performed by the OS. Such context switching can have a significant negative impact on the response time.

As the theoretical variants of the strategies listed earlier are non-preemptive, we want to discourage the operating system from preempting currently executed actions. To achieve that, we drastically change the default approach:
\begin{itemize}
    \item We limit the number of busy containers with the number of available CPU cores.
    \item We replace CPU limits based on memory requirements with fixed ones: a single container is always assigned a CPU limit of exactly one core.
\end{itemize}
As a direct consequence of such a change, we can state what follows. If the limit of busy containers is less or equal to the number of available CPU cores, we get close to a non-preemptive model. If the limit of busy containers is greater than the number of available CPU cores, then preemption is introduced by the operating system natively.

Using a limit on busy containers equal to the number of CPU cores seems more reasonable when one assumes that the executed actions are CPU-intensive. Otherwise, i.e. for I/O-intensive actions, some CPU cores may stay idle, although they could execute another function. As in the SeBS benchmark \cite{copik2021sebs} we find both CPU- and I/O-intensive functions, we will verify the impact of that experimentally.

\subsection{Implementation overview}

We modified the source code of Apache OpenWhisk in order to implement the changes described above. We stress that these modifications are orthogonal to the currently-developed changes in the action scheduling model \cite{OWSP}.

Our policies need data on recent invocations of action calls. We gather this data by extending the invoker's processing pipeline. We log the moment a request is pulled from Kafka by the invoker to calculate $r'(i)$ (required by FC, EECT, and RECT). 
Then, when the invoker gets the response from the action container, we store the processing time in a per-function fixed-size buffer (if a function has never been executed, we set its estimated execution time to 0).
We also replace the invoker's simple queue by a priority queue. The priorities are computed based on the scheduling policy selected at the start of each experiment (based on a new configuration option we added to OpenWhisk). 

To implement the CPU limits (Sect.~\ref{subsec:mbcpu}), we changed the parameters of the \emph{docker run} command used to create each action container. In our implementation, each container is always assigned a whole CPU core. Moreover, we modified the invoker's behavior, so there are no more concurrent calls than the number of available CPU cores.

%%%%%%%%%%%%%%%%%%%%%%%%%%%%%%%%%%%%%%%%%%%%%%%%%%%%%%%%%%%%%%%%%%%%%%%%%%%%%%%%
\section{Experimental setup}
\label{sec:experimental-setup}
%%%%%%%%%%%%%%%%%%%%%%%%%%%%%%%%%%%%%%%%%%%%%%%%%%%%%%%%%%%%%%%%%%%%%%%%%%%%%%%%

\begin{figure}[tb]
    \centering
    \includegraphics[width=\columnwidth,trim={0 0mm 0 0mm},clip]{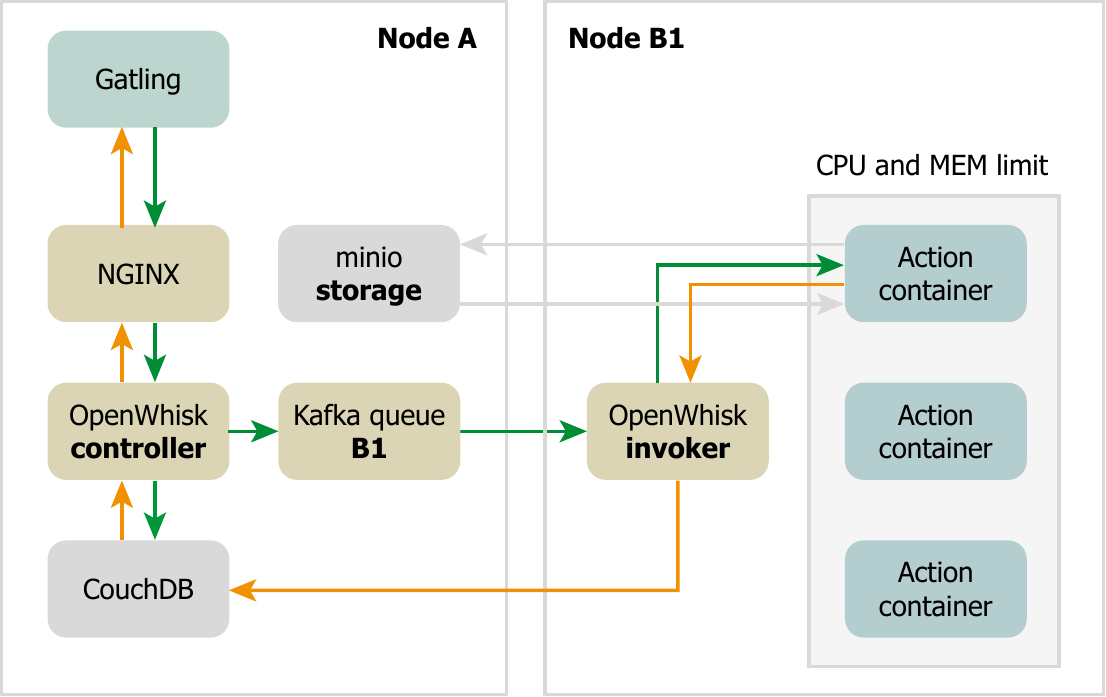}%
    \caption{A flow of a single request generated by Gatling during our experiments.}
	\label{fig:gatling-scheme}
\end{figure}

The goal of our experiments is to quantify the impact of the improved node-level scheduling policies. We thus use at least two separate OpenWhisk machines (physical in the on-premises experiments; virtual in the multi-machine experiments) --- one for the controller, and at least one for the invoker with its action containers. Fig.~\ref{fig:gatling-scheme} shows a flow of a single request in our setup, for a single invoker.

We use functions from the SeBS~\cite{copik2021sebs} benchmark.
Roughly half of these functions are computationally-intensive, while others strain I/O and network.
SeBS is designed to measure commercial cloud services providers such as AWS Lambda or Microsoft Azure Serverless. 
We extend SeBS by designing and implementing wrappers allowing us to deploy SeBS on OpenWhisk.

We simulate end-clients by generating requests with the  Gatling~\cite{Gatling} load testing tool (to reduce network noise, Gatling is deployed on the same node as the OpenWhisk controller). We define test scenarios as sequences of function requests (see Sect.~\ref{subsec:sexp}). Gatling executes such scenarios in a controlled and monitored manner. In contrast to other load testing tools, like Apache JMeter \cite{JMeter}, Gatling allows us to use Scala to create, manage and execute multiple tests.

Our on-premises experiments use two machines: each machine has 256~GB of RAM and two Intel Xeon Silver 4210R CPUs @ 2.40GHz with 20 hyper-threaded cores. Our setup runs Ubuntu 20.04 (Linux kernel 5.11.0-34-generic) with Docker 20.10.7. We changed the default CPU frequency governors of both machines to \emph{performance} to avoid CPU frequency scaling during experiments. 
We store Docker images and containers on NVMe. 
Our preliminary experiments used SSDs and the default 5.4.0-88-generic kernel, but we noticed Docker stability issues, especially for the OpenWhisk baseline, high load and 128~GB of OpenWhisk memory pool.

All FaaS requests are considered to be \emph{blocking}, i.e. the request is sent over an HTTP connection which remains open until the result is returned to the client. If the system is overloaded, this time may increase significantly --- by default OpenWhisk will return an error if it takes more than approx. 60s to execute the action.
Thus, in an overloaded node with standard time limits, the ``result'' of a scheduling strategy is a tuple --- response times and the number of failed requests. 
However, to be able to directly compare strategies, we strongly prefer to have a single measure.
We thus decided to increase the time limits high enough to have zero time-outs.
We claim this is fair as these increased limits have the same influence on all the measured strategies. 
Additionally, different FaaS providers set different timeouts on function duration (9 minutes in GCP and 15 minutes in AWS, while Azure Premium and Azure AppService allows arbitrary large timeouts).
Moreover, not every function has to produce a response that is directly delivered to the end-user -- in which case short timeouts are pointless.

\subsection{The structure of an experiment}
\label{subsec:sexp}

Our goal is to measure and compare the effectiveness of different scheduling policies fairly.
The general structure of a single experiment is as follows.

First, we warm up the action containers by issuing a certain number of function calls. These calls initialize the action containers, thus reducing the influence of cold starts on the results (see Sect.~\ref{sec:cont-evic} for a detailed discussion). If the number of available CPU cores is equal to $c$, then at most $c$ containers for each function can be used simultaneously. This is so as the number of busy containers is limited by the number of available CPU cores (c.f. Sect.~\ref{subsec:mbcpu}). Thus, we issue $c$ parallel calls for each function. Note that we do not measure the response times of these warm-up calls. After the containers are warmed-up, we are ready to start the actual experiment.

We generate the measured load as a burst of requests that are all uniformly issued in a 60-second time window.
These 60 seconds simulate a ``difficult'' overload scenario. With shorter (e.g. 10-second) bursts, scheduling inefficiencies will have a short-term influence on the response time and thus will not be that noticeable by the end-user. Conversely, longer bursts are best handled by a horizontal autoscaler adding more nodes. However, it takes at least dozens of seconds to set up a new, cold worker node --- and then seconds to warm up the action containers.

After 60 seconds, no new requests are issued. Otherwise, e.g., if the intensity decreased gradually, it would be harder to measure the influence of the precise load intensity on the performance of the scheduling strategies.

After sending the last call, Gatling waits until all the responses are returned. 
We report response times as measured by Gatling.
We monitor the response errors to make sure that no (or very few) response errors are generated. In our initial experiments, such errors generally hinted to various system-level problems (e.g.: I/O bottlenecks on an SSD or Docker's inability to maintain many containers).

We use functions from SeBS~\cite{copik2021sebs}, Table~\ref{tab:sebs-functions}. We considered all the functions defined in the benchmark except the 3 Node.js implementations (we use their Python alternatives) and the network microbenchmarks (as they measure network latency).
For each function, we defined a separate address of the HTTP endpoint used to invoke it, and specified its call parameters.
To approximate function's processing time $p(i)$ for stretch-related metrics, we benchmarked each function in an idle on-premises setup: 
we warmed up the corresponding containers, and then we called this function $50$ times.
In our experiments, we do not want to measure the latency of the network, but the capacity of the FaaS node. On the other hand, we are unable to directly measure the duration of the execution of an action call on the invoker level. Thus, in our stretch metrics, instead of the processing time (denoted earlier by $p(i)$), we use the median response time measured on the level of the Gatling client (see Table~\ref{tab:sebs-functions}). Although it lets us compare the performance of the system in the context of different functions, such a change may result in stretch values that are less than $1$.

\begin{table}
    \caption{Functions from the SeBS benchmark tool measured on the client side, on-premises setup. The measurements include ca. 10\,ms Kafka overhead.}
    \label{tab:sebs-functions}
    \begin{tabularx}{\columnwidth}{Xrrr}
    \toprule
                  & \multicolumn{3}{c}{Response time}\\
                  \cmidrule(lr){2-4}
    Function name & 5th perc. & Median & 95th perc.\\
    \midrule
    \texttt{dna-visualisation} & 8 415\,ms & \textbf{8 552\,ms} & 8 847\,ms\\
    \texttt{sleep} (1000\,ms) & 1 020\,ms & \textbf{1 022\,ms} & 1 026\,ms\\
    \texttt{compression} & 793\,ms & \textbf{807\,ms} & 832\,ms\\
    \texttt{video-processing} & 586\,ms & \textbf{593\,ms} & 605\,ms\\
    \texttt{uploader} & 184\,ms & \textbf{192\,ms} & 405\,ms\\
    \texttt{image-recognition} & 117\,ms & \textbf{121\,ms} & 237\,ms\\
    \texttt{thumbnailer} & 112\,ms & \textbf{118\,ms} & 124\,ms\\
    \texttt{dynamic-html} & 18\,ms & \textbf{19\,ms} & 22\,ms\\
    \texttt{graph-pagerank} & 11\,ms & \textbf{12\,ms} & 15\,ms\\
    \texttt{graph-bfs} & 11\,ms & \textbf{12\,ms} & 13\,ms\\
    \texttt{graph-mst} & 11\,ms & \textbf{12\,ms} & 13\,ms\\
    \bottomrule
    \end{tabularx}
\end{table}

\subsection{Hyperparameters: load intensity and CPU cores}

Two configuration options significantly influence the results: (1) the available resources --- in our case, the number of available CPU cores; (2) the load --- in our case, the amount of requests issued during the 60-second window. Roughly, by doubling the amount of resources (e.g., increasing the number of cores from 10 to 20), the system should double the load it can handle. Thus, to meaningfully compare the impact of increased resources with constant load --- or increased load with constant resources --- we introduce the notion of \emph{intensity} of a scenario: a multiplicative factor regulating the load (keeping the available resources constant).

Our experiments are designed so that each action is called the same number of times in a 60-second window.
As we consider 11 functions, in a scenario of intensity $\emph{v}$, denoting as $c$ the number of CPU cores for action containers, we generate exactly $1.1\cdot c\cdot \emph{v}$ requests (to simplify notation, we only consider intensities that are multiples of 10). For example, if there are $20$ CPU cores and the intensity is $30$, we generate a total of $660$ requests distributed uniformly in the 60-second window.

The average response time for the function selected uniformly from Table~\ref{tab:sebs-functions} is \textasciitilde$1.042$s. Thus, we state that for intensity 30, the CPU executes the function calls for roughly 50\% of the time.
Consequently, higher intensities correspond to higher actual utilization, and progressively more overloaded systems. Intensity 40 should make the processor execute the functions for 65\% of time, and we additionally consider the intensity of 60 (and, in the appendix~[p. \pageref{appendix}], intensities 90 and 120). Our estimations of CPU utilization do not take into account overheads of container creation and management. In fact, intensity 30 may result in full utilization of the processor, if managing container executing the function requires more time, on average per call, than executing the function itself.

Even with the same intensity, the number of CPU cores may still impact the performance. For example, some strategies may behave more flexibly when $660$ requests are executed on $20$ cores compared to when $330$ requests are executed on $10$ cores, although in both cases the intensity is equal to $30$. With the increasing number of cores, we observe the effects of the regression to the mean utilization on a single core, i.e. the load can be balanced between cores more efficiently. For this reason, we perform our experiments on 10 and 20 cores. In the appendix~[p. \pageref{appendix}], we also present the results on 5 cores.

For each pair of hyperparameters: numbers of CPUs and intensity, we generate $5$ different random sequences of calls. 

As the baseline, we use OpenWhisk with a single configuration change, the extended time limit. We set the same CPU and memory limits on the baseline workers as on our variants.

\section{Cold starts, evictions and the impact on the response time}
\label{sec:cont-evic}

Theoretical considerations on scheduling policies often assume that the functions can be executed exactly when needed (cf. \cite{Bap2021}). Unfortunately, this is not always a case in practice. Although some requests are processed by idle containers that were already initialized (\emph{warm start}), others may --- and sometimes even must --- be served by a new container (\emph{cold start}). It takes $500$\,ms on the average \cite{shahrad_architectural_2019} (and, in our measurements, up to $2$\,s) to fully initialize a new container, and thus the response time of a cold started request is always longer than a warm started one.
On the other hand, reducing the total processing time by maintaining an unlimited number of containers is impossible due to limited memory, so it is profitable to evict (remove) unused containers.

Decisions on which containers to create and which to remove significantly impact the response time. However, these are orthogonal to what we measure in this paper --- the impact of the scheduling policy on the response times of individual requests.
With evictions, even small, quasi-random changes in the scheduling policies can be randomly amplified by evictions, which would introduce noise to our results.
We decided to diminish this effect by reducing the number of evictions (and thus, cold starts) to zero or almost zero.

Our policies upper-bound the maximum number of containers in use by the number of functions and the CPU cores.
Thus, we expect that with increasing memory capacity, the number of evictions will eventually approach zero, and thus the number of measured cold starts will be almost zero (as we warm the containers before the measurement starts). 
On the other hand, OpenWhisk is greedy:
a pending request that has no free, warm container will trigger initialization of a new container.
Thus, we expect that with increasing memory capacity, the number of cold starts will decrease, but only slowly.

\begin{figure*}[tb]
    \centering
    \subfloat[original approach to node-level scheduling]{{\includegraphics[width=0.4\textwidth,clip]{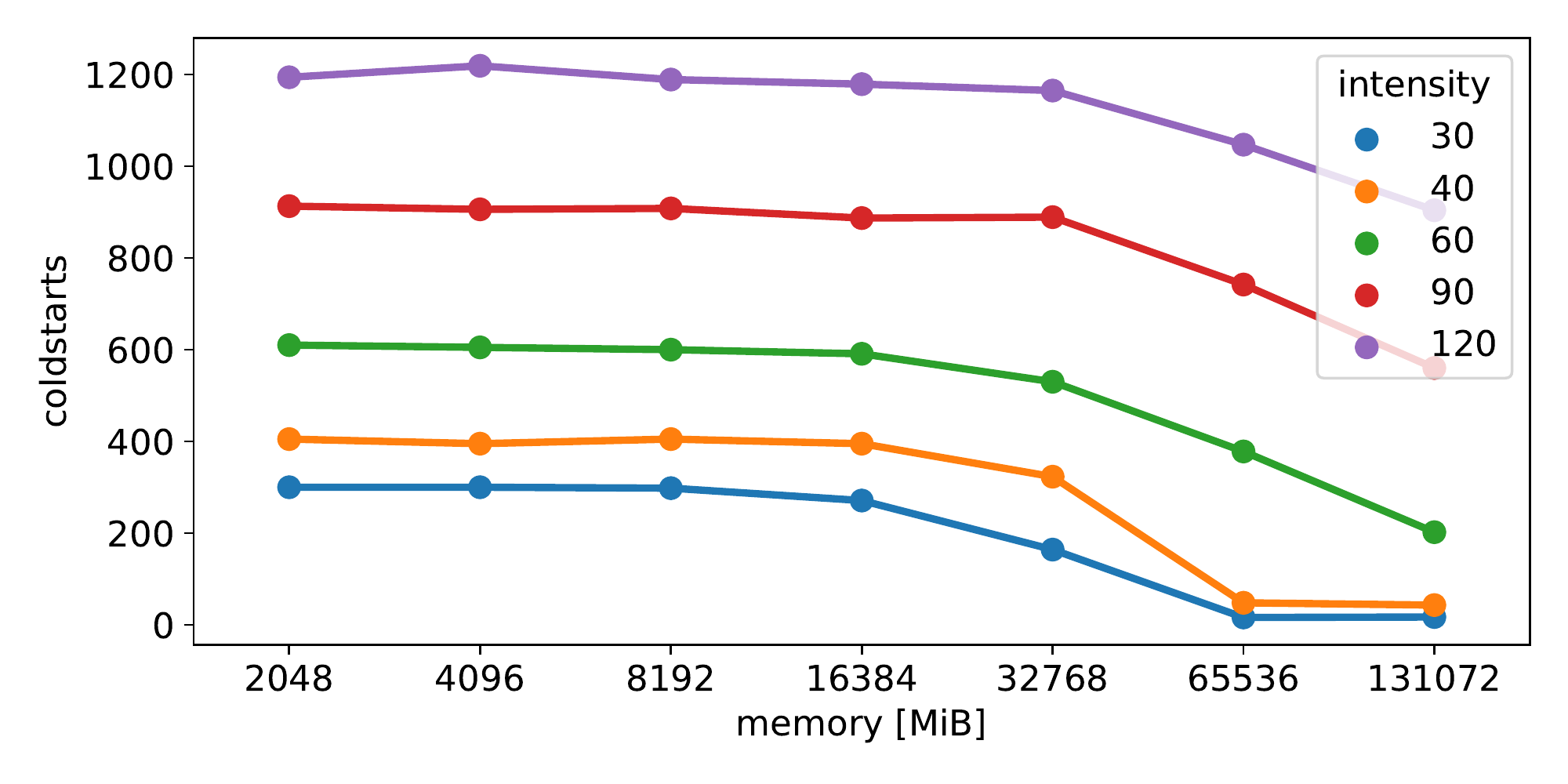}}}%
    \subfloat[our approach to node-level scheduling (FIFO variant)]{{\includegraphics[width=0.4\textwidth,clip]{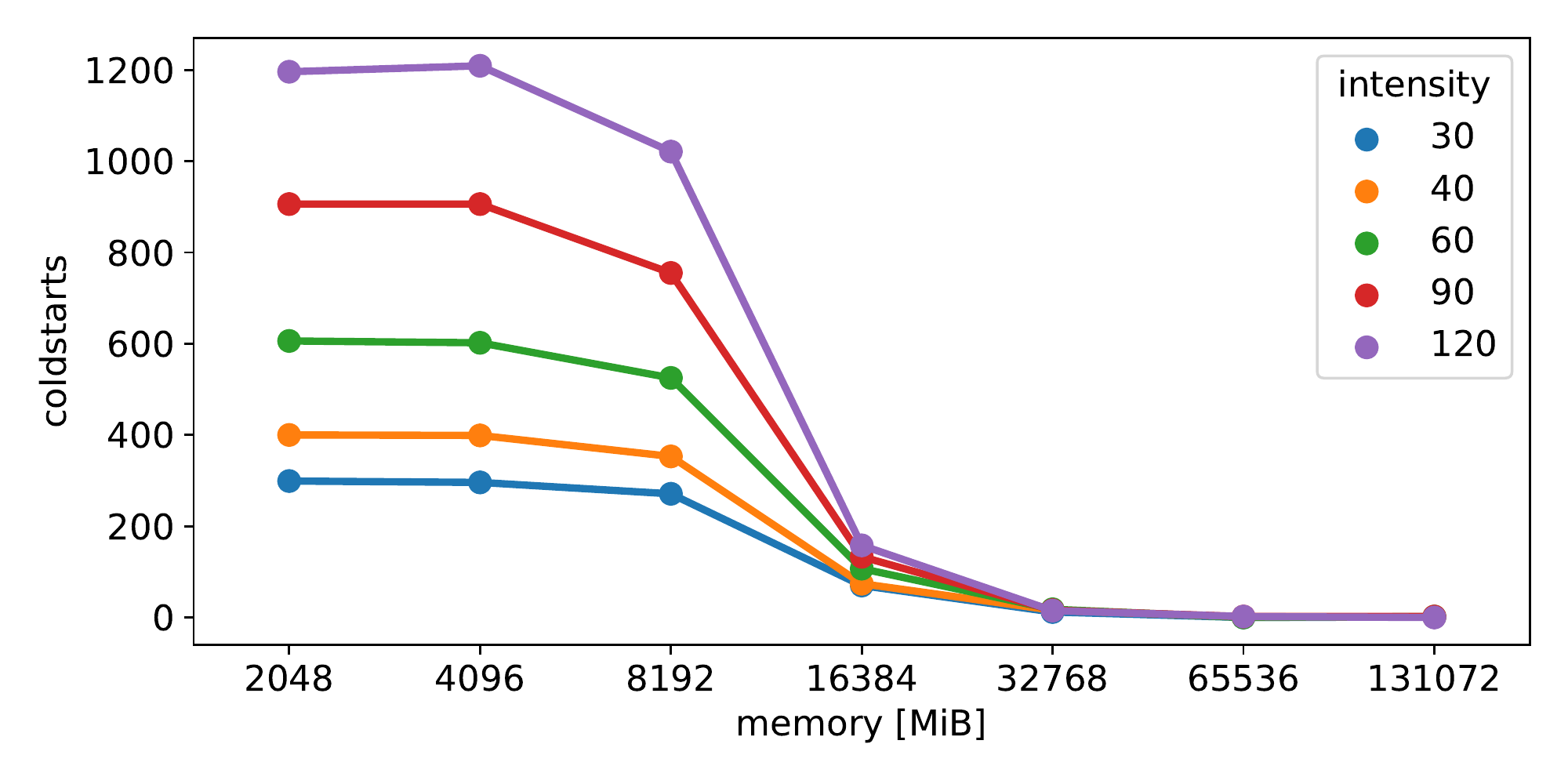}}}%
    \caption{Comparison of the number of cold starts on 10 CPU cores depending on intensity and amount of available memory.}
    \label{fig:coldstarts_vs_ram}
\end{figure*}

We verified these intuitions experimentally. We measured the number of cold starts for different load intensities and for increasing memory (following the setup as in Sect.~\ref{subsec:sexp}). The results are presented in Fig.~\ref{fig:coldstarts_vs_ram}.
For the original approach, Fig.~\ref{fig:coldstarts_vs_ram}(a), the number of cold starts strongly depends on the intensity, but less so on the memory. In fact, for load intensity 120, the node processes 1\,320 requests, and the number of cold starts exceeds 1\,100 with almost no dependency on the amount of available memory. This means that over 80\% of requests were processed on newly-created containers, each with the cold start overhead of almost 2 seconds. Moreover, for 128~GB of RAM, the number of concurrently-running containers was so large that Docker had problems running them --- and OpenWhisk responded with a barrage of errors. 

In contrast, in our FIFO policy, Fig.~\ref{fig:coldstarts_vs_ram}(b), starting from 32~GB, the number of cold starts does not change --- which suggests that RAM is no longer a constraint and thus almost no evictions happen.

Our goal was to determine the memory threshold starting from which evictions can be neglected --- and then run the rest of our experiments with this setup.  Based on the above, we perform all further experiments with the OpenWhisk memory pool restricted to 32 GB.

%%%%%%%%%%%%%%%%%%%%%%%%%%%%%%%%%%%%%%%%%%%%%%%%%%%%%%%%%%%%%%%%%%%%%%%%%%%%%%%%
\section{Influence of scheduling policies}
\label{sec:experiments}
%%%%%%%%%%%%%%%%%%%%%%%%%%%%%%%%%%%%%%%%%%%%%%%%%%%%%%%%%%%%%%%%%%%%%%%%%%%%%%%%

We start with an aggregated view over all considered load intensities and CPU cores count (Sect.~\ref{subsec:inf-st}). We then analyze the influence of the increasing load intensities (Sect.~\ref{subsec:inf-int}) and CPU cores (Sect.~\ref{subsec:inf-cpu}).
Finally, we change the relative call frequencies to highlight the fairness provided by the Fair-Choice (FC) strategy (Sect.~\ref{subsec:flm}).
Results presented here and in the following section aggregate over 5 repetitions with different call sequences (the variance between repetitions is small); our on-line appendix~[p. \pageref{appendix}] shows results for each of the 5 experiments, as well as additional aggregates and comparisons.

\begin{figure*}[tb]
    \centering
    \subfloat[10 CPU cores, intensity 30]{%
        {\includegraphics[width=0.33\textwidth,clip]{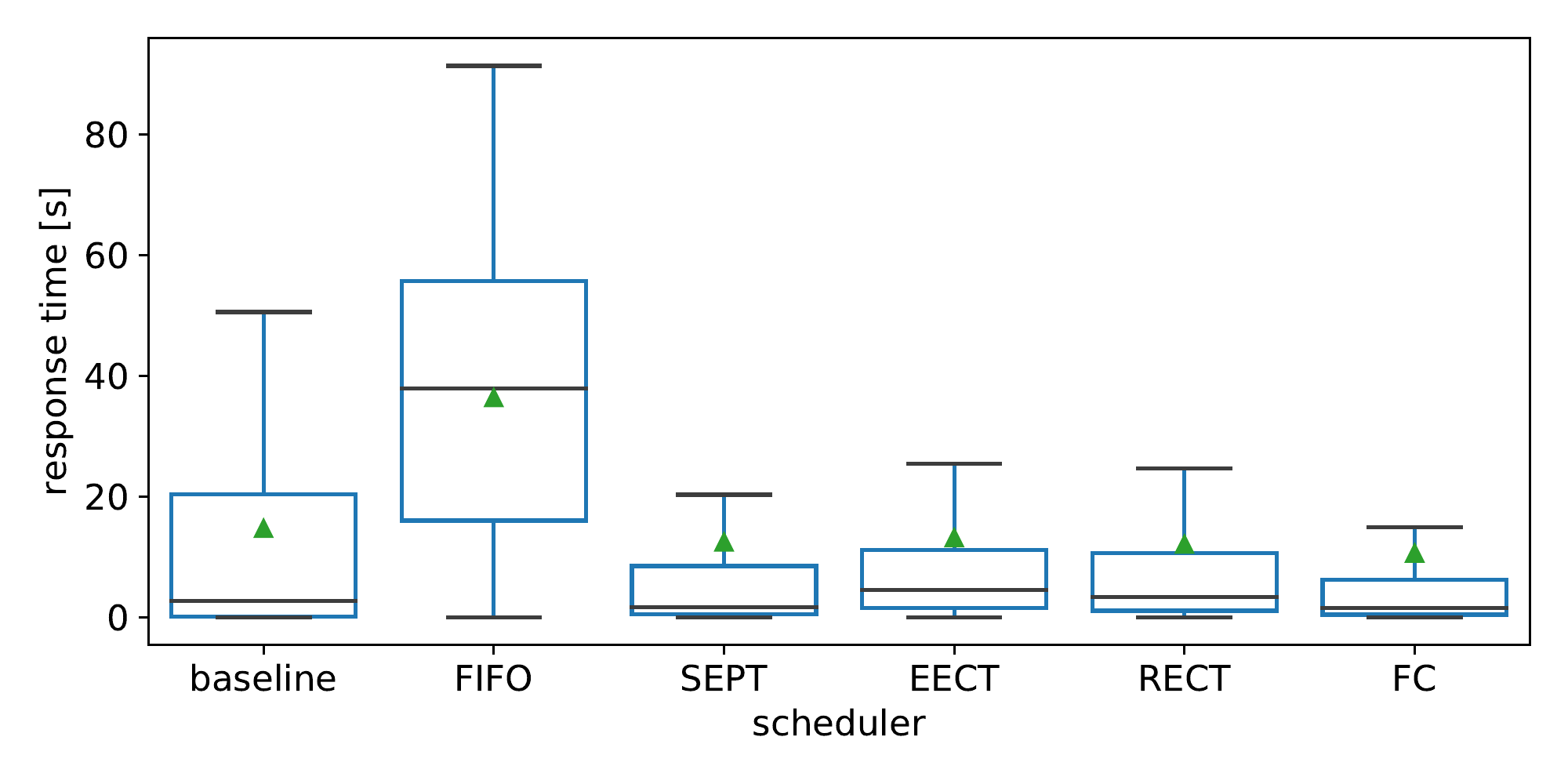}}%
    }%
    \subfloat[10 CPU cores, intensity 40]{%
        {\includegraphics[width=0.33\textwidth,clip]{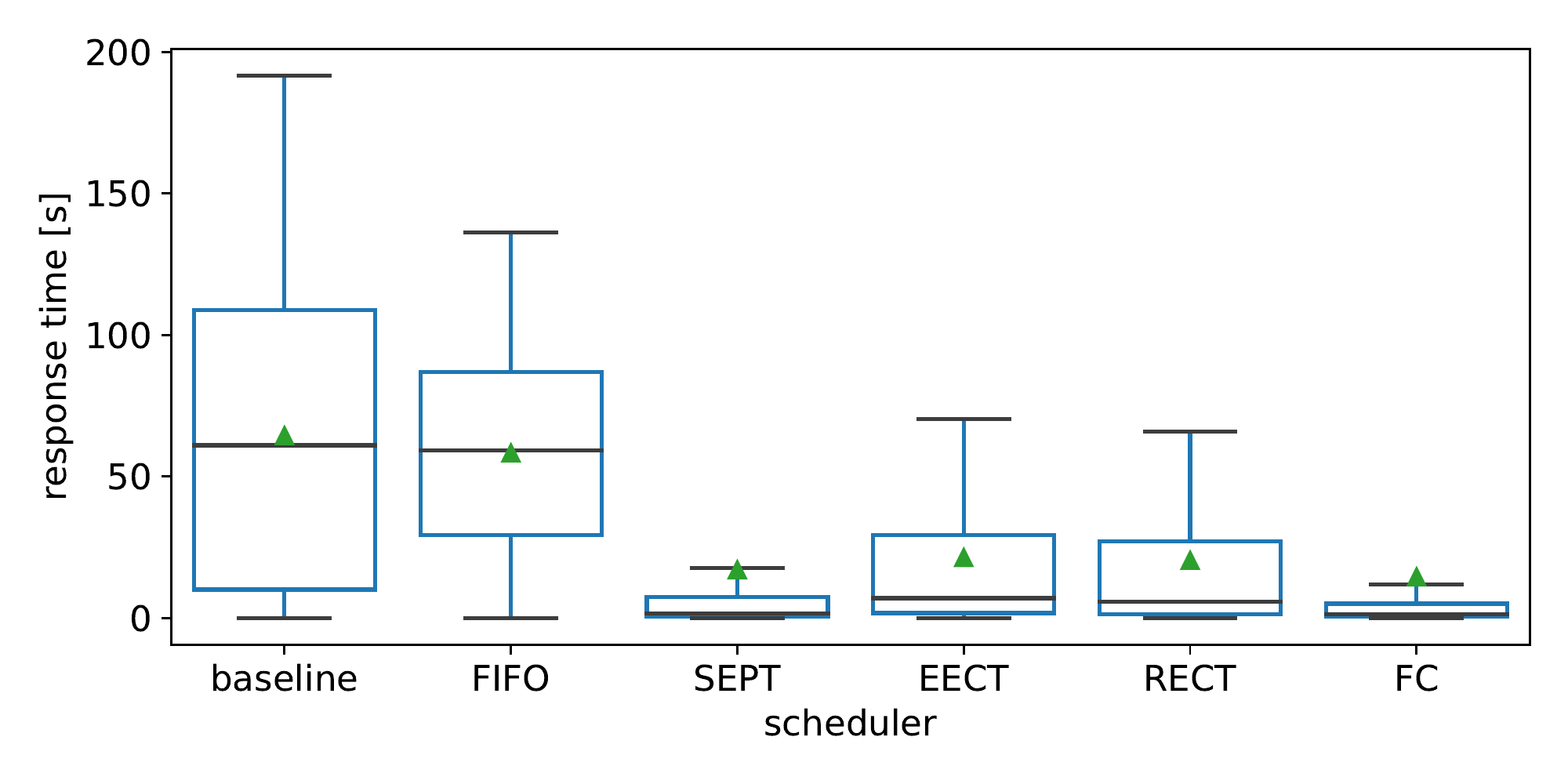}}%
    }%
    \subfloat[10 CPU cores, intensity 60]{%
        {\includegraphics[width=0.33\textwidth,clip]{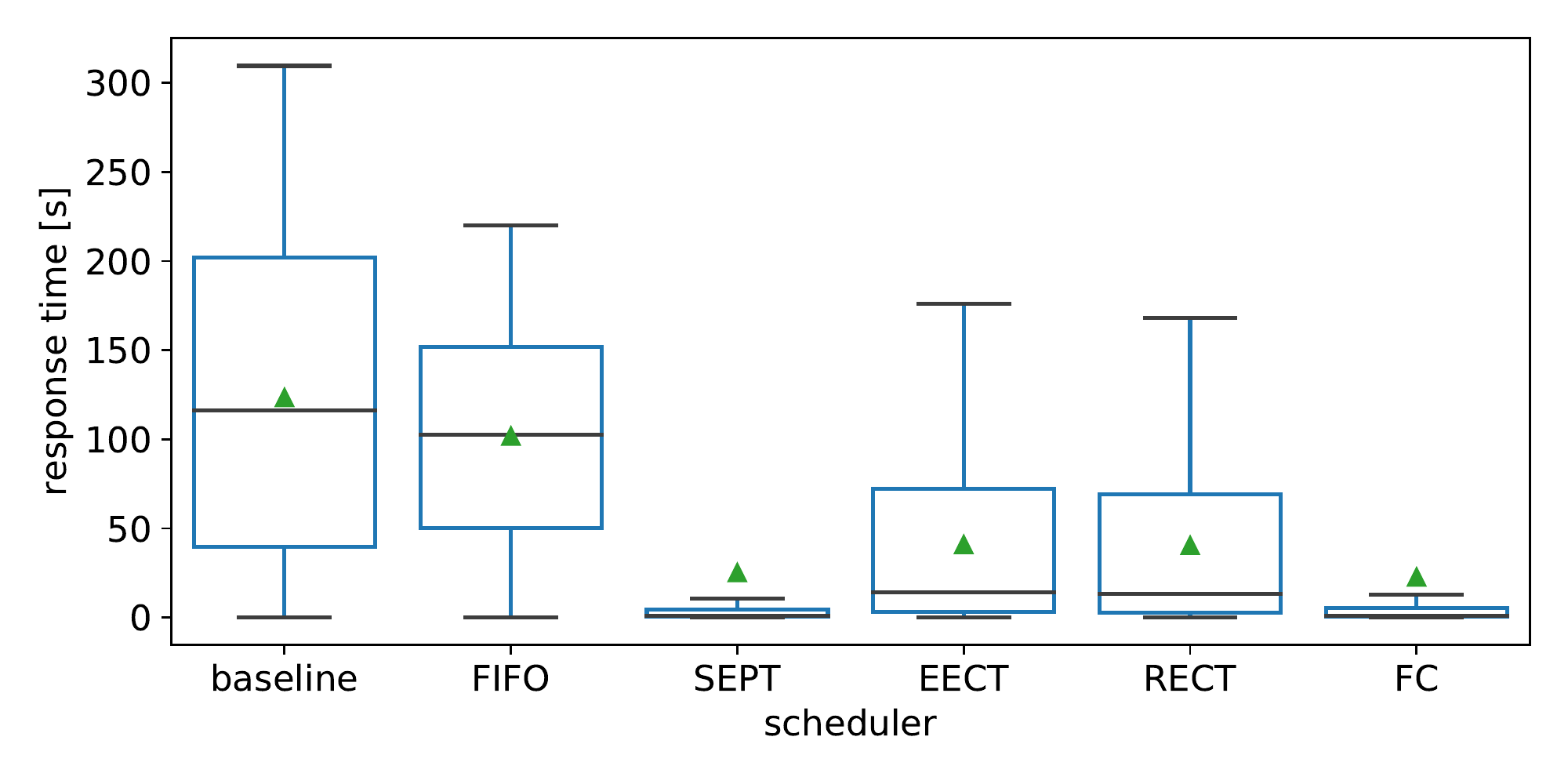}}%
    }%
    \\%
    \subfloat[20 CPU cores, intensity 30]{%
        {\includegraphics[width=0.33\textwidth,clip]{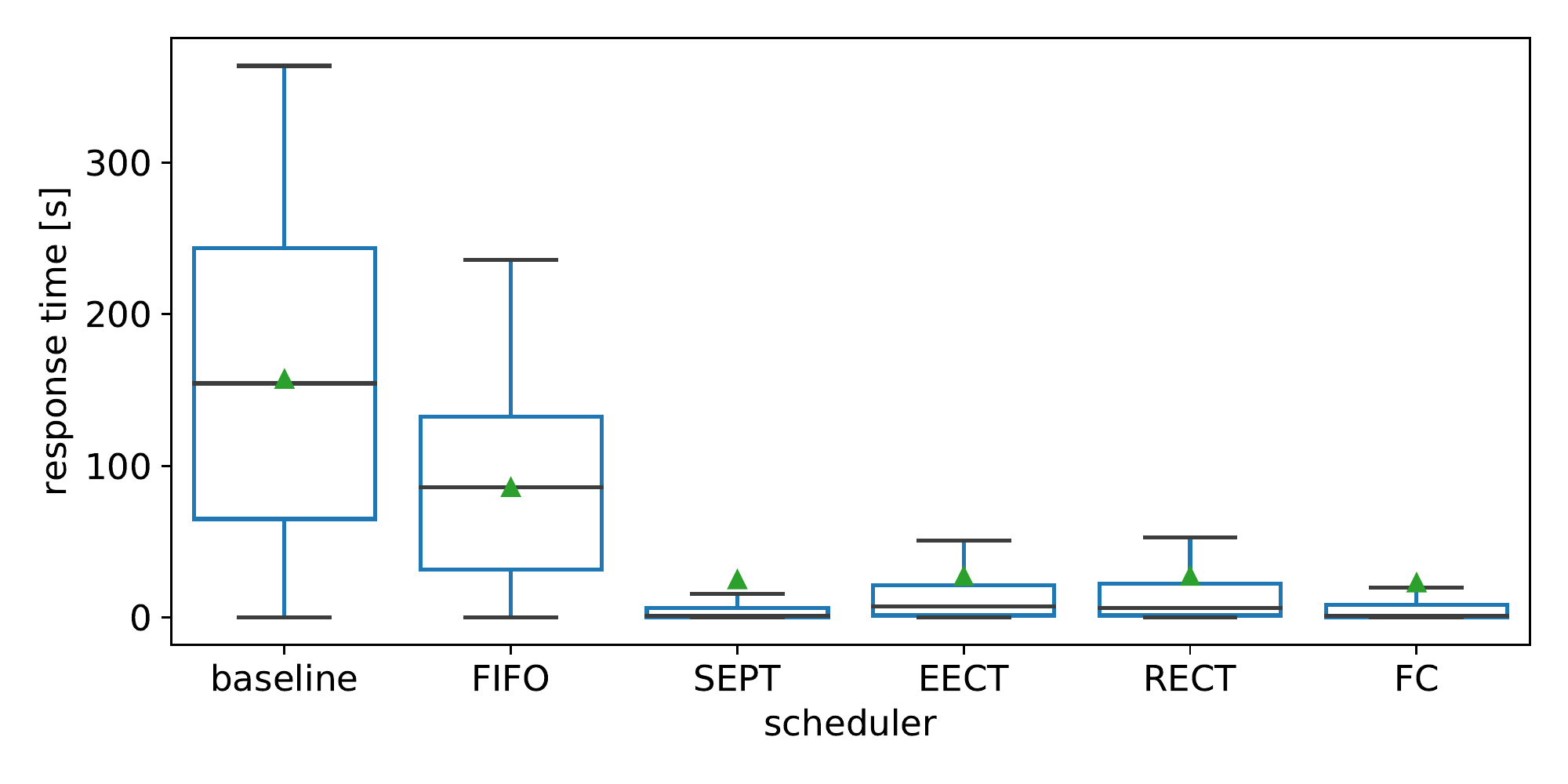}}%
    }%
    \subfloat[20 CPU cores, intensity 40]{%
        {\includegraphics[width=0.33\textwidth,clip]{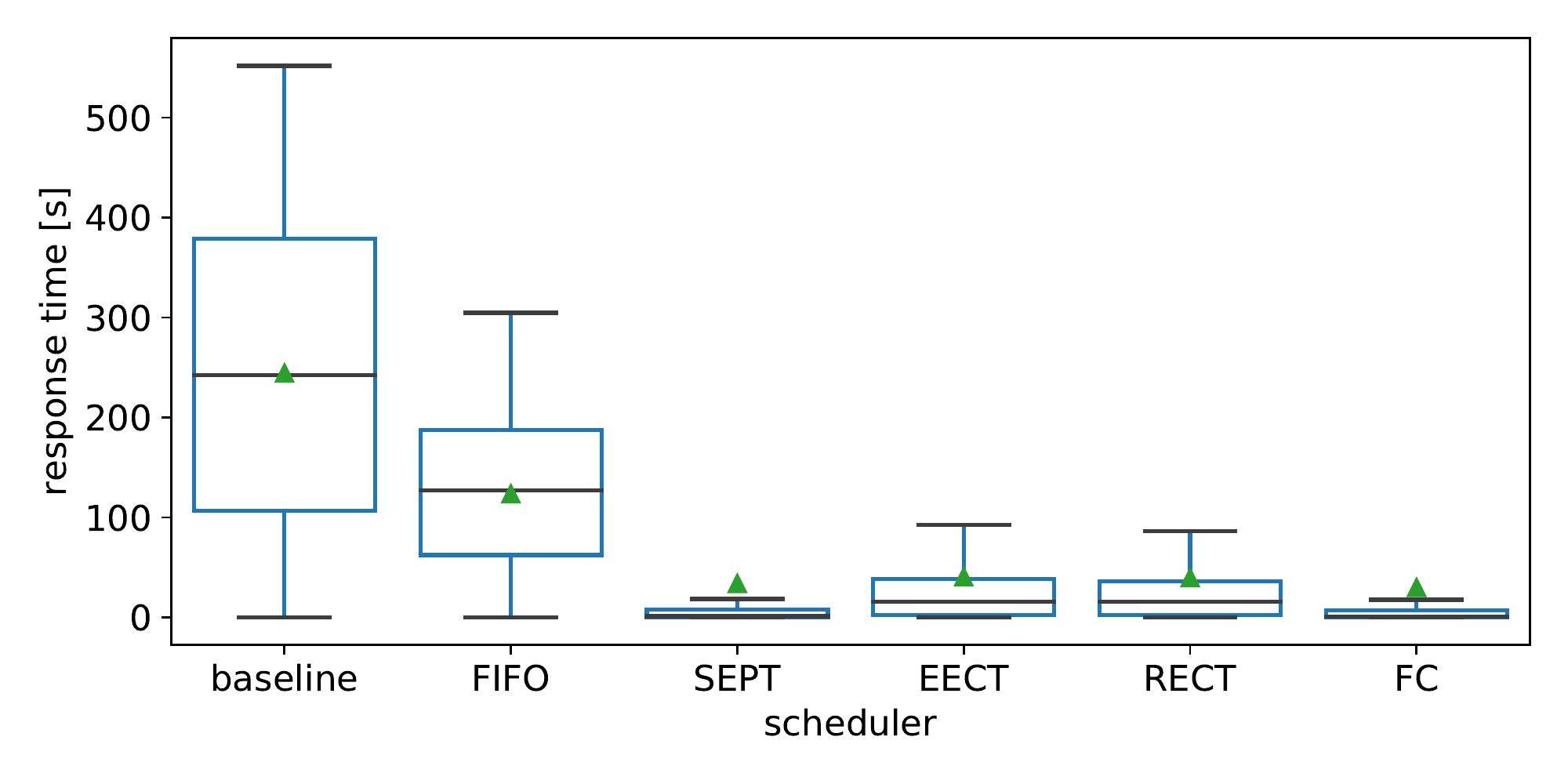}}%
    }%
    \subfloat[20 CPU cores, intensity 60]{%
        {\includegraphics[width=0.33\textwidth,clip]{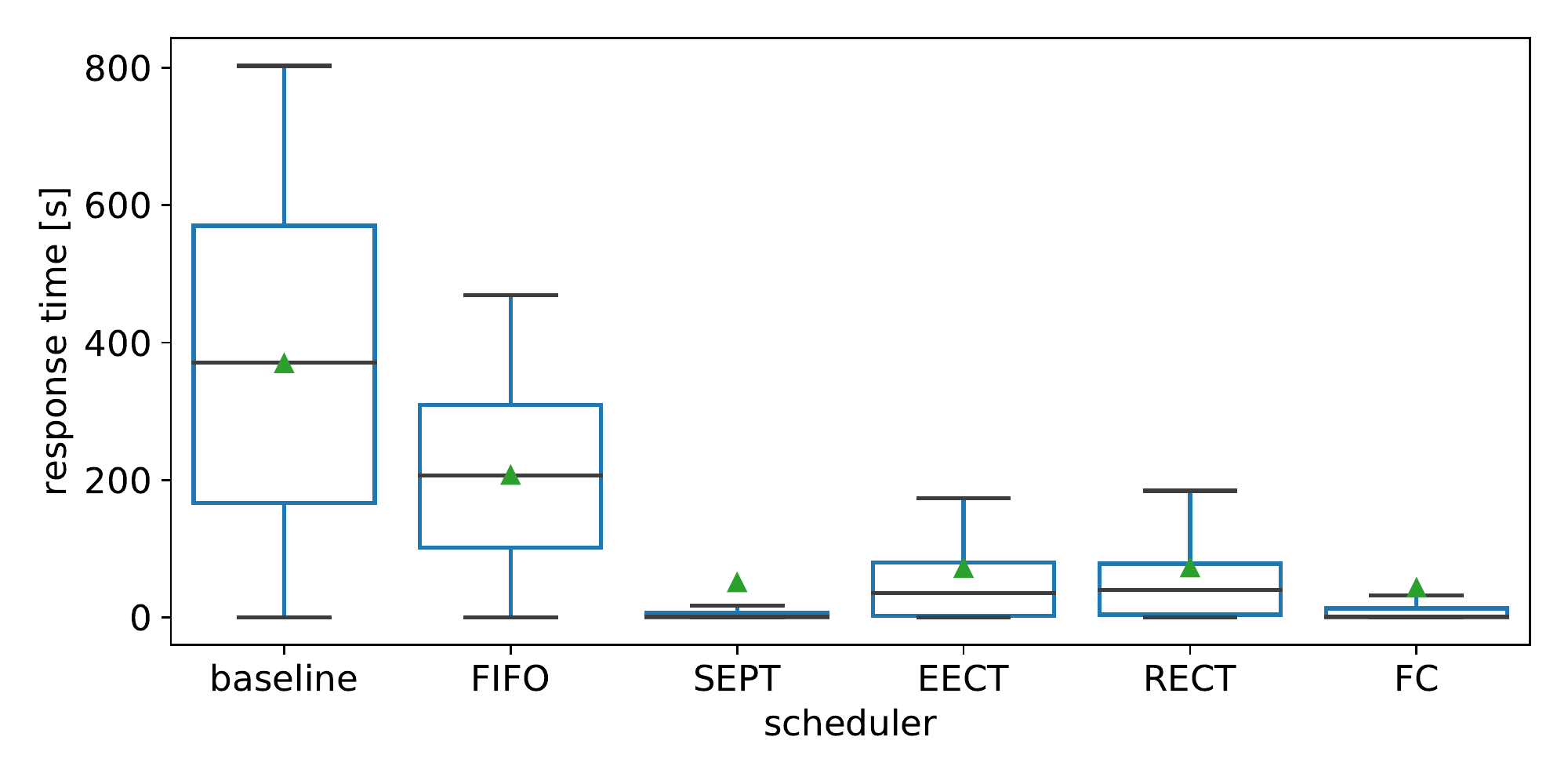}}%
    }%
    \\
    \caption{Response time for different scheduling policies, the number of CPU cores available for action containers (rows), and load intensity (columns). On-premise infrastructure. Here and in remaining plots, unless noted otherwise, each box aggregates results from all individual calls from all 5 sequences of calls. Thus, e.g., in (l), each box shows statistics over $5\cdot 2640$ individual calls. We use standard box plots with boxes spanning from the 25th to the 75th percentile (the IQR); the black line denotes the median; the green triangle denotes the average; and the whiskers span to the most extreme data point within $1.5 \cdot IQR$.}
	\label{fig:cmp_response_time}
\end{figure*}

\begin{figure*}[tb]
    \centering
    \subfloat[10 CPU cores, intensity 30]{%
        {\includegraphics[width=0.33\textwidth,clip]{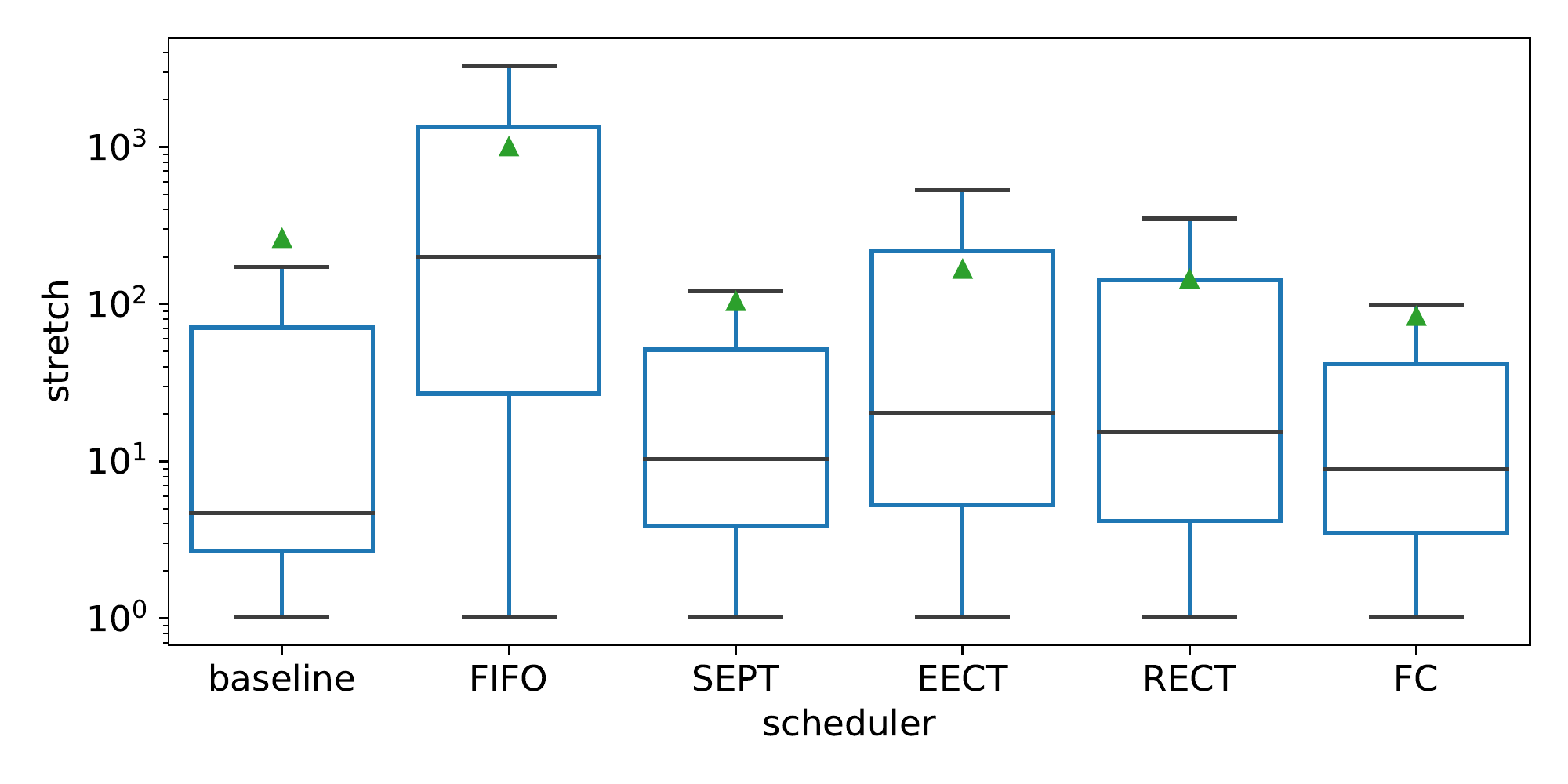}}%
    }%
    \subfloat[10 CPU cores, intensity 40]{%
        {\includegraphics[width=0.33\textwidth,clip]{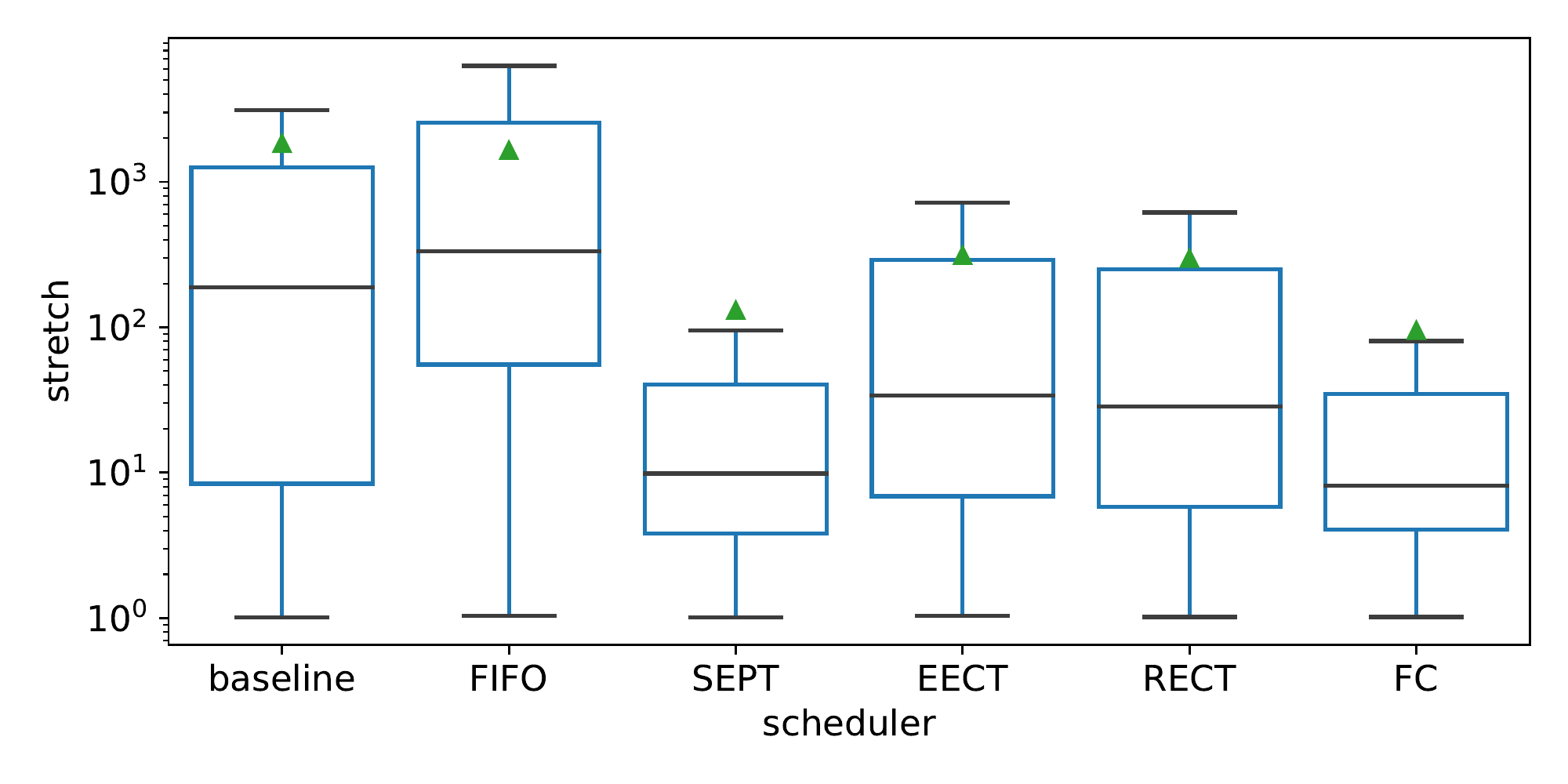}}%
    }%
    \subfloat[10 CPU cores, intensity 60]{
        {\includegraphics[width=0.33\textwidth,clip]{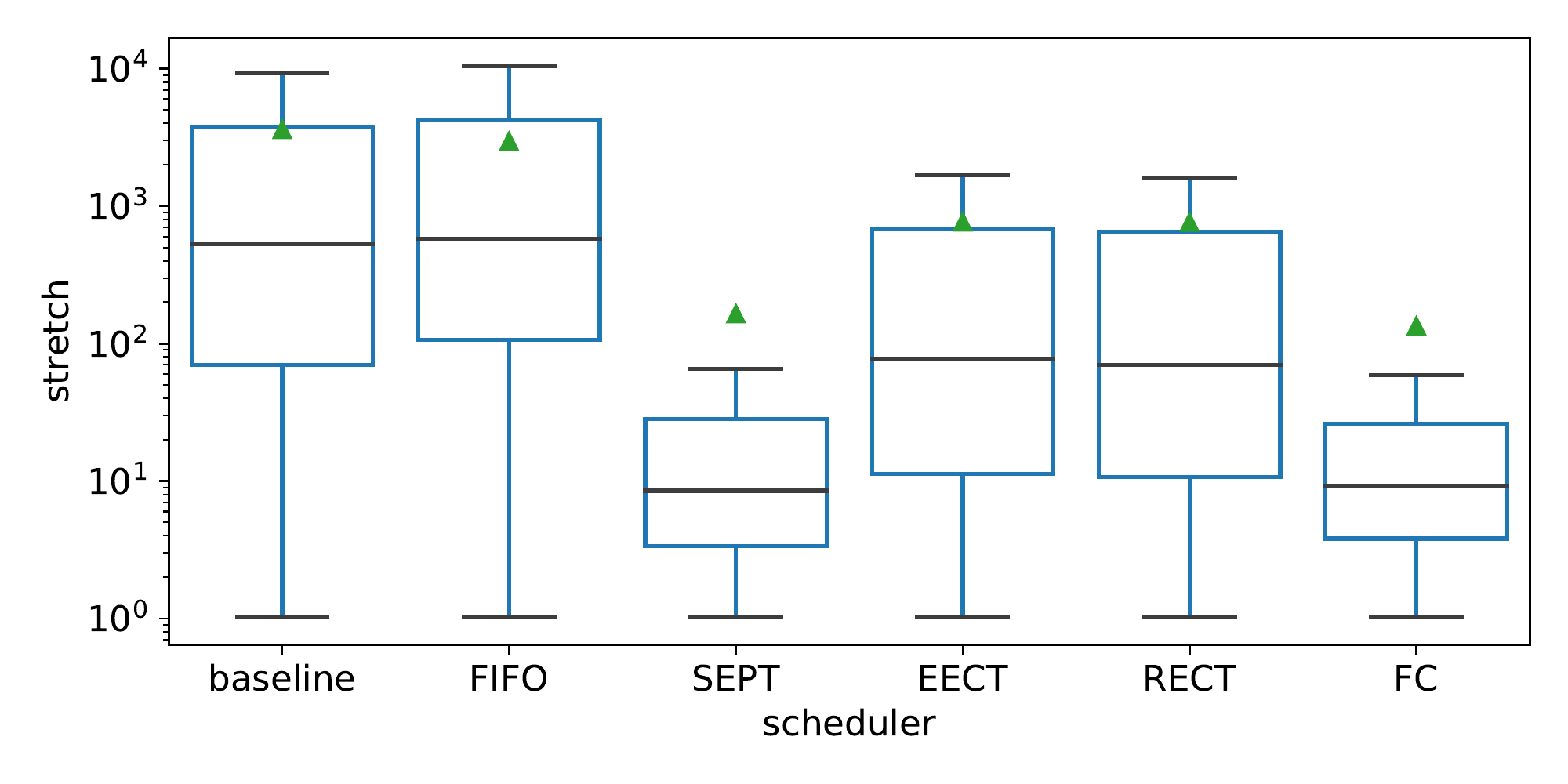}}
    }%
    \\
    %%%%%%%%%%%%%%%%%%%%%
    \subfloat[20 CPU cores, intensity 30]{%
        {\includegraphics[width=0.33\textwidth,clip]{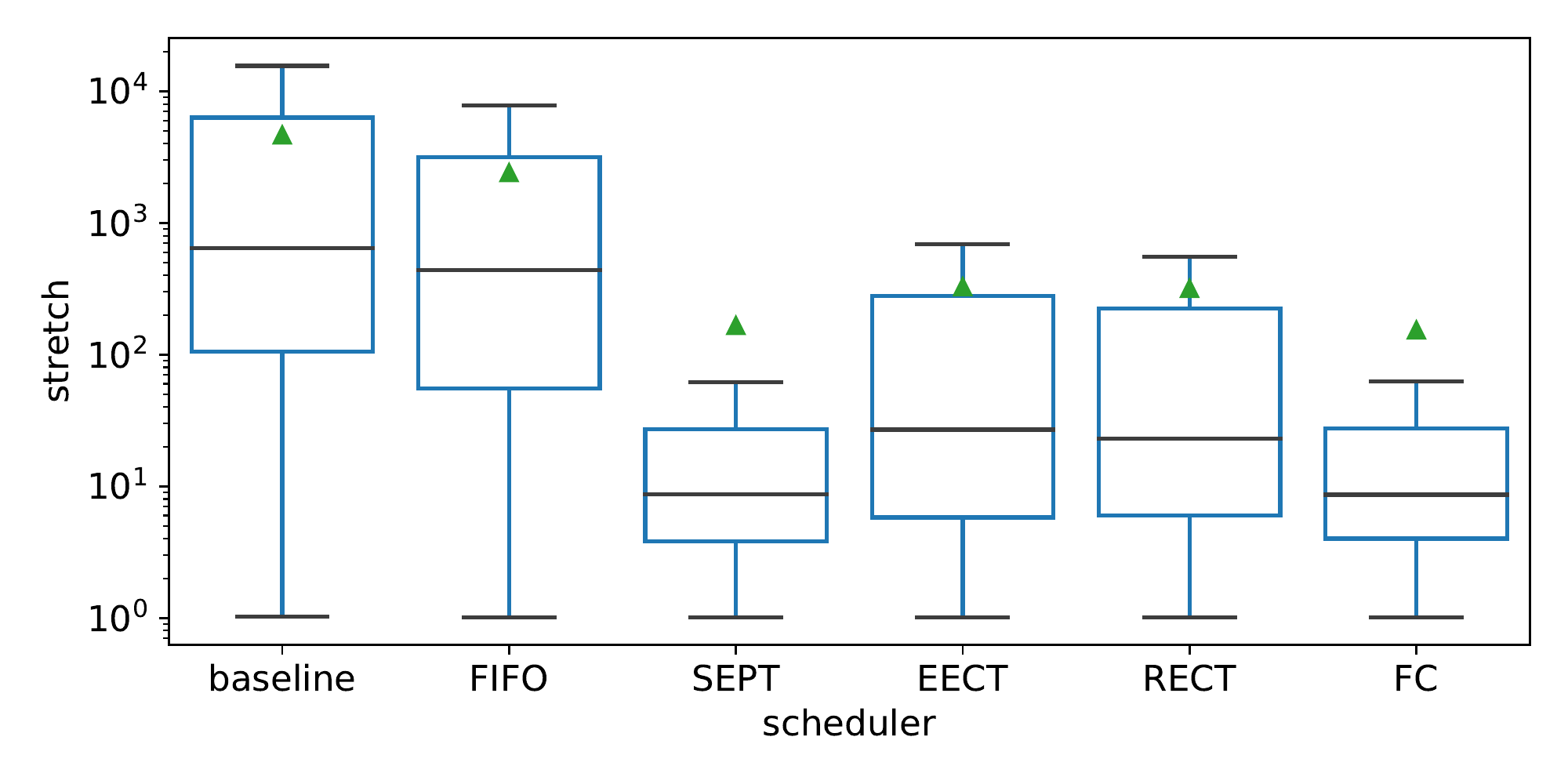}}%
    }%
    \subfloat[20 CPU cores, intensity 40]{%
        {\includegraphics[width=0.33\textwidth,clip]{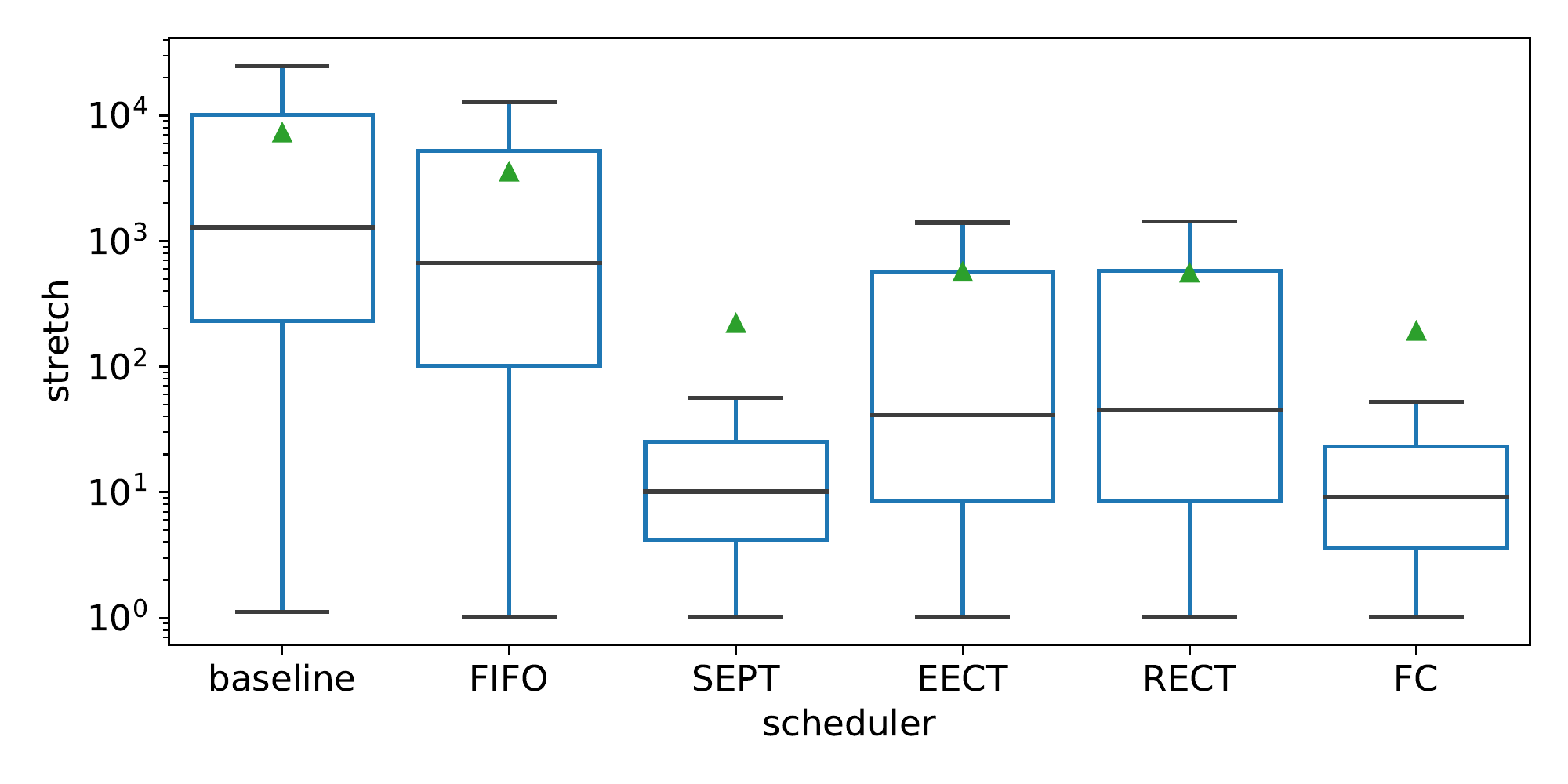}}%
    }%
    \subfloat[20 CPU cores, intensity 60]{
        {\includegraphics[width=0.33\textwidth,clip]{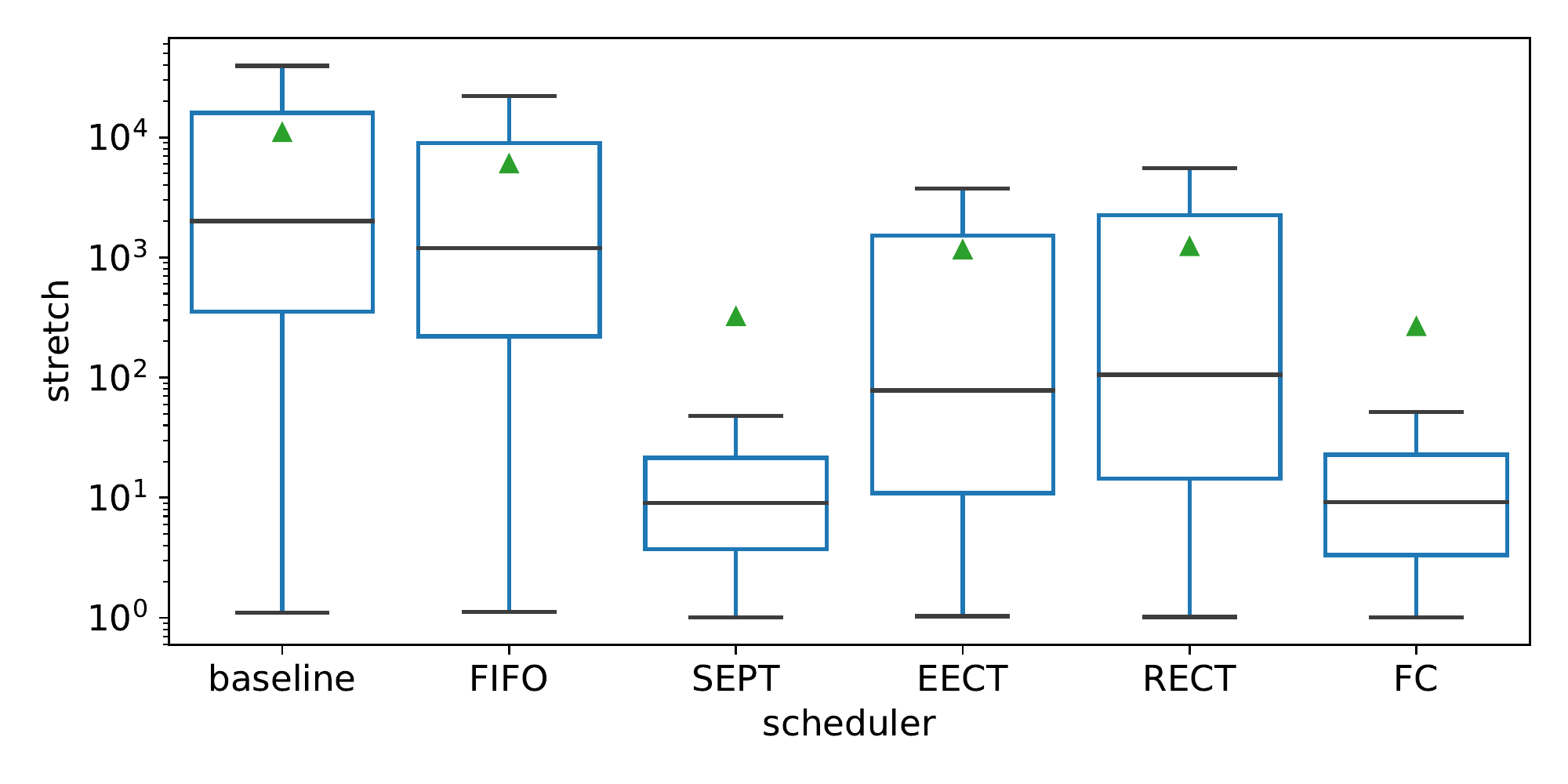}}
    }\\
    \caption{Stretch for different scheduling policies, the number of CPU cores available for action containers (rows), and load intensity (columns). On-premise infrastructure. Average stretch presented as a green triangle.}
	\label{fig:cmp_stretch}
\end{figure*}

Figures~\ref{fig:cmp_response_time} and \ref{fig:cmp_stretch} aggregate response times and stretches from all 5 experiments related to each combination of CPU cores and load intensity. Each row shows results for the same number of cores; each column --- for the same intensity. To calculate stretch, we used the median response time of an idle system (Table~\ref{tab:sebs-functions}). Thus, the stretch can be smaller than 1.

\subsection{Influence of our strategies}
\label{subsec:inf-st}

We start by comparing the five proposed scheduling strategies against each other and against the baseline approach of OpenWhisk by computing the baseline-to-us ratio, i.e., the relative improvement of a given metric (response time, stretch or the completion time) that comes from applying a given strategy --- and this over all considered CPU counts and load intensities (unless otherwise noted).

We start by analyzing our FIFO policy. In FIFO, the sequencing of calls is similar to the baseline OpenWhisk --- but our FIFO effectively eliminates preemption (see Sect.~\ref{subsec:mbcpu} for details) and limits cold starts (Sect.~\ref{sec:cont-evic}). 
This impact can be best measured by comparing the time needed to process all the requests. Each test case processes exactly the same (base) load --- thus, the differences in observed completion time are mostly caused by context switches and cold starts. Table~\ref{tab:cmax-str} shows that our FIFO always reduces the completion time when there are 20 CPU cores involved; the improvement varies from 22\% (intensity 30) to 45\% (intensity 120). For 10 CPU cores, the improvement varies from -28\% (which is actually a deterioration) to 34\%.

\begin{table}[b]
\caption{Maximum request completion times. For each pair of parameters (CPU cores/intensity) 
we show FIFO to baseline ratios (over 5 experiments). 
Configurations for which our method improves upon FIFO in each experiment in bold.}
\label{tab:cmax-str}
\begin{tabularx}{\columnwidth}{X|ccccc}
    %\toprule
    int.~$\rightarrow$ & & & & &\\
    cores~$\downarrow$ & 30 & 40 & 60 & 90 & 120 \\
    \midrule
    5  & 1.14--1.20 & 1.10--1.13 & 0.98--1.05 & 0.97--1.02 & \textbf{0.90--0.98} \\
    10 & 1.11--1.28 & \textbf{0.76--0.90} & \textbf{0.74--0.90} & 0.92--1.04 & \textbf{0.66--0.70} \\
    20 & \textbf{0.67--0.78} & \textbf{0.59--0.66} & \textbf{0.60--0.64} & \textbf{0.57--0.60} & \textbf{0.55--0.58} \\
    %\bottomrule
\end{tabularx} 
\end{table}

We continue by aggregating the response time statistics of Figures~\ref{fig:cmp_response_time} and~\ref{fig:cmp_stretch}. Comparing FIFO to baseline, the average relative improvement of the average response time is 1.39. However, this improvement strongly depends on both the number of cores and load intensity, increasing with both parameters --- with 10 CPUs, the improvement factor varies from 0.41 for intensity of 30 (i.e., an increase of the response time by the factor of over 2), to 1.21 for intensity of 60. With 20 CPUs, however, the improvement ranges from 1.79 to 1.98. For stretch, the relative improvements are almost the same (ranging from 0.26 to 1.22 for 10 CPUs, and from 1.82 to 2.05 for 20 CPUs). When it comes to the response time tails, these values are also decreased --- the improvement factor for 10 CPUs varies between 1.02 and 1.41 for the 95th percentile of the response time, and between 0.98 and 1.44 for the 99th percentile. In case of 20 CPUs, these values range between 1.85 and 2.02, and 1.70 and 1.90, respectively.

Our remaining strategies show the additional benefit of smarter queuing. Overall, both SEPT and Fair-Choice improve upon FIFO in all experiments: the average relative response time improvement of SEPT is 3.59 and of FC is 4.10; while the average relative stretch improvement of SEPT is 14.89 and of FC is 18.02. Although one can see a relative degradation of the 99th percentile of response time compared to FIFO, there is still a visible improvement compared to the baseline: the average improvement factor for SEPT is 1.16, and FC is 1.40. EECT and RECT also show an advantage over FIFO (with the average relative response time improvement of 2.88).
However, both these strategies prevent calls from starving.

\subsection{Influence of the intensity}
\label{subsec:inf-int}

With the same number of CPU cores available for action containers, when load intensity increases, the queue of action calls waiting to be processed gets longer. Thus, the choice of the sequencing strategy is expected to have a more significant impact. We thus anticipate that the advantage of SEPT and FC over FIFO will increase with increasing load intensity.

Even with intensity 30, the advantage of SEPT and FC over FIFO, EECT and RECT can be clearly seen both for response time and for stretch. SEPT, FC, EECT and RECT  perform better than FIFO in terms of the average response time (with the ratio of FIFO-to-other varying from 2.5 to 4.8) and the stretch (ratio from 3.9 to 22.6). Improvements further increase for intensities above 60~[p. \pageref{appendix}].

Consider the case of 10 CPU cores. For the intensity of 30, the average response time from our FIFO and SEPT strategies is 36.42\,s and 12.52\,s (with a ratio of 2.9). For intensity 40, these values are equal to 58.29\,s and 17.01\,s (with a ratio of 3.4); and for intensity 60, to 101.76\,s and 25.14s (with a ratio of 4.0). At the same time, the FIFO-to-SEPT ratio of the median response time varies from 22.0, through 38.7, to 95.9. This quick growth can be explained as follows: for large load intensity, we may almost freely choose short functions from the queue. This drastically reduces the median response time.
Similarly, when it comes to stretch and intensity 30, the averages provided by our FIFO and SEPT strategies are 1000 and 104 (with a ratio of 9.6). For intensity 40, it is 1647 and 130 (with a ratio of 12.6), and for intensity 60, it is 2959 and 164 (with a ratio of 18.0), respectively. At the same time, the FIFO-to-SEPT ratio of median stretch increases from 19.3, through 33.7, to 68.0. This phenomenon can be explained in the same way.

The significant influence of the scheduling strategy can be concluded from the fact that the difference between the baseline approach and our approaches remains stable in terms of the order of magnitude. For 20 CPU cores and for intensities 30, 40, 60, 90 and 120, the ratio of average response time obtained for the baseline and for our FIFO implementation is 1.8, 2.0, 1.8, 1.8 and 1.9, respectively. Similarly, the ratio of average stretch obtained for the baseline and for our FIFO implementation is 1.9, 2.0, 1.8, 1.8 and 1.9.

\subsection{Influence of the number of CPU cores}
\label{subsec:inf-cpu}

When the number of CPU cores available for action containers increases while the intensity remains the same, the load --- although the same, on average, for each core --- can be better balanced between cores. On the other hand, higher load is processed on more cores that share parts of their caches. Thus, we expect that with the increase of CPU cores, the improvements of our strategies will change slowly. On the other hand, we expect the baseline approach to behave worse for a higher number of CPU cores due to its container management policies. Our results confirm this intuition, especially for higher intensities.

Consider the intermediate load intensity 40. If 10 CPU cores are used, the average response times for our FIFO and SEPT implementations are 58.29\,s and 17.01\,s, with their ratio equal to 3.4. When the number of CPU is 20, however, the average values for FIFO and SEPT are 123.64s and 33.92s, with their ratio equal to 3.6. At the same time, the baseline-to-FIFO ratio of the average response time increases, from 1.1 to 2.0.

The same observations can be made for stretch. For 10 CPU cores, the average stretches for our FIFO and SEPT implementations are 1647.40 and 130.87, with their ratio equal to 12.6. When the number of CPU cores is 20, these values are 3538.65 and 220.89. As expected, the ratio of the average stretch for the baseline and for our FIFO implementation increases, from 1.1 to 2.0.

For FIFO and with fixed intensity (40 or more), the median response time (and thus, also stretch) increases almost linearly with the increase of the number of CPU cores (we confirmed that by analyzing additional results for 5 CPU cores~[p. \pageref{appendix}]). If we increase the number of cores twice, the median response time and stretch will double. Our experiments thus show that the system overheads (related to container management) have a significant impact on the overall performance. At the same time, in the case of the baseline, doubling the number of cores triples the medians. Surprisingly, for the same core-level intensity, the best performance is presented by nodes that have lower numbers of cores.

\subsection{Function-level metrics}
\label{subsec:flm}

\begin{figure*}[t]
    \centering
    \subfloat[All functions]{%
        {\includegraphics[width=0.33\textwidth,clip]{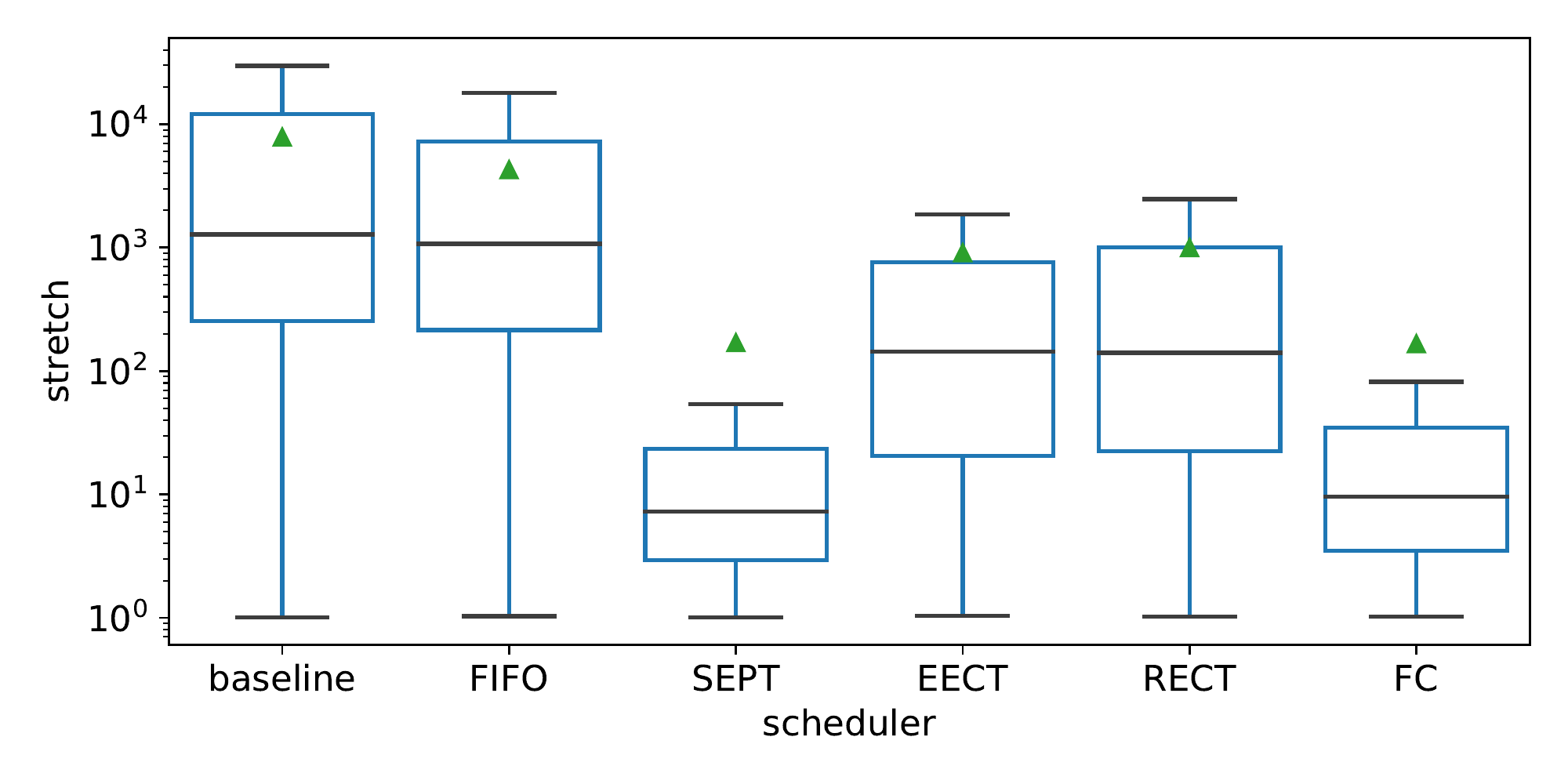}}% 
    }%
    \subfloat[\texttt{dna-visualisation} (1\% of all calls)]{%
        {\includegraphics[width=0.33\textwidth,clip]{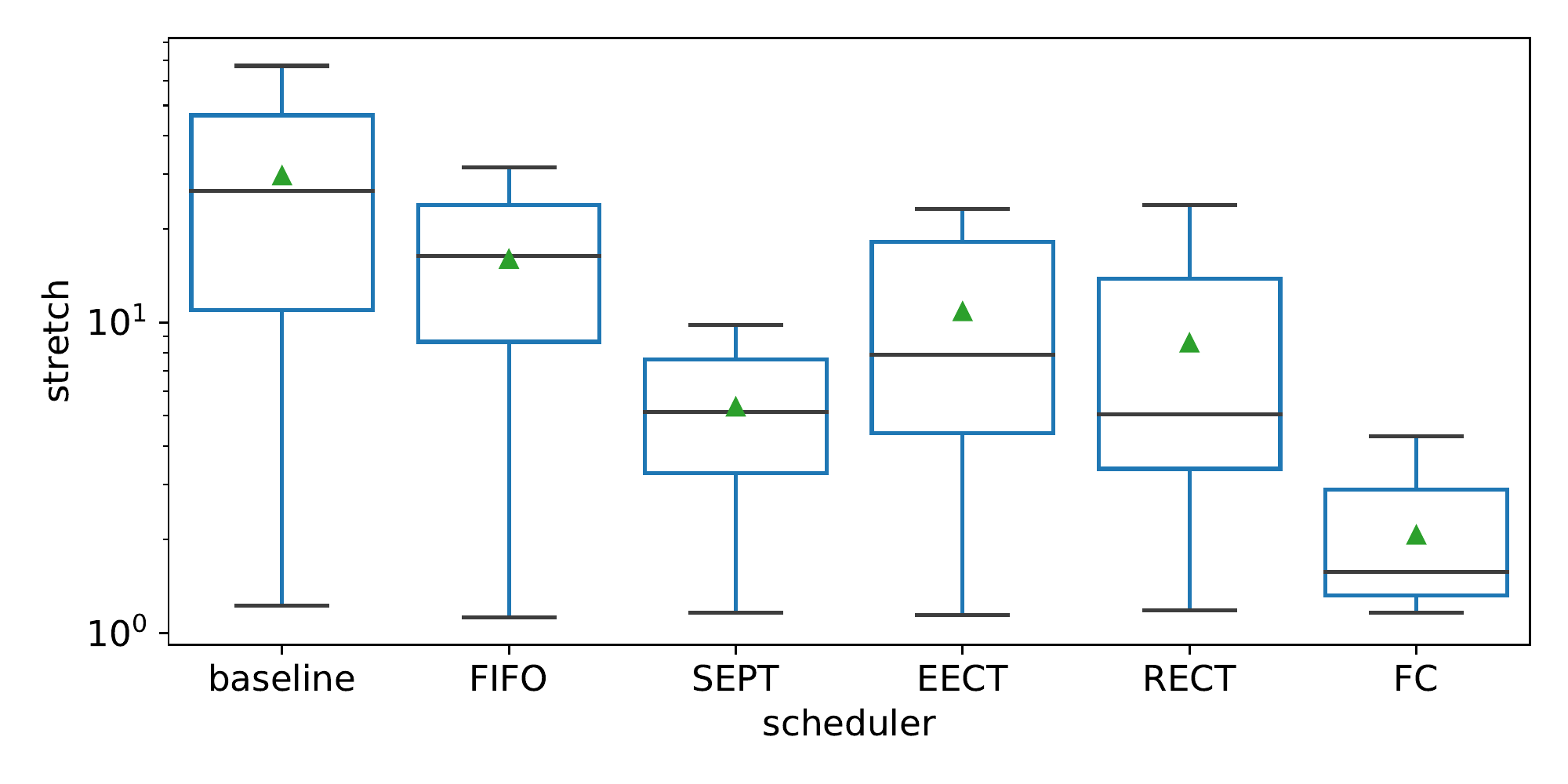}}%
    }%
    \subfloat[\texttt{graph-bfs} (9.9\% of all calls)]{
        {\includegraphics[width=0.33\textwidth,clip]{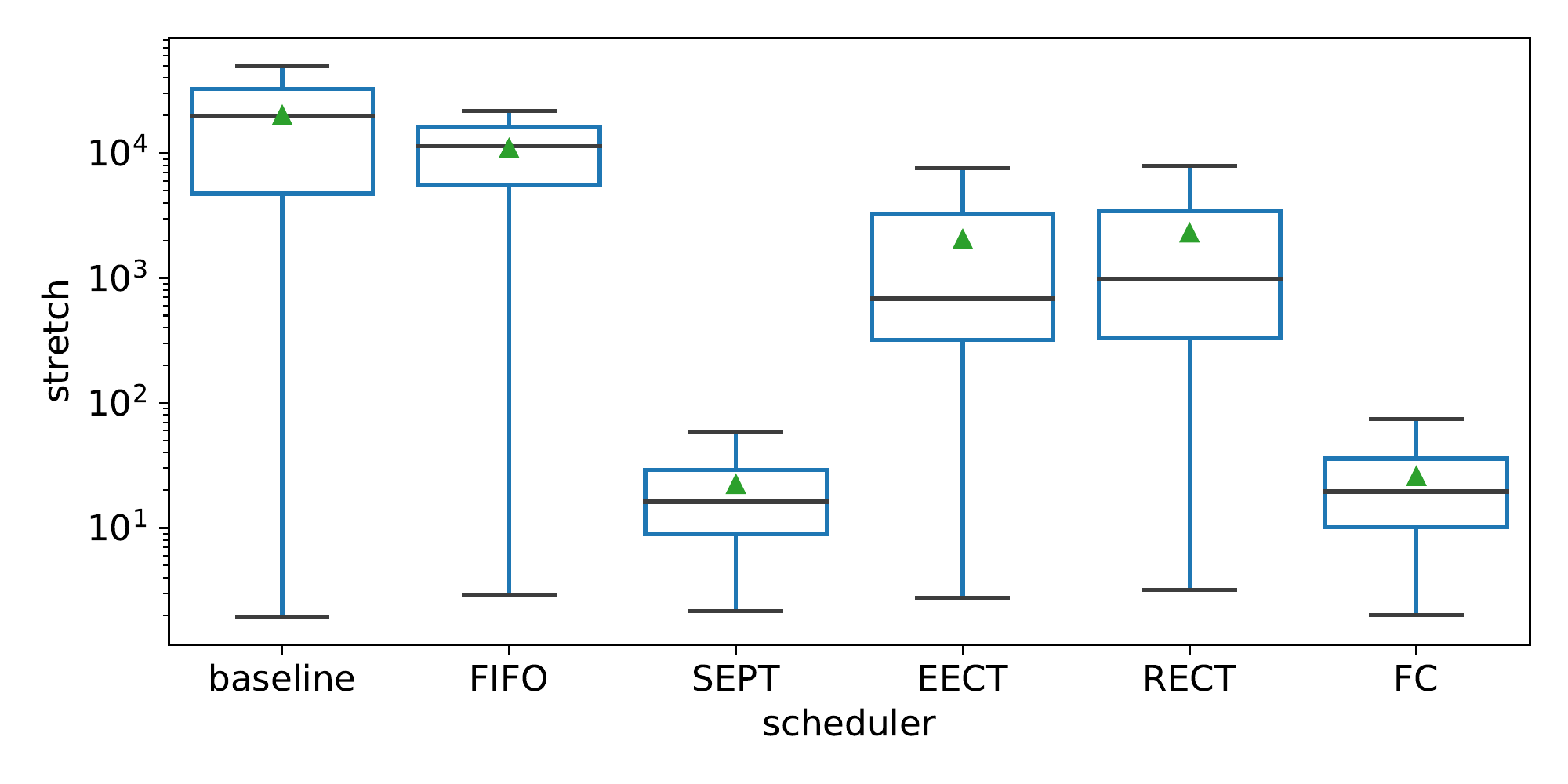}}
    }%
    \caption{Comparison of aggregated stretch for different scheduling approaches and functions, 10 CPU cores and intensity 90. The \texttt{graph-bfs} function is short (12\,ms) and thus we expect larger values of stretch compared to the long (8\,552\,ms) \texttt{dna-visualisation} function.}
	\label{fig:cmp_fn_stretch}
\end{figure*}

The SEPT strategy will always prioritize short calls, independently of how often corresponding functions are called. As a consequence, if a long function is called relatively rare, it will be discriminated against. Our Fair-Choice (FC) strategy (Sect.~\ref{sec:nlsp}) prioritizes calls based on the total resource consumption in a moving window. In order to quantify the fairness of the FC strategy, we performed 5 additional groups of experiments for the intermediate configuration with 10 CPU cores and intensity 90. In these experiments, exactly 10 calls corresponded to the relatively-long \texttt{dna-visualisation} function. Other calls were uniformly distributed among other functions, including a very short \texttt{graph-bfs} function. In contrast to our previous experiments, this time we made no assumptions on partial-uniformity of the call distribution. Fig.~\ref{fig:cmp_fn_stretch} presents the aggregated stretch.

In Fig.~\ref{fig:cmp_fn_stretch}(a) we observe the distribution of stretch among all the calls. Although the call frequencies differ, Fig.~\ref{fig:cmp_fn_stretch}(a) is almost identical to Fig.~\ref{fig:cmp_stretch}(c). The fairness of FC is apparent when we show just the results of the infrequent, long \texttt{dna-visualisation} function, Fig.~\ref{fig:cmp_fn_stretch}(b). 
FC reduces the average stretch from 5.3 (SEPT) to 2.1; the median stretch is reduced from 5.2 (SEPT) to 1.6, which hints that often the call is started almost immediately. 
These gains are not for free, though, Fig.~\ref{fig:cmp_fn_stretch}(c): FC increases the average stretch for a short, often-called \texttt{graph-bfs} function to 25.8 (compared with SEPT's 22.2).

\section{Multiple worker nodes}
\label{sec:multiple-workers}

Encouraged by our single-node results, we evaluate our approach in a multi-node environment. We measure the performance of a setup in which up to 4 workers operate in parallel. To do so, we created 5 virtual machines in our on-premises cloud running under Proxmox 7.1 \cite{Proxmox} on physical machines equipped with AMD EPYC 7402P CPUs @ 2.80GHz with 24 hyper-threaded cores and 128 GB or RAM. Our controller is running on a VM limited to 4 CPU cores and 16 GB of RAM. Each of the four VMs executing OpenWhisk workers were assigned 20 CPU cores and 40 GB of RAM.
The software setup of these workers is similar to the on-premises setup: two cores are reserved for system and invoker processes, while 18 were reserved for the action containers.

\begin{figure}[t]
    \centering
    \includegraphics[width=0.8\columnwidth,clip]{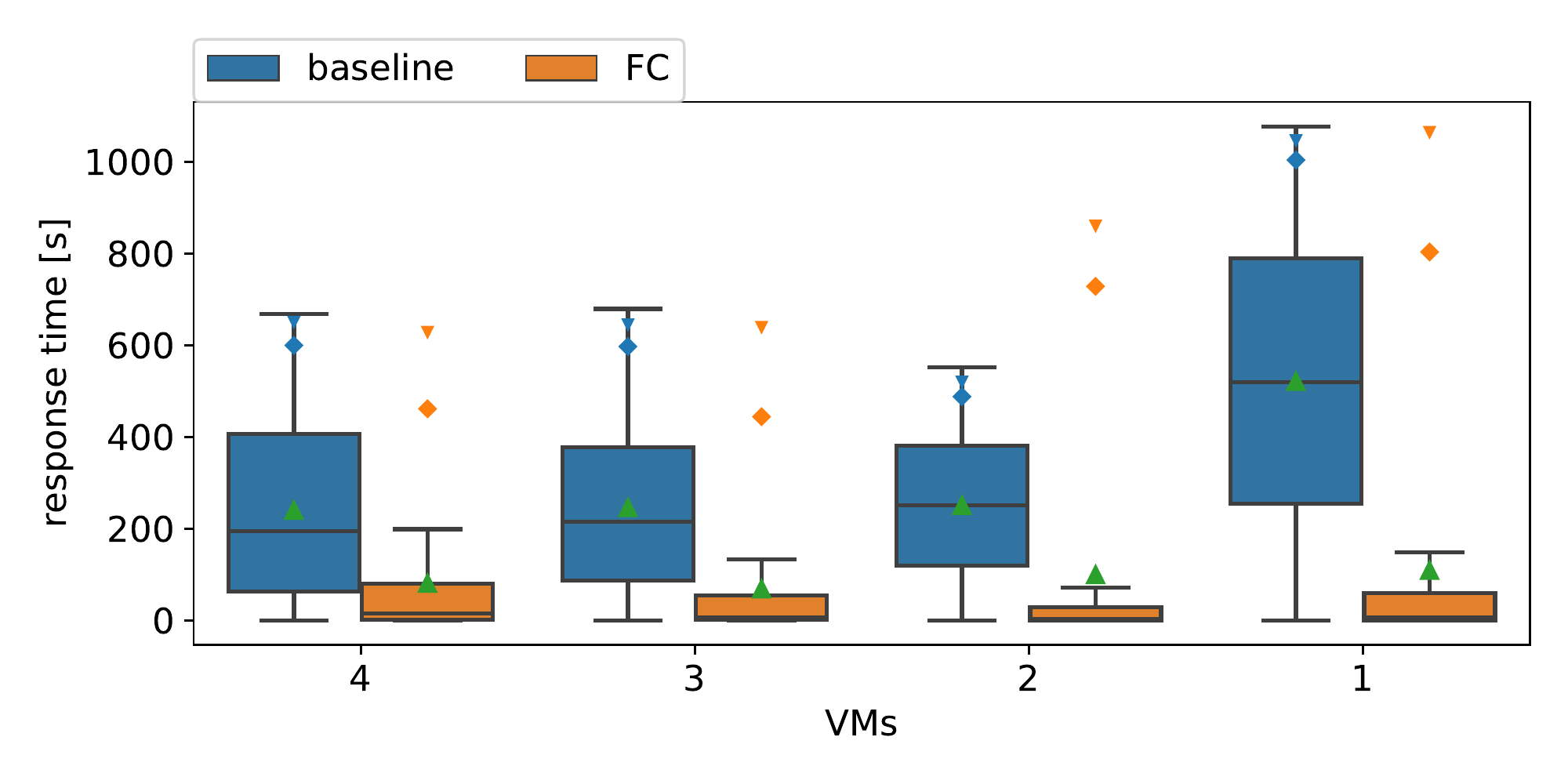}
    \caption{Response times in a multi-node environment. Each configuration processes the same sequence of 2376 requests. $\triangle$ is the average; $\diamond$ the 95th percentile; $\triangledown$ the 99th percentile.}
	\label{fig:cmp_proxmox}
\end{figure}

In each multi-node experiment, we send the same sequence of 2376 requests uniformly during a 60-second time-window. With 4 workers, 18-CPU nodes, 2376 corresponds to the core-level intensity of 30. We test the baseline and FC, but in different runs we reduce the number of available worker nodes from 4 up to 1. With 3 workers operating in parallel, 2376 requests correspond to the core-level intensity of 40; 2 workers correspond to the intensity of 60; and 1 worker to 120. As in the previous section, we repeated the experiment 5 times with different request sequences.
Fig.~\ref{fig:cmp_proxmox} shows the results.

\emph{With 3 machines, our FC strategy provides better quality of service than the baseline using 4 machines.} 
For the baseline on 4 VMs, the average response time is 240 seconds (with the 75th percentile of 406 seconds, the 95th percentile of 600 seconds, and the 99th percentile of 649 seconds). 
Although our FC strategy processed the same load on just 3 VMs, 
we reduce the average response time to 68 seconds (71\% less than the baseline on 4 VMs), its 75th percentile of 54 seconds (97\% less), the 95th percentile of 443 seconds (26\% less), and even the 99th percentile of 638 seconds (2\% less).

Surprisingly,
even on 2 VMs,
some metrics are still better for our FC strategy 
compared to the baseline operating on 4 VMs. 
FC reduces the average response time by 58\% and its 75th percentile by 93\%. Although FC worsens the 95th percentile by 21\% and the 99th percentile by 32\%, we argue that 95th percentile response time during a load peak corresponds to a much higher percentile in a longer time window.

%%%%%%%%%%%%%%%%%%%%%%%%%%%%%%%%%%%%%%%%%%%%%%%%%%%%%%%%%%%%%%%%%%%%%%%%%%%%%%%%
\section{Related work}
%%%%%%%%%%%%%%%%%%%%%%%%%%%%%%%%%%%%%%%%%%%%%%%%%%%%%%%%%%%%%%%%%%%%%%%%%%%%%%%%

Serverless computing is a rapidly evolving cloud computing model. Recent overviews are~\cite{jonas2019cloud,Castro2019}; a recent survey is~\cite{hassan_survey_2021}. Below, we focus on papers that optimize resource management; we propose a classification based on the core problem they focus on.
The first three groups (optimizing assignments of calls to nodes, reducing cold starts, and reducing resource requirements) are largely orthogonal to our contribution; while the node level problems are directly related.

\textbf{Assigning calls to nodes.}
In \cite{Kaffes2019}, a central scheduler assigns requests to individual cores (rather than nodes, as in standard OpenWhisk). 
FaaSRank~\cite{yufaasrank} optimizes the assignment of the invoker to a function call through a neural-network approach. 
Further optimizations are possible for complex FaaS invocations (in which a function calls another function forming a chain of invocations).
Fifer~\cite{10.1145/3423211.3425683} optimizes execution of such chains while avoiding cold starts by reactive scaling. \cite{zuk2020schedulingfaas,DBLP:journals/jpdc/ZukR22} schedule chain and DAG invocations taking into account future preparation of the environment.
\cite{carver2020wukong} analyzes DAGs and shows a decentralized scheduler reducing DAG processing time.
Faastlane~\cite{kotni_faastlane_nodate} speeds up processing of workflows by executing multiple steps within a single executor in parallel, while providing isolation between steps through Intel Memory Protection Keys (MPK).
Executed functions may interact with the environment by modifying shared state. In current FaaS platforms, interruption of a function call may result in inconsistent state. \cite{sreekanti_fault-tolerance_2020} presents shim enabling fault-tolerance and benchmarks its impact on the performance.

\textbf{Reducing cold starts.}
\cite{du2020catalyzer} reduces cold start time by restoring a function instance from a predefined state (with an implementation on Google gVisior~\cite{gvisior} sandbox).
SEUSS~\cite{cadden_seuss_2020} uses unikernel snapshots to reduce time of initialization by over an order of magnitude and additionally saves memory by sharing pages between instances.
\cite{8752939} decouples libraries and other dependencies from function packages; libraries are cached on worker nodes, and call scheduling is aware of the packages available on a node. 
\cite{li_pagurus_2021} analyzes functions with similar dependencies --- this allows a function call to use an environment dedicated for another, related function.
\cite{agarwal2021reinforcement} uses reinforcement learning to  identify function call patterns and starting containers in advance.
All these approaches require provider-side modifications. In contrast, \cite{bermbach_using_2020} proposes middleware to be deployed with function code that can trigger container creation in advance.

\textbf{Reducing resource requirements.}
Photons~\cite{dukic2020photons} share context (runtime, libraries, etc.) between multiple concurrent calls, thus reducing the memory footprint.
OFC~\cite{mvondo_ofc_2021}~uses ML to predict the true memory usage of an invocation; and then overcommits the rest to act as a cache for remote data stores. 
SAND~\cite{akkus_sand_2018} distinguishes between applications and functions; multiple functions of the same application share a common container.
\cite{sreekanti_cloudburst_2020}~extends FaaS to handle stateful functions natively, rather than through a remote data store. 

\textbf{Node-level scheduling.}
Our paper focuses on node-level scheduling policies. Two recently-published papers are~\cite{Bap2021} and ETAS~\cite{banaei_etas_2021}. In \cite{Bap2021}, a number of scheduling strategies is proposed, including SEPT and theoretical variants of Fair-Choice. These strategies are evaluated based on the Azure function trace~\cite{Shahrad2020}; it is demonstrated, among others, that reliable estimates of execution times can be obtained from just the 10 most recent invocations. In contrast to these theoretical considerations, we implement these strategies in OpenWhisk and evaluate them with a standard benchmark. We additionally propose EECT and RECT policies, and discuss the impact of cold starts on the performance of the system. ETAS~\cite{banaei_etas_2021} proposes a queuing policy similar to our EECT. Although this policy is also implemented in OpenWhisk, the impact of OS-level preemption is not discussed, and thus the potential of reducing their number is untapped. In contrast, we propose container-management methods that address preemption. In \cite{banaei_etas_2021}, the policy is tested in an underloaded system against a set of $4$ custom functions (with no I/O-intensive functions), rather than a standard benchmark we use. Also, no fairness of function executions is considered (in contrast to our FC policy).

%%%%%%%%%%%%%%%%%%%%%%%%%%%%%%%%%%%%%%%%%%%%%%%%%%%%%%%%%%%%%%%%%%%%%%%%%%%%%%%%
\section{Conclusions}
\label{sec:conclusions}
%%%%%%%%%%%%%%%%%%%%%%%%%%%%%%%%%%%%%%%%%%%%%%%%%%%%%%%%%%%%%%%%%%%%%%%%%%%%%%%

In this paper, we introduce new methods of allocating computational resources on a single FaaS node. Our aim is to reduce the response time and the stretch in a temporarily overloaded system. We modify OpenWhisk and introduce a number of sequencing policies. We take advantage of the fact that in FaaS each function is usually called repeatedly. This way, we are able to implement strategies that make use of historical data on the executions of a specific function. These strategies include adaptations of FIFO and SEPT, together with three new policies: EECT, RECT and FC. These new policies use both the expected processing time of a function call and the frequency of calls of the same function in the recent past.

Compared to a baseline approach, the current version of OpenWhisk, our proposed algorithms show general advantages. Moreover, compared to other analyzed strategies, SEPT and FC significantly decrease the response time --- median, average and the tail percentiles. At the same time, the FC strategy prioritizes jobs based on their previous usage, introducing inter-function fairness in the system.
We also show that all the response-time metrics for our implementation of the FC strategy are better on 3 VMs compared to the baseline running on 4 VMs. This means that our solution allows us to reduce the number of machines by a factor of at least 25\% without decreasing the quality of the service. In general, we expect this improvement to remain stable even for larger production clusters.

\section*{Acknowledgements}

The authors thank the Faculty of Mathematics and Computer Science, Adam Mickiewicz University in Poznan, Poland, for the access to its computational resources for the multi-node experiments, and Maciej Prill for his invaluable technical help during those experiments.

This research is supported by a Polish National Science Center grant Opus (UMO-2017/25/B/ST6/00116).

\balance

\bibliographystyle{IEEEtran}
\bibliography{article}

\appendix
\label{appendix}

This supplementary material presents detailed information on the results of our experiments. Its aim is to present all the data that was aggregated in the paper.

\subsection{Single-node experiments}

In Table~\ref{tab:num-premises}, we present numerical results aggregated from all five experiments performed for each configuration. Then, in Table~\ref{tab:num-premises-exp}, we present the same results in the context of each single experiment.

\begin{figure*}[h]
    \centering
    \chart{Experiment 1}{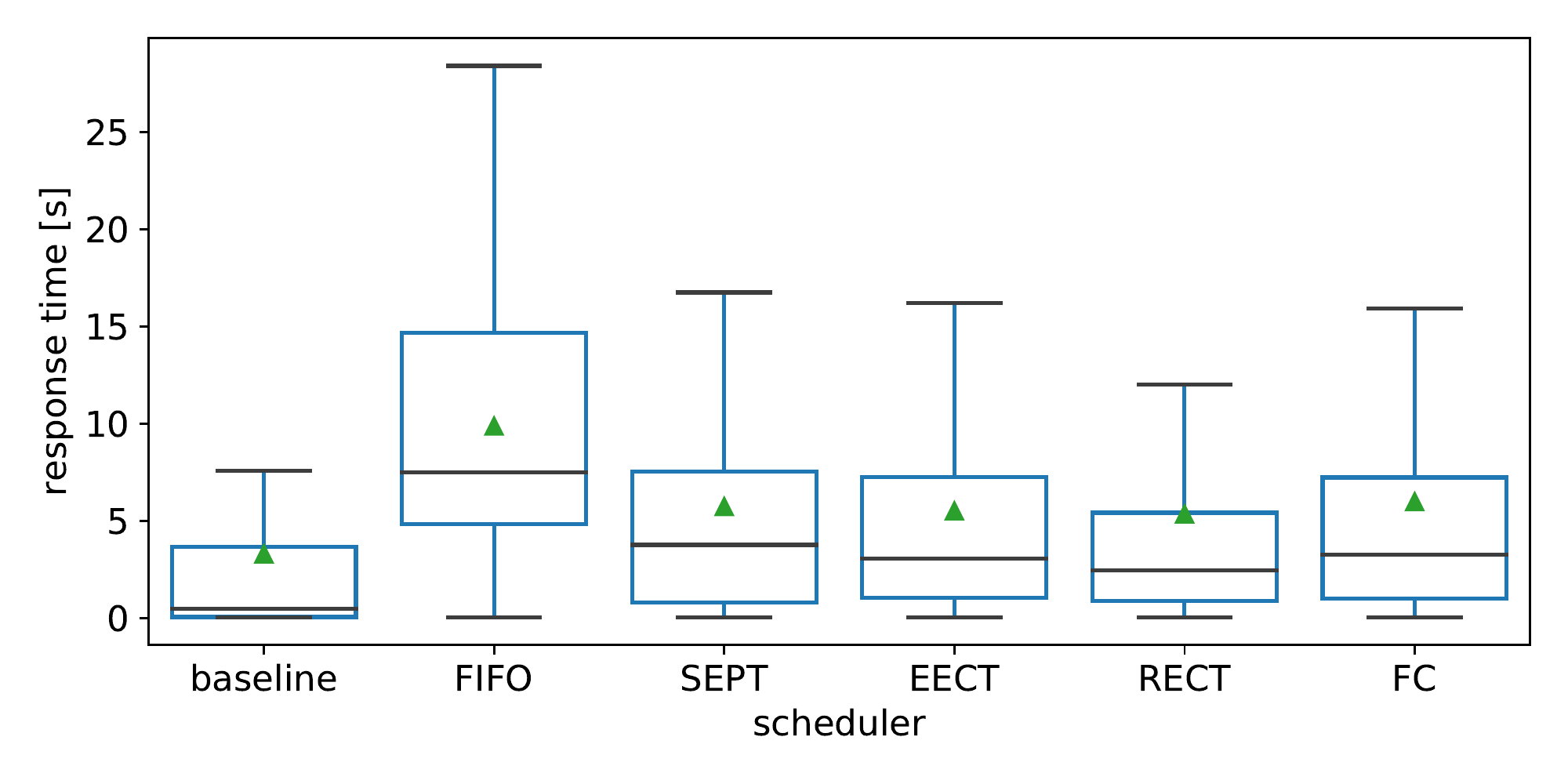}
    \chart{Experiment 2}{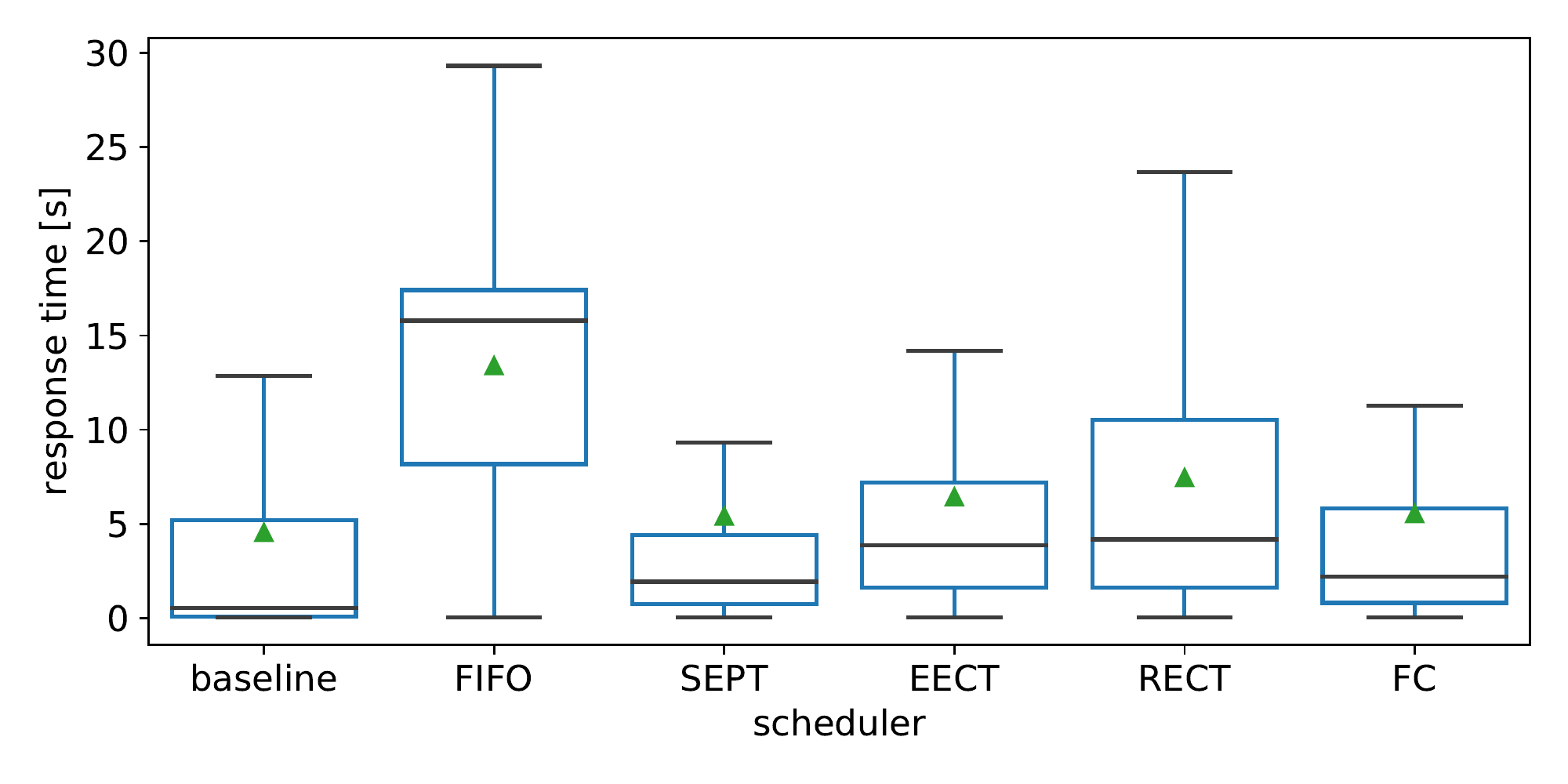}
    \chart{Experiment 3}{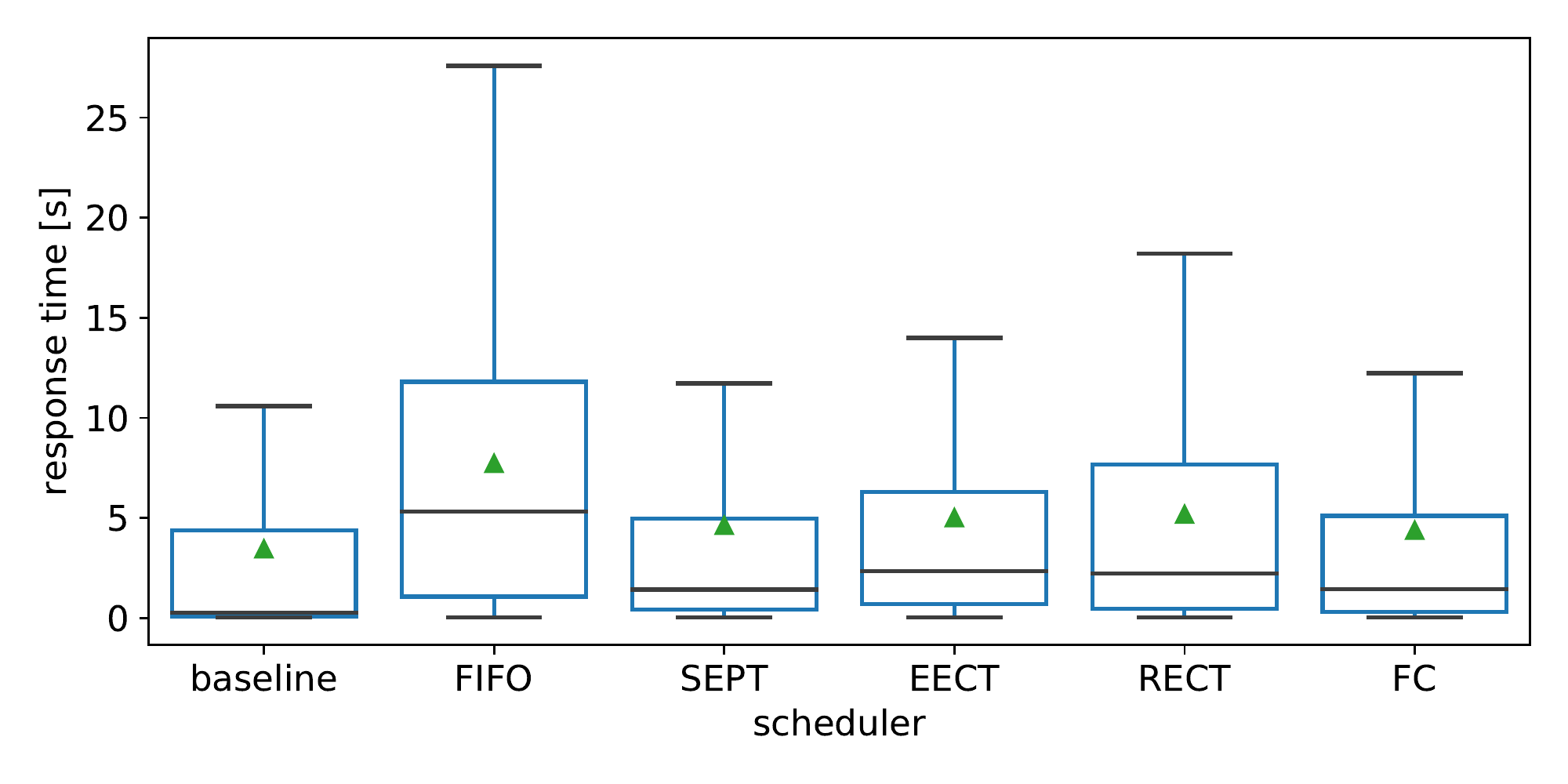}
    \\
    \chart{Experiment 4}{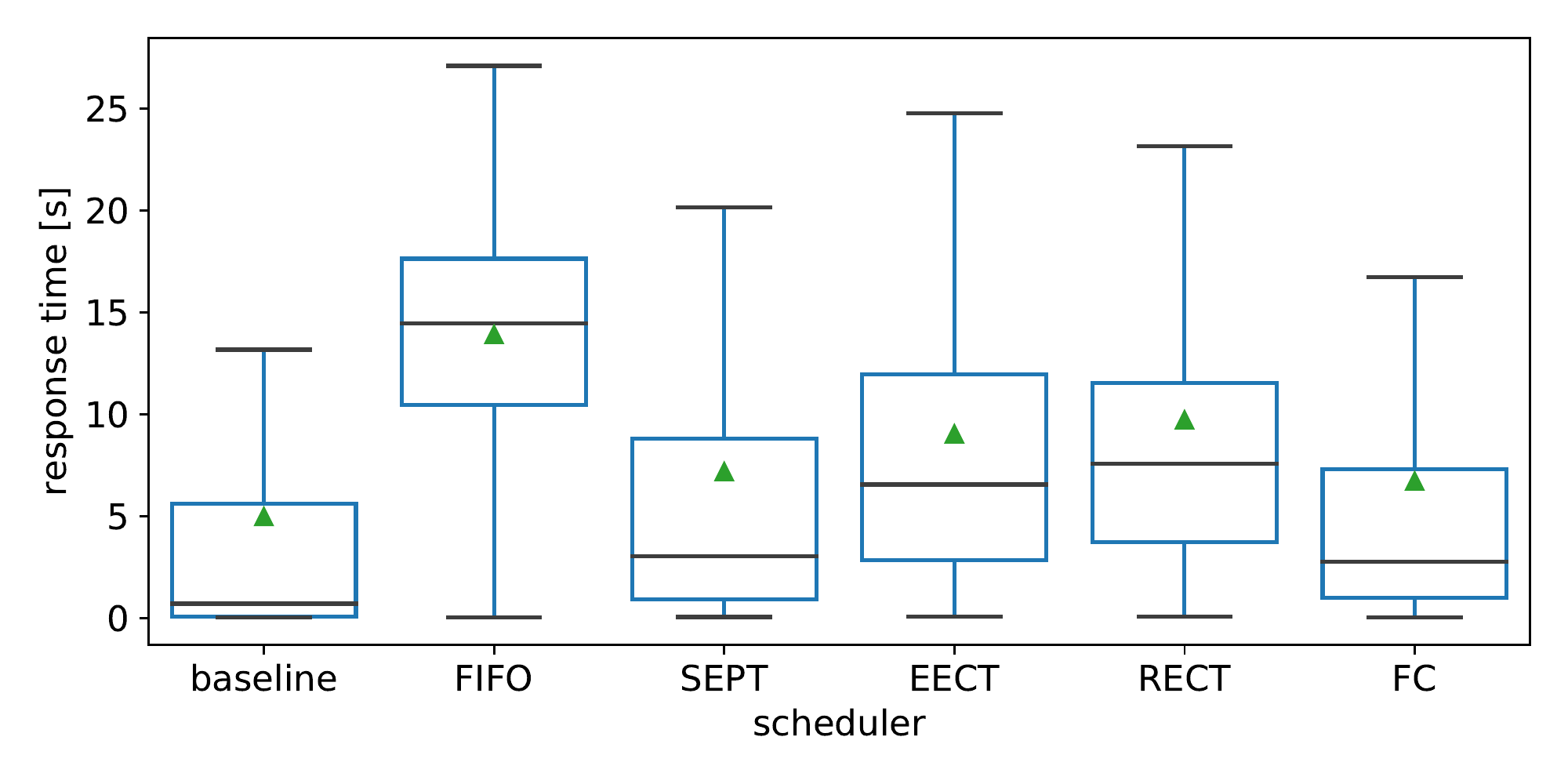}
    \chart{Experiment 5}{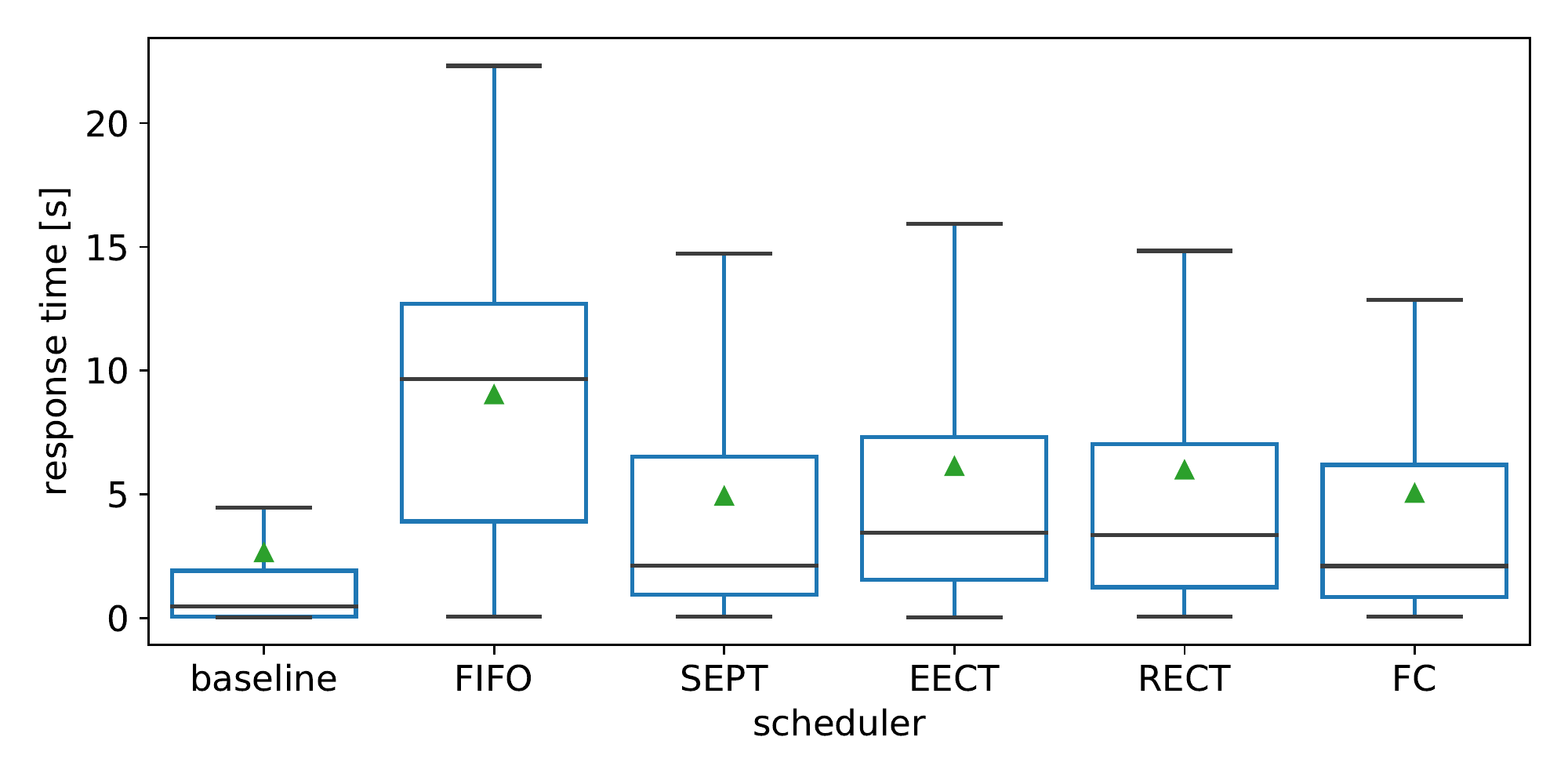}
    \\
    \caption{Response time for different scheduling policies. Here, we present results for 5 CPUs and intensity 30. On-premise infrastructure. Average response time presented as a green triangle.}
    \label{fig:response_time_5_30}
\end{figure*}

\begin{figure*}[h]
    \centering
    \chart{Experiment 1}{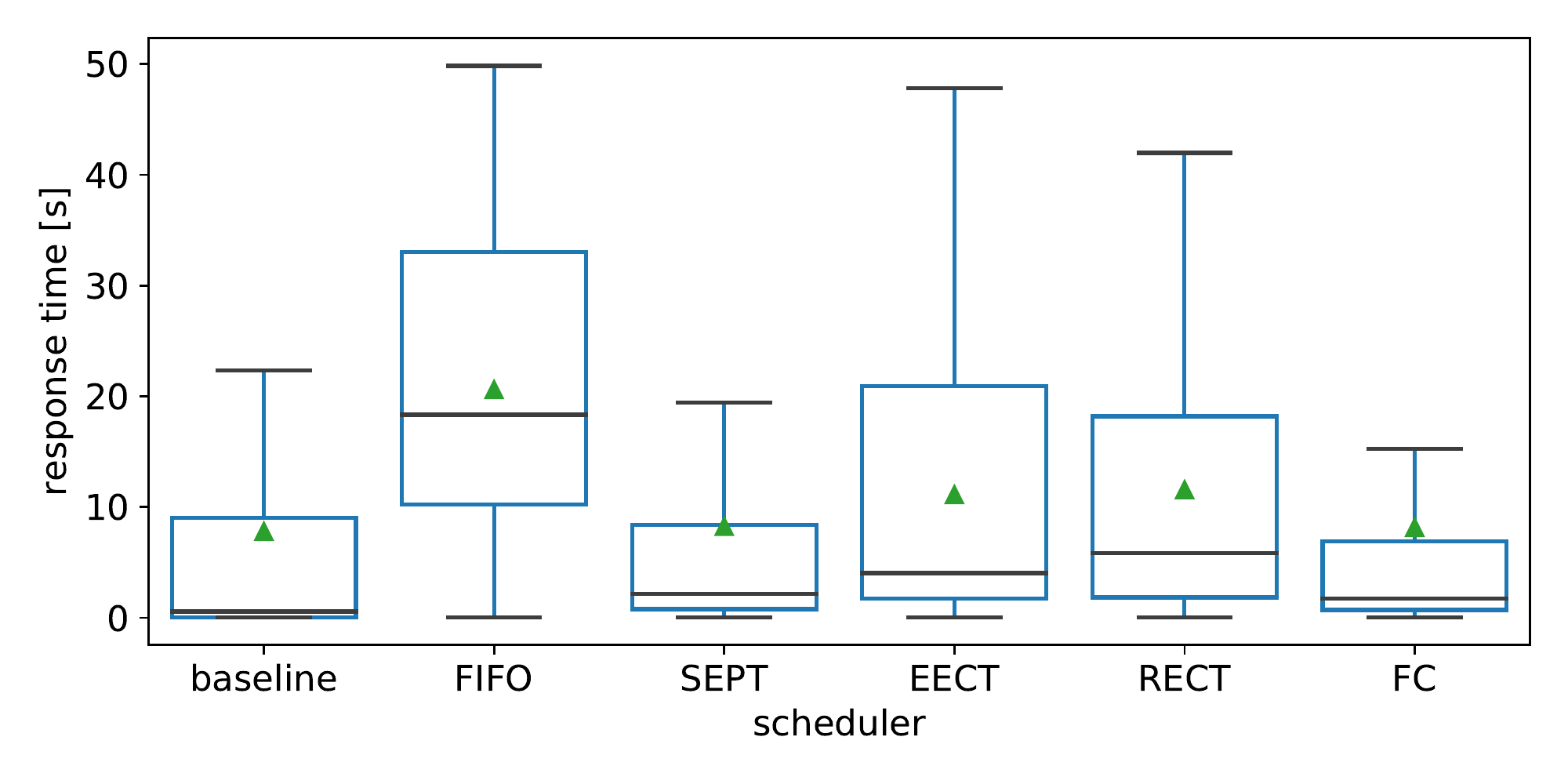}
    \chart{Experiment 2}{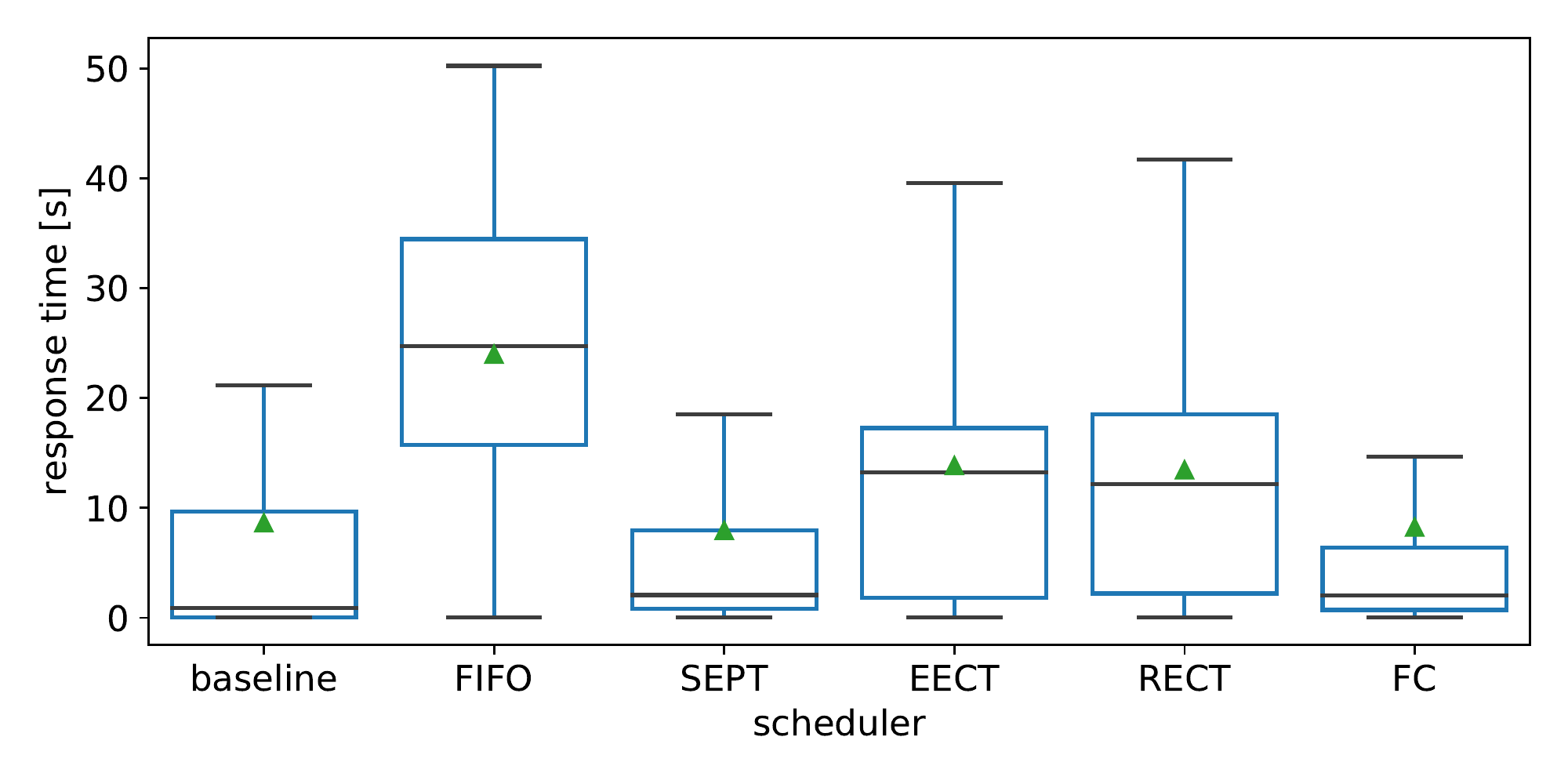}
    \chart{Experiment 3}{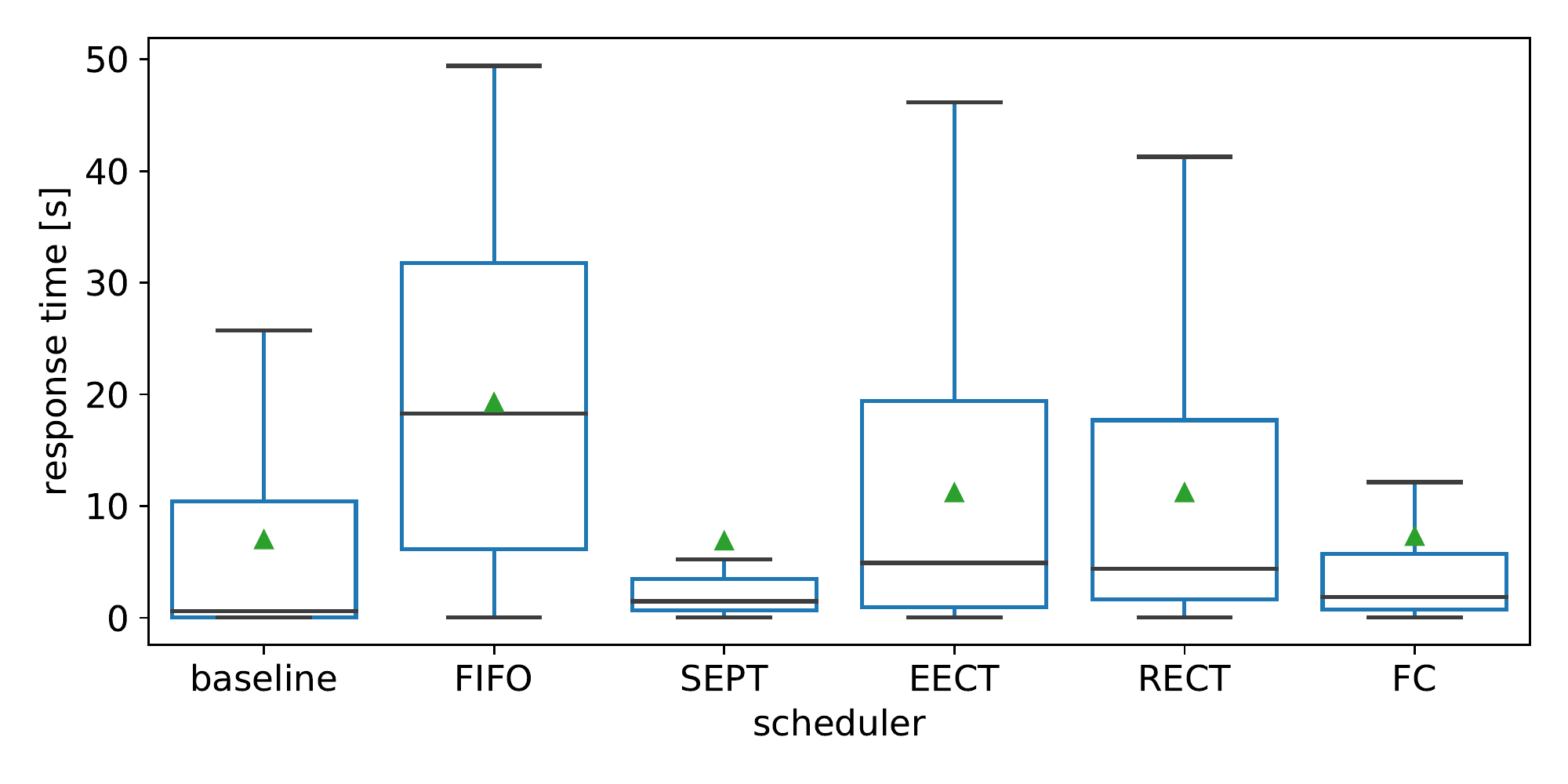}
    \\
    \chart{Experiment 4}{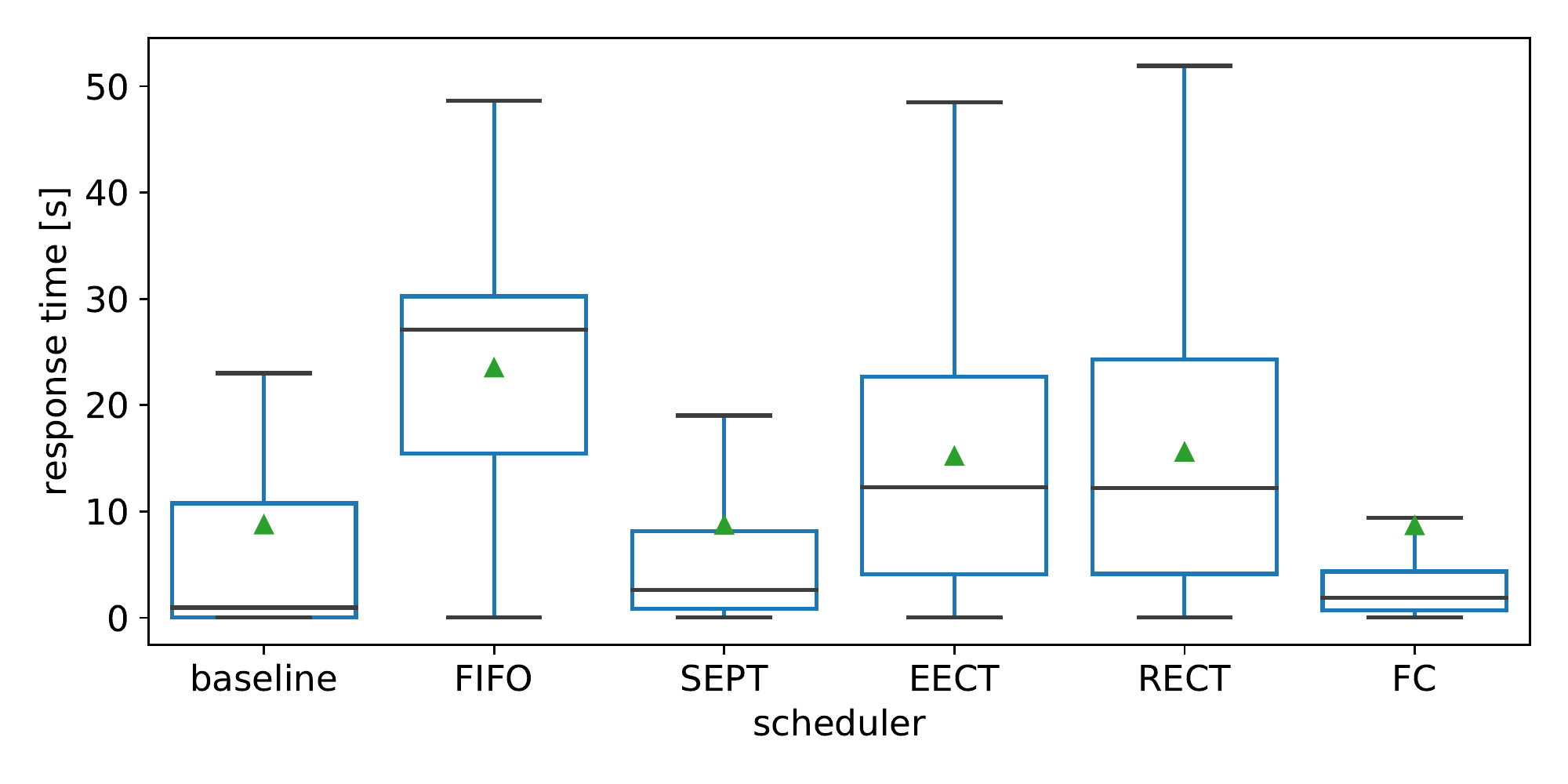}
    \chart{Experiment 5}{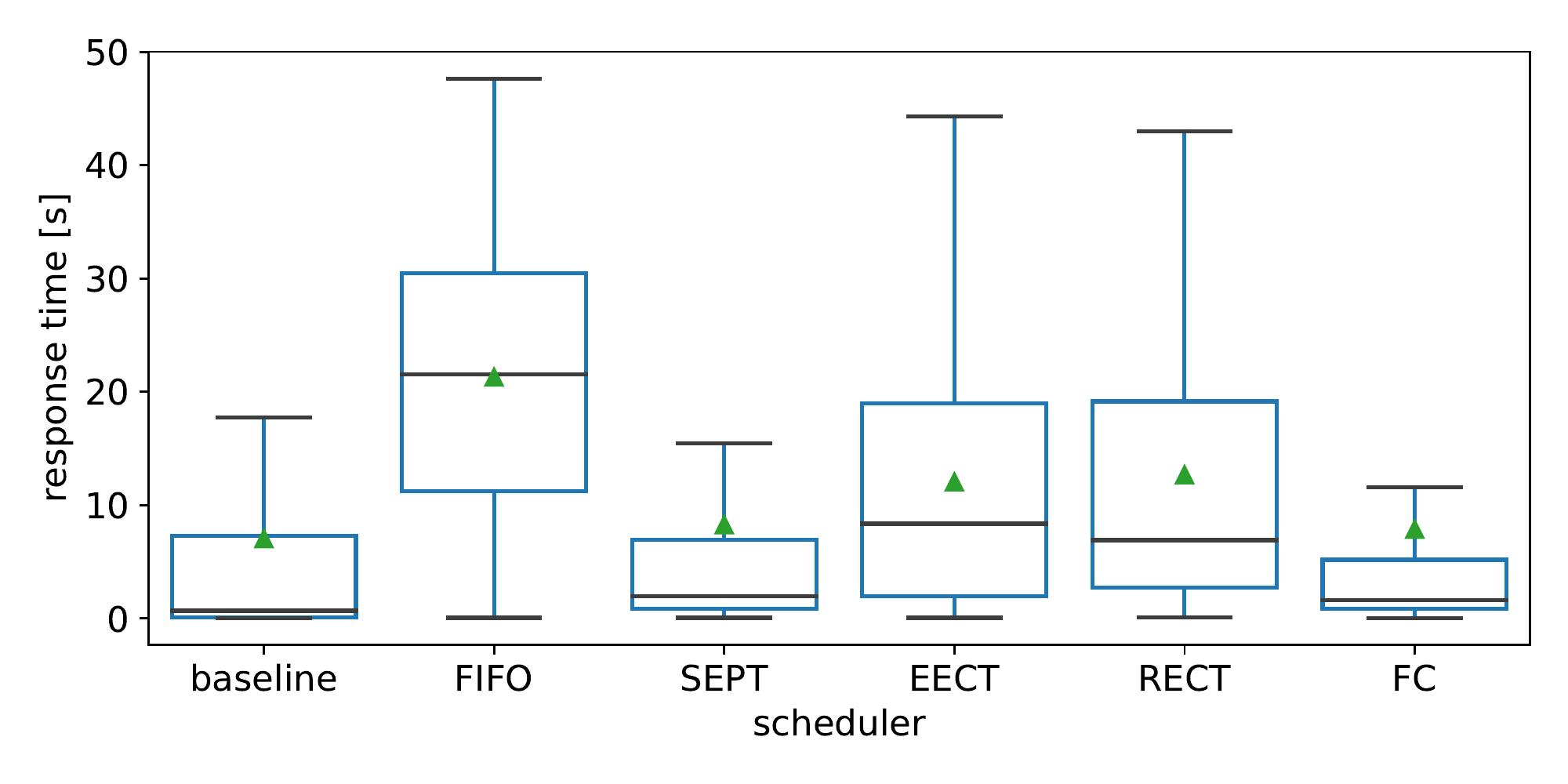}
    \\
    \caption{Response time for different scheduling policies. Here, we present results for 5 CPUs and intensity 40. On-premise infrastructure. Average response time presented as a green triangle.}
    \label{fig:response_time_5_40}
\end{figure*}

\begin{figure*}[h]
    \centering
    \chart{Experiment 1}{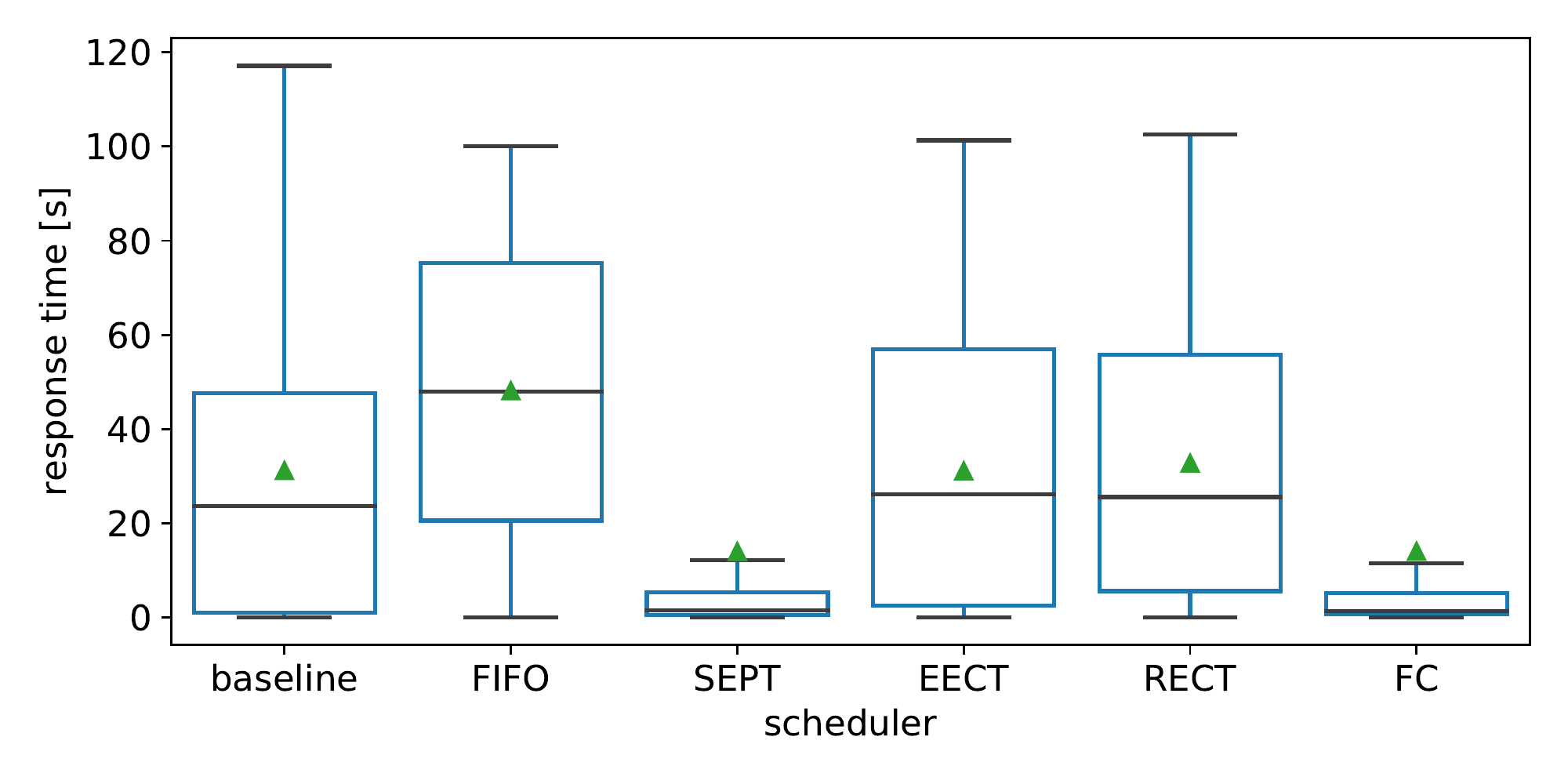}
    \chart{Experiment 2}{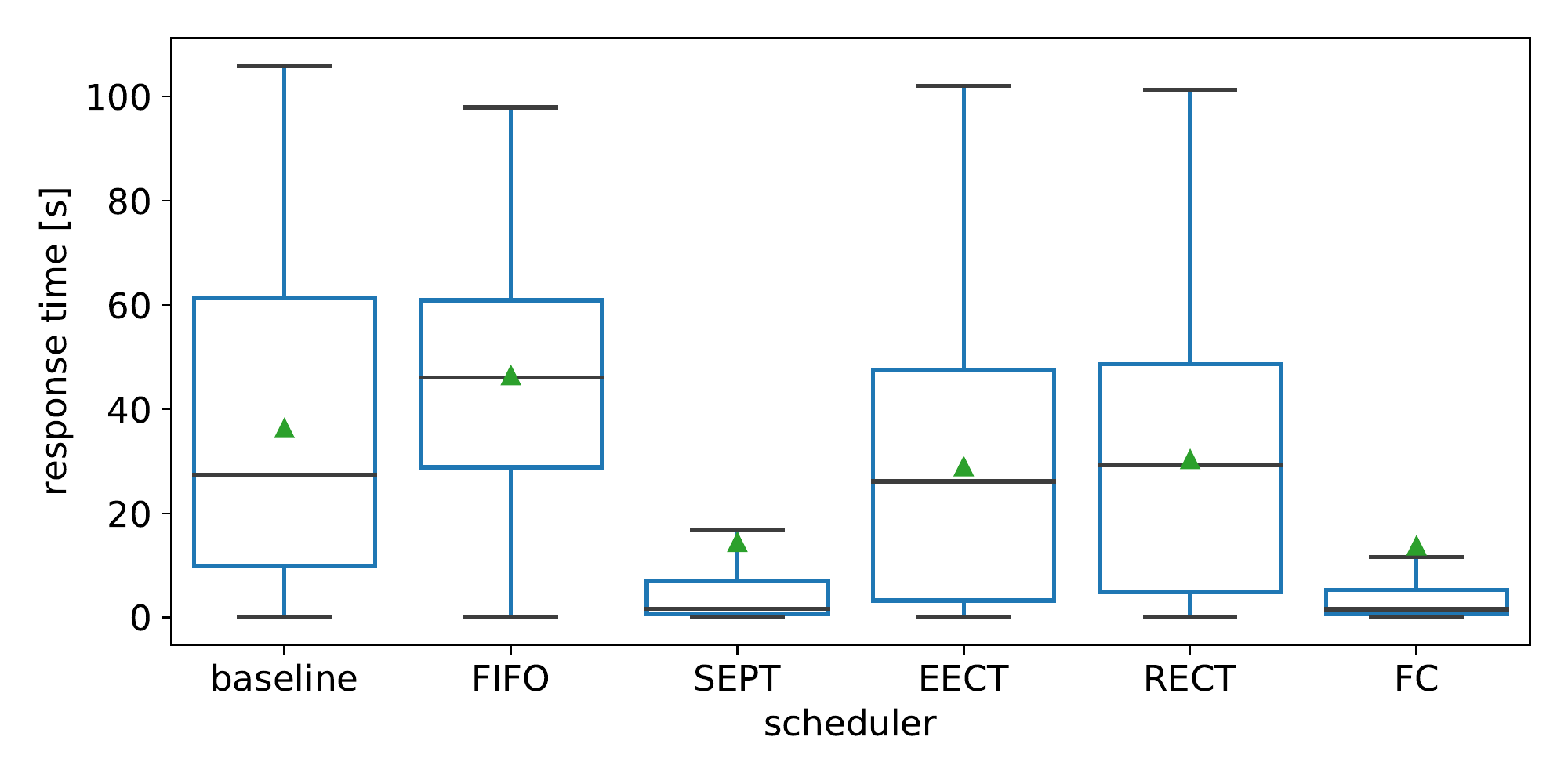}
    \chart{Experiment 3}{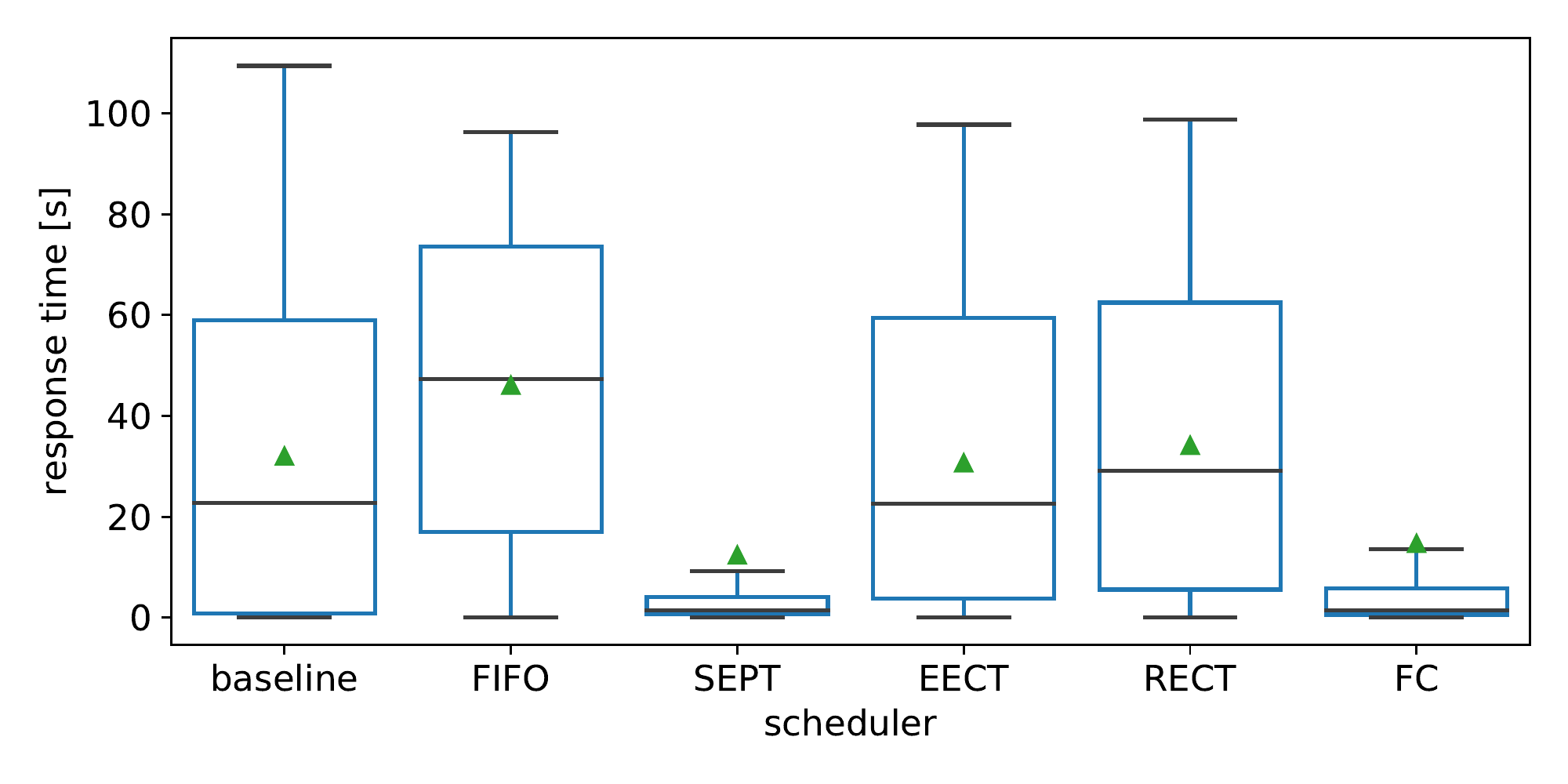}
    \\
    \chart{Experiment 4}{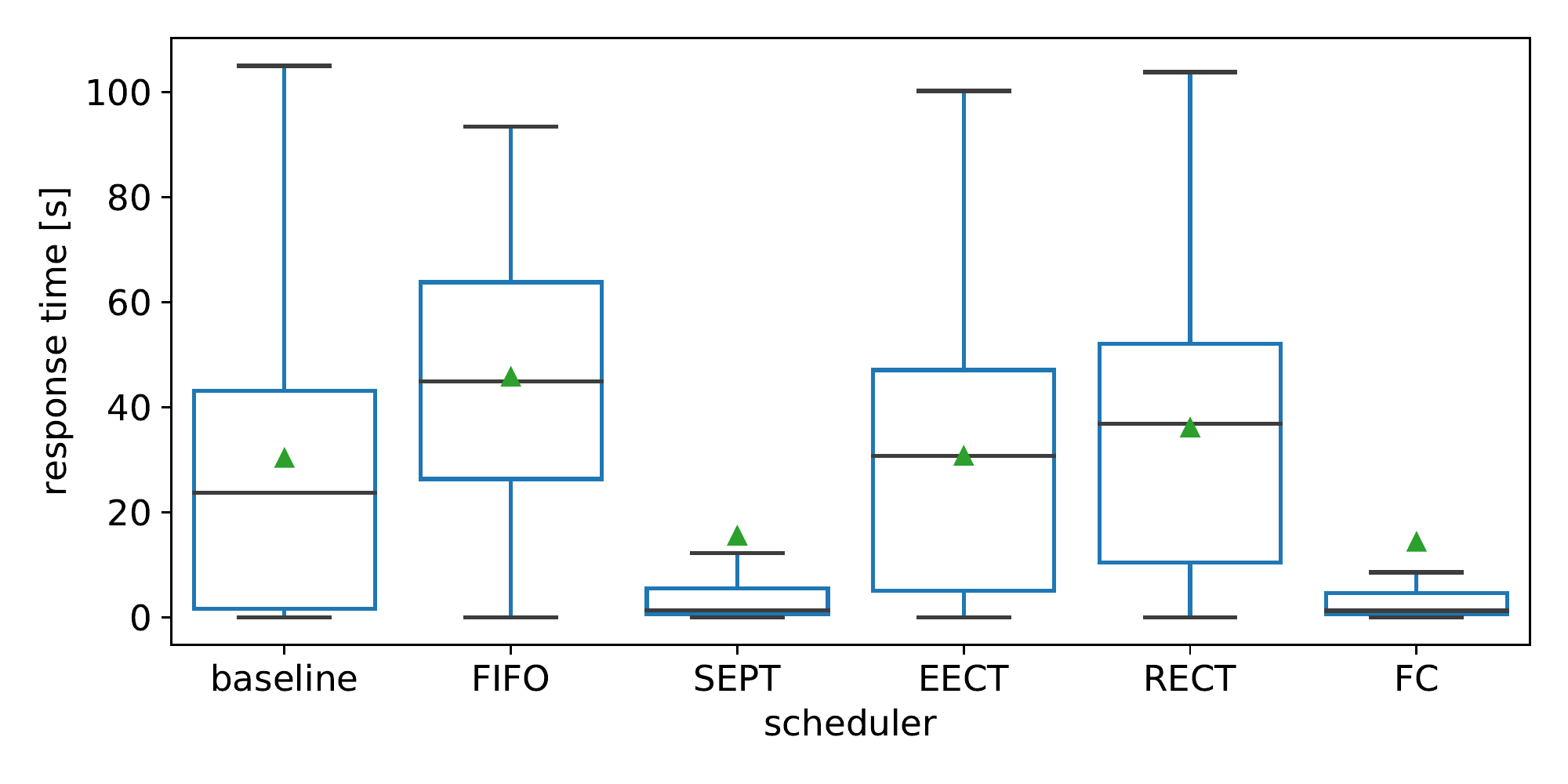}
    \chart{Experiment 5}{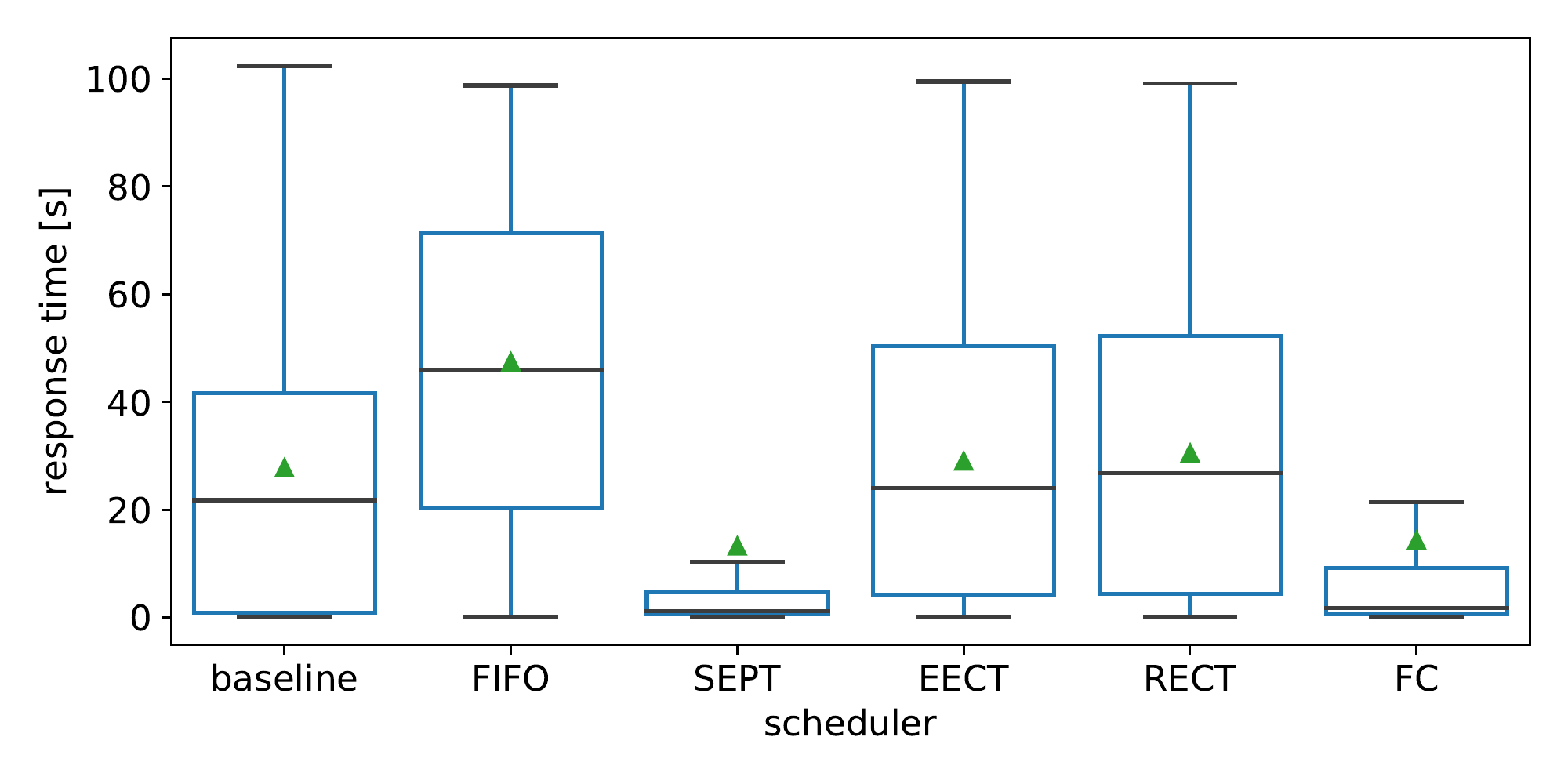}
    \\
    \caption{Response time for different scheduling policies. Here, we present results for 5 CPUs and intensity 60. On-premise infrastructure. Average response time presented as a green triangle.}
    \label{fig:response_time_5_60}
\end{figure*}

\begin{figure*}[h]
    \centering
    \chart{Experiment 1}{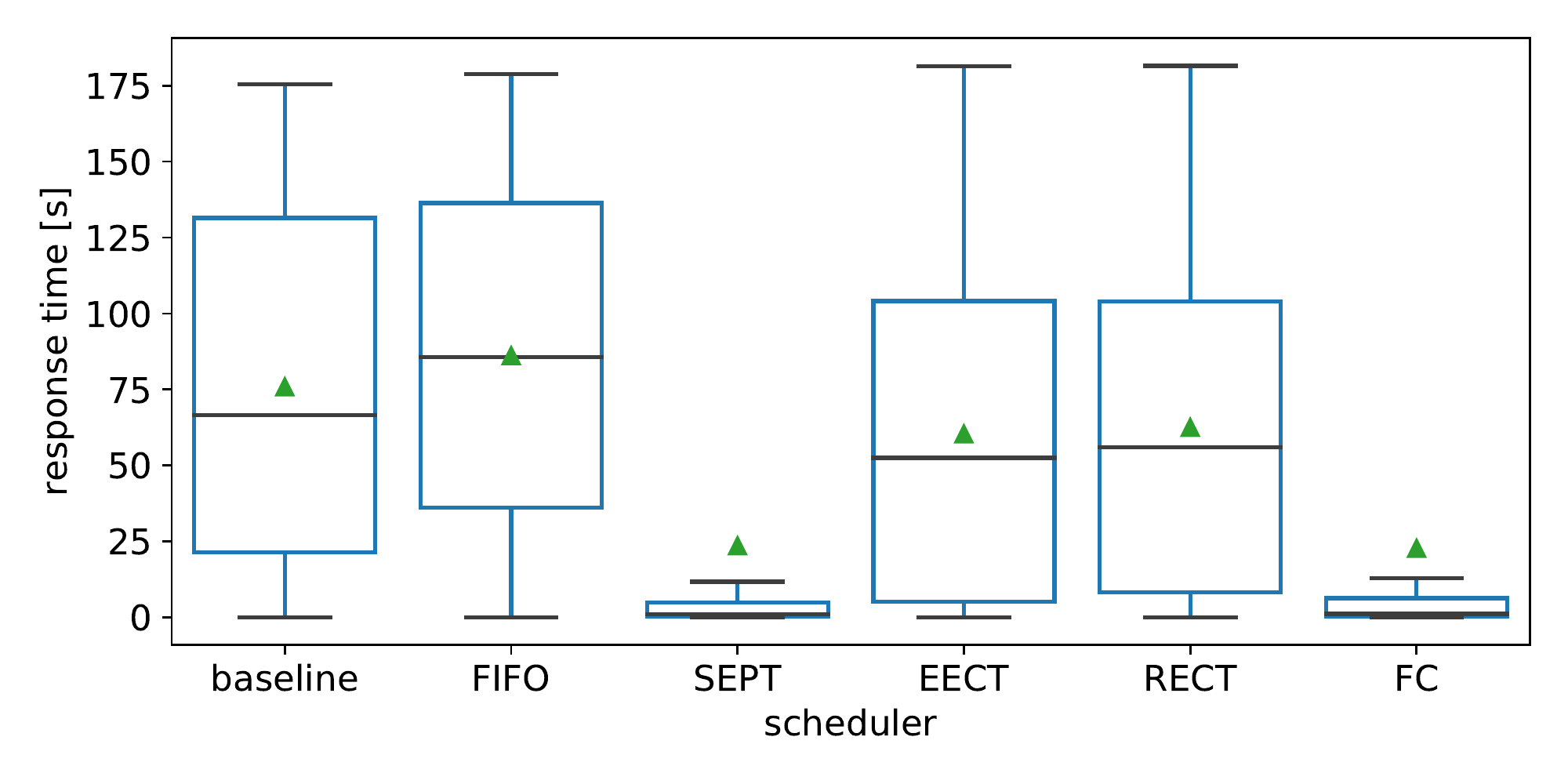}
    \chart{Experiment 2}{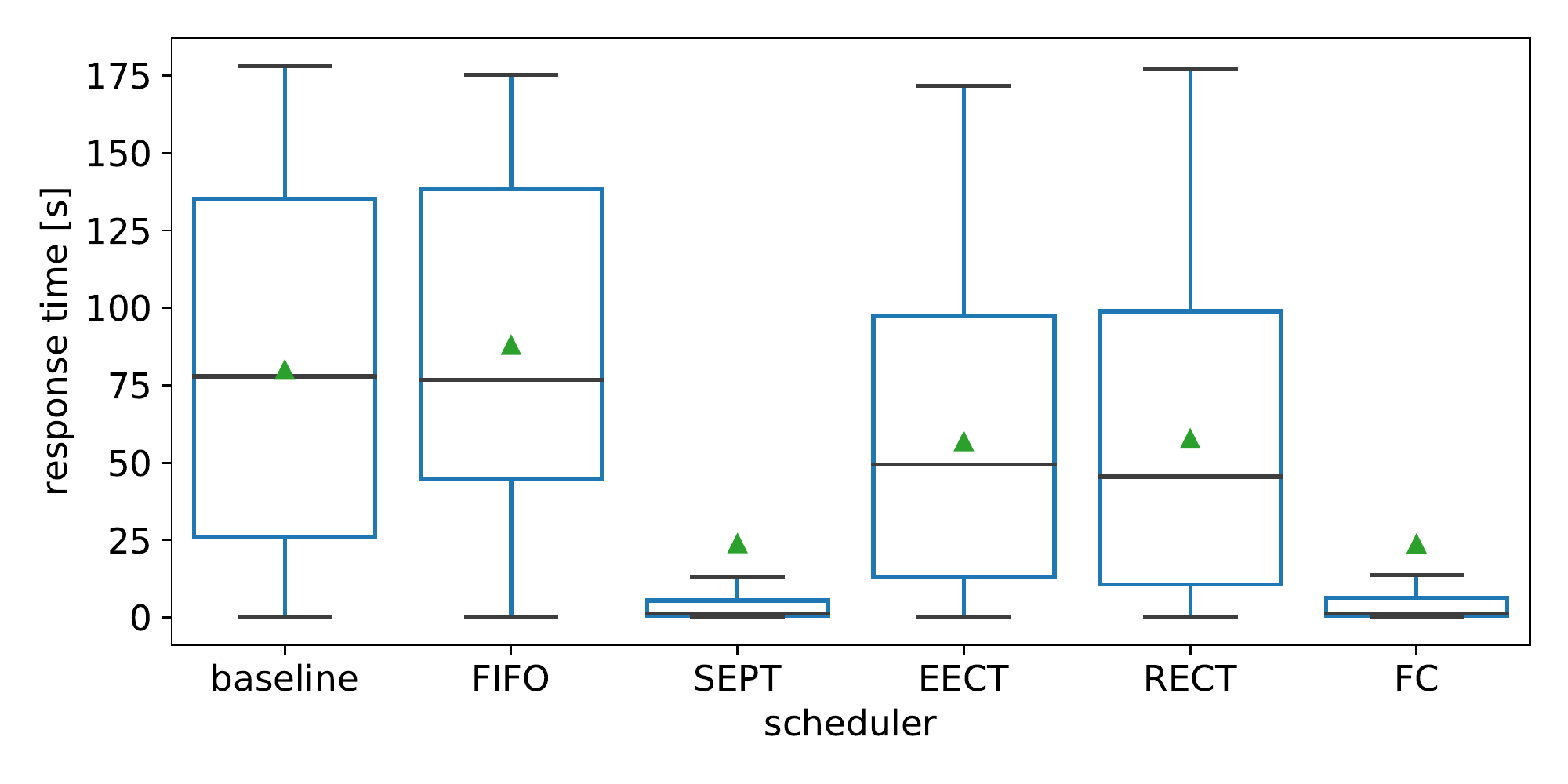}
    \chart{Experiment 3}{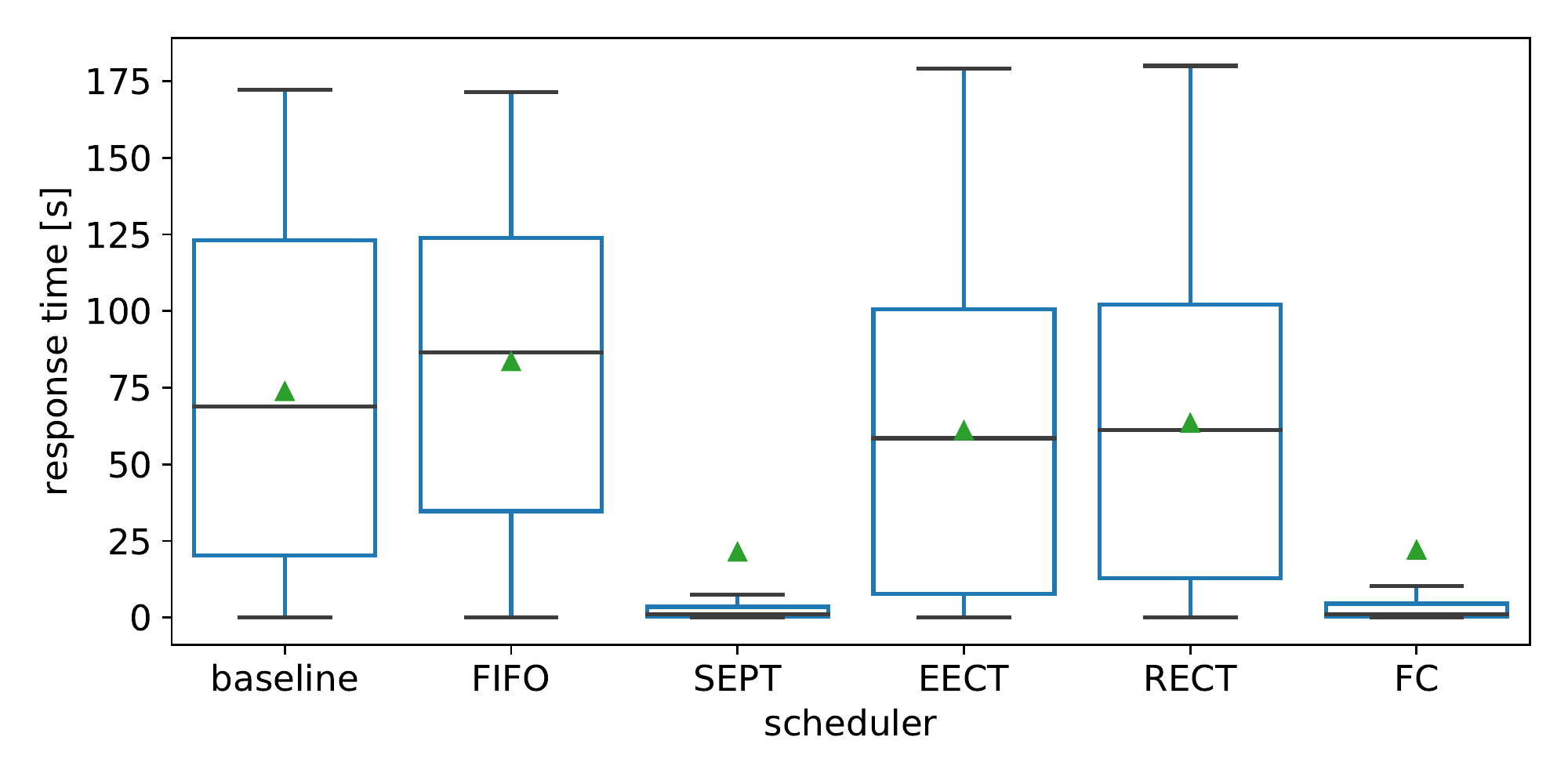}
    \\
    \chart{Experiment 4}{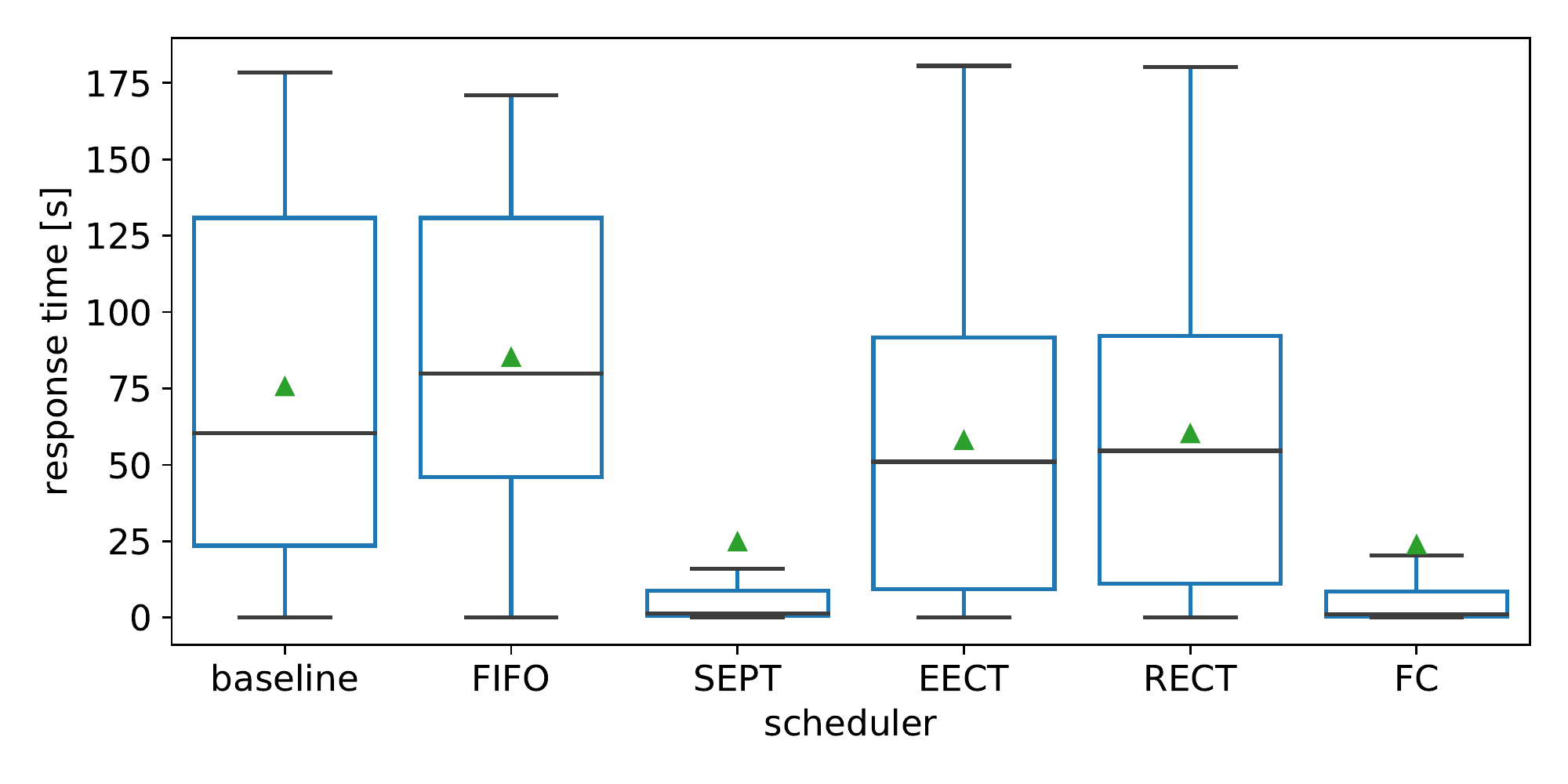}
    \chart{Experiment 5}{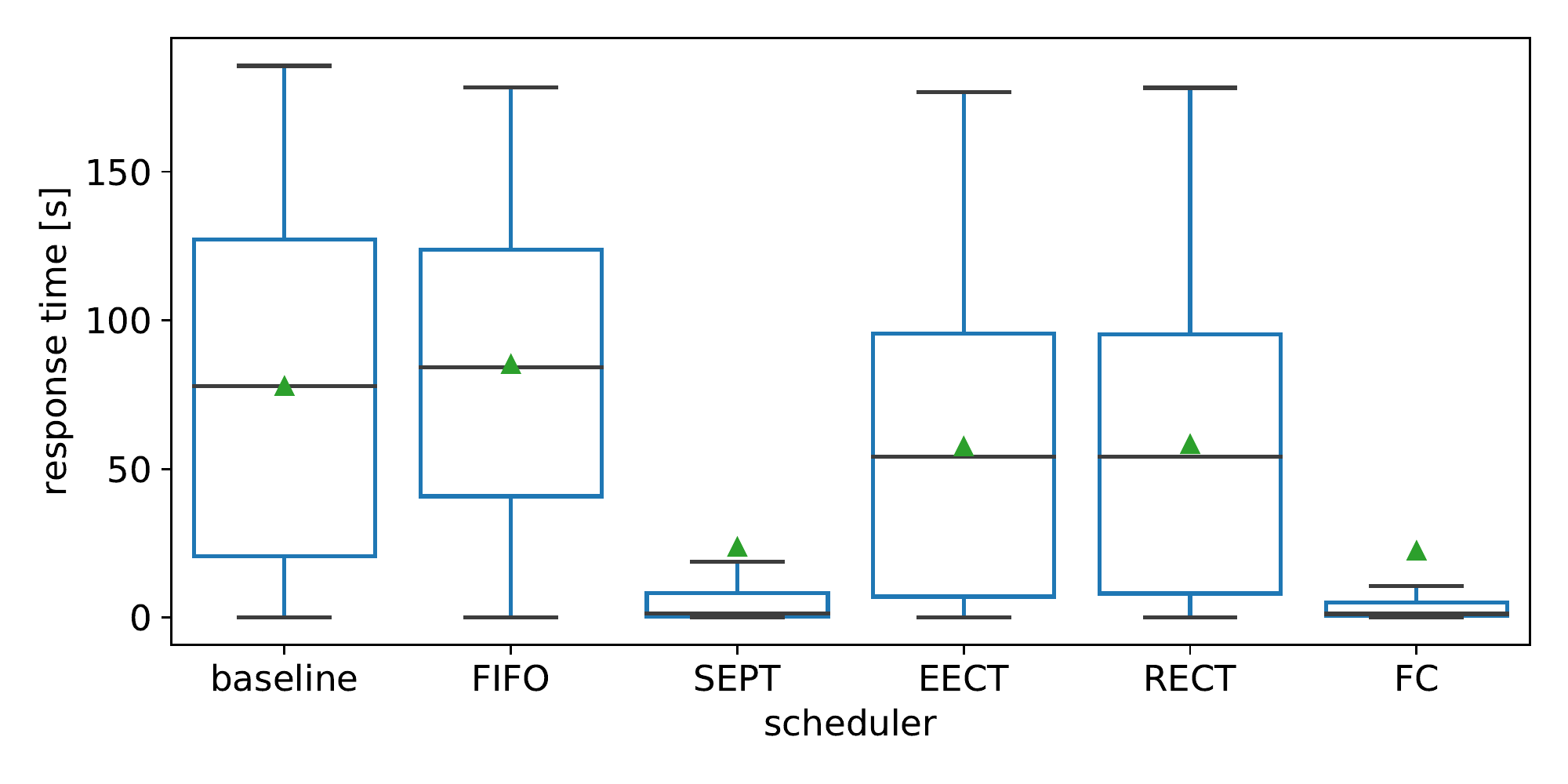}
    \\
    \caption{Response time for different scheduling policies. Here, we present results for 5 CPUs and intensity 90. On-premise infrastructure. Average response time presented as a green triangle.}
    \label{fig:response_time_5_90}
\end{figure*}

\begin{figure*}[h]
    \centering
    \chart{Experiment 1}{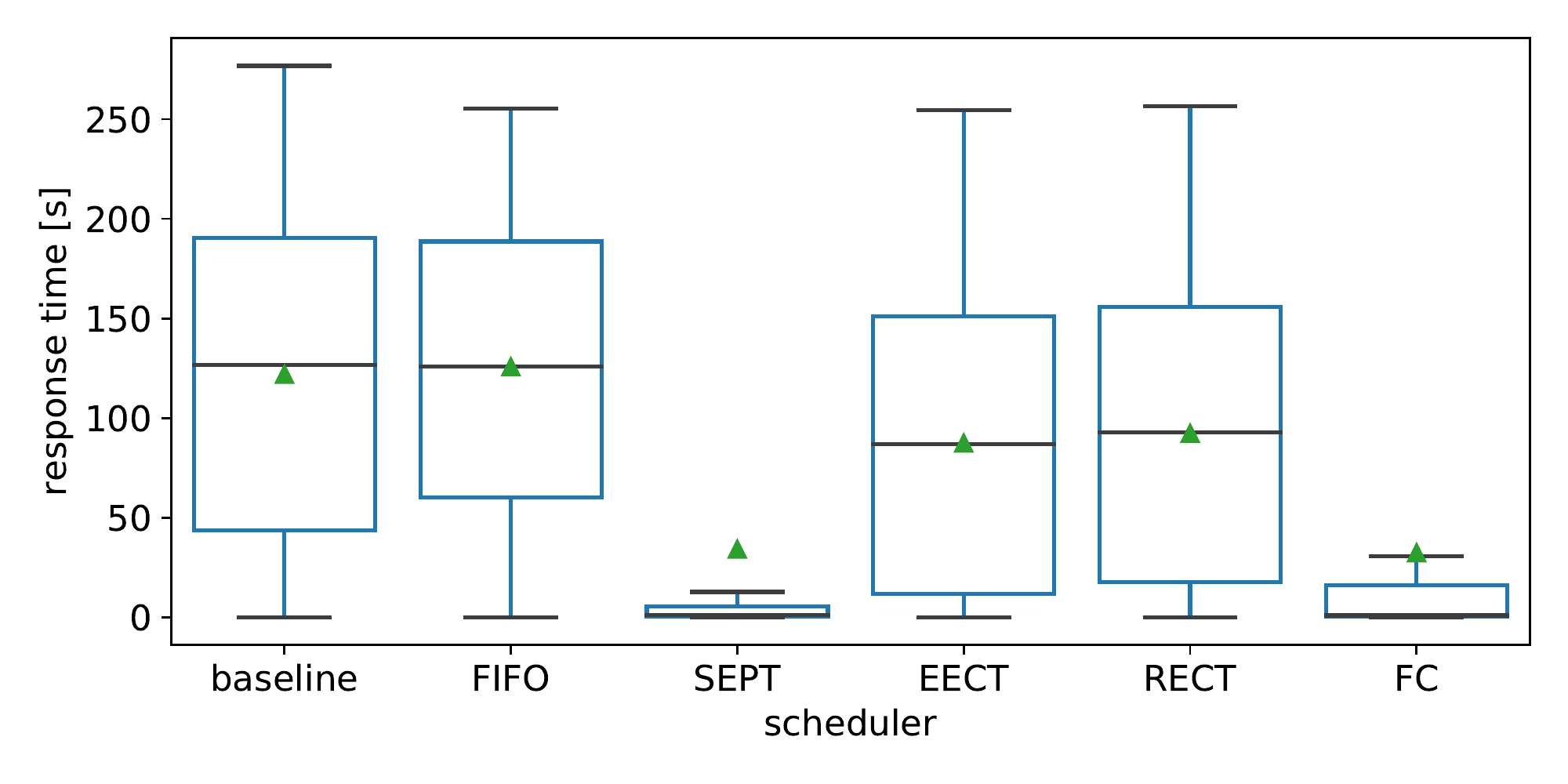}
    \chart{Experiment 2}{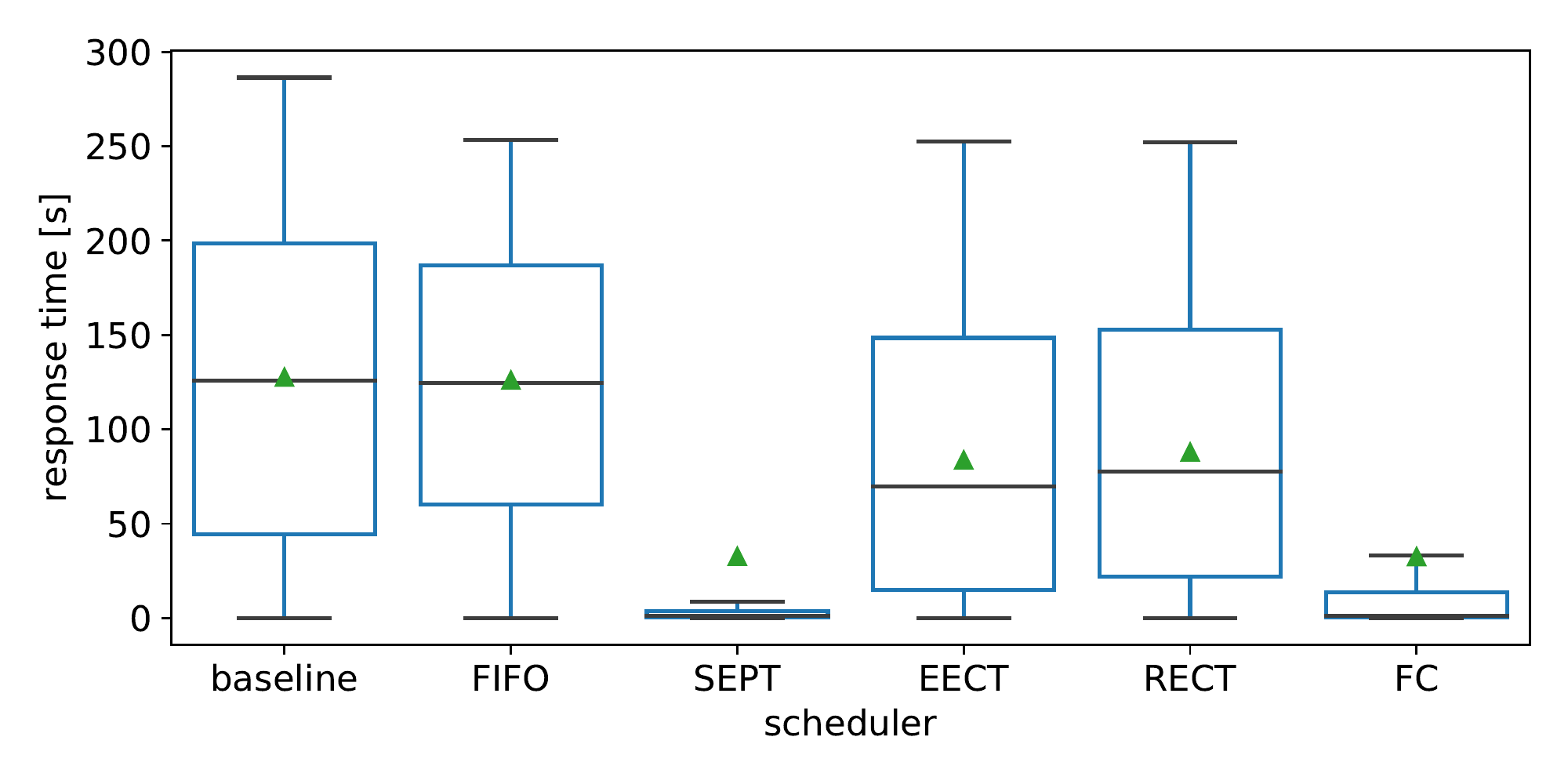}
    \chart{Experiment 3}{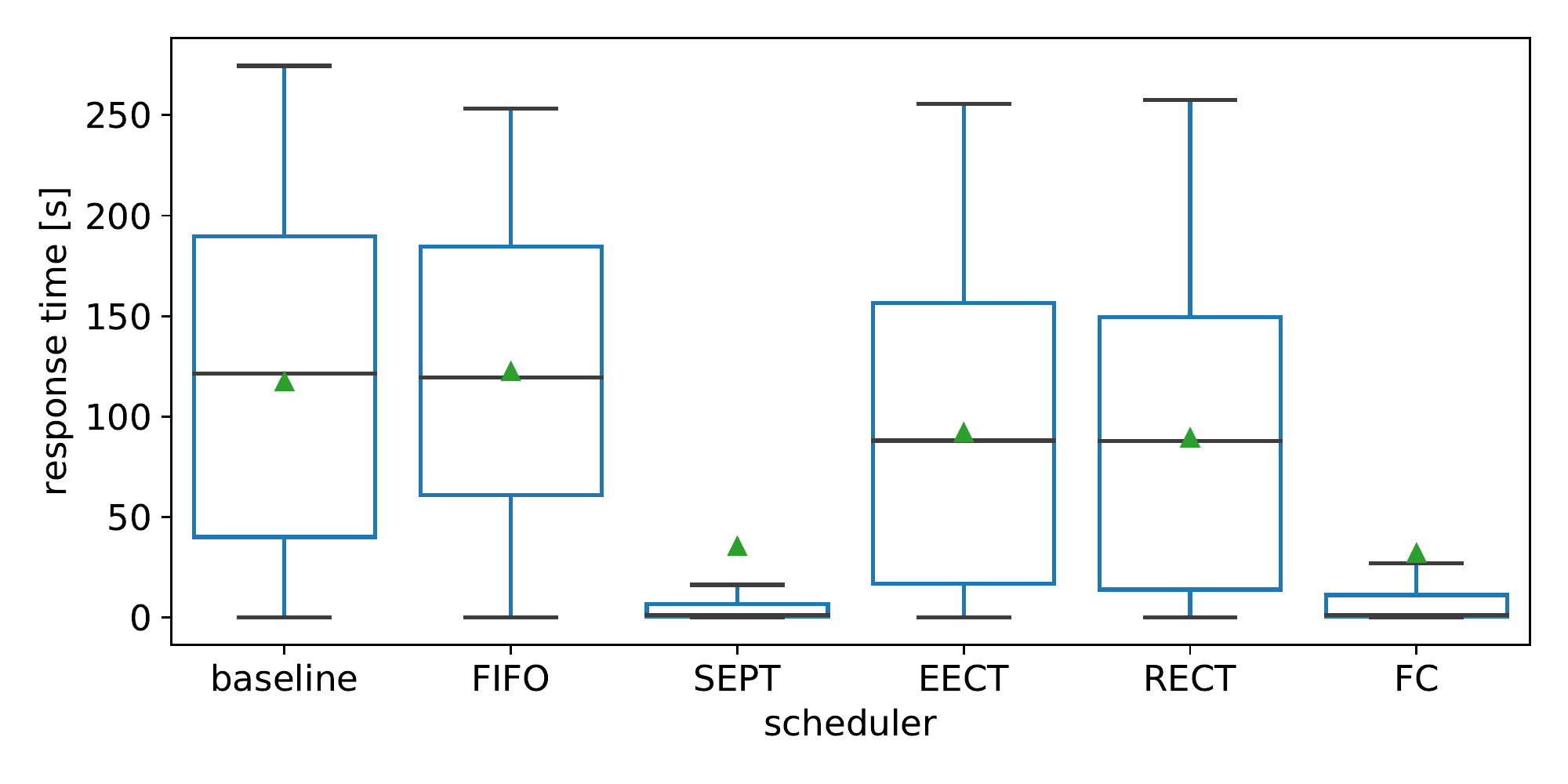}
    \\
    \chart{Experiment 4}{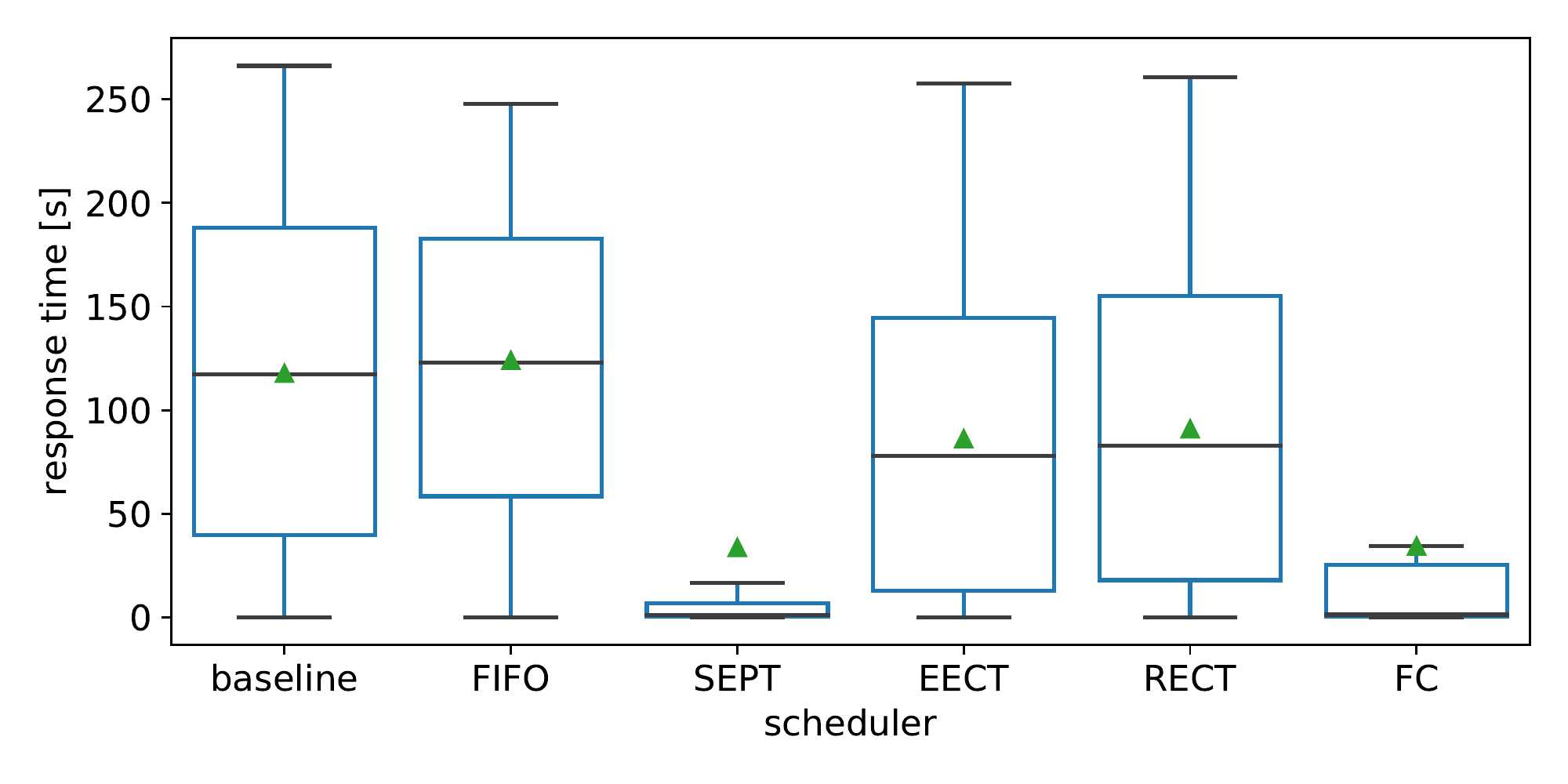}
    \chart{Experiment 5}{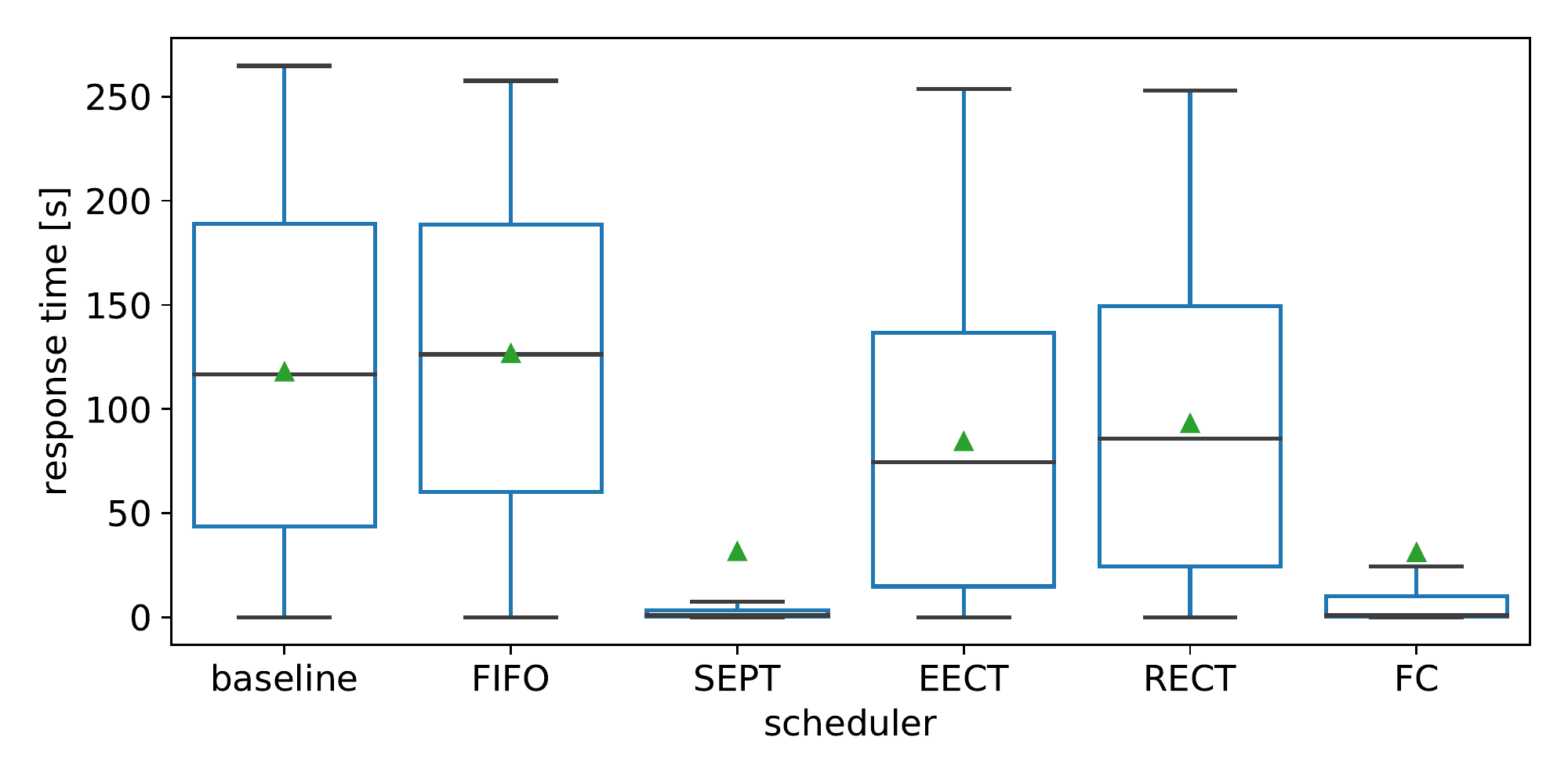}
    \\
    \caption{Response time for different scheduling policies. Here, we present results for 5 CPUs and intensity 120. On-premise infrastructure. Average response time presented as a green triangle.}
    \label{fig:response_time_5_120}
\end{figure*}

\begin{figure*}[h]
    \centering
    \chart{Experiment 1}{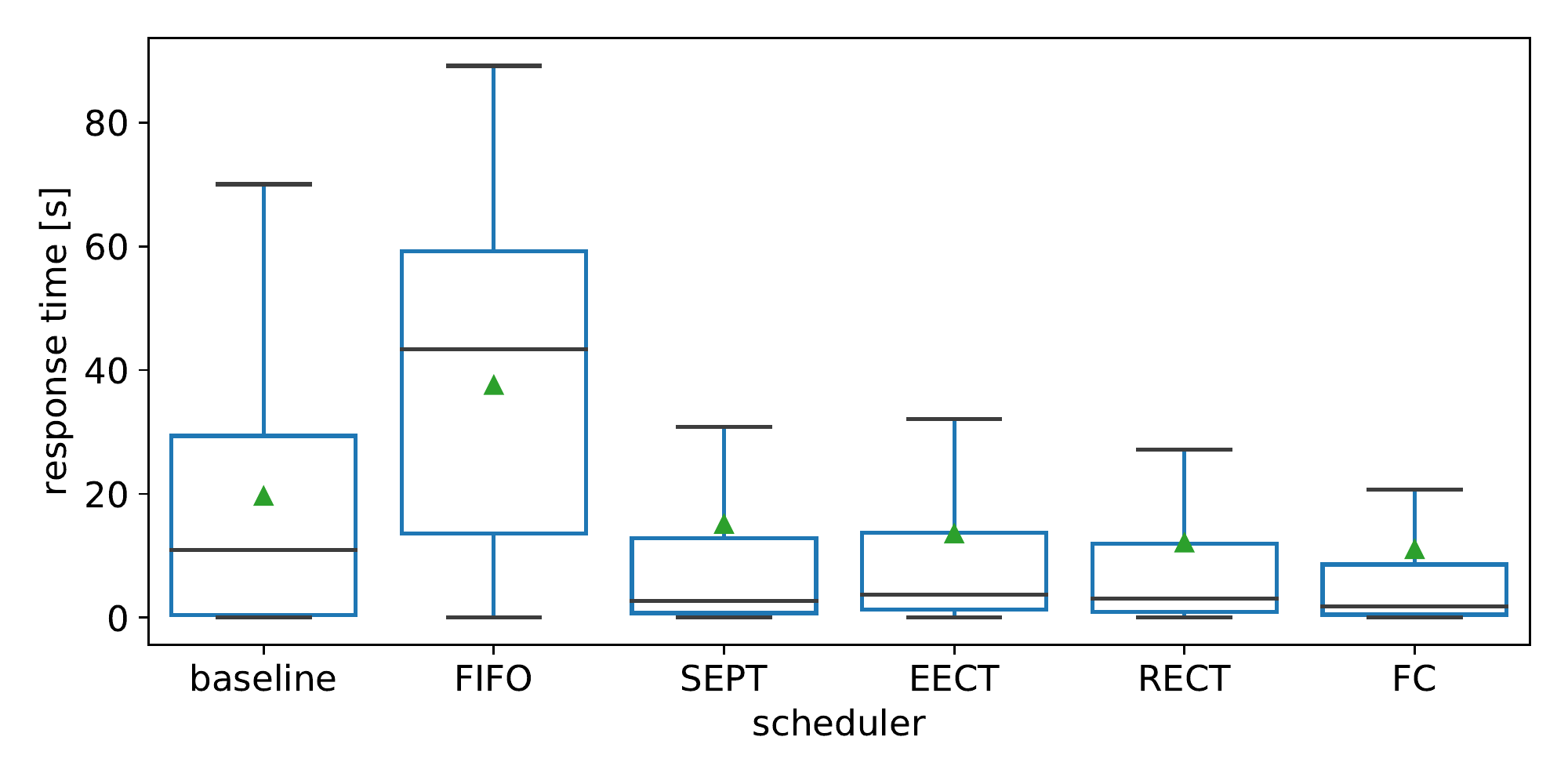}
    \chart{Experiment 2}{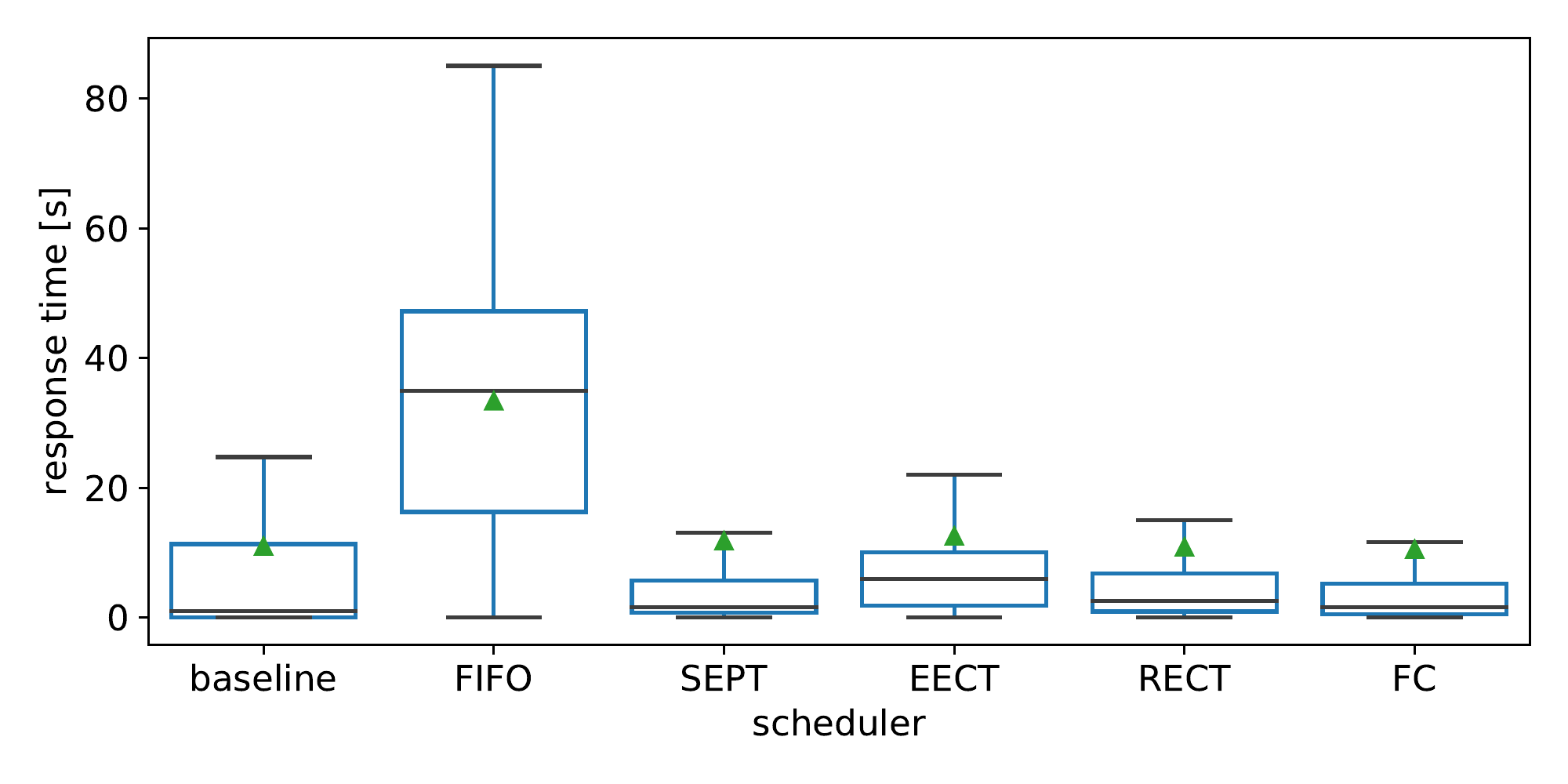}
    \chart{Experiment 3}{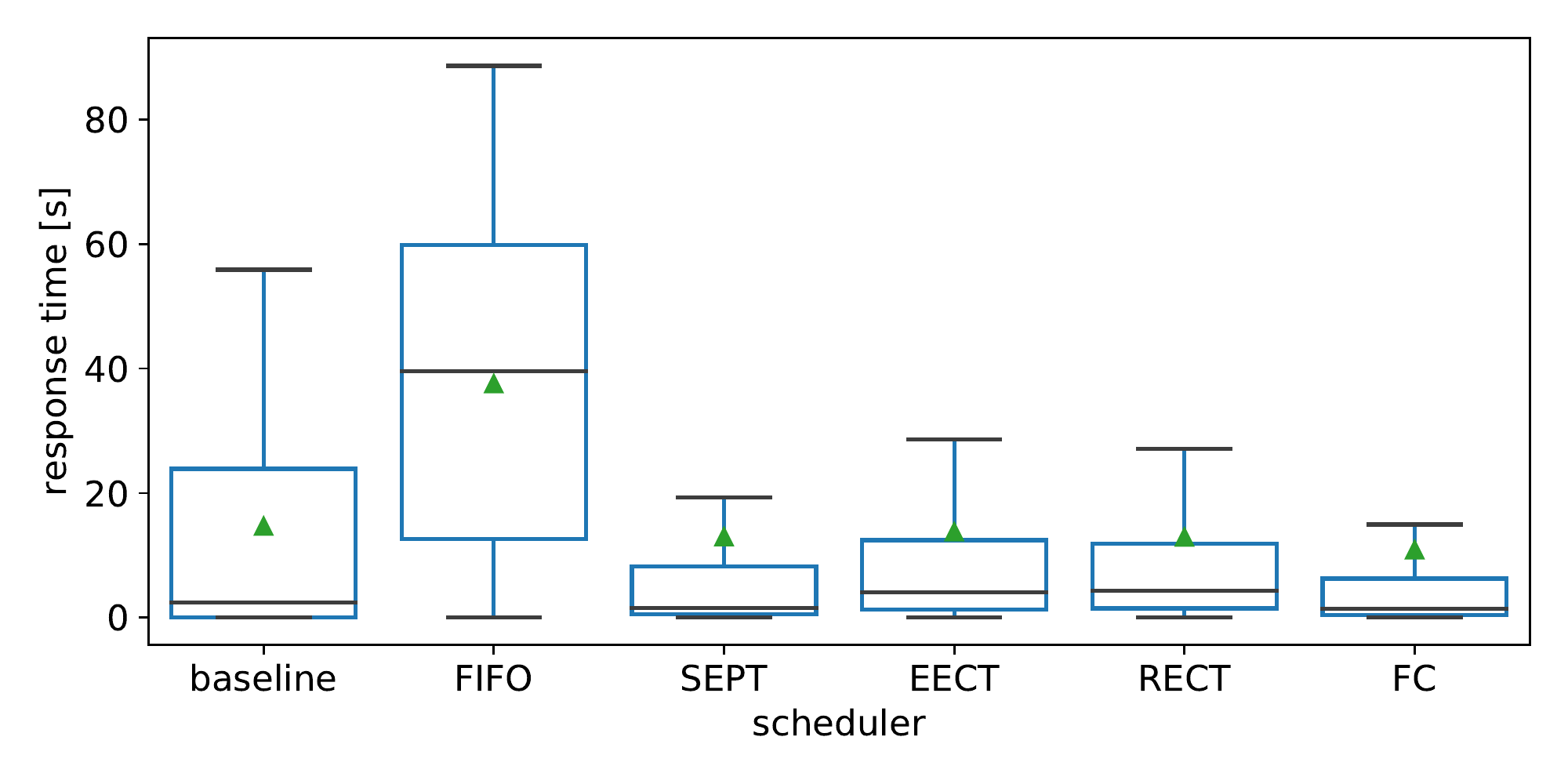}
    \\
    \chart{Experiment 4}{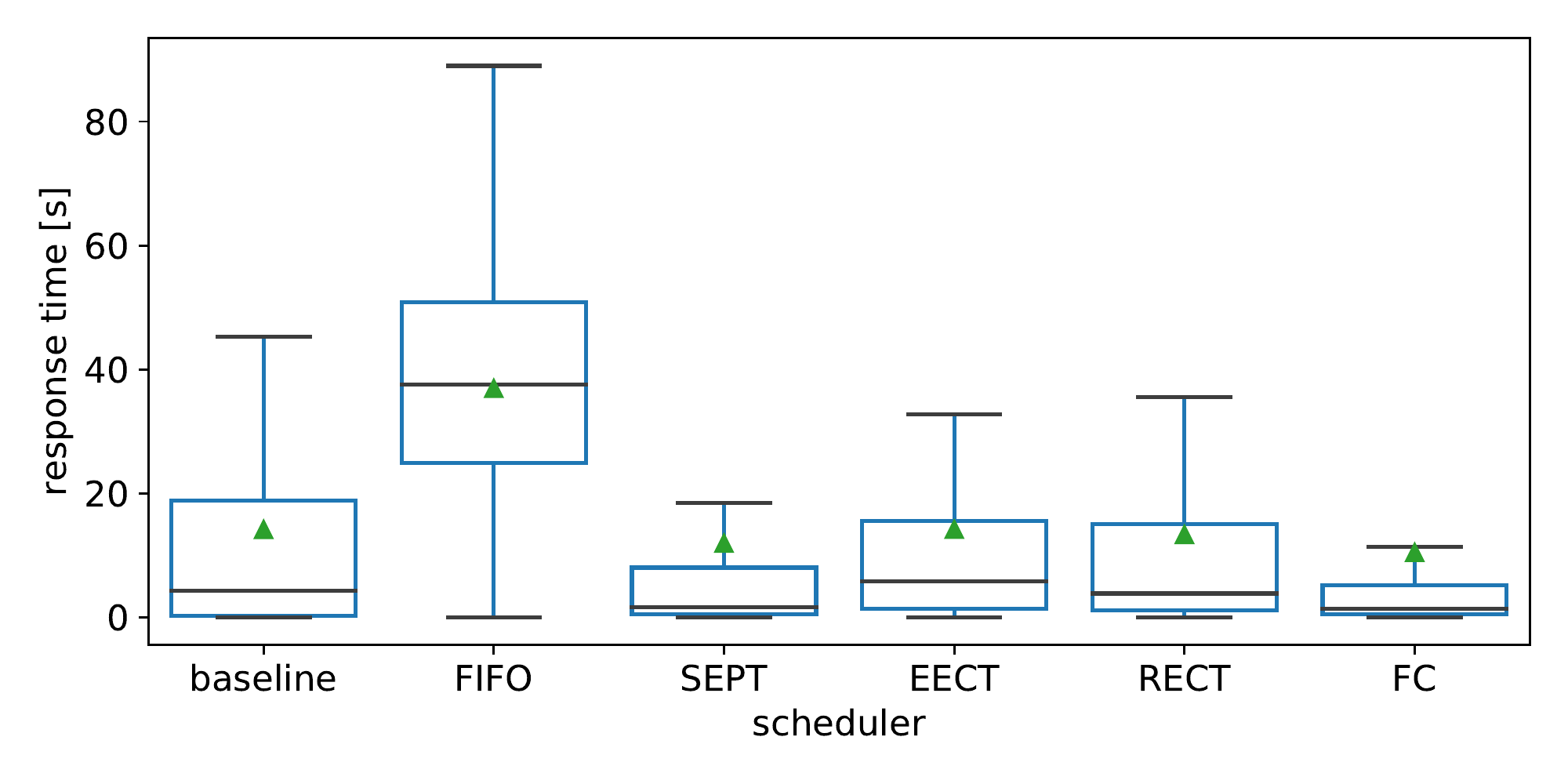}
    \chart{Experiment 5}{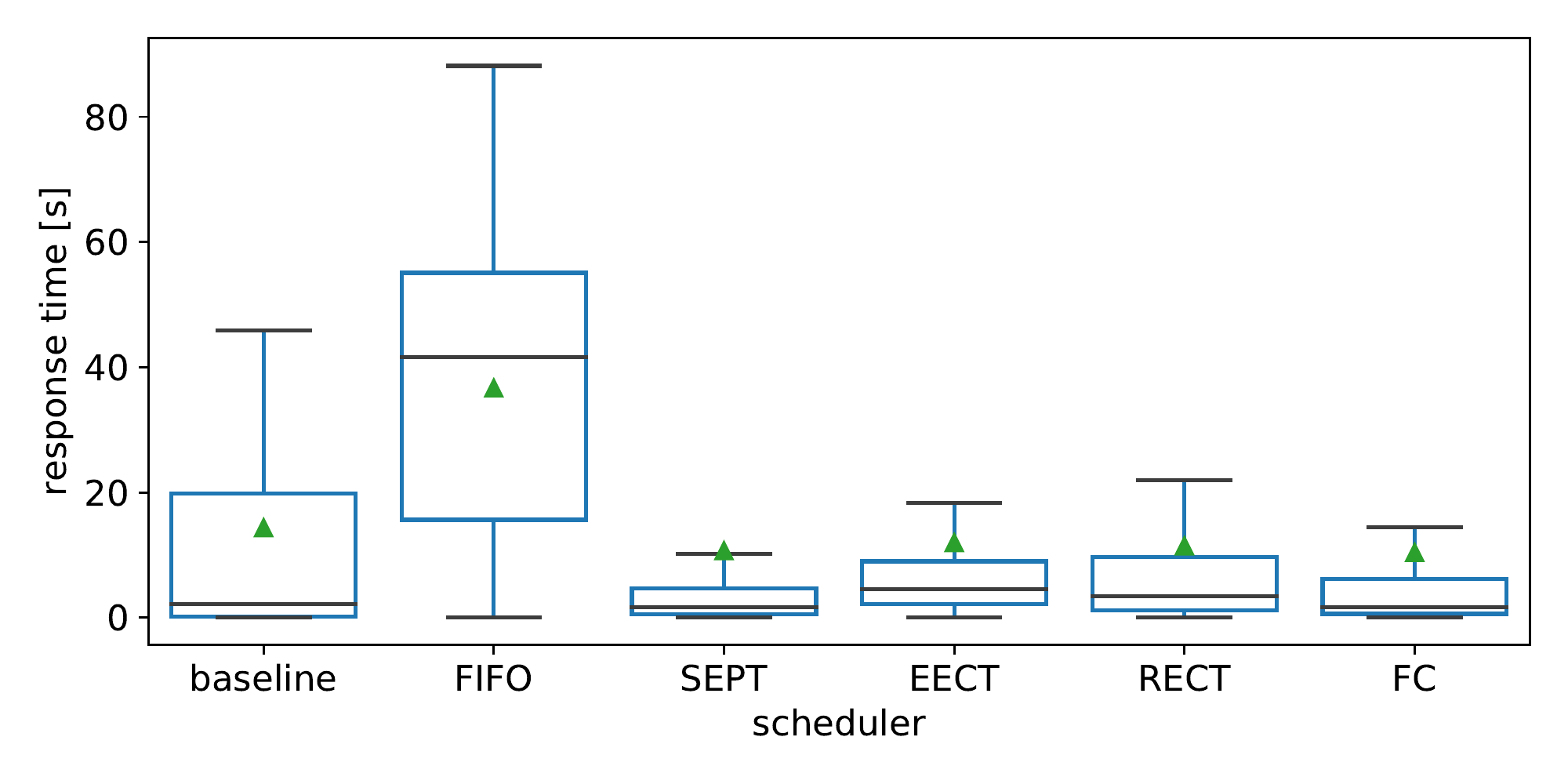}
    \\
    \caption{Response time for different scheduling policies. Here, we present results for 10 CPUs and intensity 30. On-premise infrastructure. Average response time presented as a green triangle.}
    \label{fig:response_time_10_30}
\end{figure*}

\begin{figure*}[h]
    \centering
    \chart{Experiment 1}{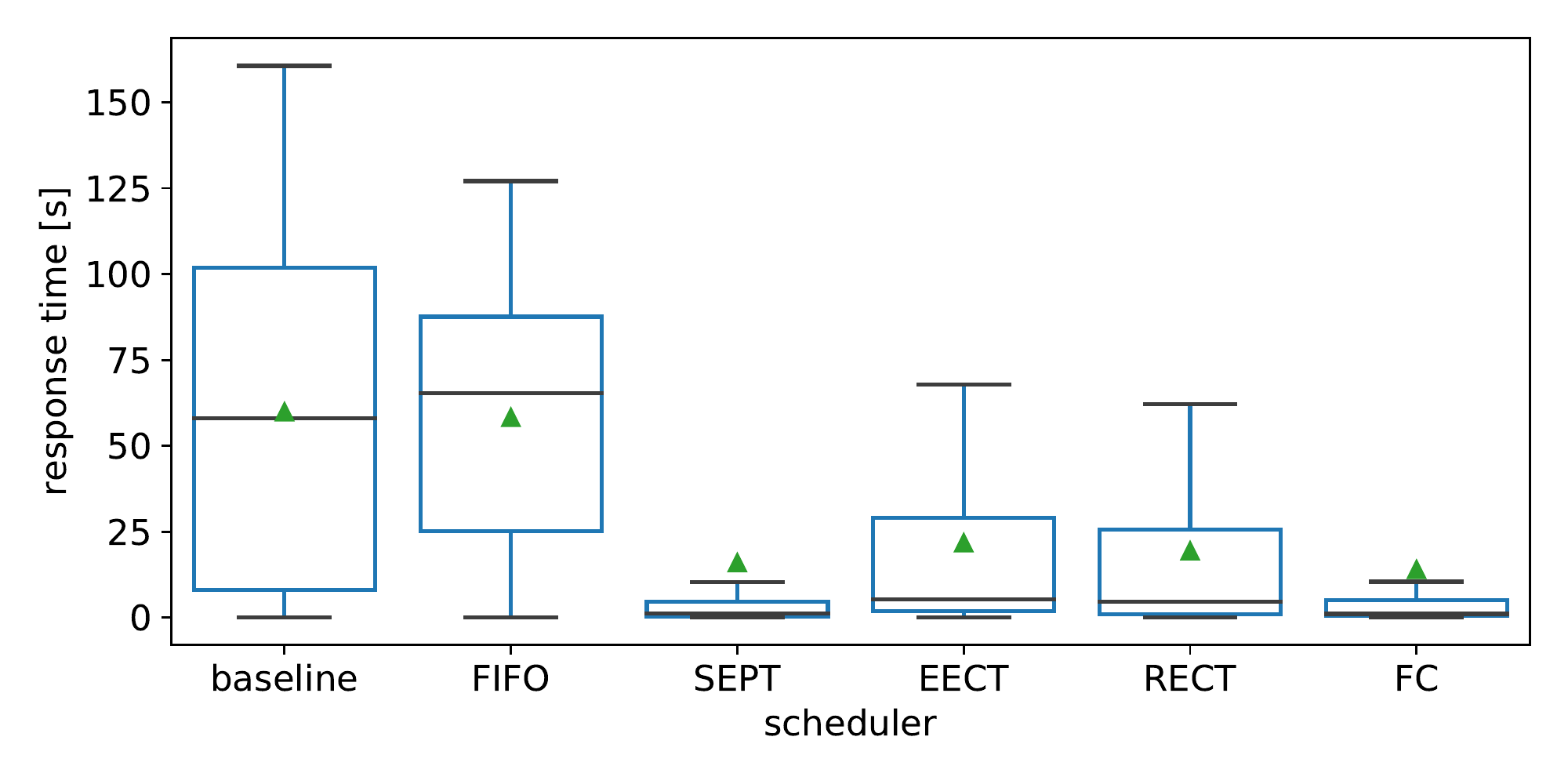}
    \chart{Experiment 2}{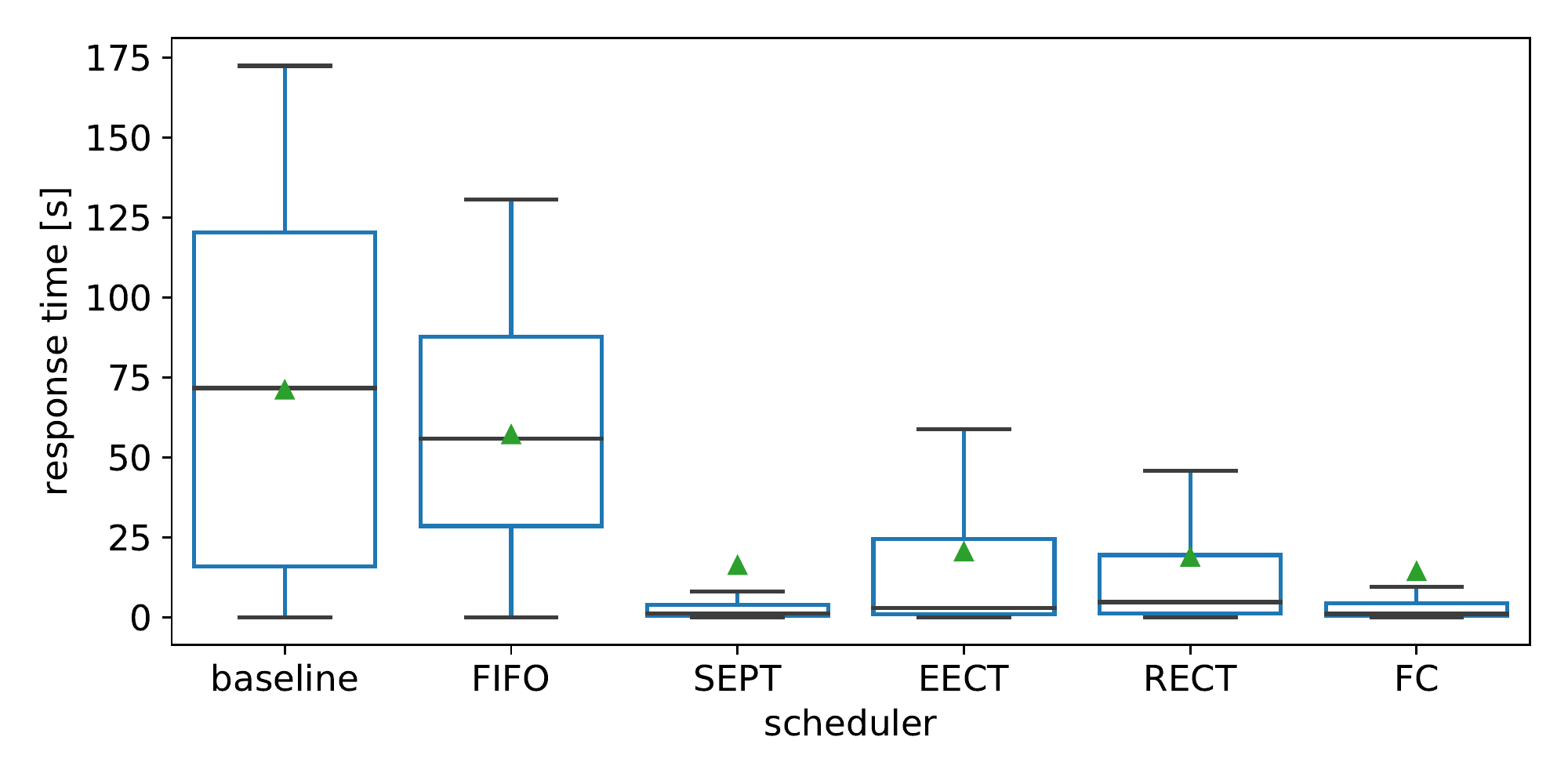}
    \chart{Experiment 3}{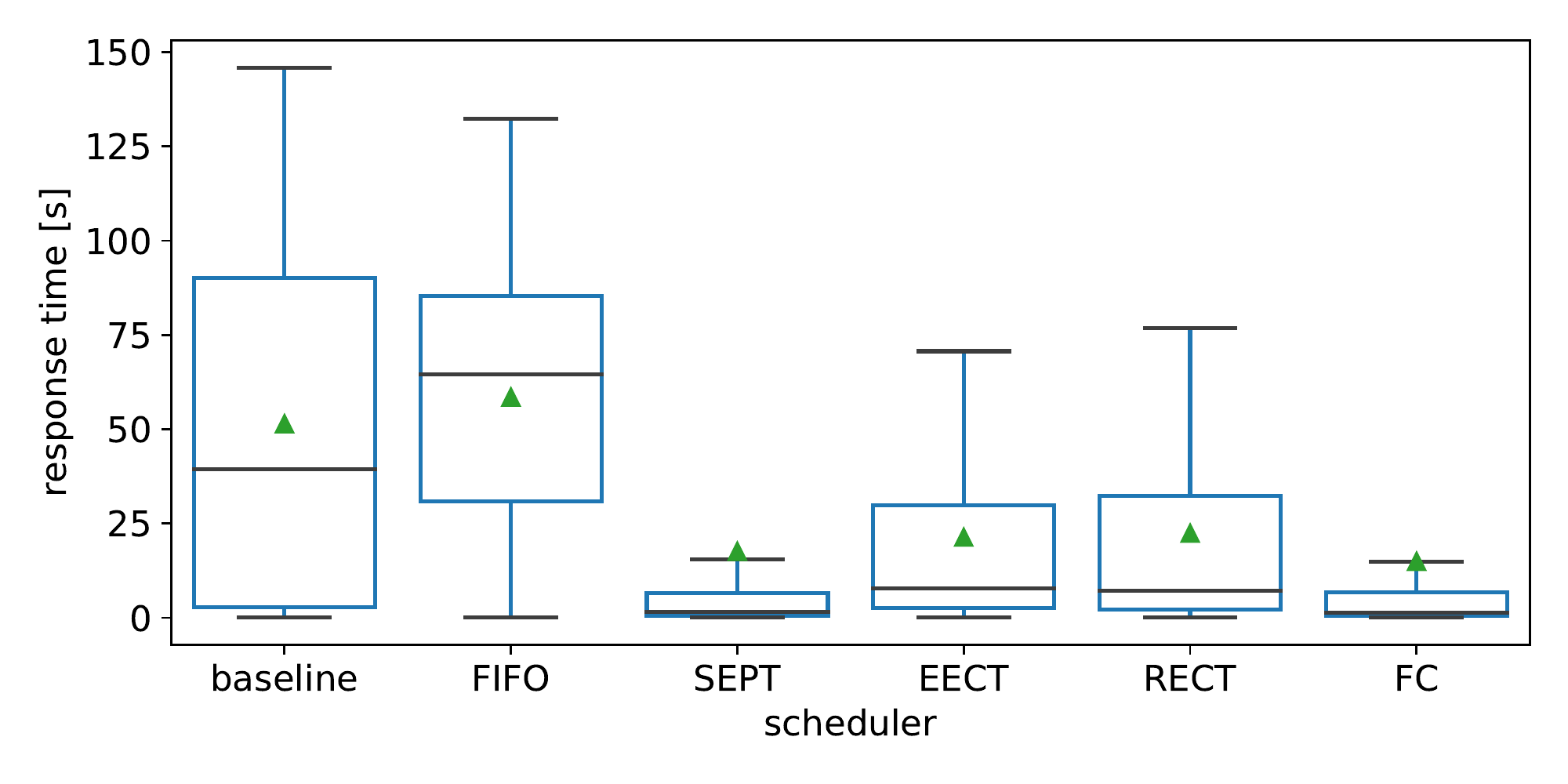}
    \\
    \chart{Experiment 4}{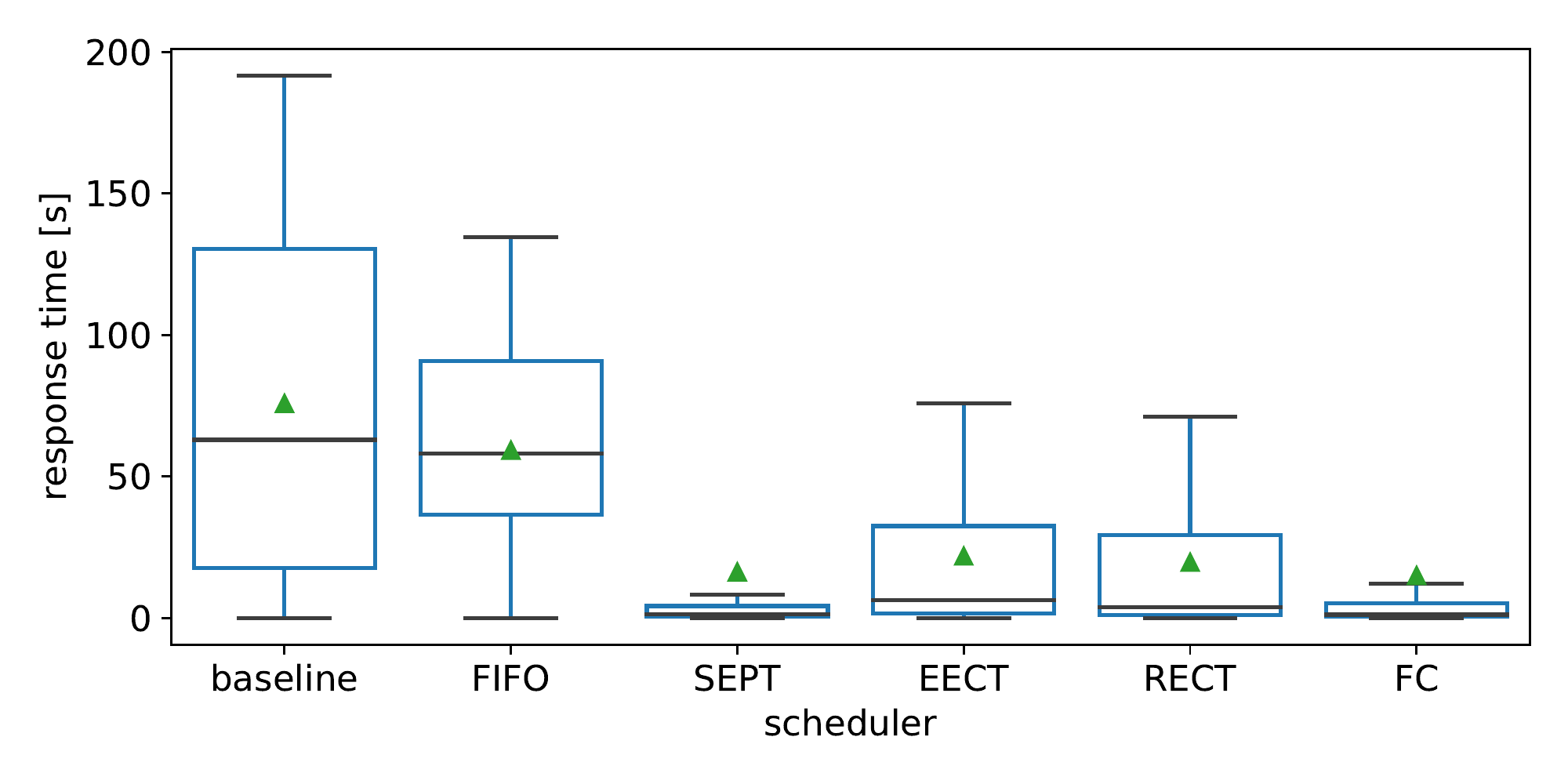}
    \chart{Experiment 5}{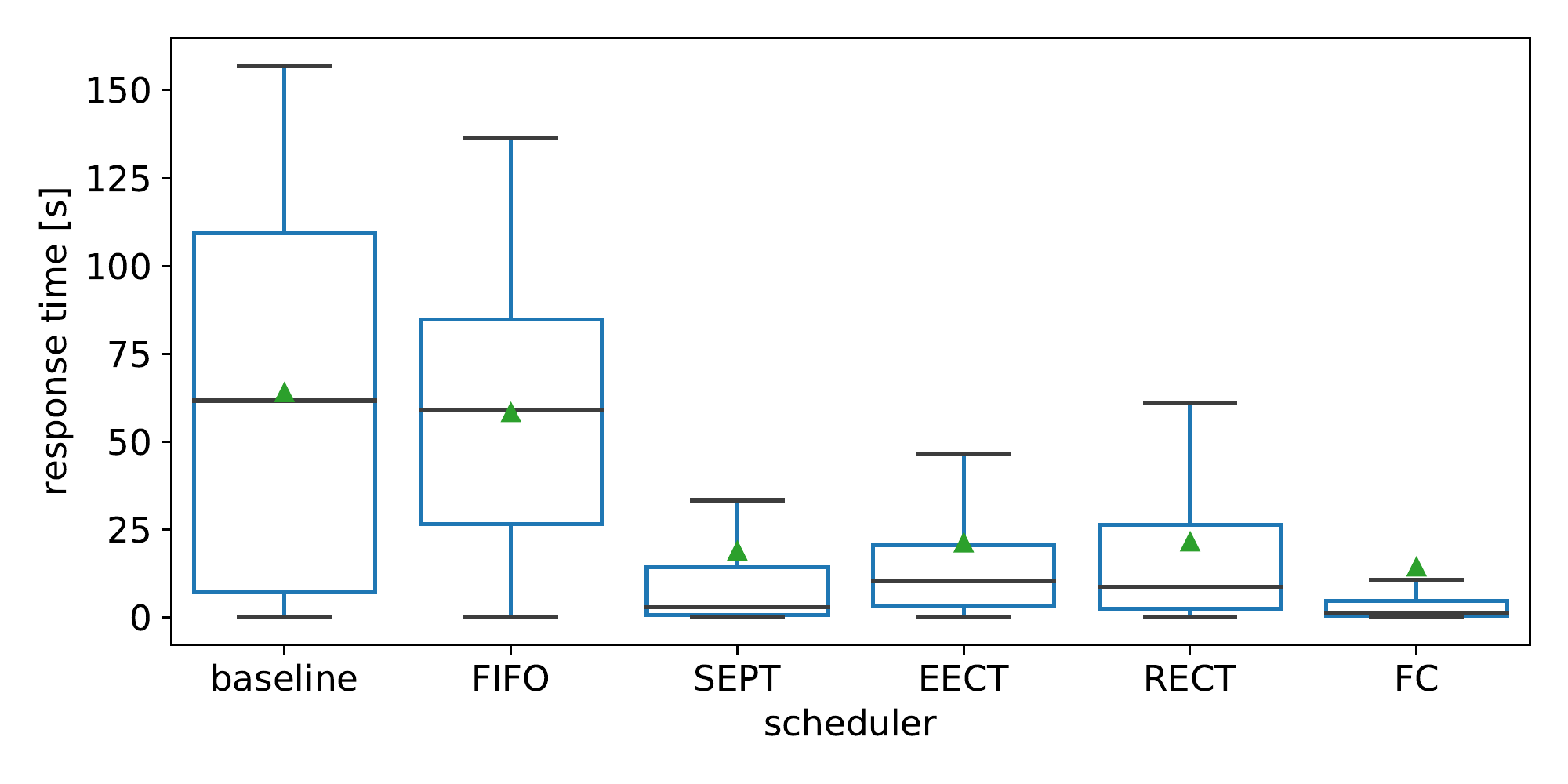}
    \\
    \caption{Response time for different scheduling policies. Here, we present results for 10 CPUs and intensity 40. On-premise infrastructure. Average response time presented as a green triangle.}
    \label{fig:response_time_10_40}
\end{figure*}

\begin{figure*}[h]
    \centering
    \chart{Experiment 1}{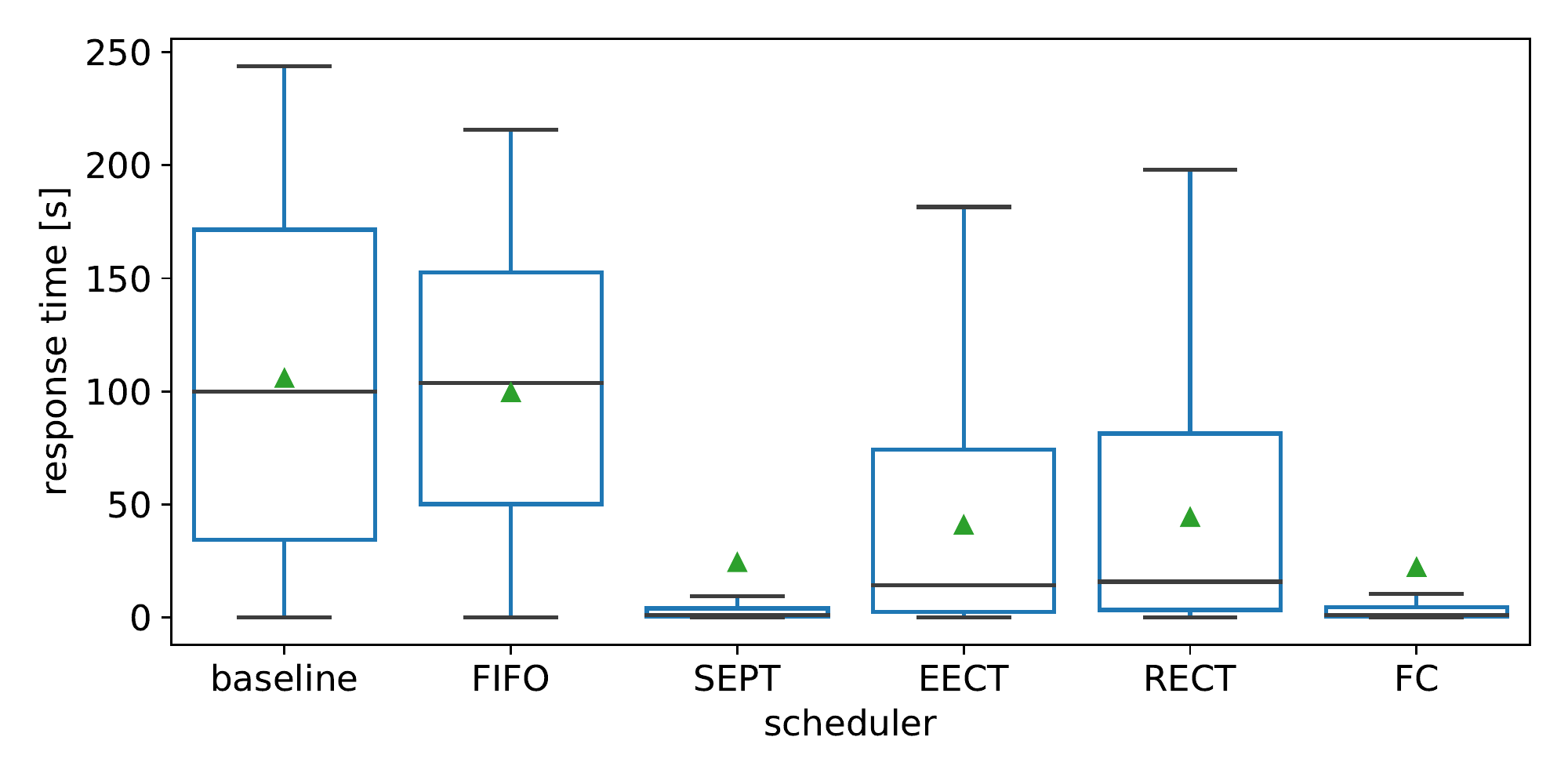}
    \chart{Experiment 2}{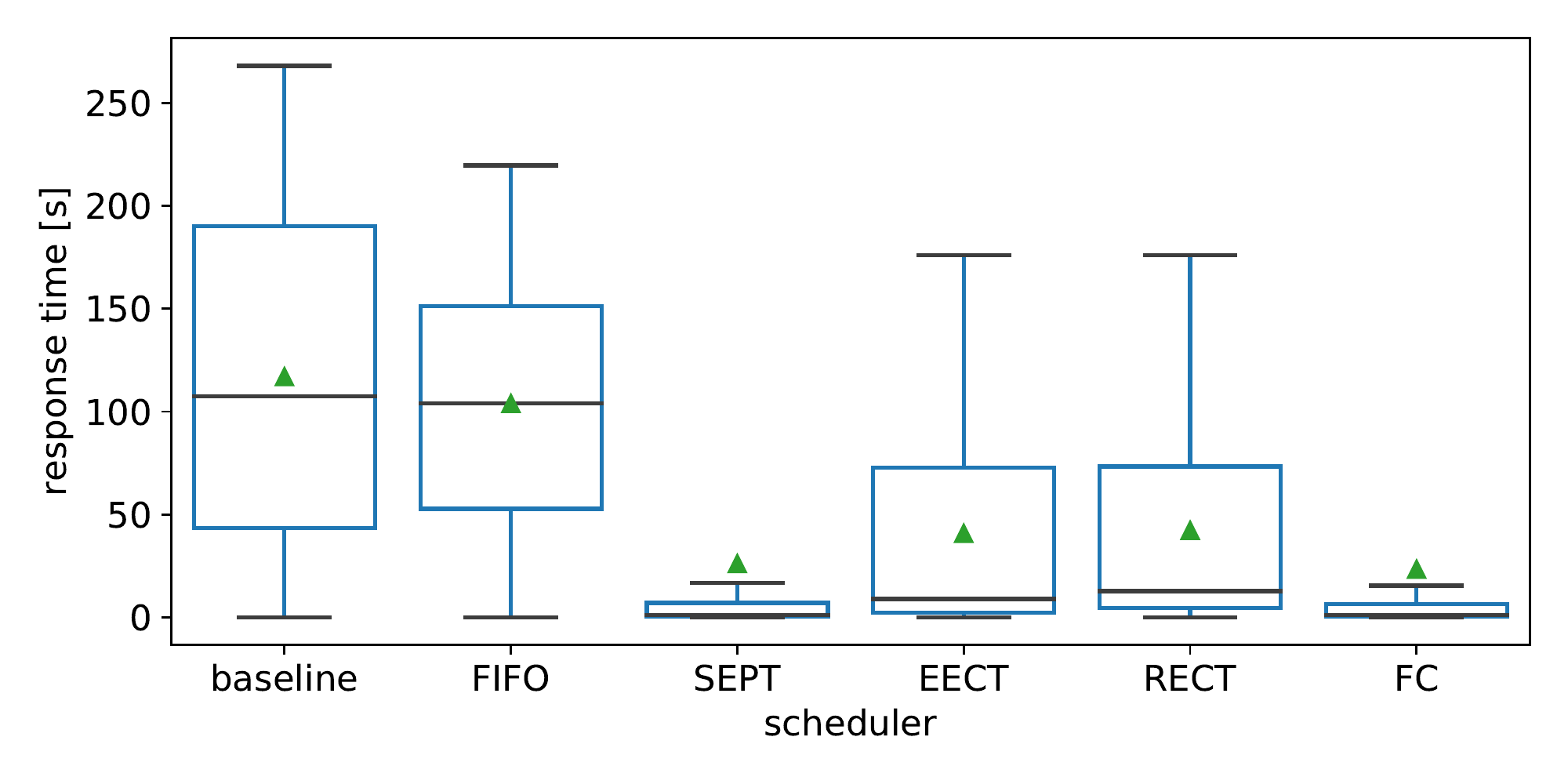}
    \chart{Experiment 3}{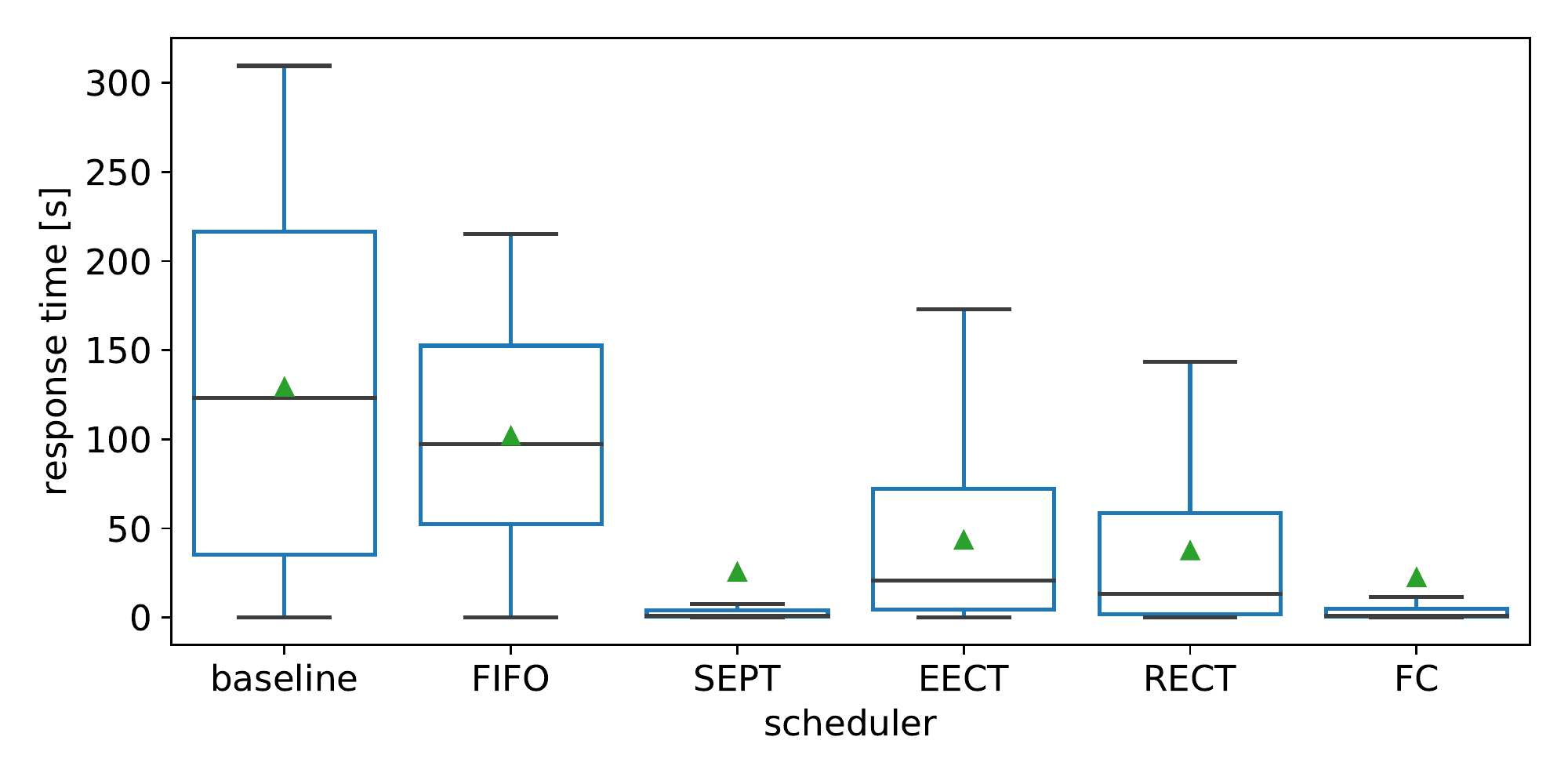}
    \\
    \chart{Experiment 4}{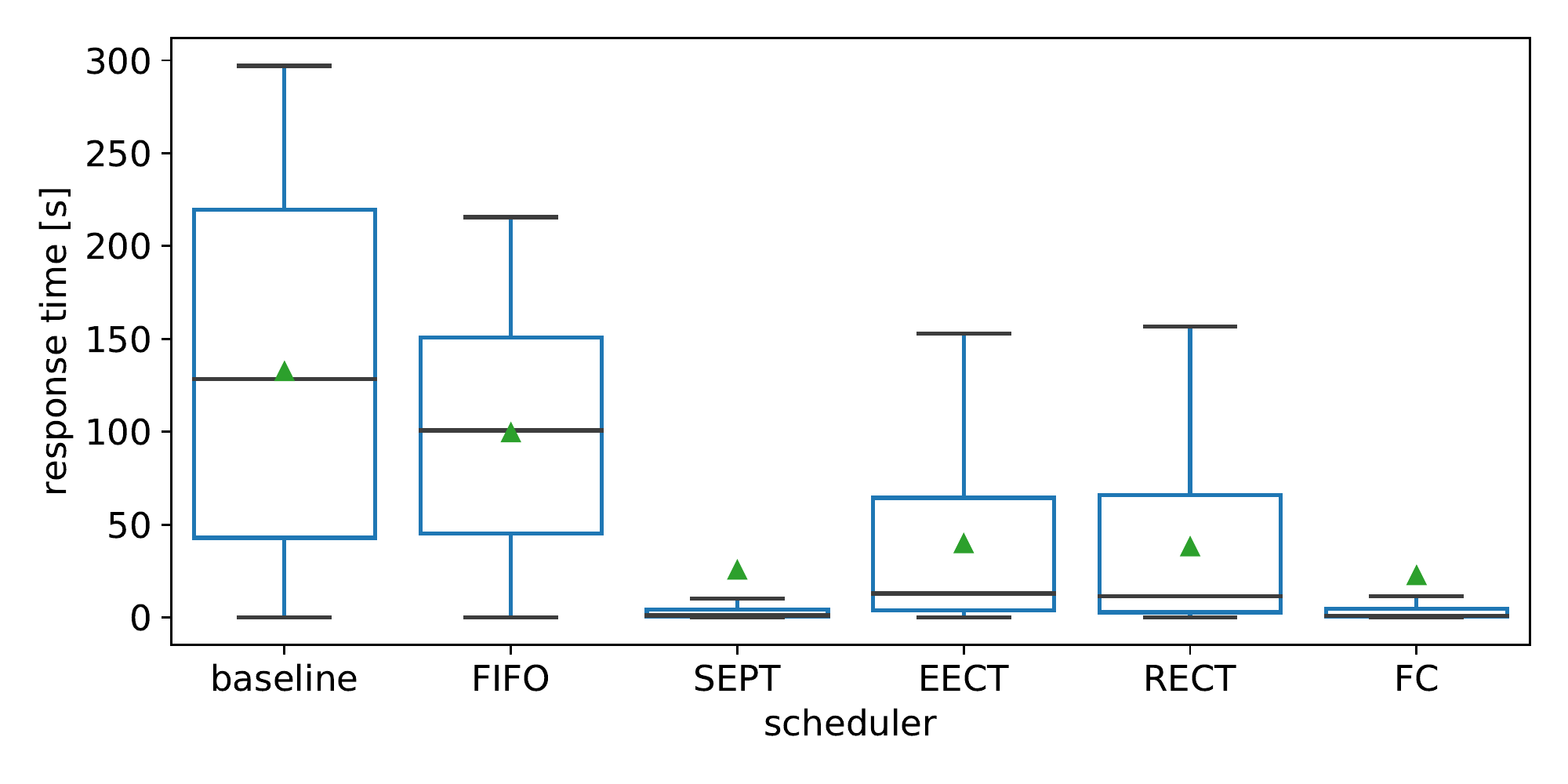}
    \chart{Experiment 5}{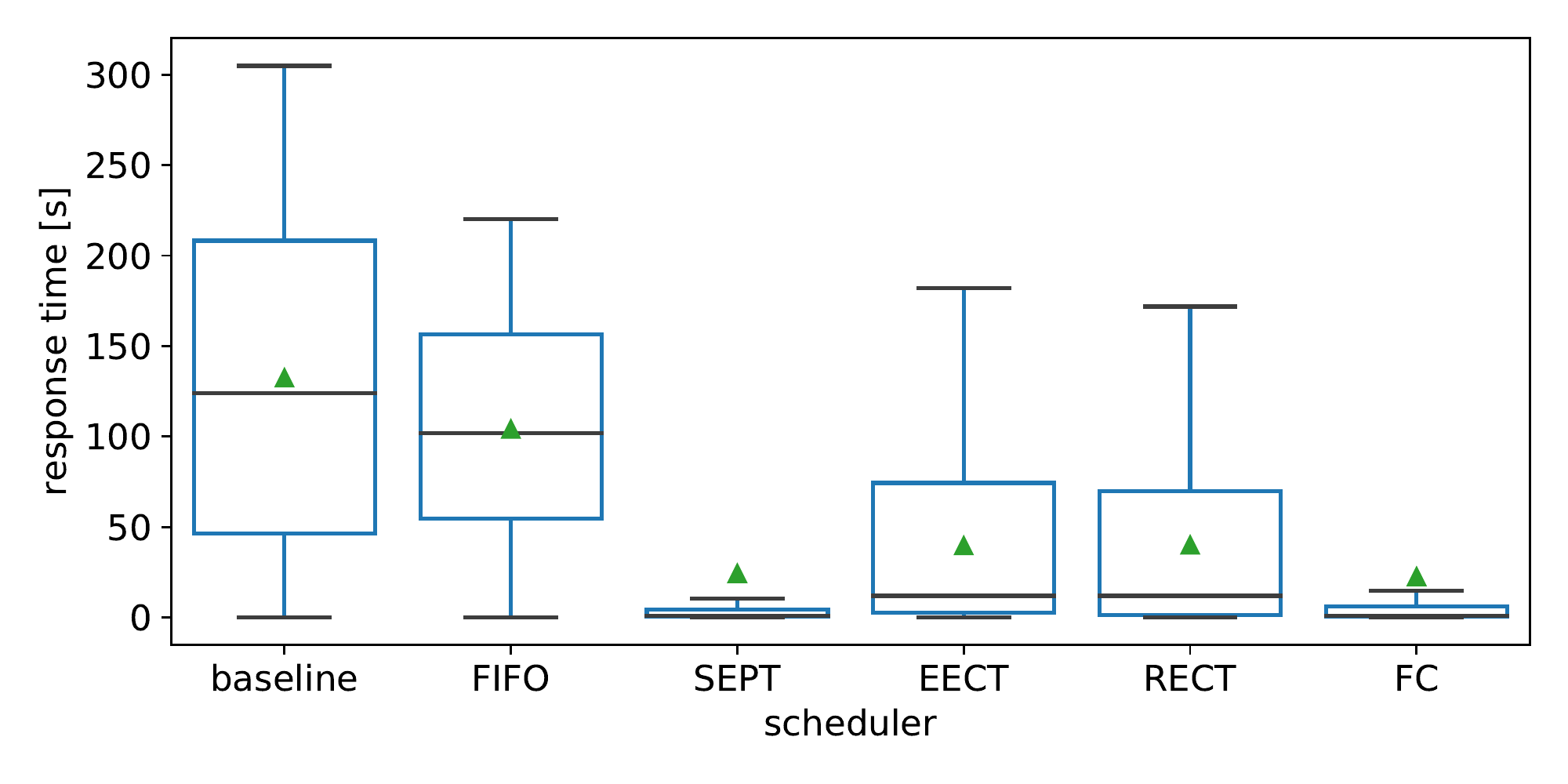}
    \\
    \caption{Response time for different scheduling policies. Here, we present results for 10 CPUs and intensity 60. On-premise infrastructure. Average response time presented as a green triangle.}
    \label{fig:response_time_10_60}
\end{figure*}

\begin{figure*}[h]
    \centering
    \chart{Experiment 1}{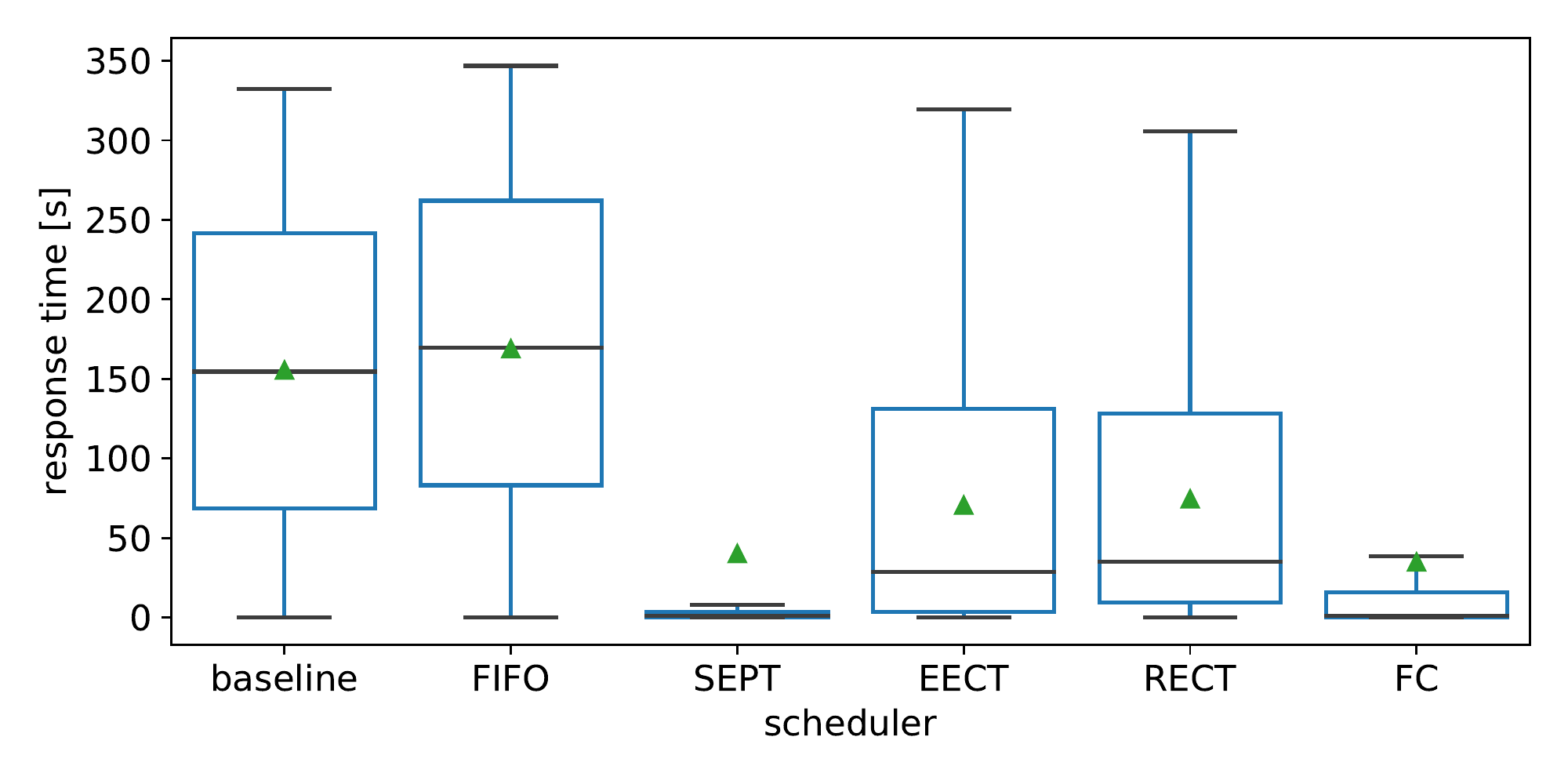}
    \chart{Experiment 2}{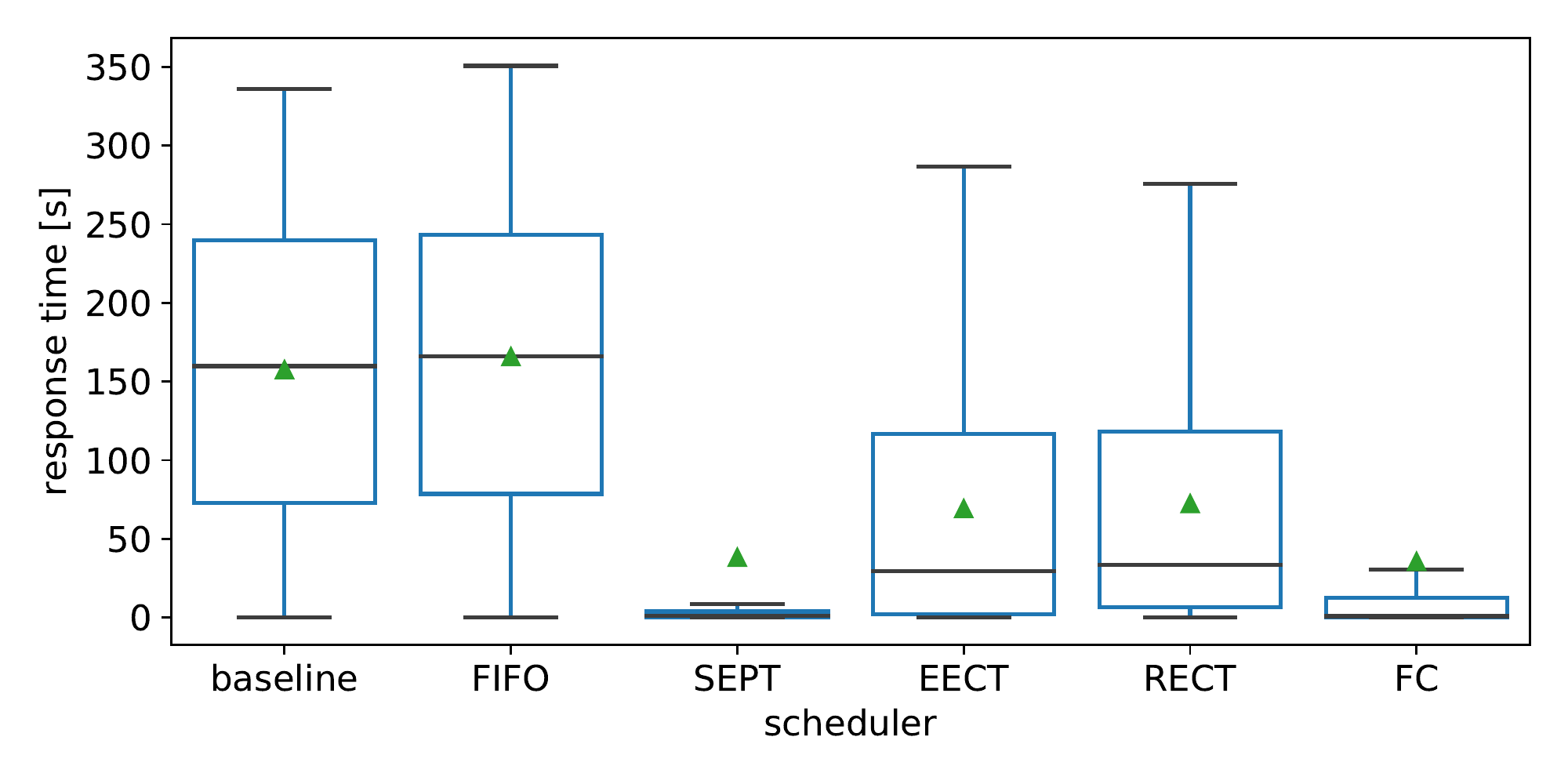}
    \chart{Experiment 3}{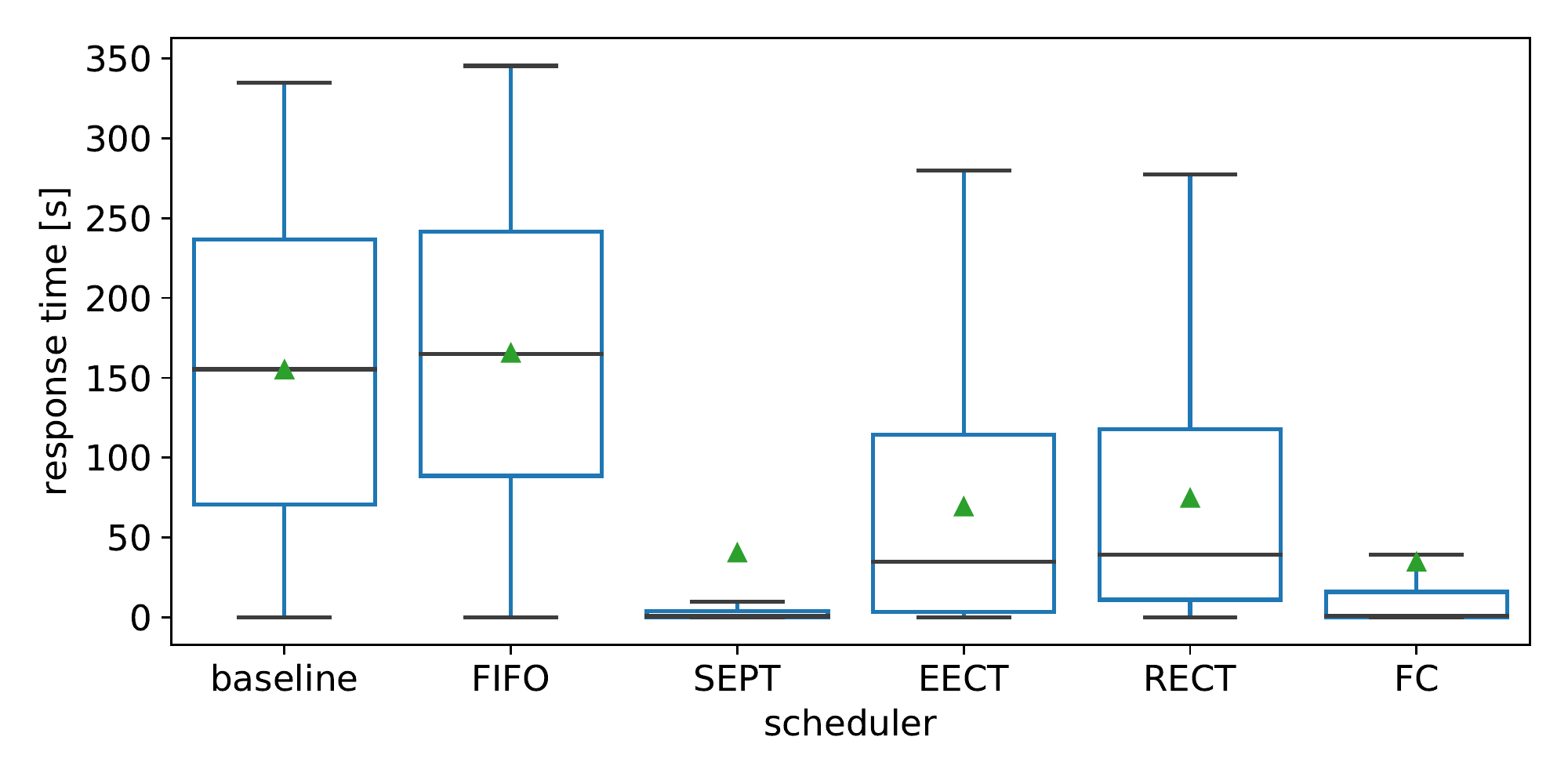}
    \\
    \chart{Experiment 4}{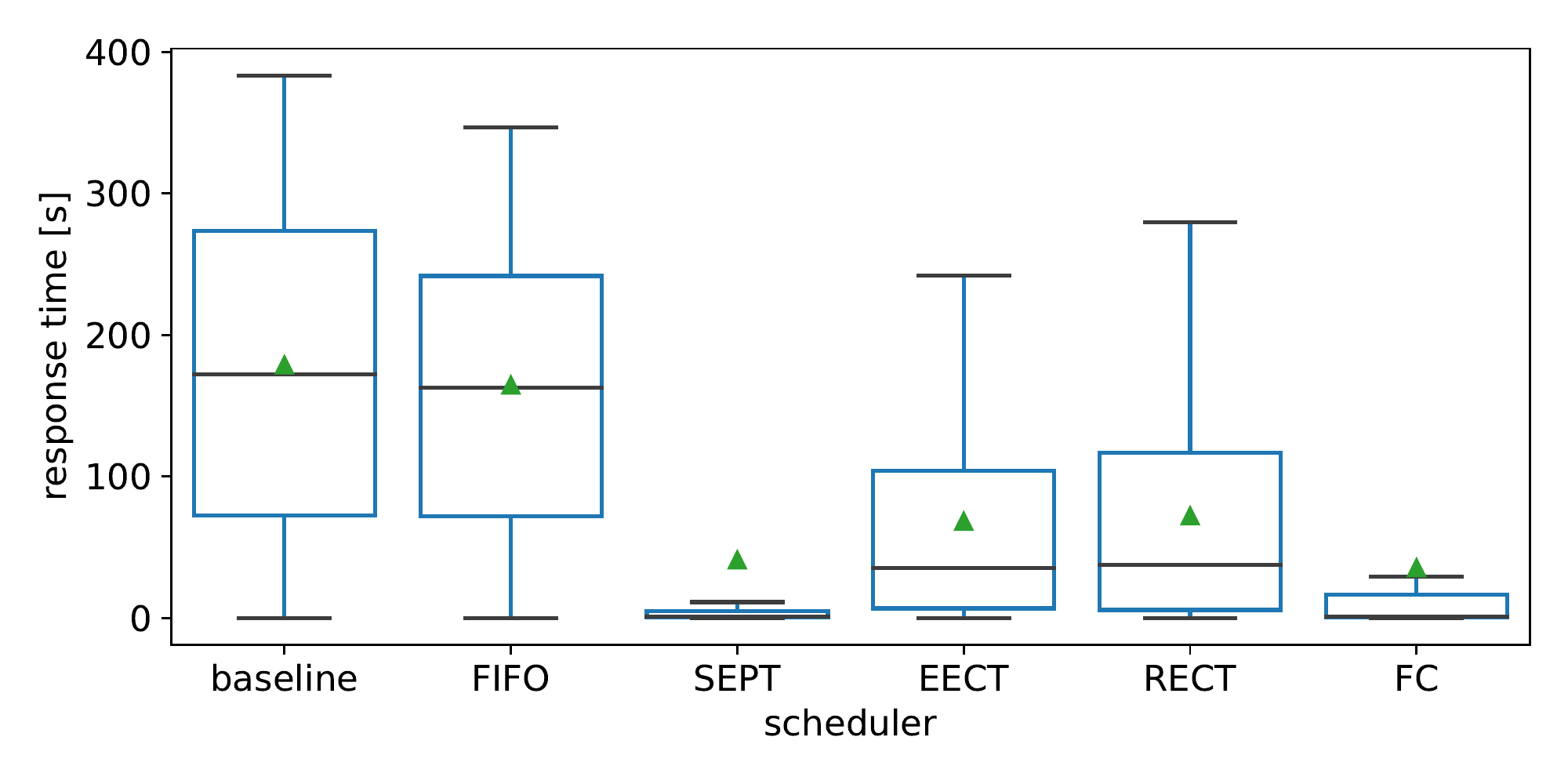}
    \chart{Experiment 5}{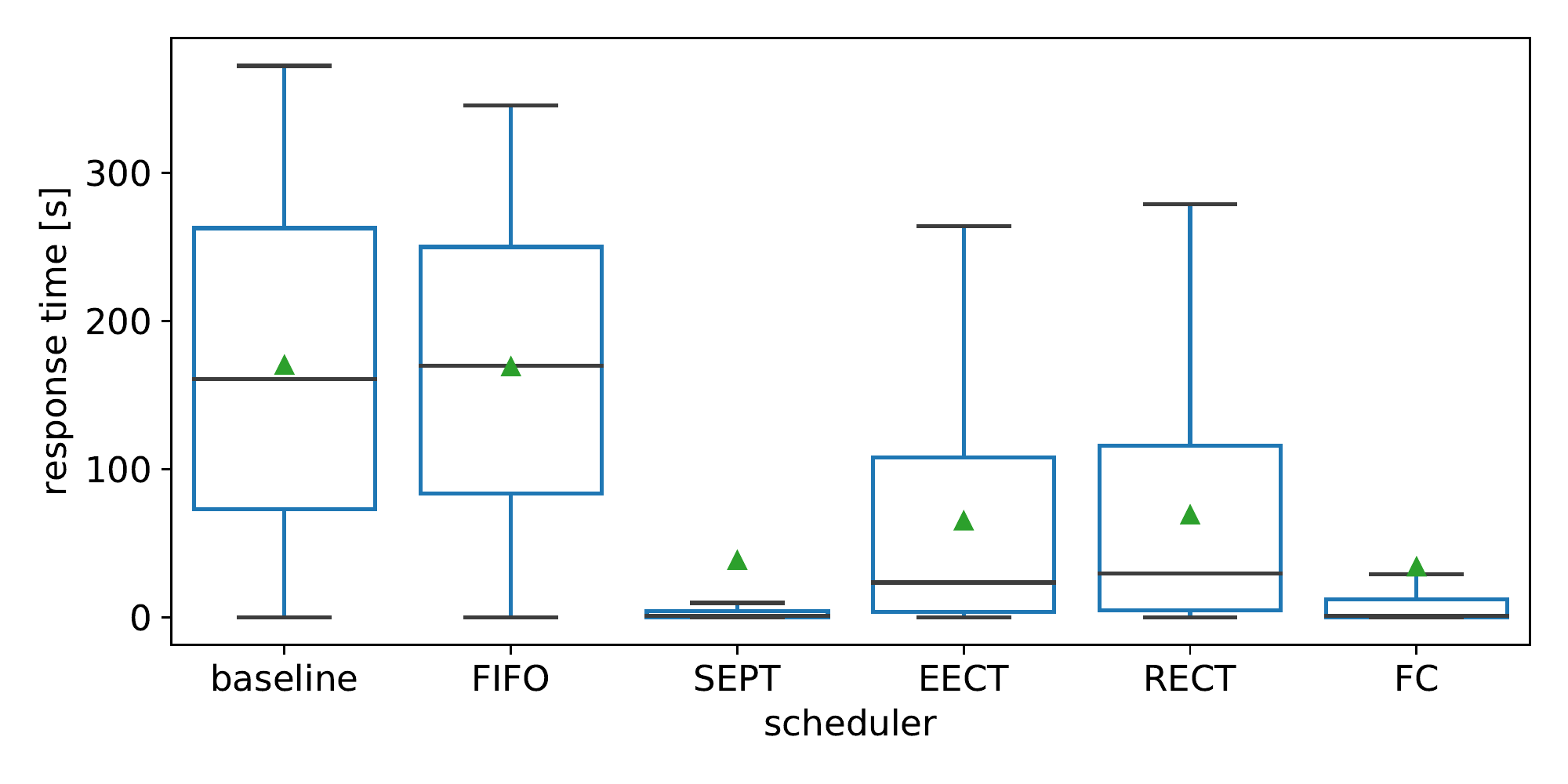}
    \\
    \caption{Response time for different scheduling policies. Here, we present results for 10 CPUs and intensity 90. On-premise infrastructure. Average response time presented as a green triangle.}
    \label{fig:response_time_10_90}
\end{figure*}

\begin{figure*}[h]
    \centering
    \chart{Experiment 1}{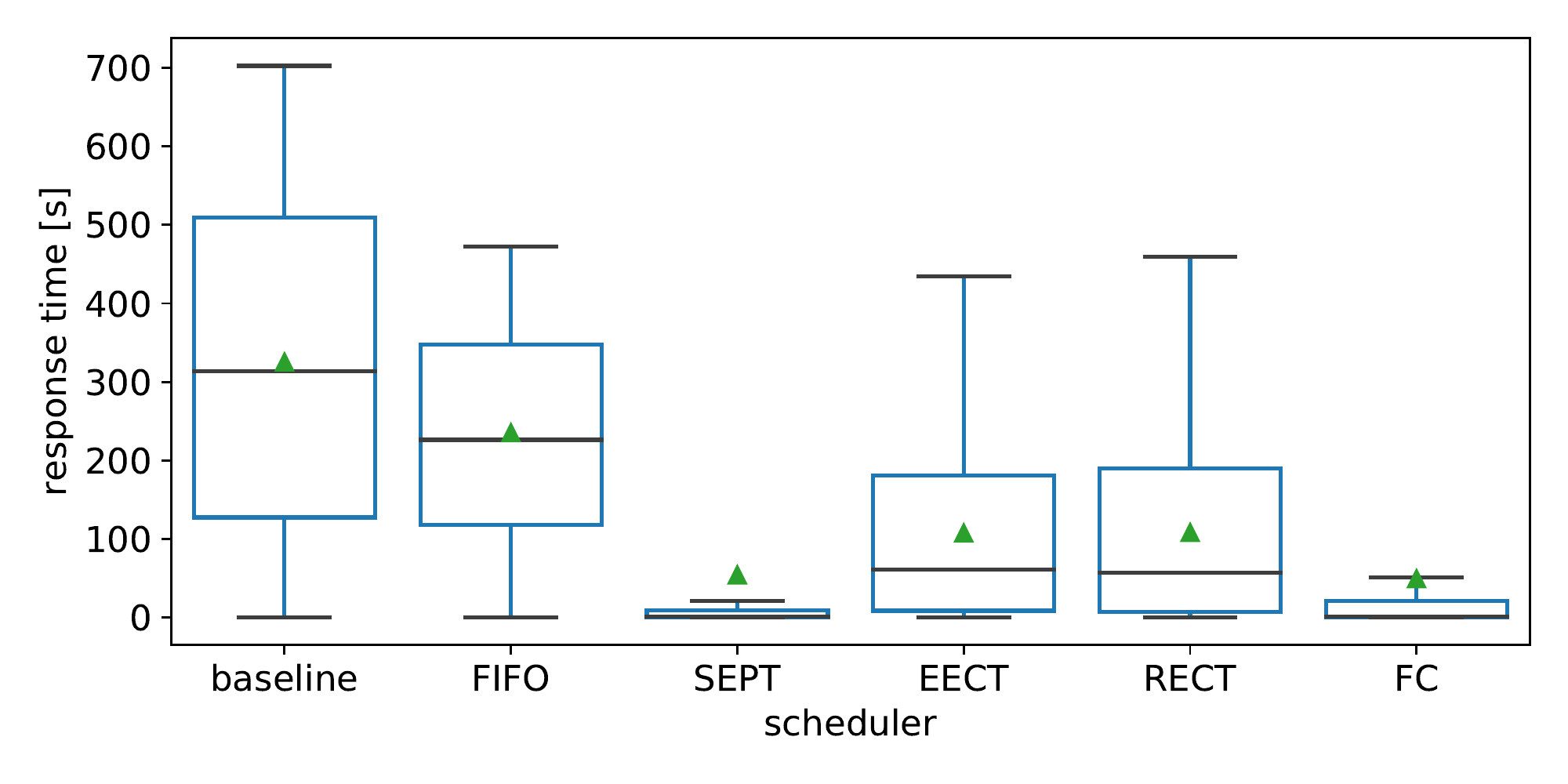}
    \chart{Experiment 2}{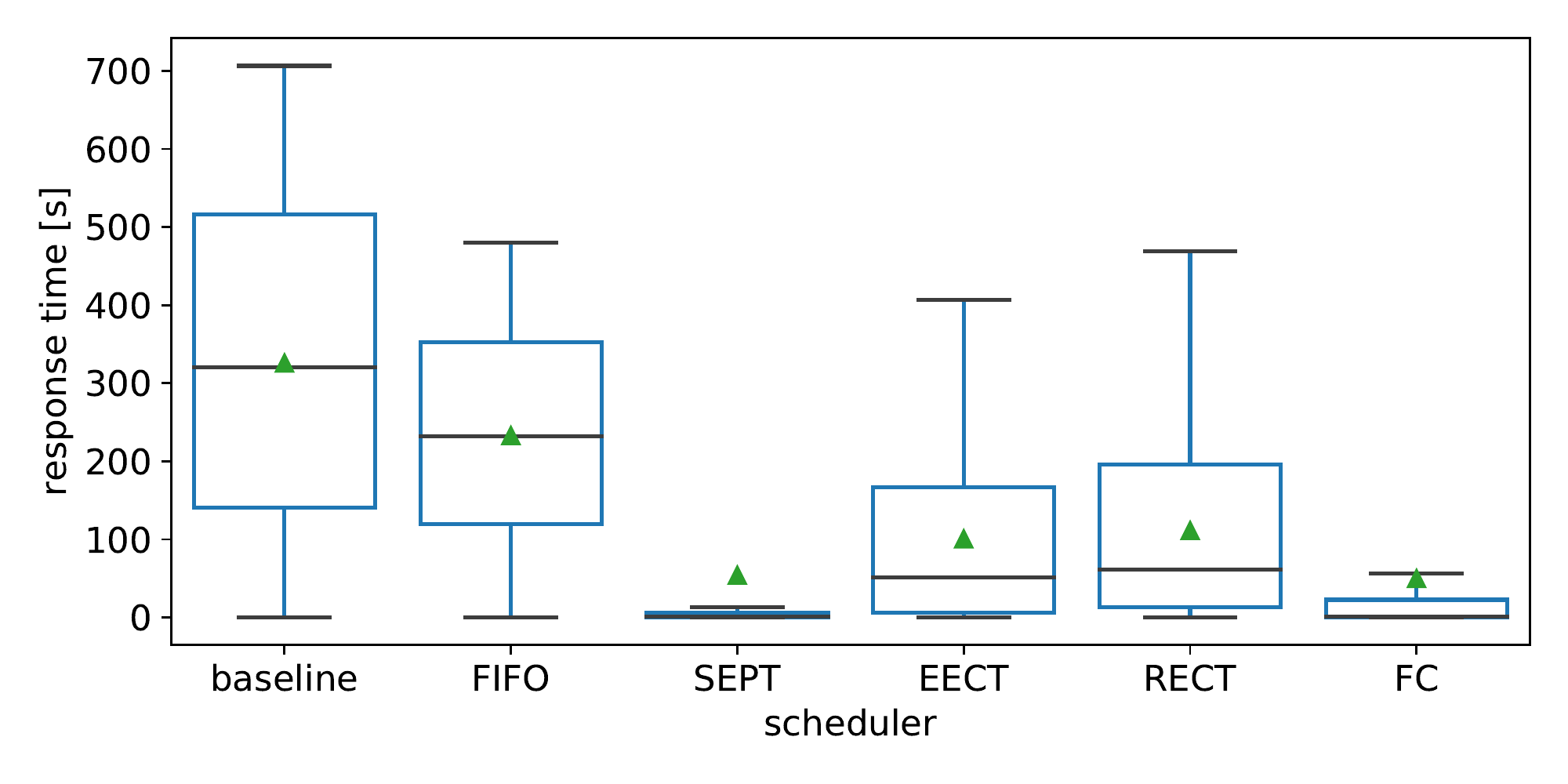}
    \chart{Experiment 3}{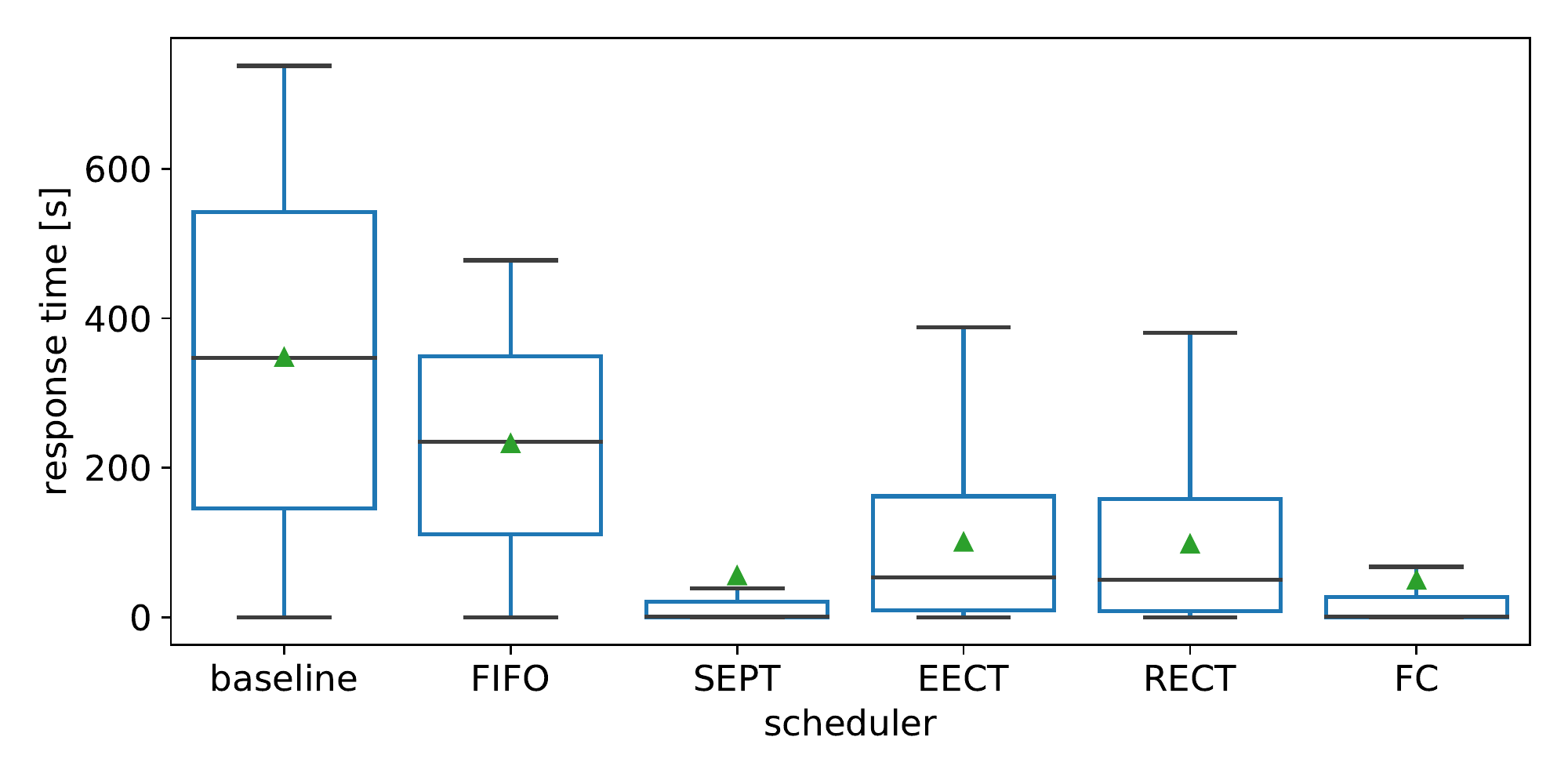}
    \\
    \chart{Experiment 4}{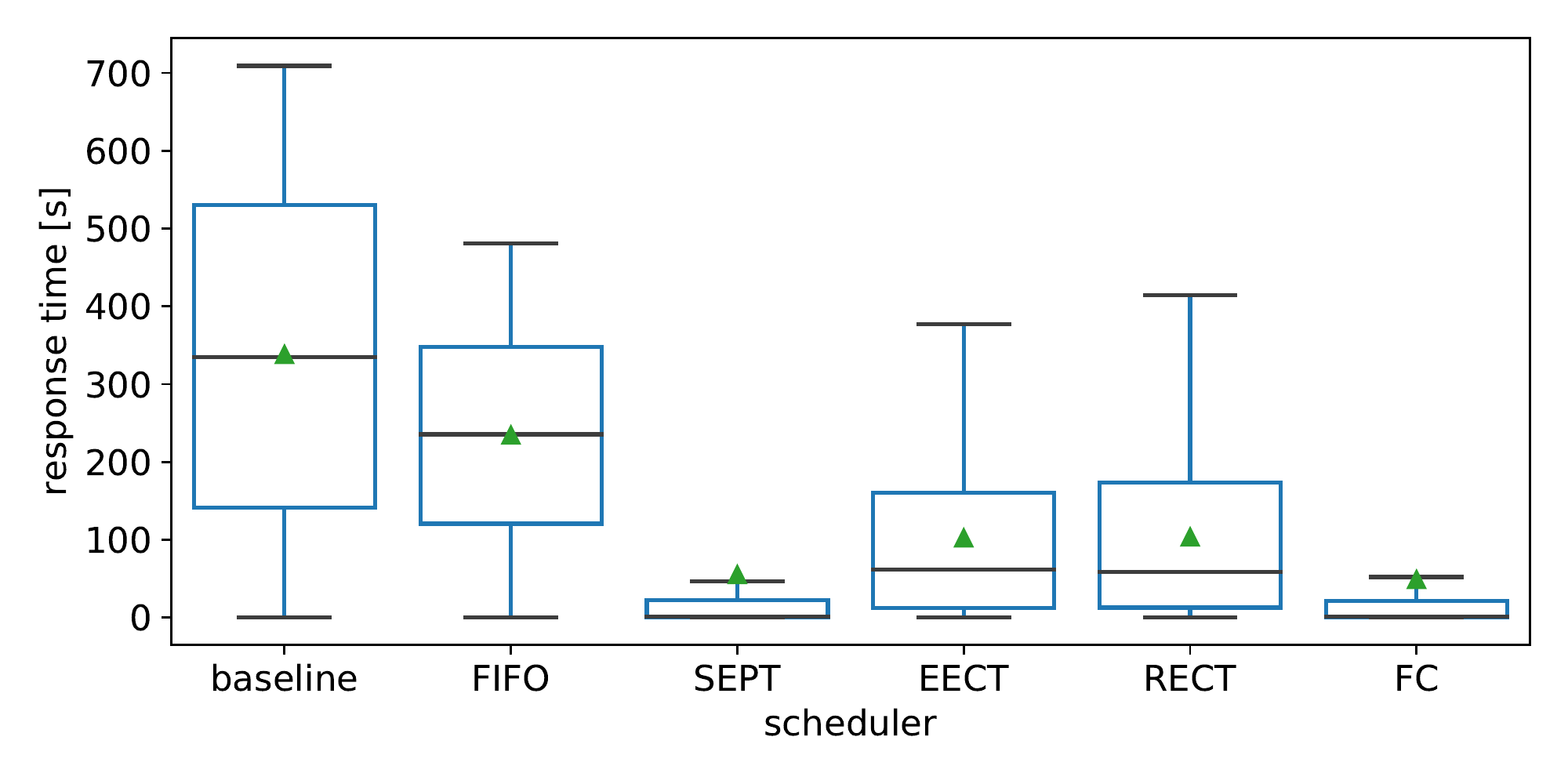}
    \chart{Experiment 5}{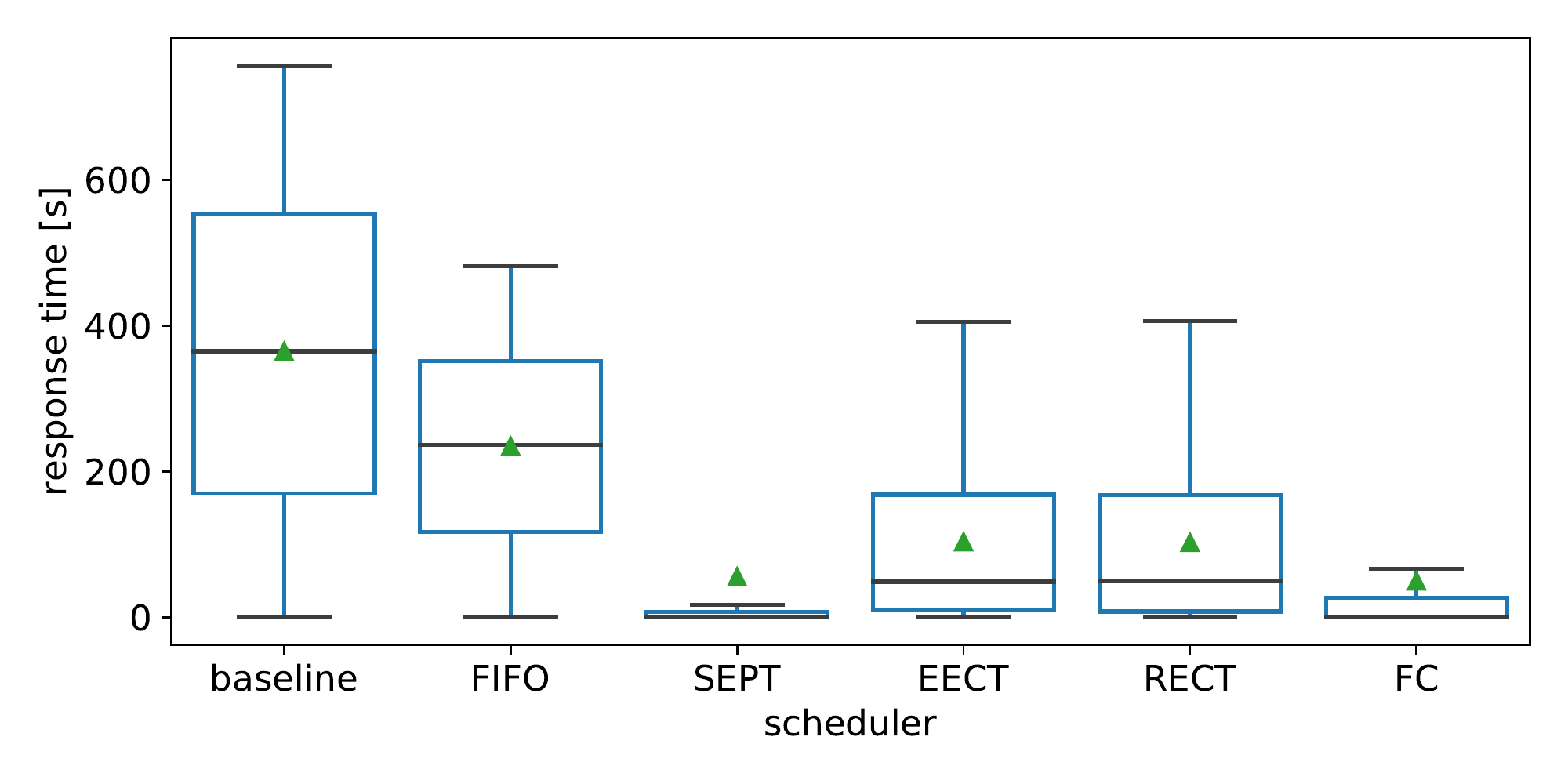}
    \\
    \caption{Response time for different scheduling policies. Here, we present results for 10 CPUs and intensity 120. On-premise infrastructure. Average response time presented as a green triangle.}
    \label{fig:response_time_10_120}
\end{figure*}

\begin{figure*}[h]
    \centering
    \chart{Experiment 1}{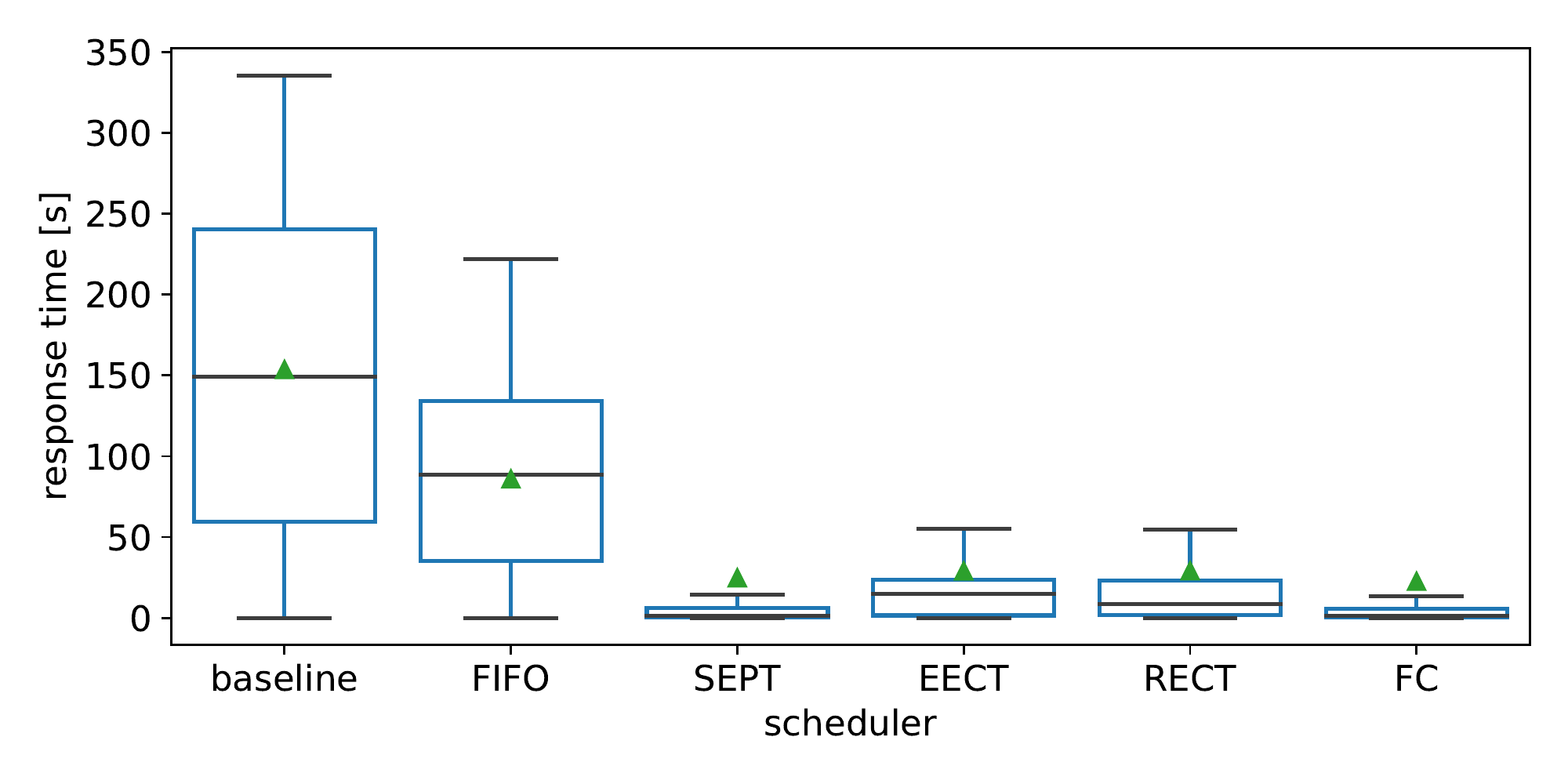}
    \chart{Experiment 2}{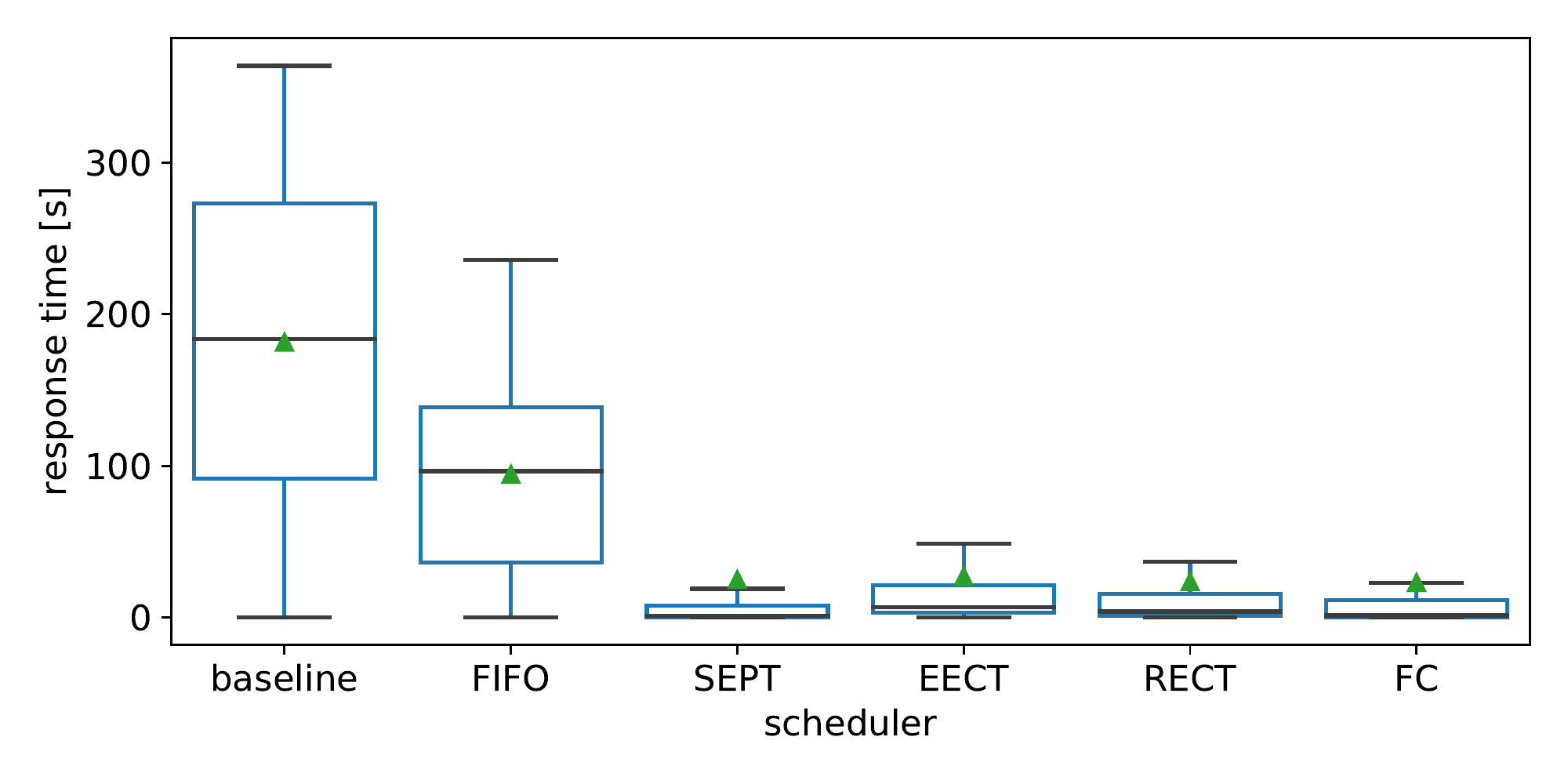}
    \chart{Experiment 3}{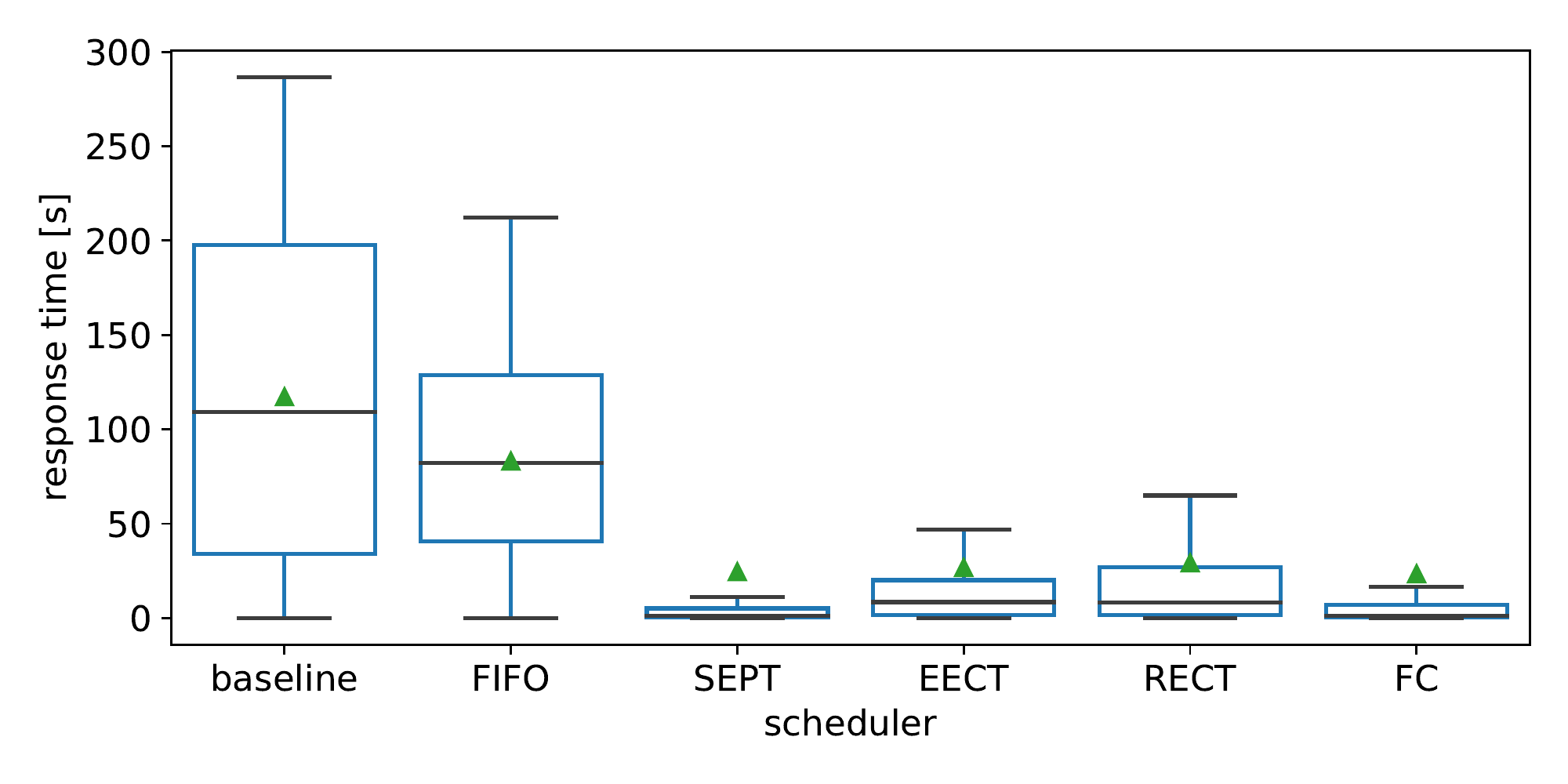}
    \\
    \chart{Experiment 4}{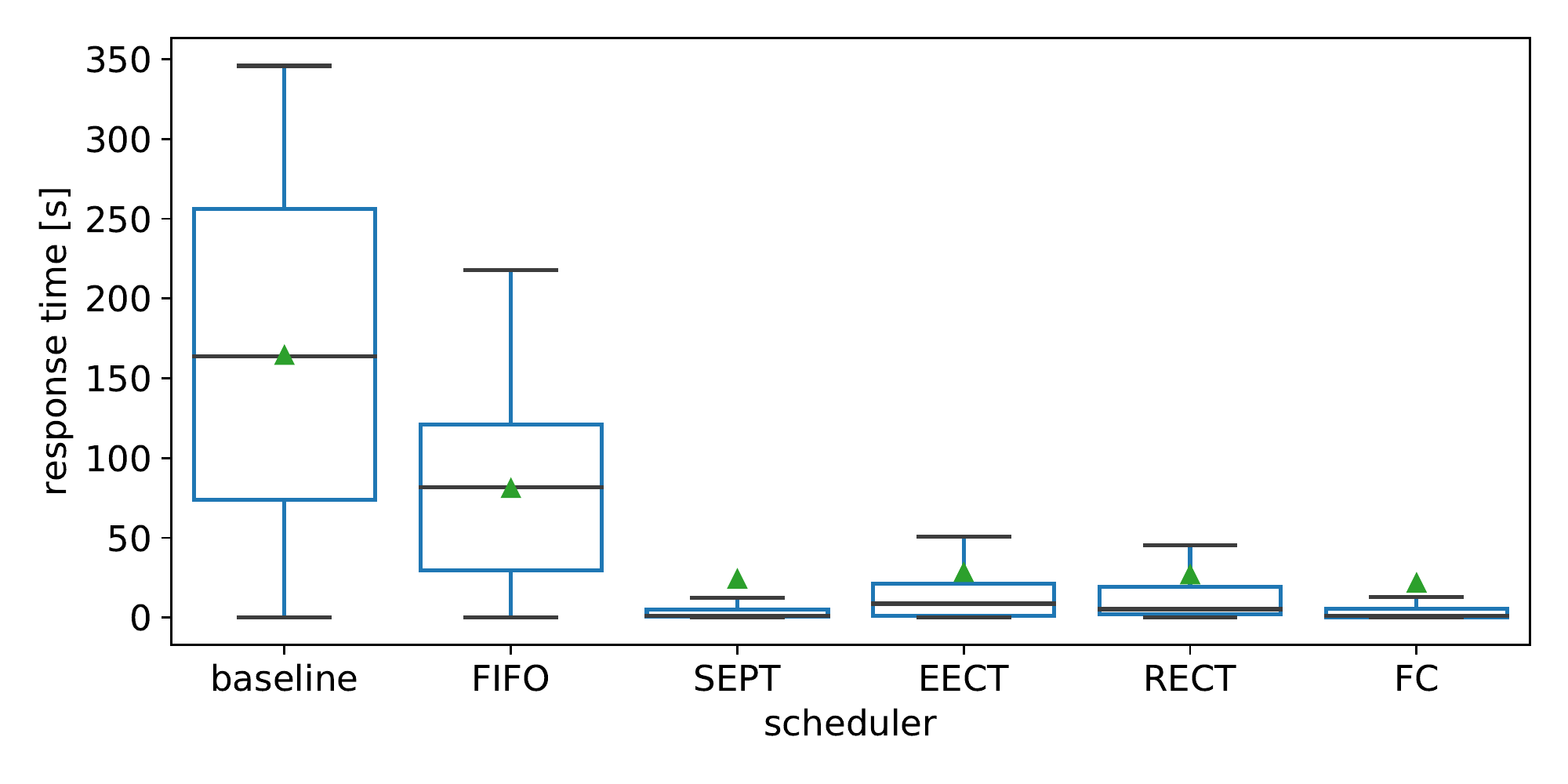}
    \chart{Experiment 5}{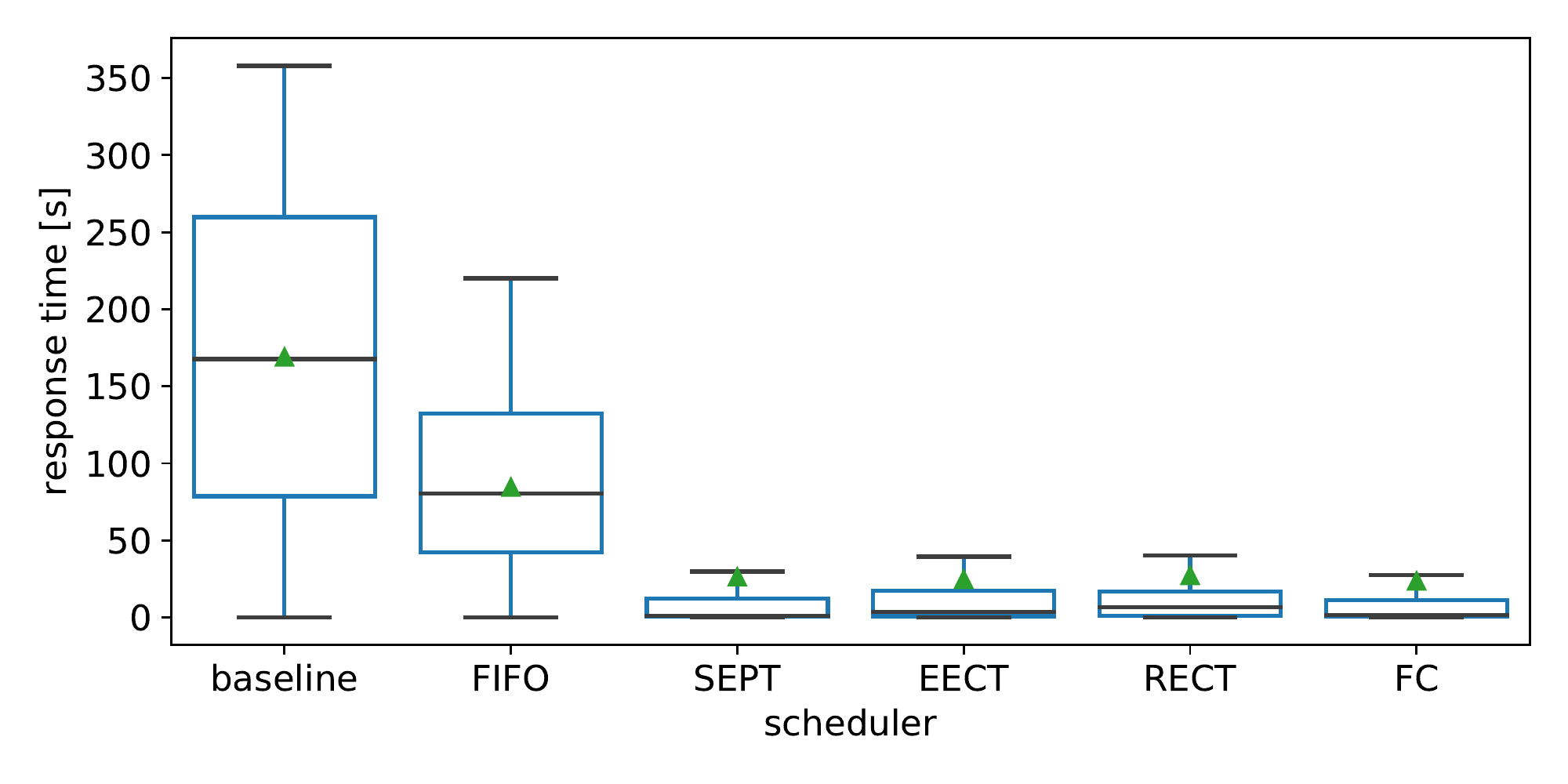}
    \\
    \caption{Response time for different scheduling policies. Here, we present results for 20 CPUs and intensity 30. On-premise infrastructure. Average response time presented as a green triangle.}
    \label{fig:response_time_20_30}
\end{figure*}

\begin{figure*}[h]
    \centering
    \chart{Experiment 1}{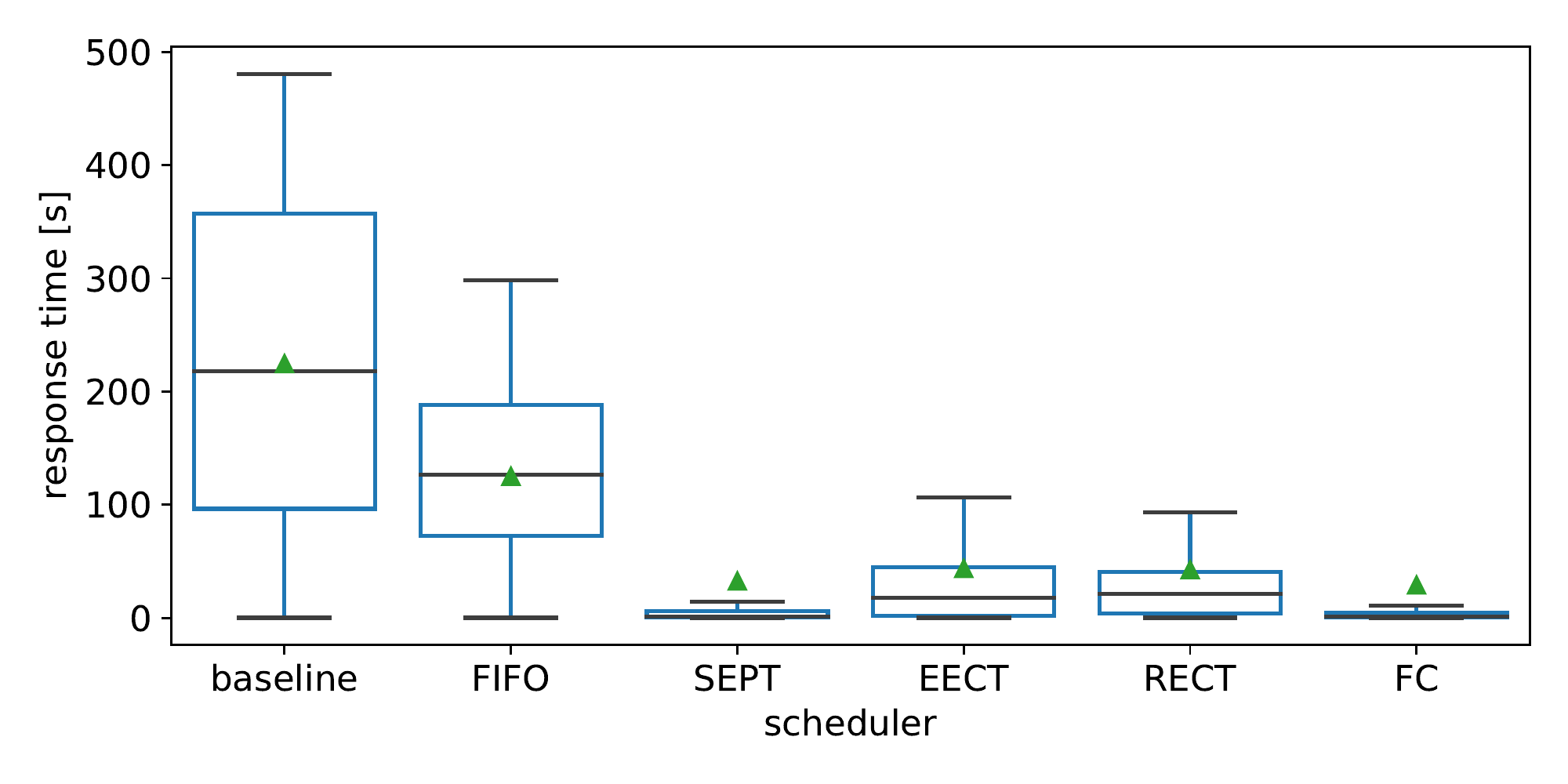}
    \chart{Experiment 2}{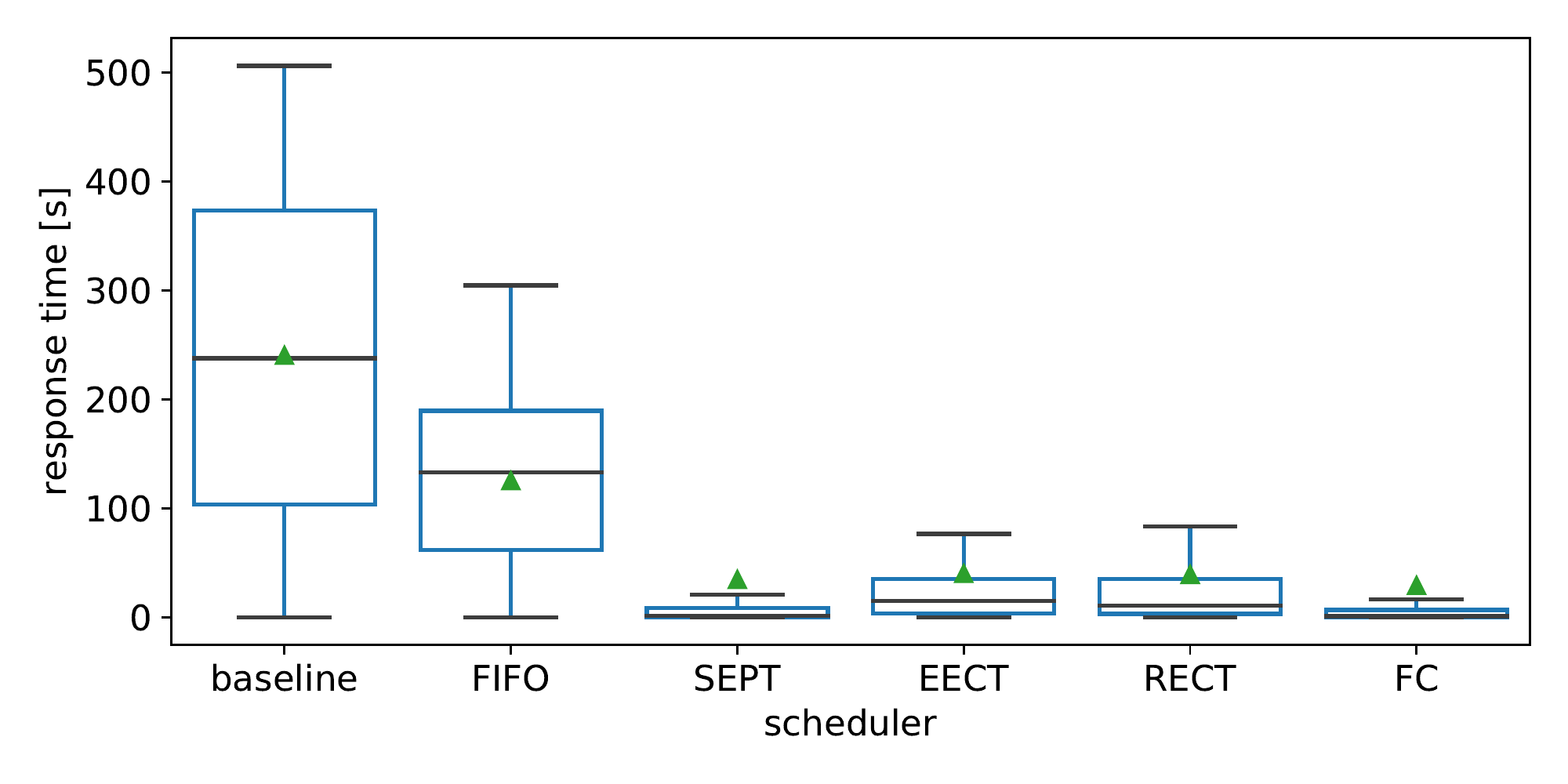}
    \chart{Experiment 3}{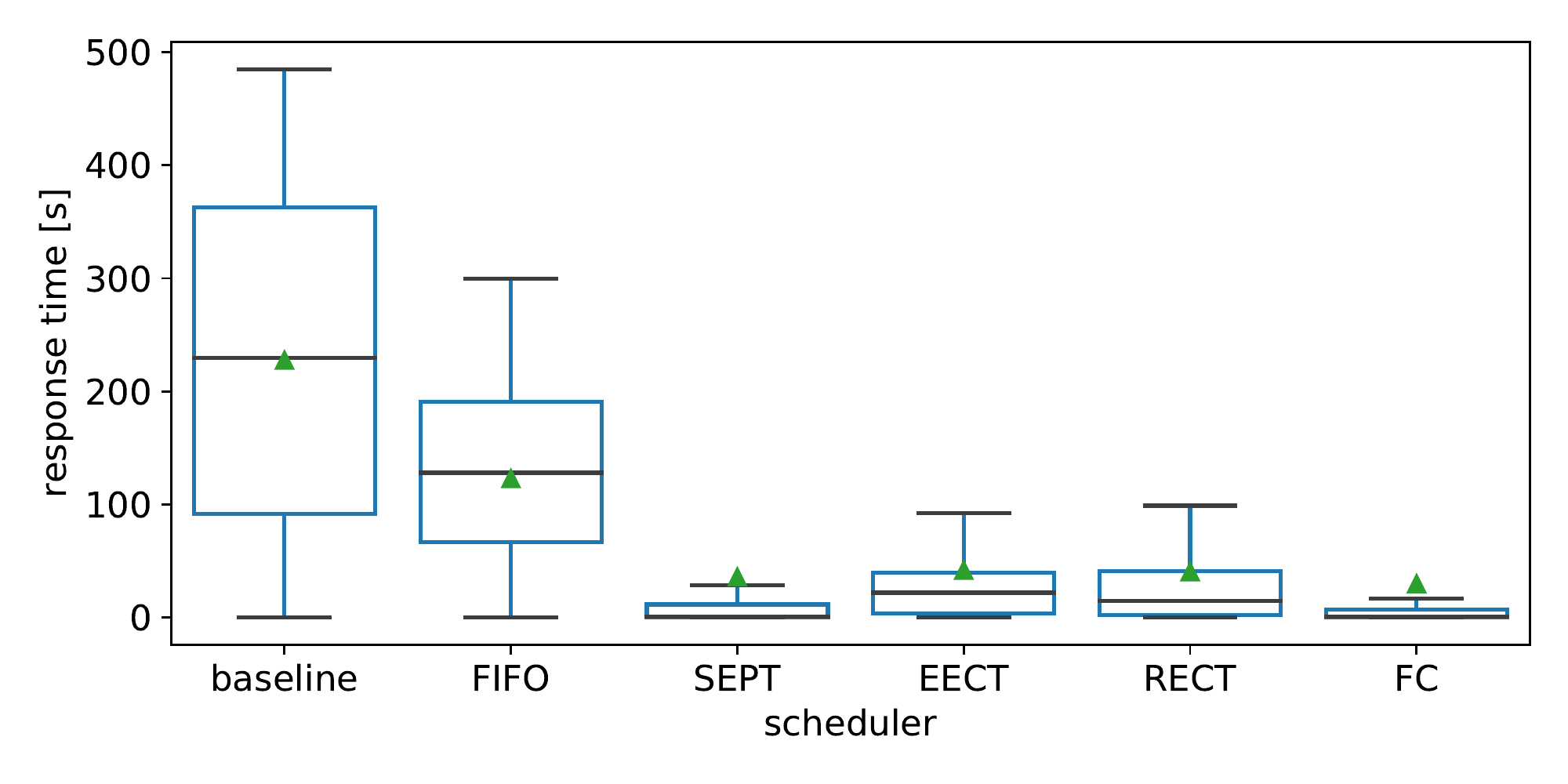}
    \\
    \chart{Experiment 4}{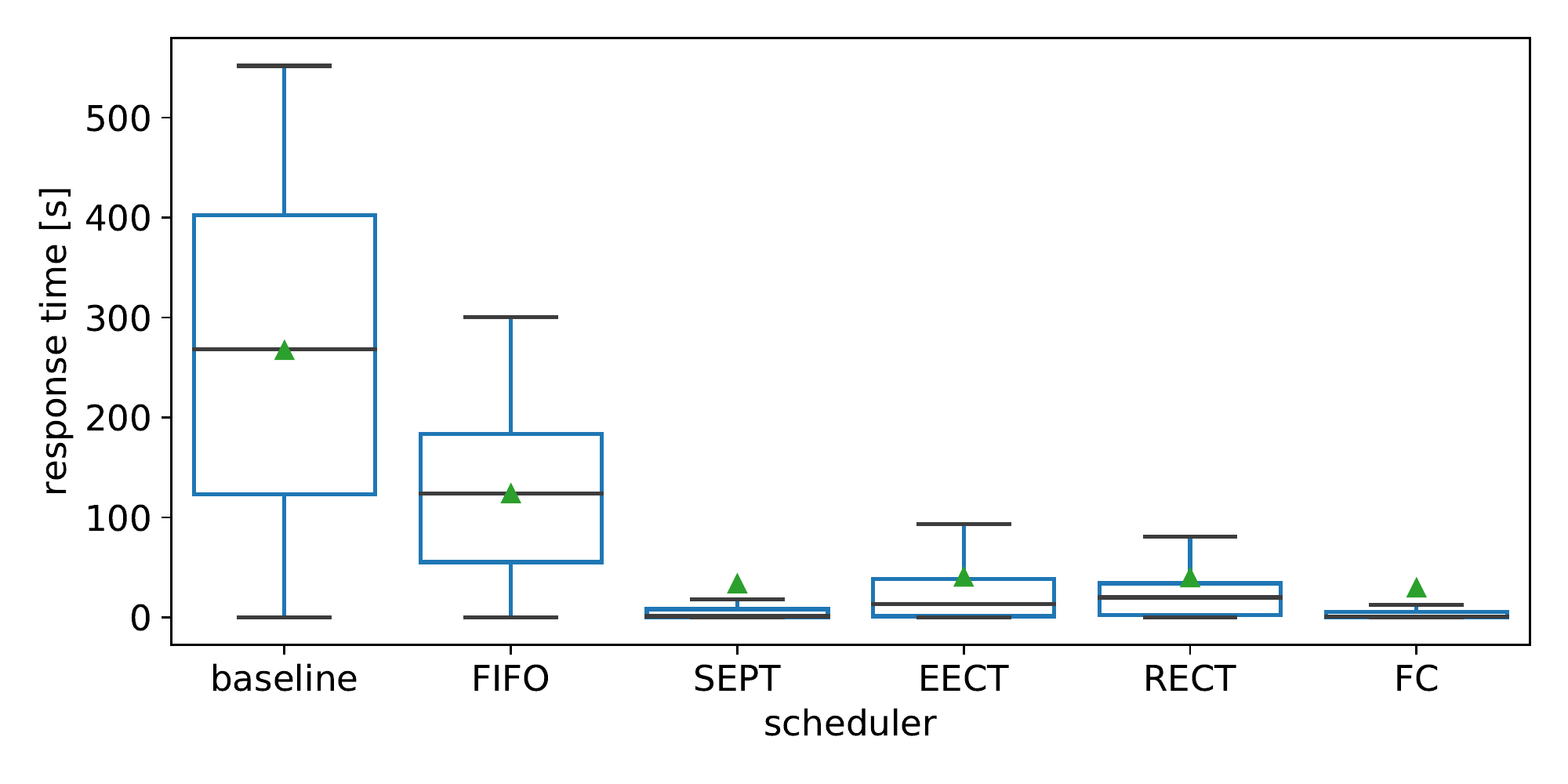}
    \chart{Experiment 5}{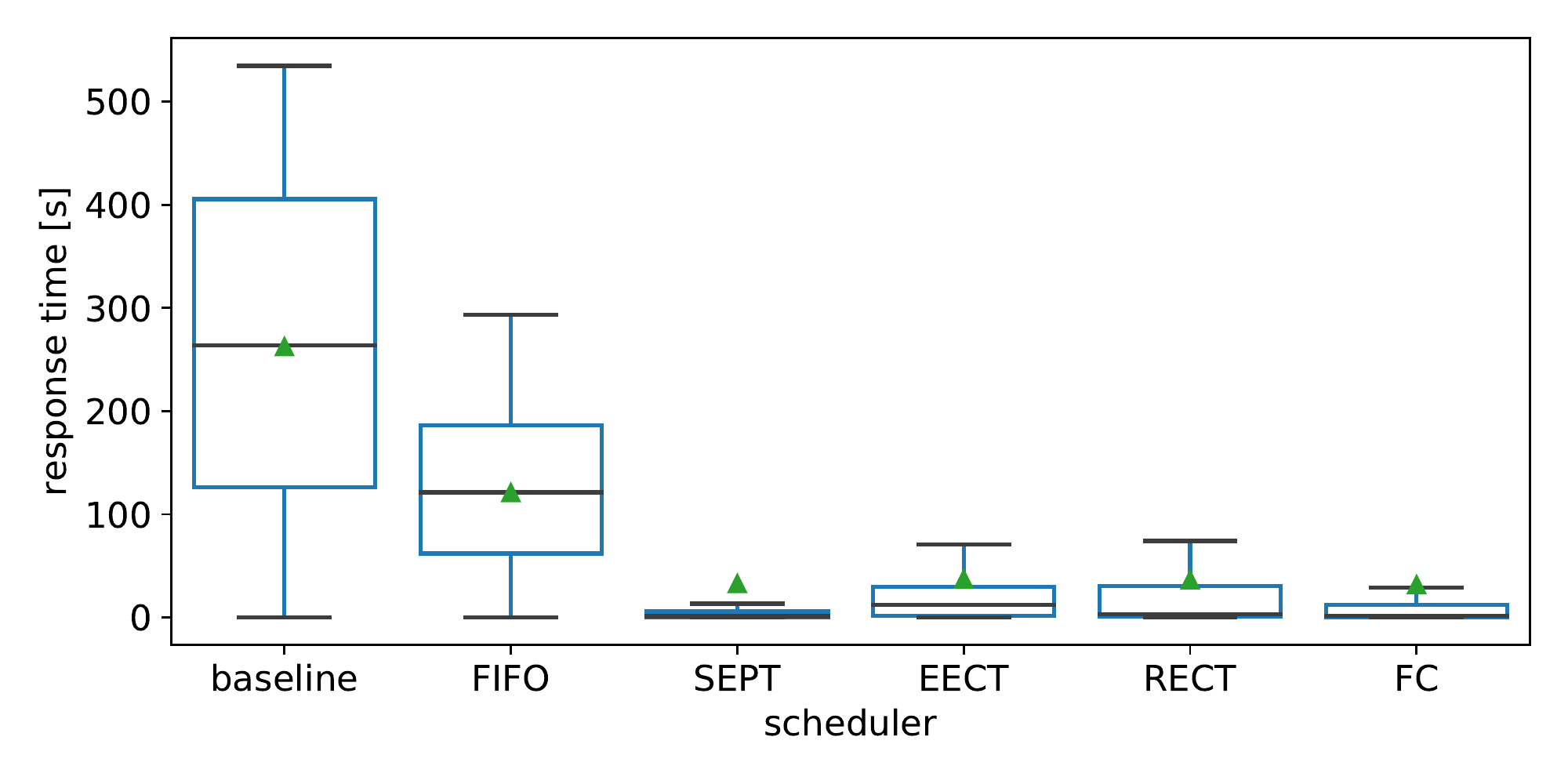}
    \\
    \caption{Response time for different scheduling policies. Here, we present results for 20 CPUs and intensity 40. On-premise infrastructure. Average response time presented as a green triangle.}
    \label{fig:response_time_20_40}
\end{figure*}

\begin{figure*}[h]
    \centering
    \chart{Experiment 1}{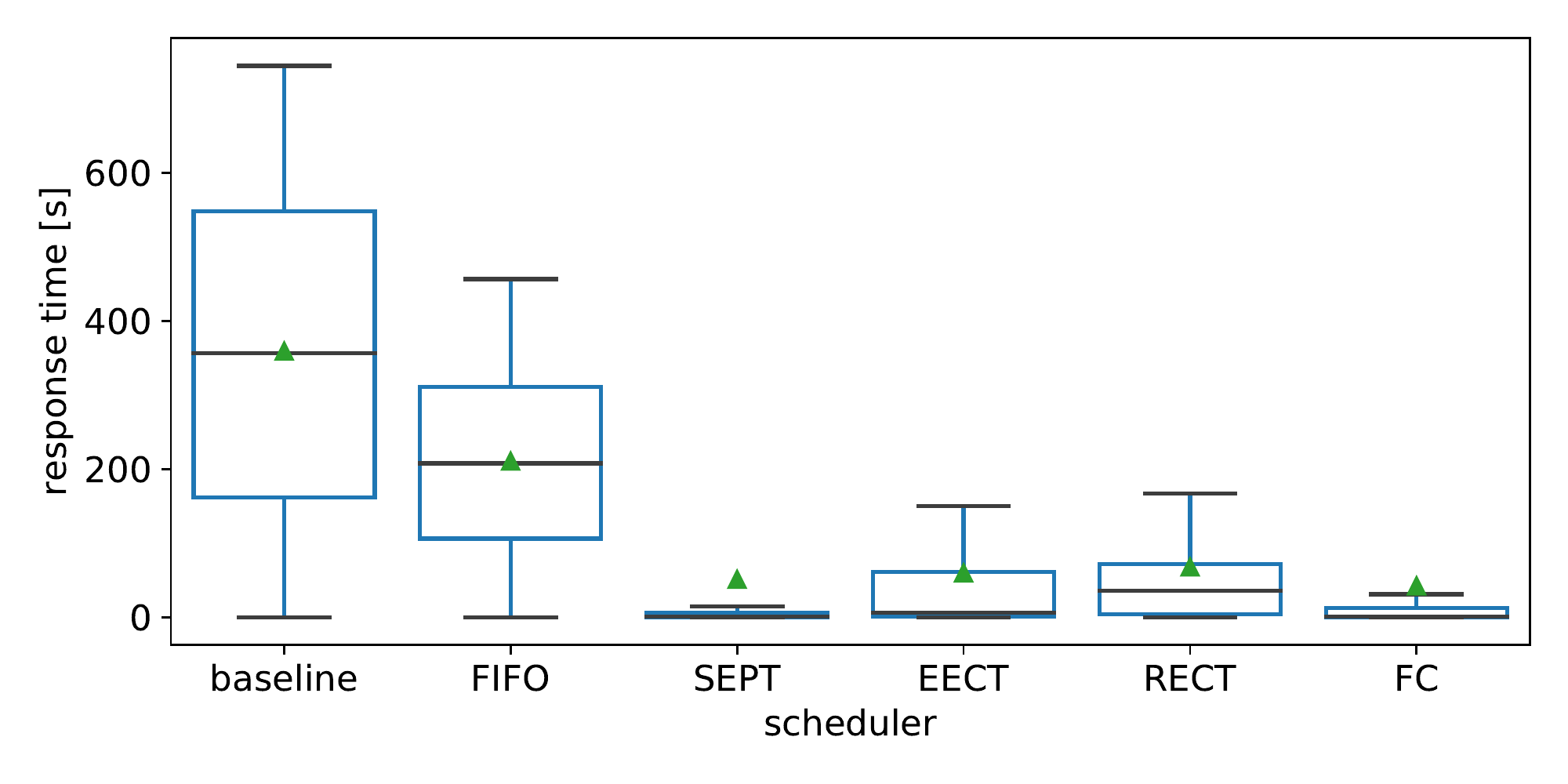}
    \chart{Experiment 2}{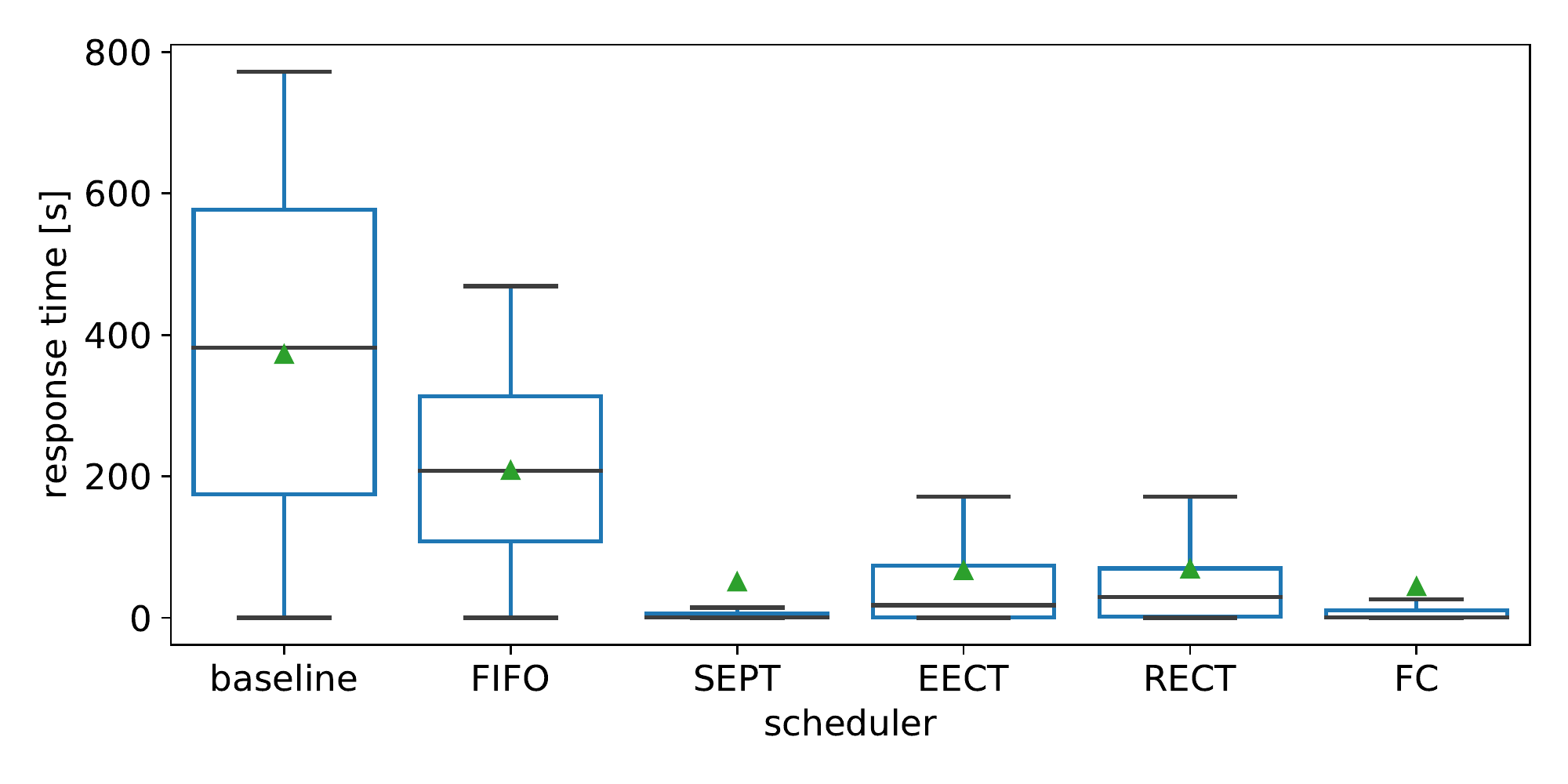}
    \chart{Experiment 3}{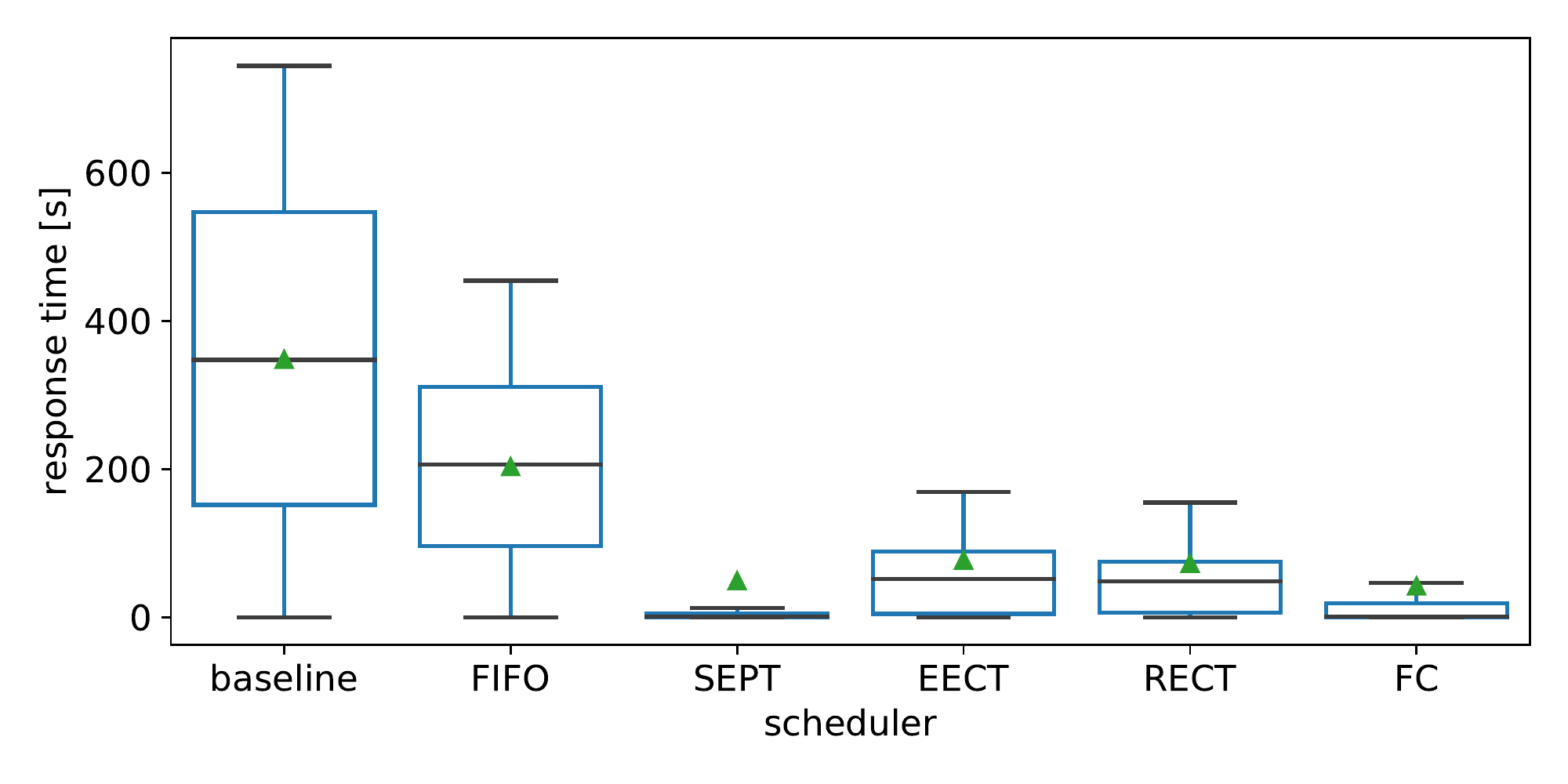}
    \\
    \chart{Experiment 4}{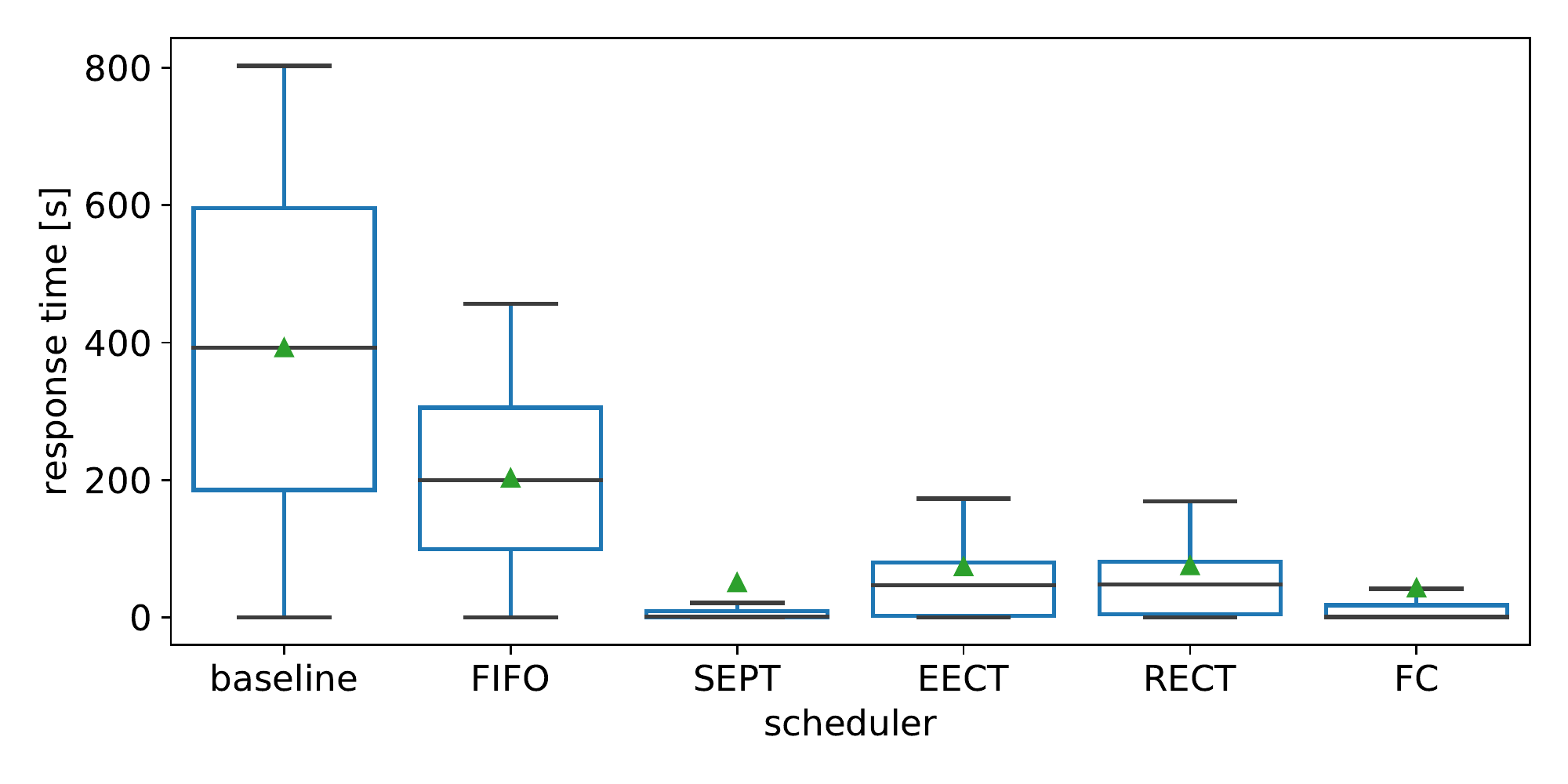}
    \chart{Experiment 5}{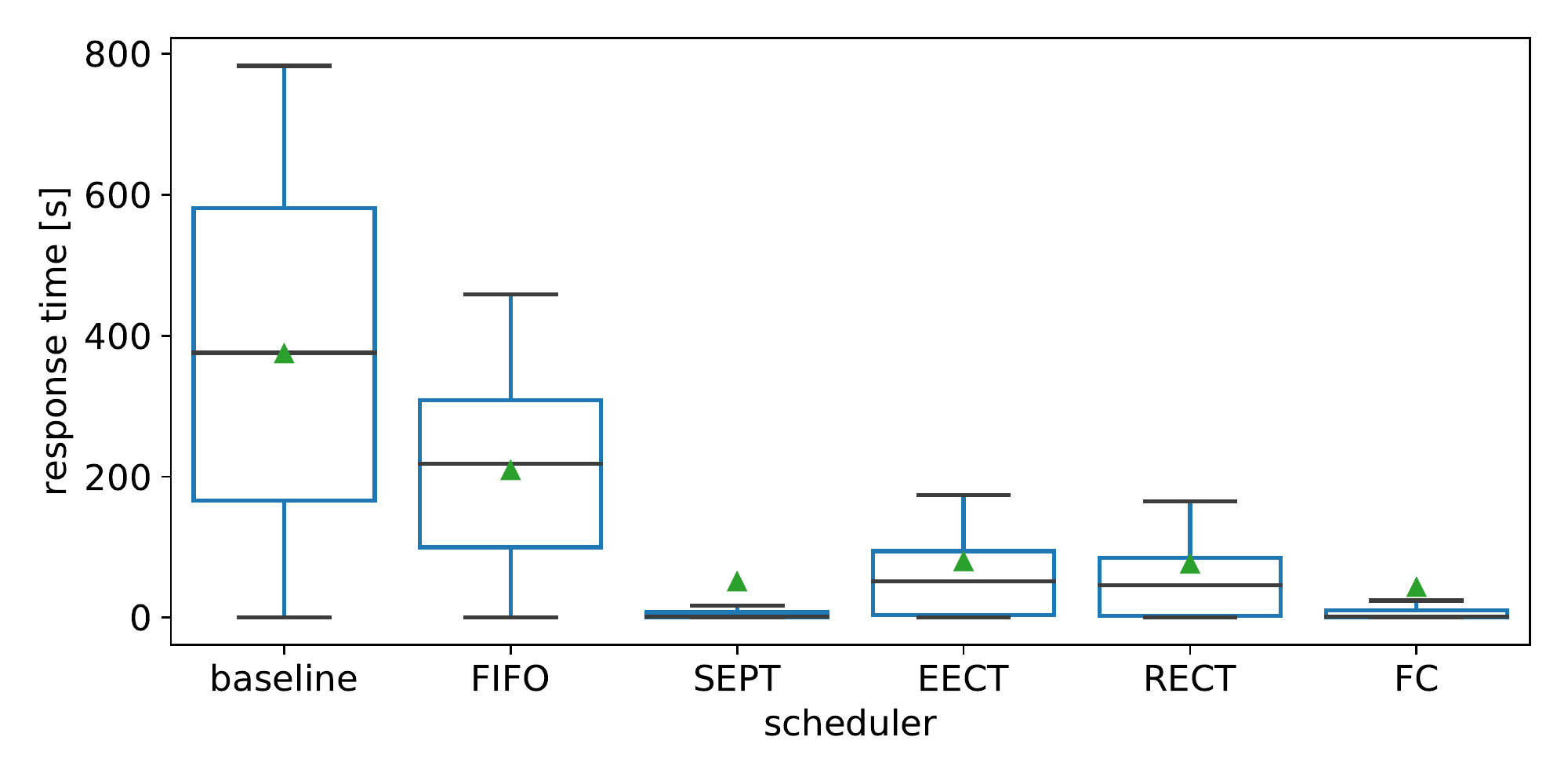}
    \\
    \caption{Response time for different scheduling policies. Here, we present results for 20 CPUs and intensity 60. On-premise infrastructure. Average response time presented as a green triangle.}
    \label{fig:response_time_20_60}
\end{figure*}

\begin{figure*}[h]
    \centering
    \chart{Experiment 1}{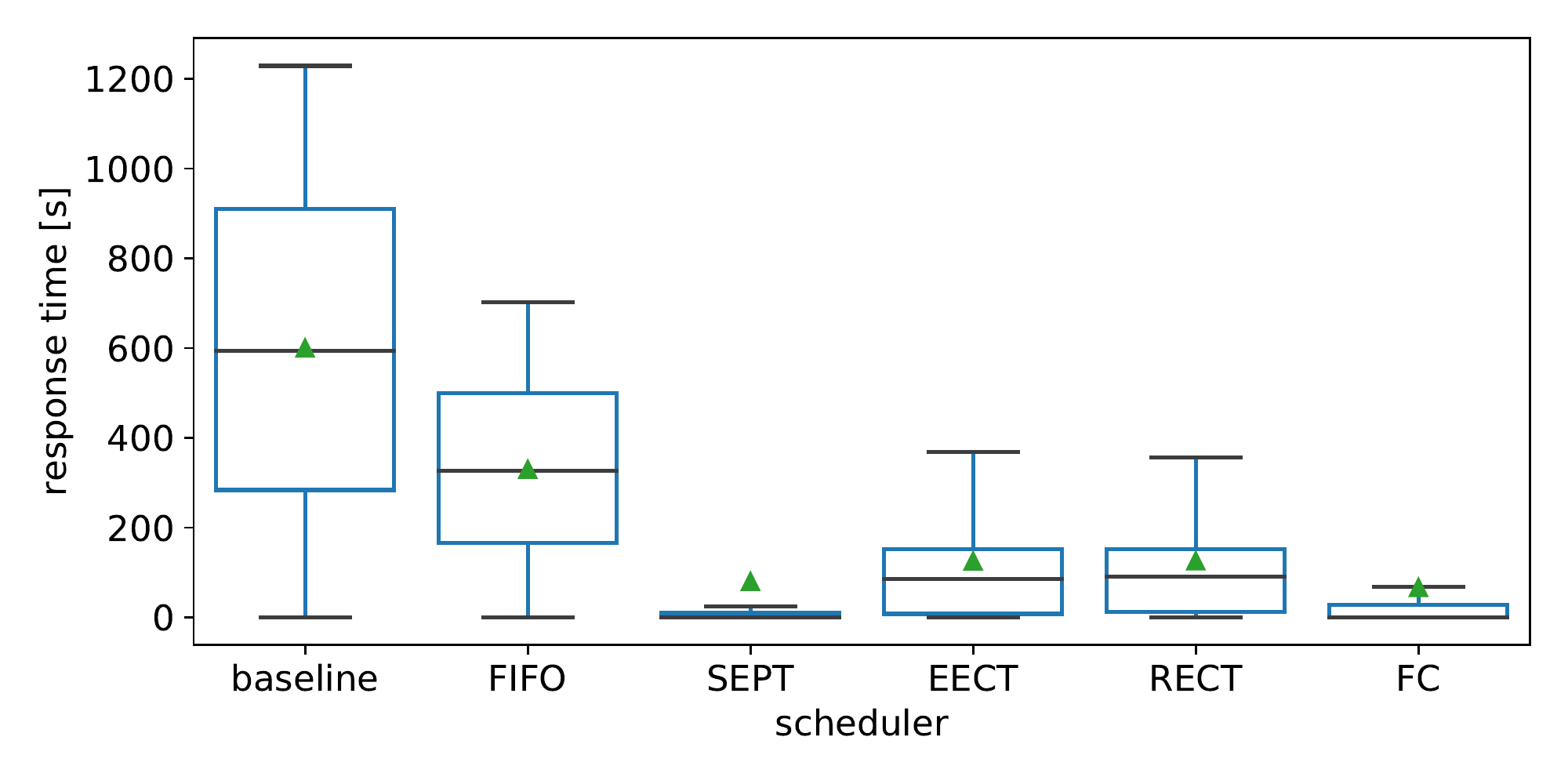}
    \chart{Experiment 2}{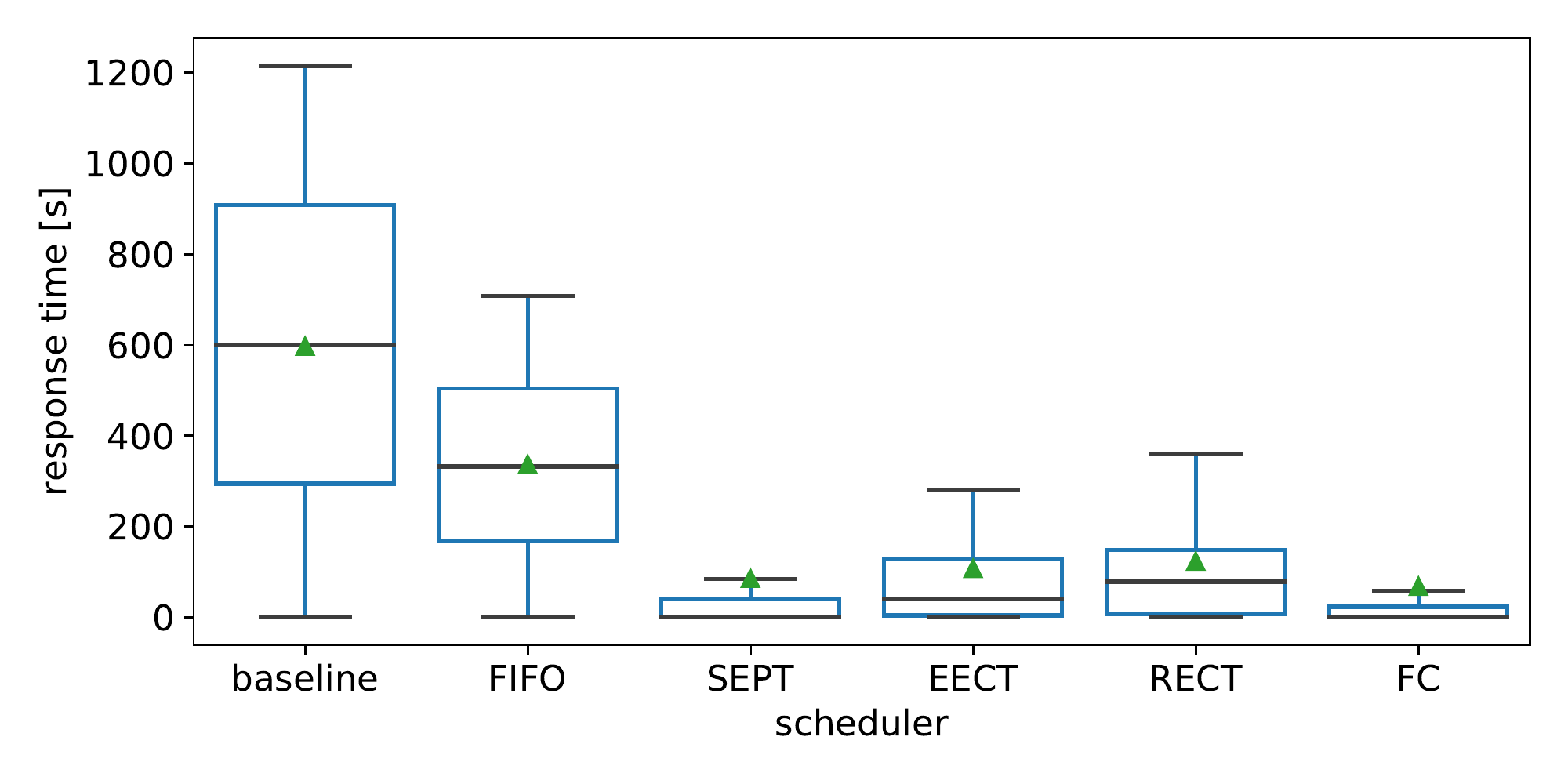}
    \chart{Experiment 3}{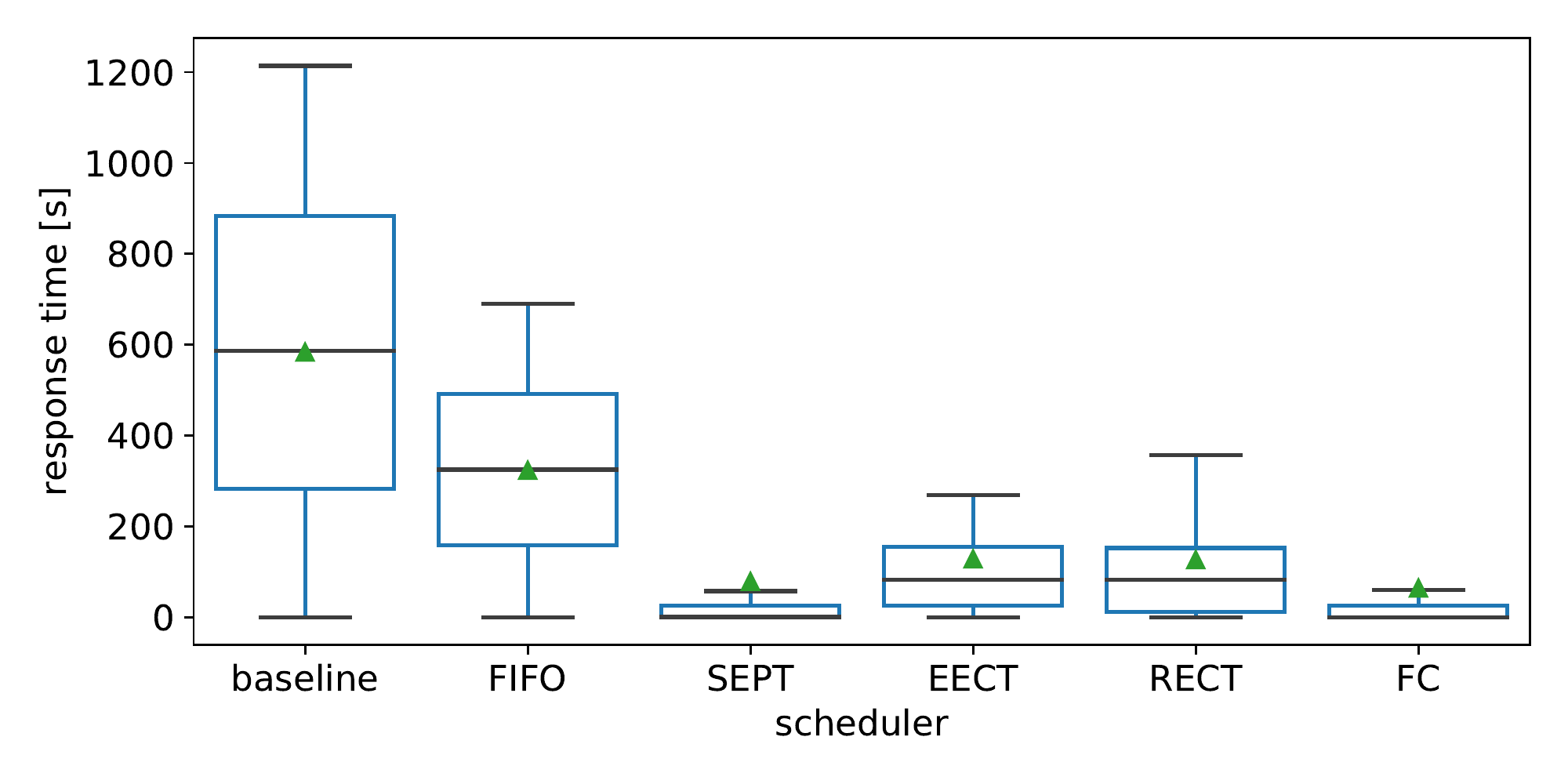}
    \\
    \chart{Experiment 4}{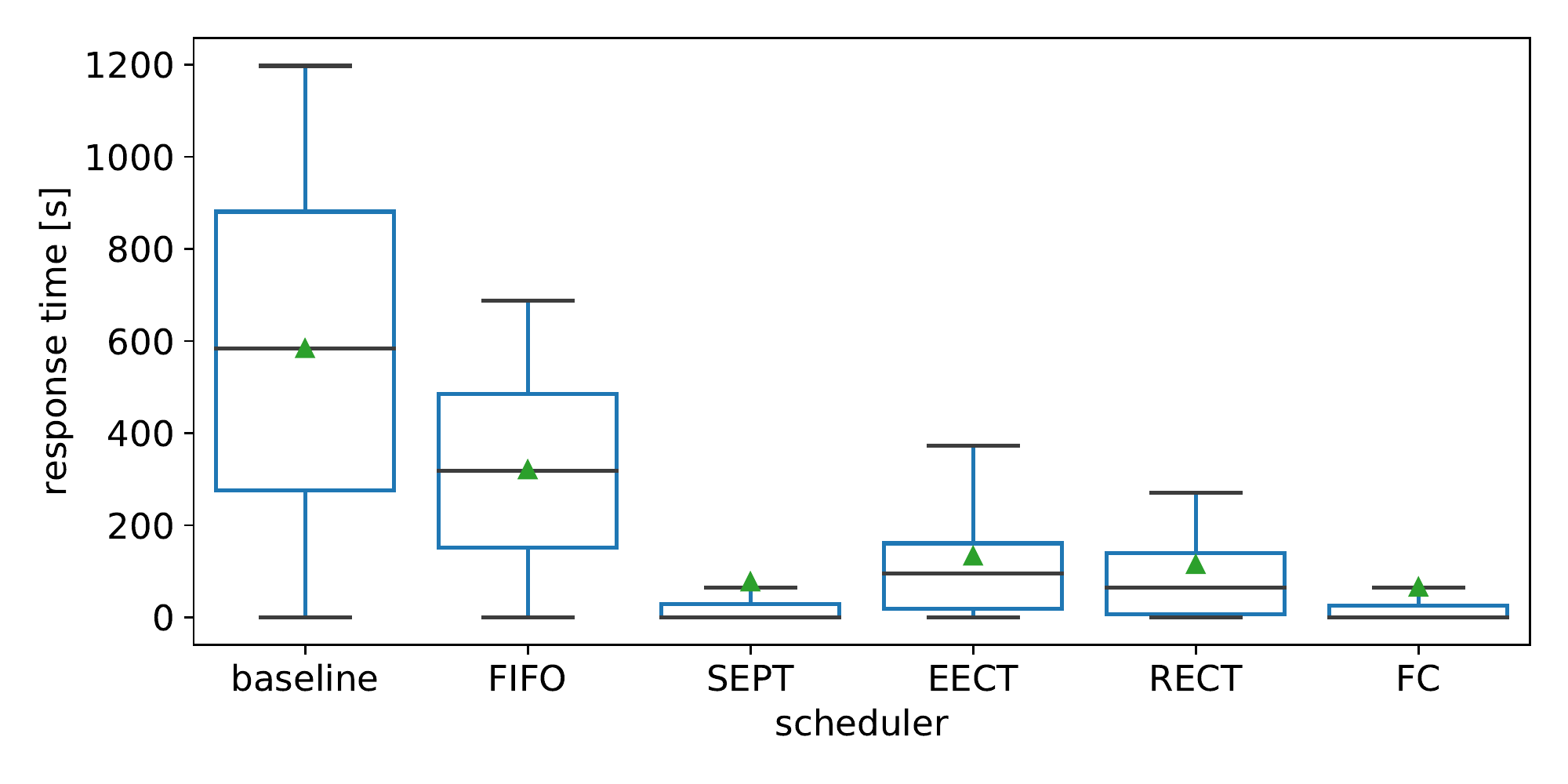}
    \chart{Experiment 5}{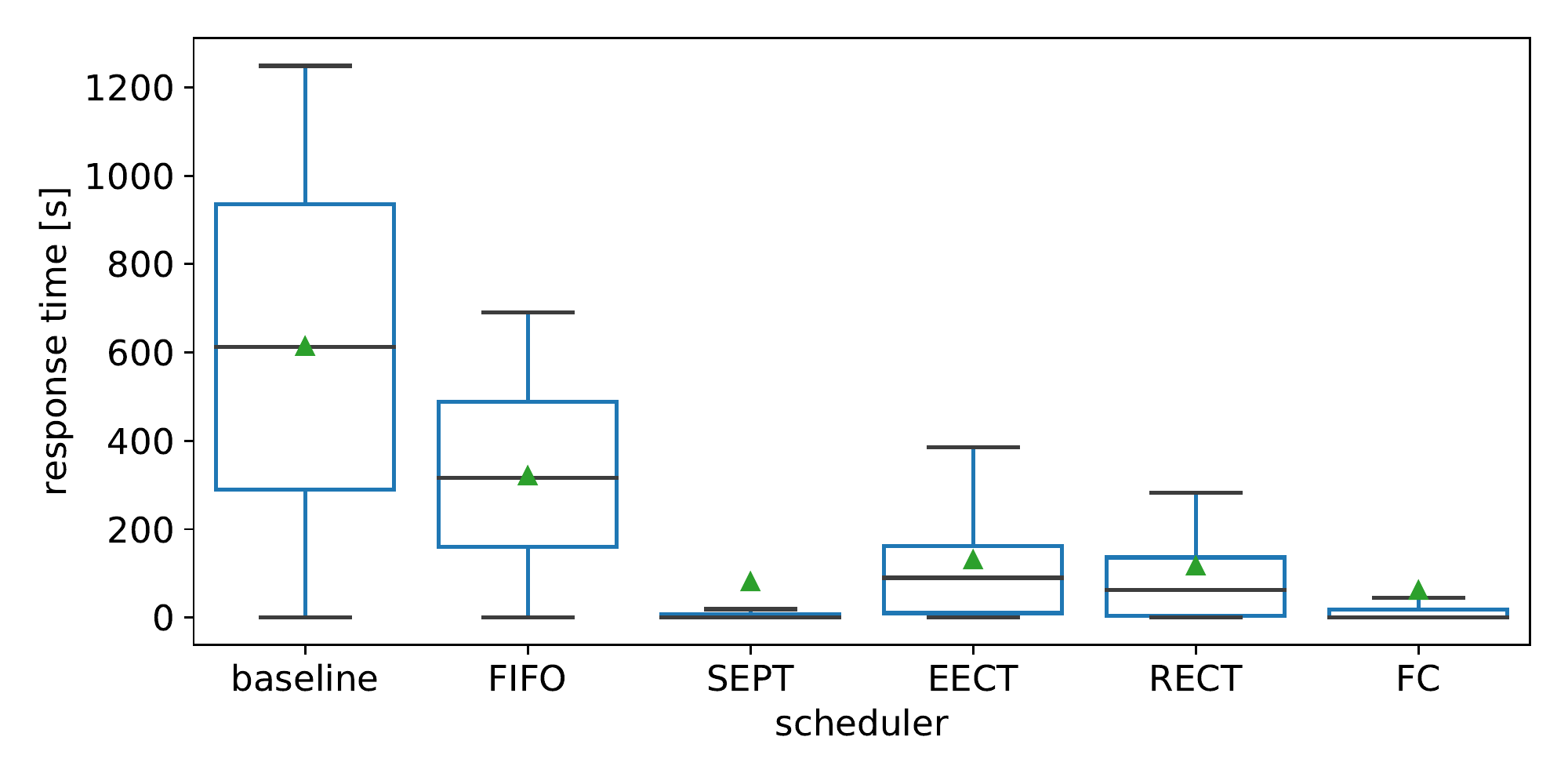}
    \\
    \caption{Response time for different scheduling policies. Here, we present results for 20 CPUs and intensity 90. On-premise infrastructure. Average response time presented as a green triangle.}
    \label{fig:response_time_20_90}
\end{figure*}

\begin{figure*}[h]
    \centering
    \chart{Experiment 1}{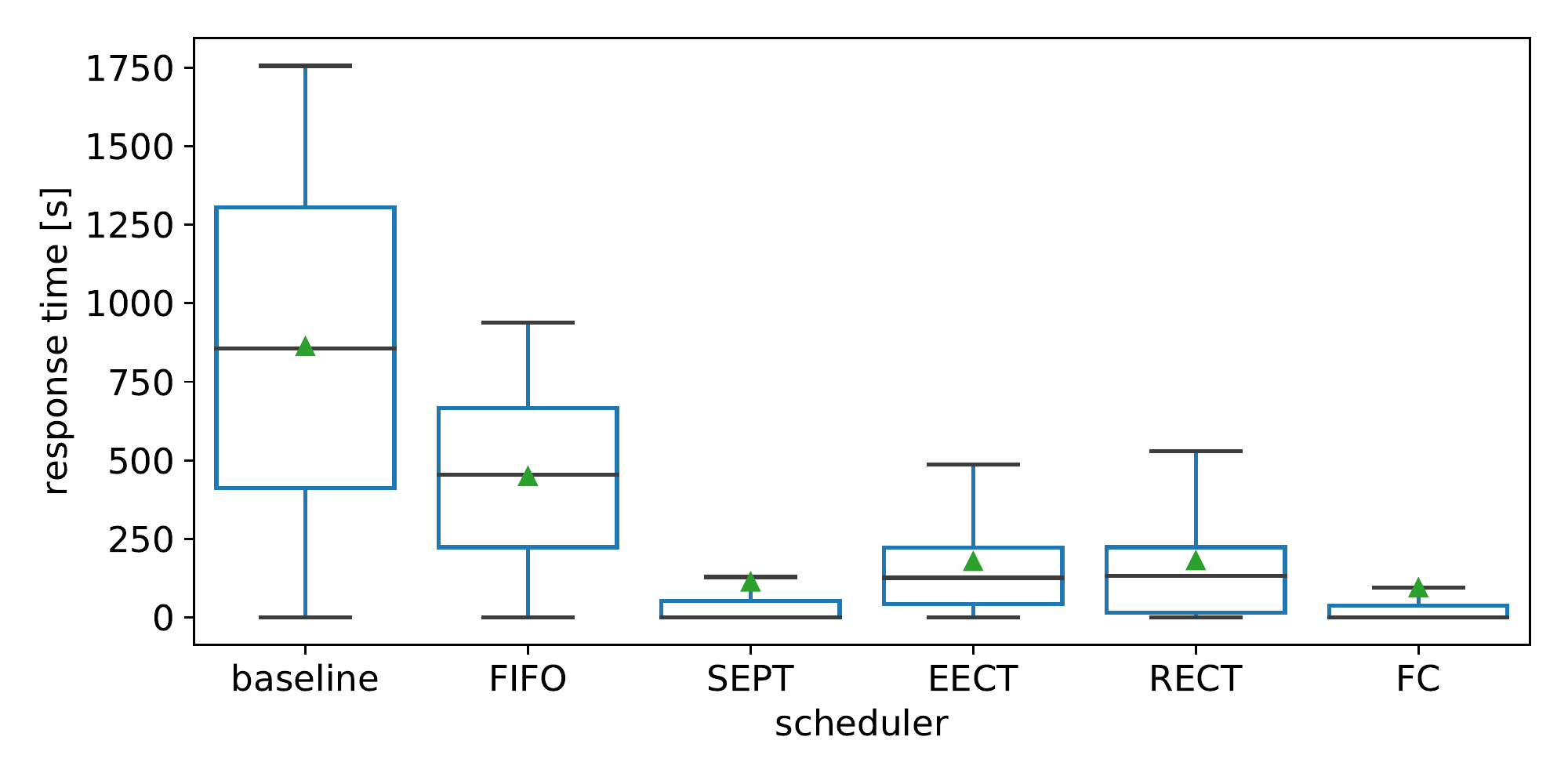}
    \chart{Experiment 2}{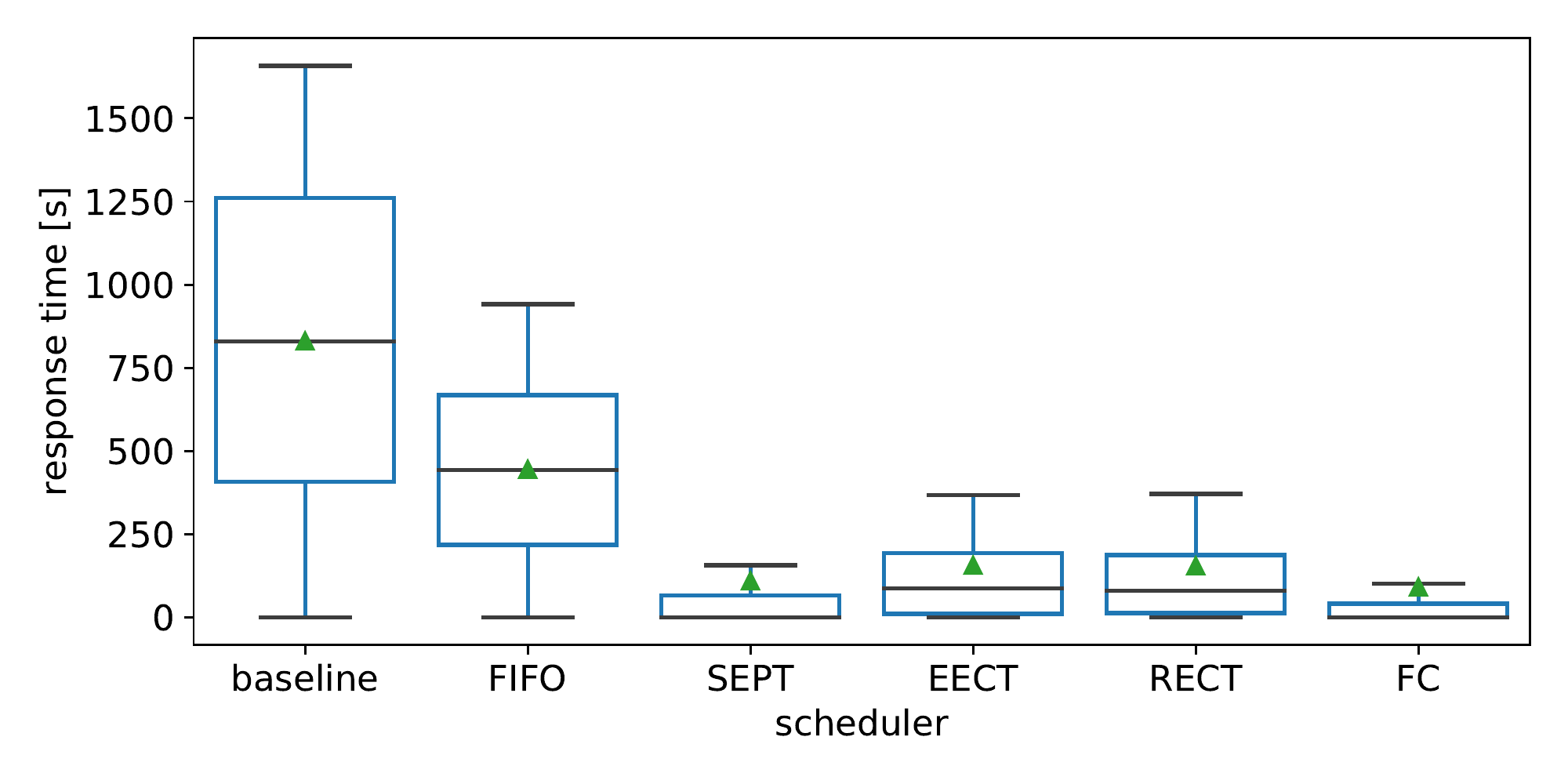}
    \chart{Experiment 3}{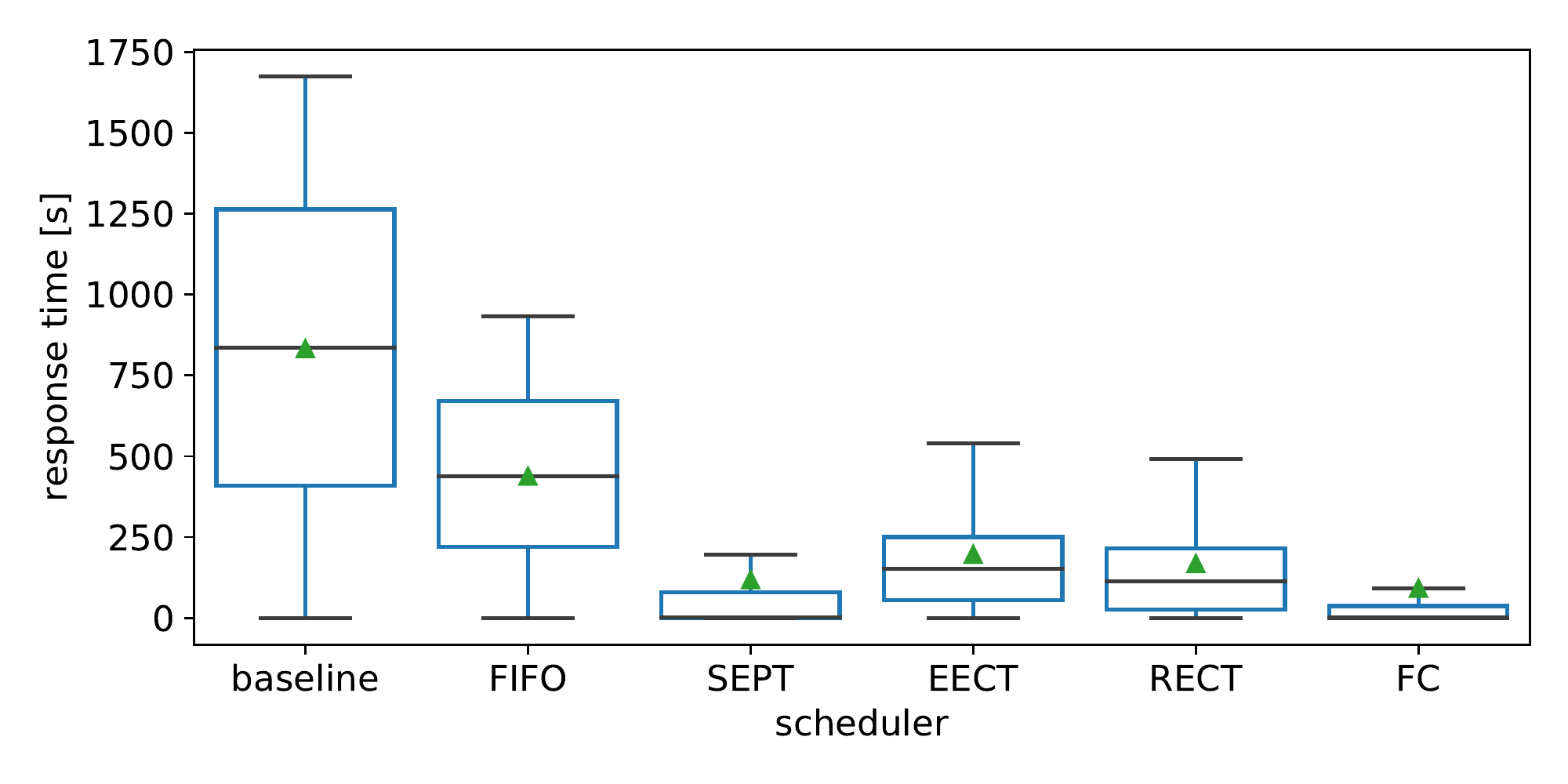}
    \\
    \chart{Experiment 4}{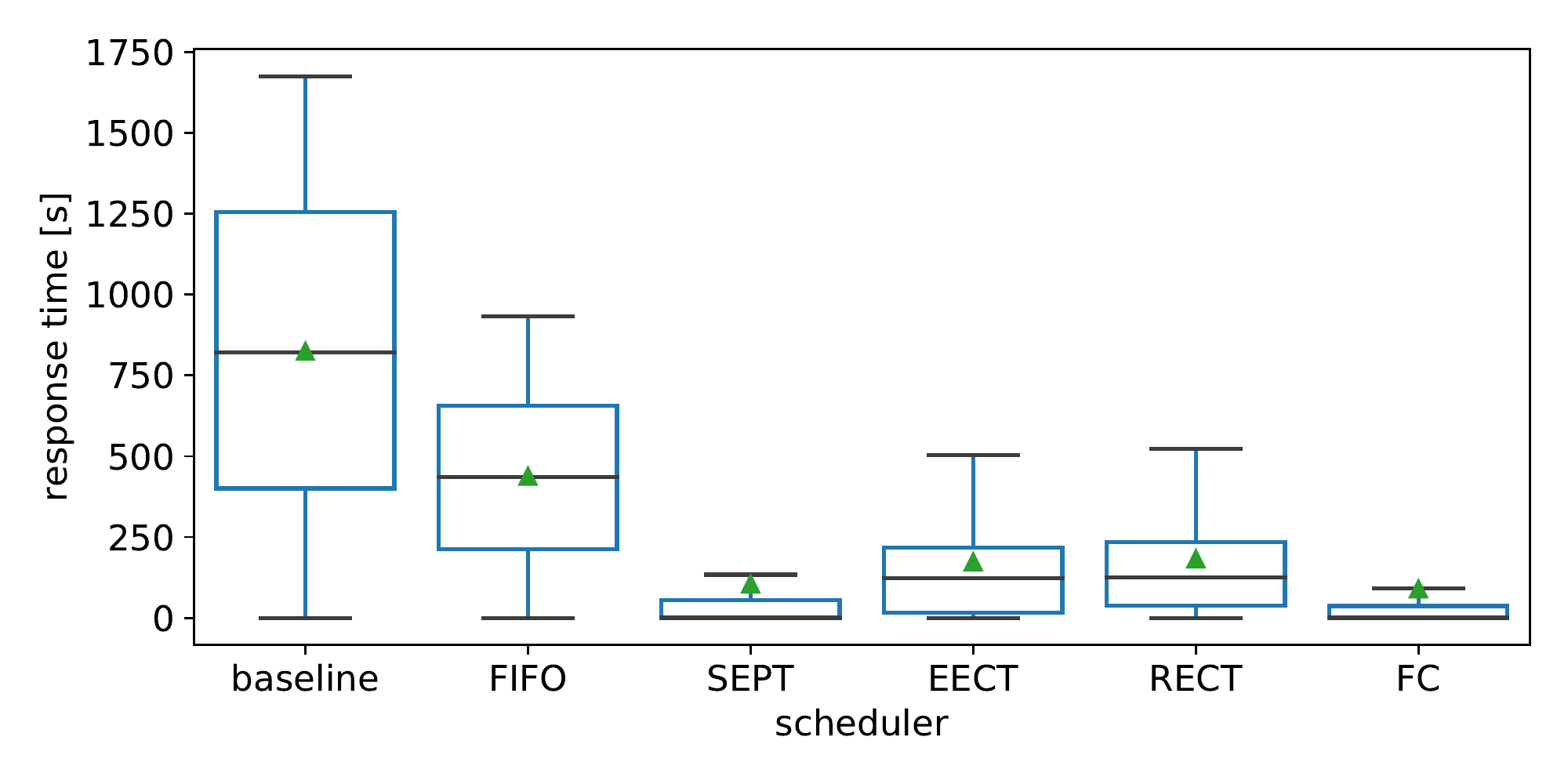}
    \chart{Experiment 5}{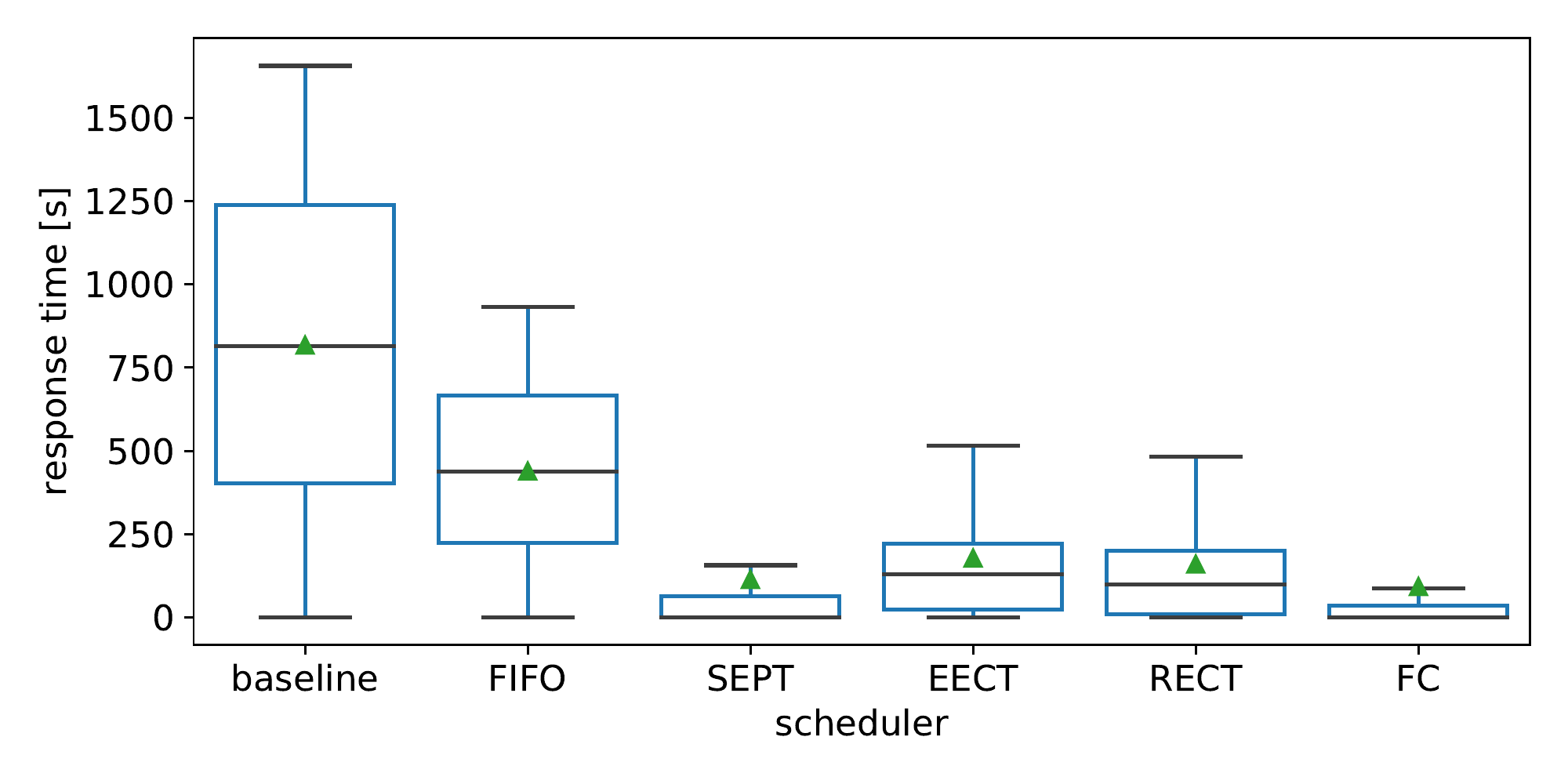}
    \\
    \caption{Response time for different scheduling policies. Here, we present results for 20 CPUs and intensity 120. On-premise infrastructure. Average response time presented as a green triangle.}
    \label{fig:response_time_20_120}
\end{figure*}

\begin{figure*}[h]
    \centering
    \chart{Experiment 1}{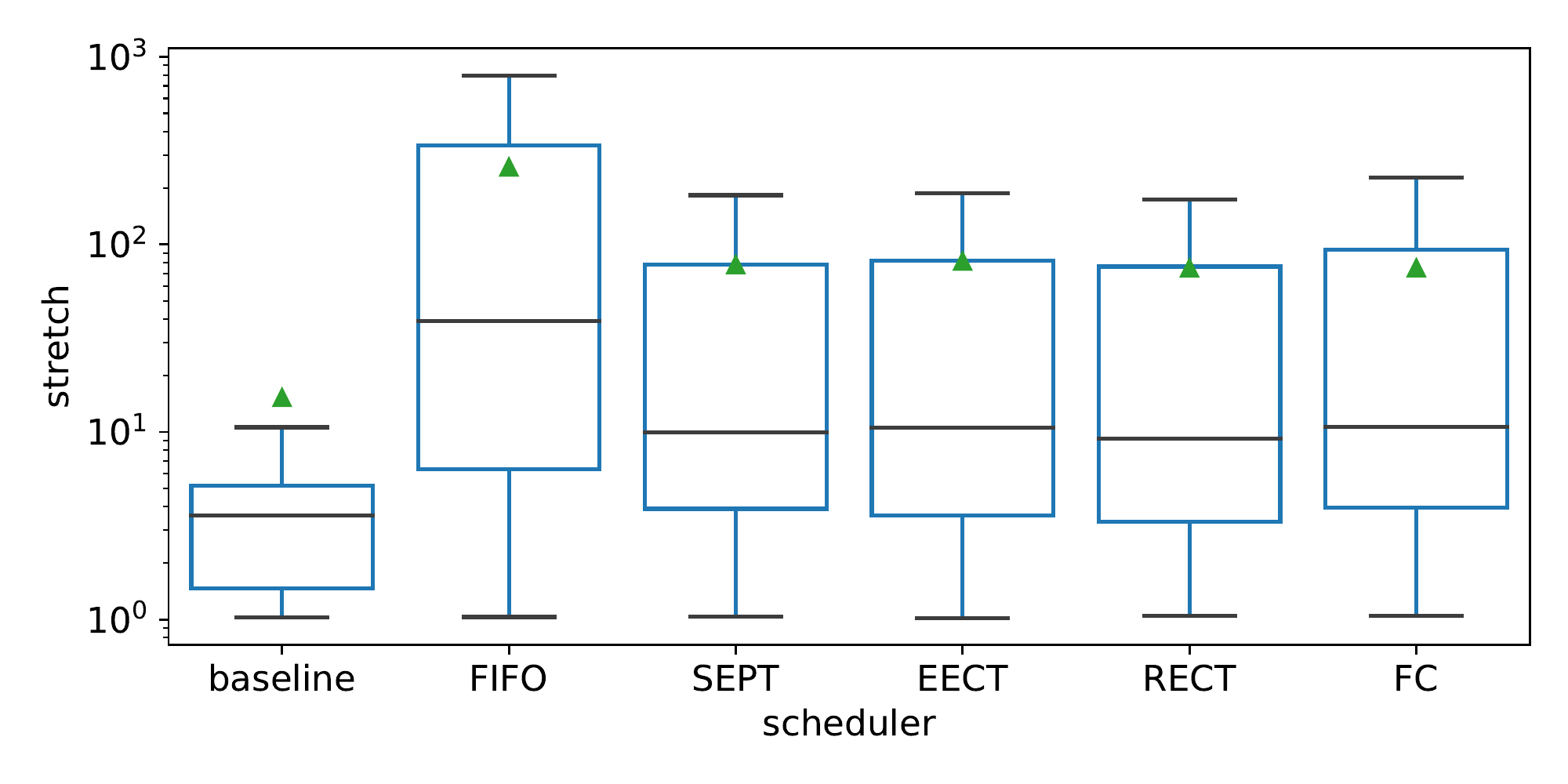}
    \chart{Experiment 2}{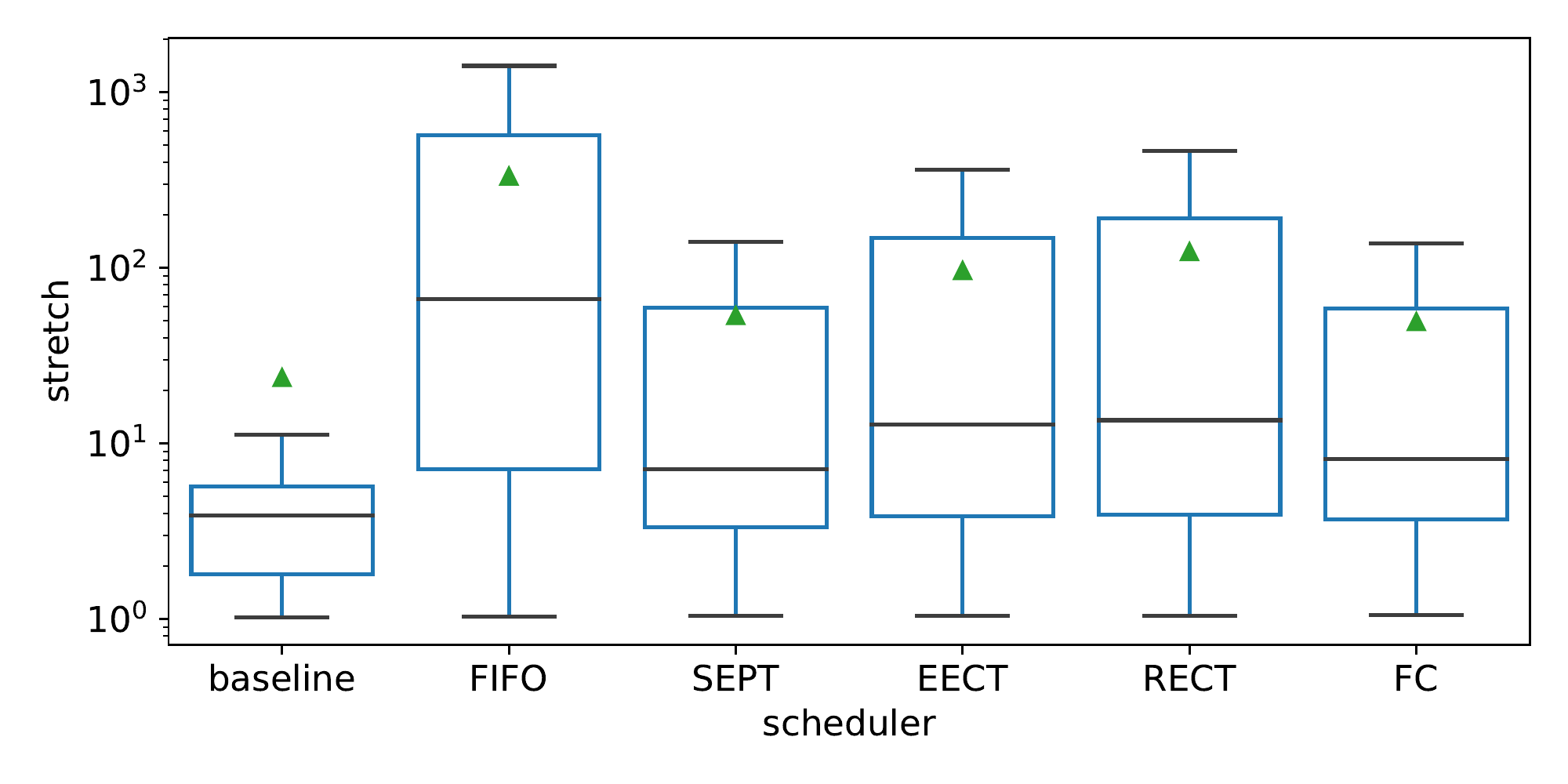}
    \chart{Experiment 3}{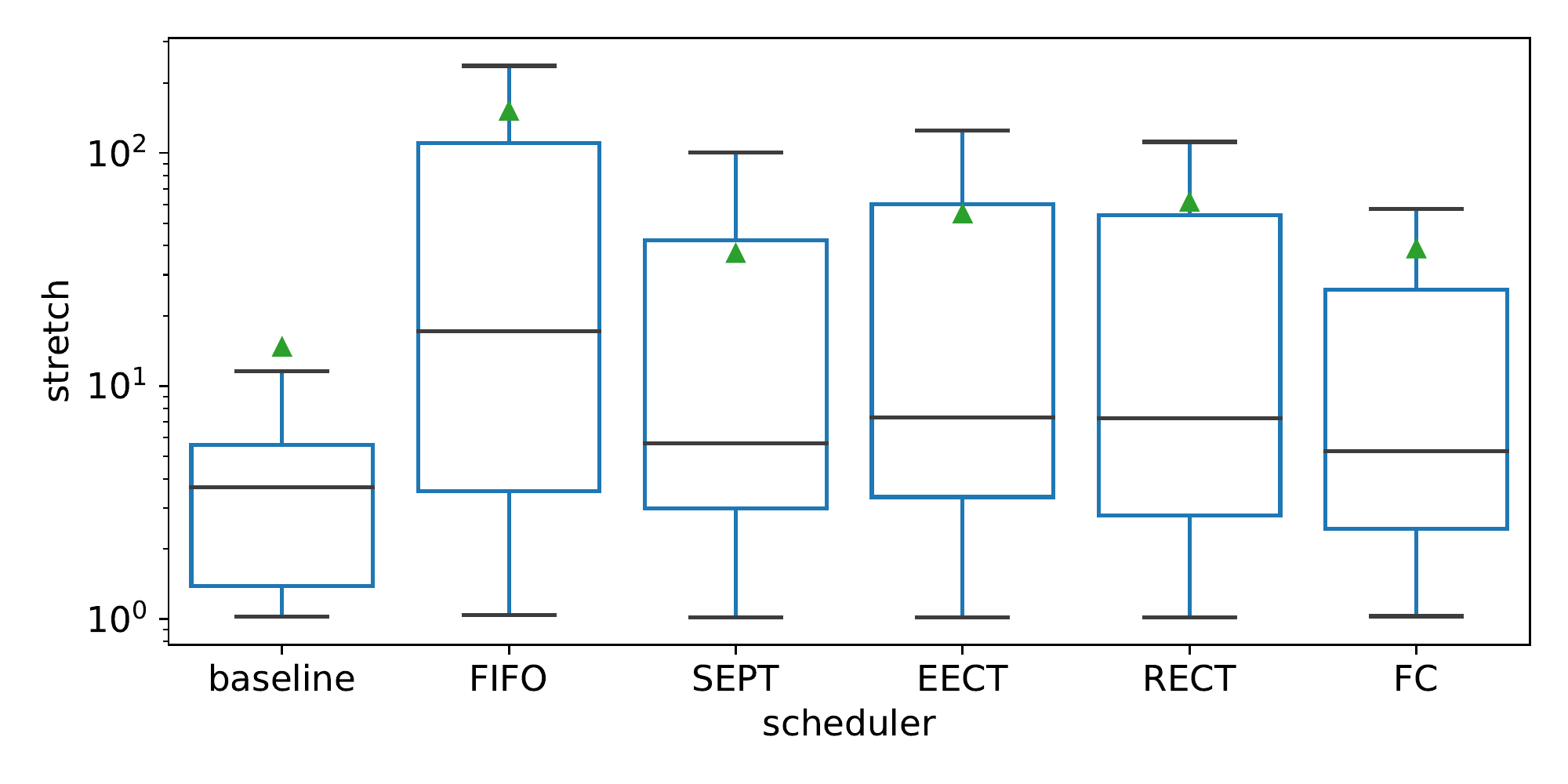}
    \\
    \chart{Experiment 4}{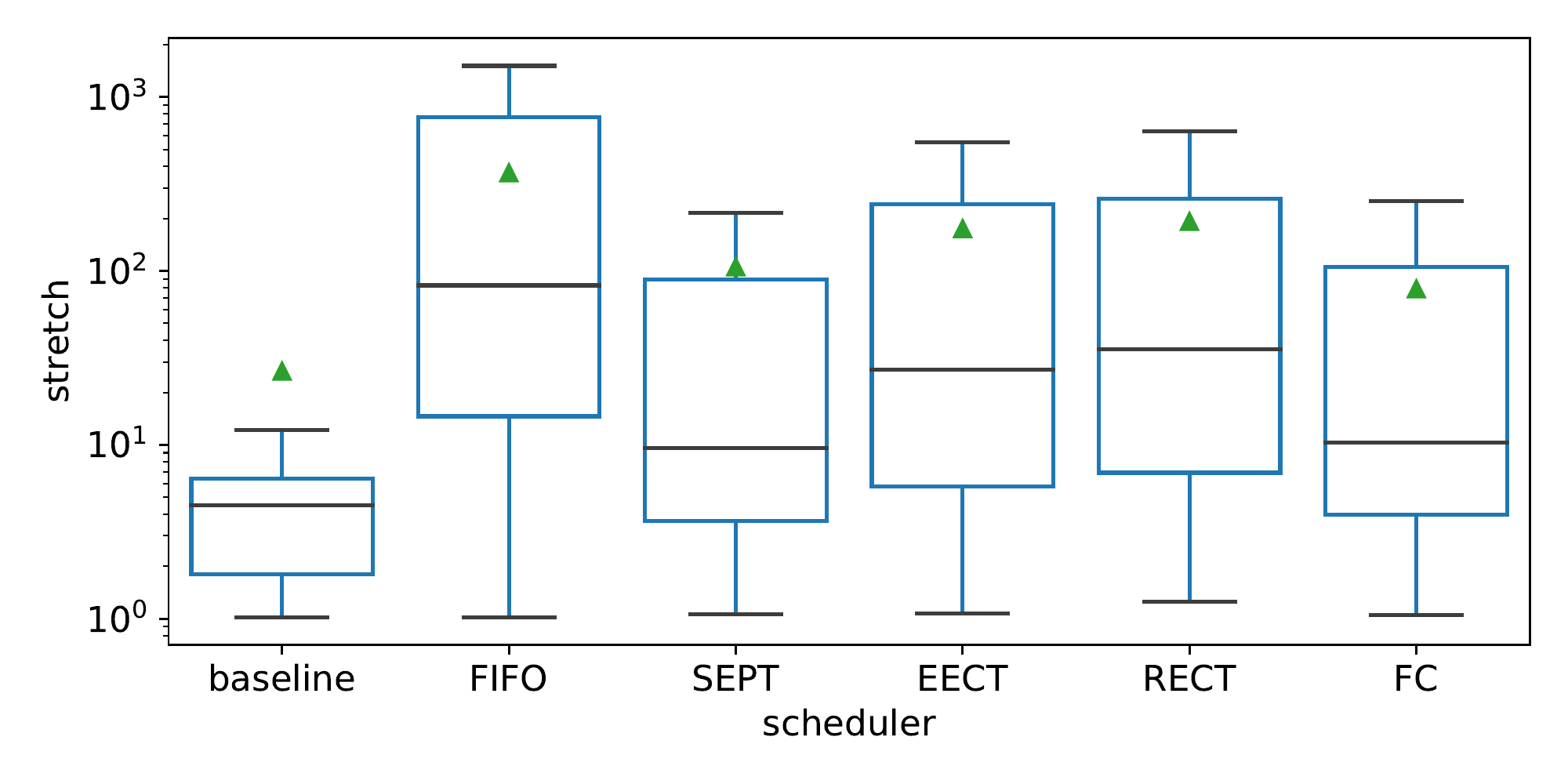}
    \chart{Experiment 5}{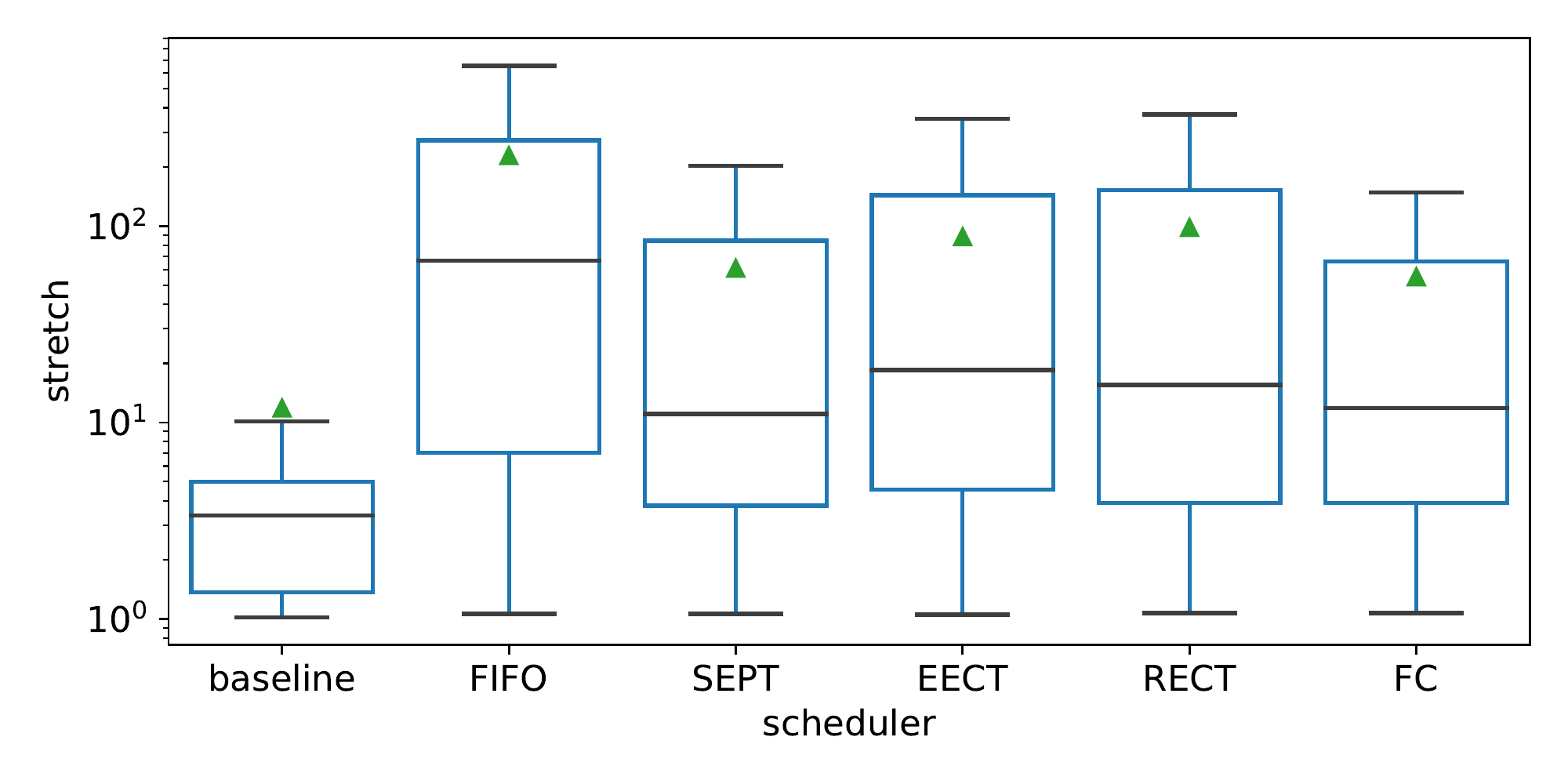}
    \\
    \caption{Stretch for different scheduling policies. Here, we present results for 5 CPUs and intensity 30. On-premise infrastructure. Average stretch presented as a green triangle.}
    \label{fig:stretch_5_30}
\end{figure*}

\begin{figure*}[h]
    \centering
    \chart{Experiment 1}{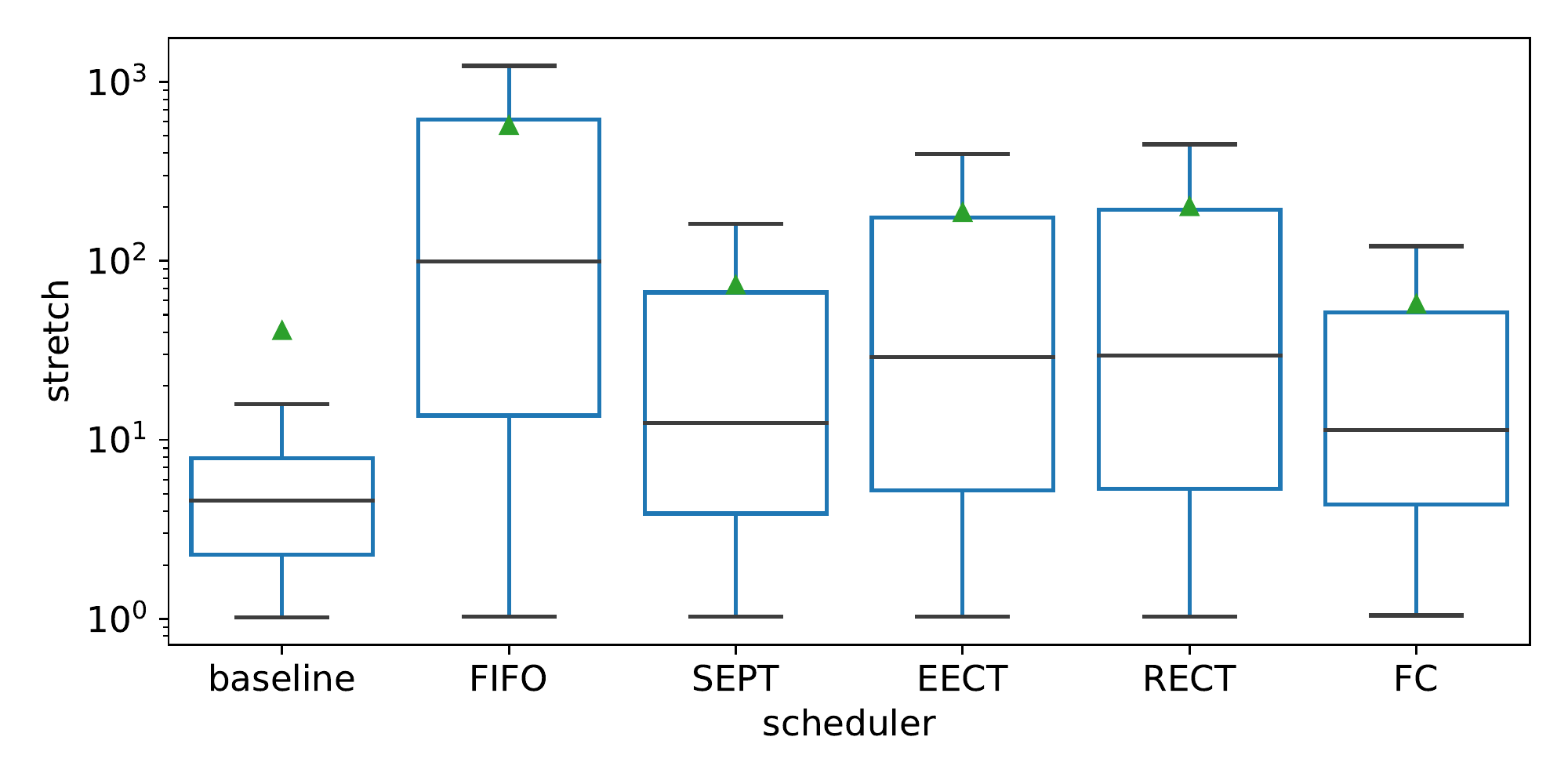}
    \chart{Experiment 2}{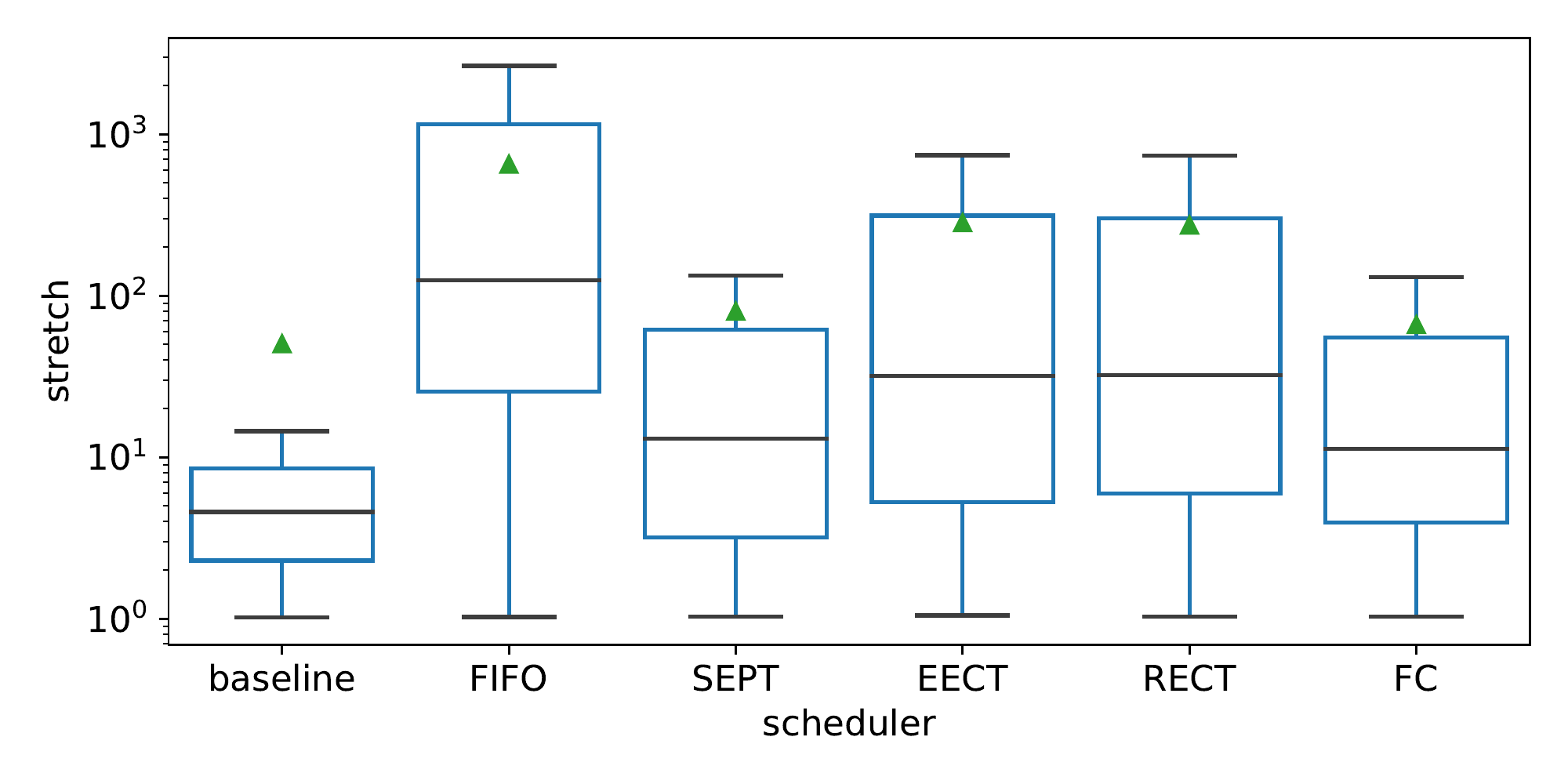}
    \chart{Experiment 3}{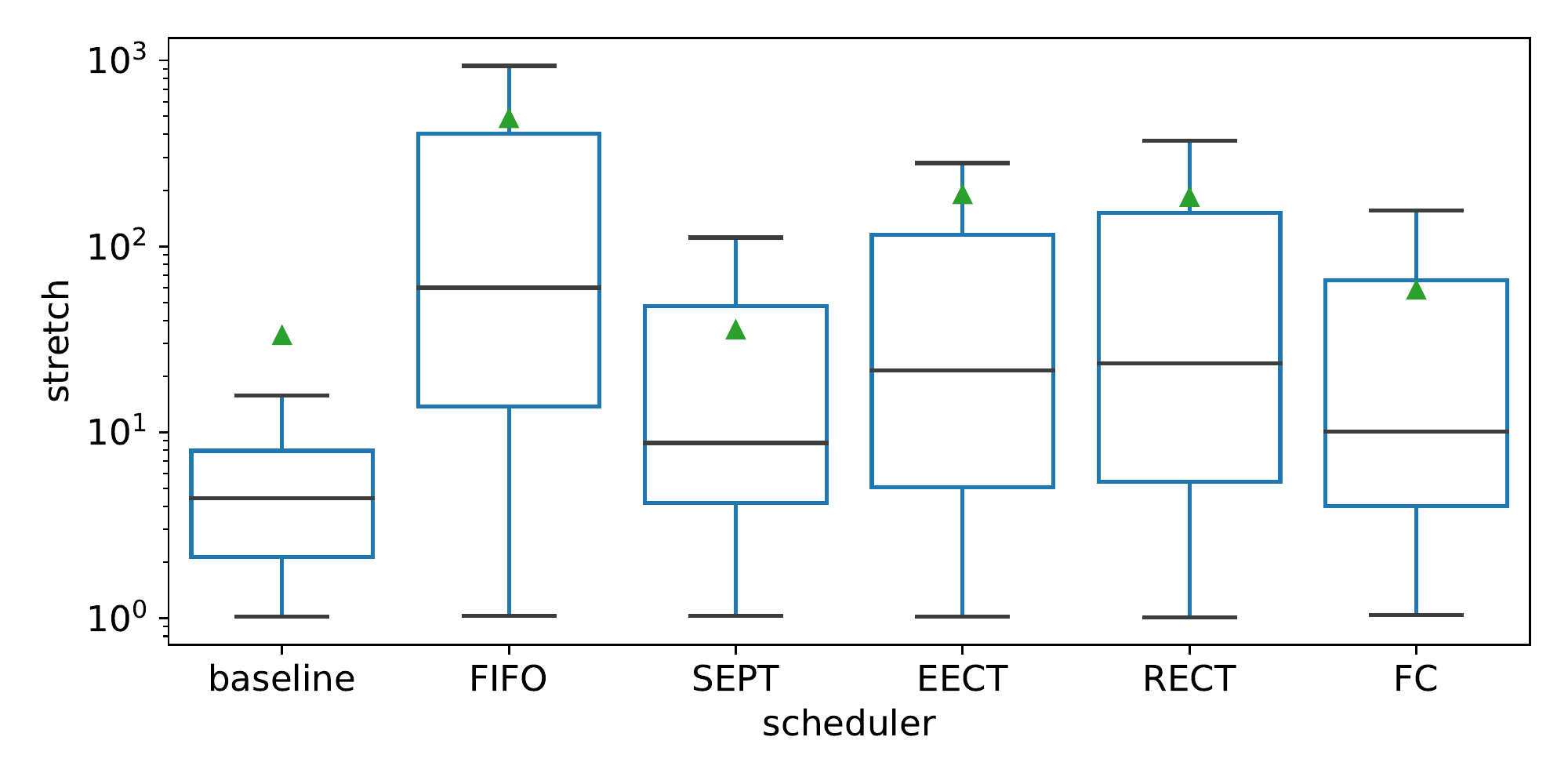}
    \\
    \chart{Experiment 4}{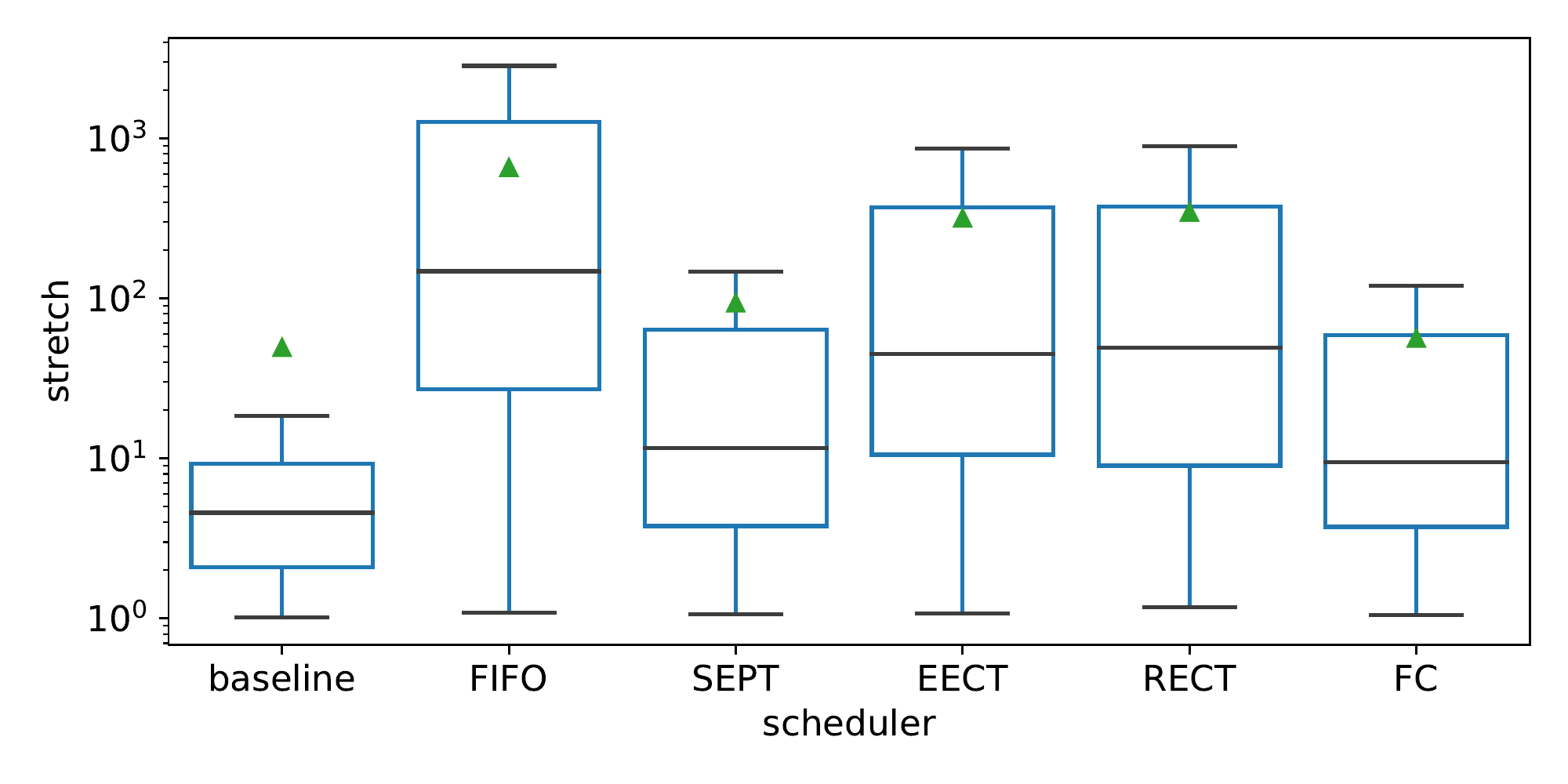}
    \chart{Experiment 5}{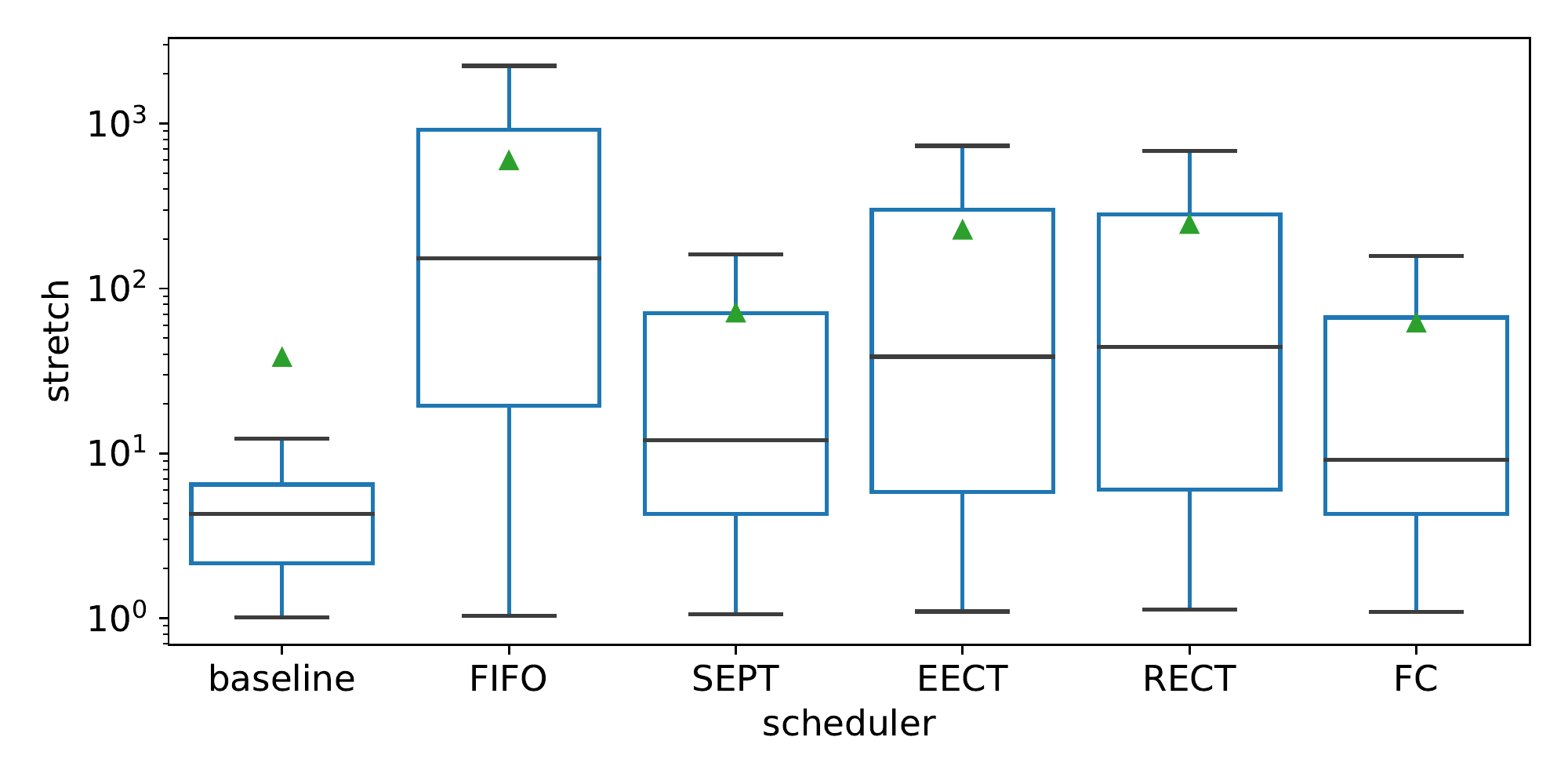}
    \\
    \caption{Stretch for different scheduling policies. Here, we present results for 5 CPUs and intensity 40. On-premise infrastructure. Average stretch presented as a green triangle.}
    \label{fig:stretch_5_40}
\end{figure*}

\begin{figure*}[h]
    \centering
    \chart{Experiment 1}{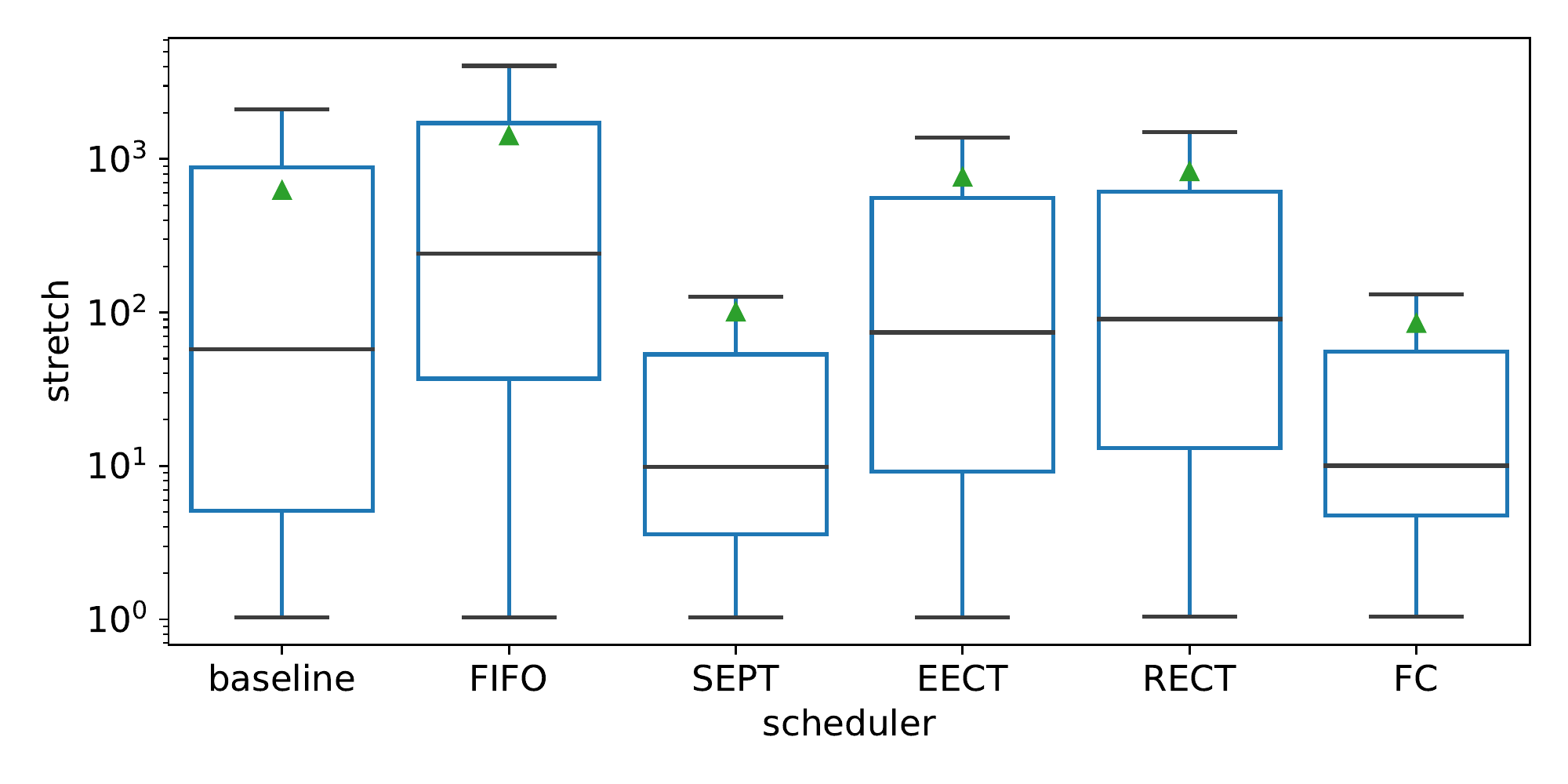}
    \chart{Experiment 2}{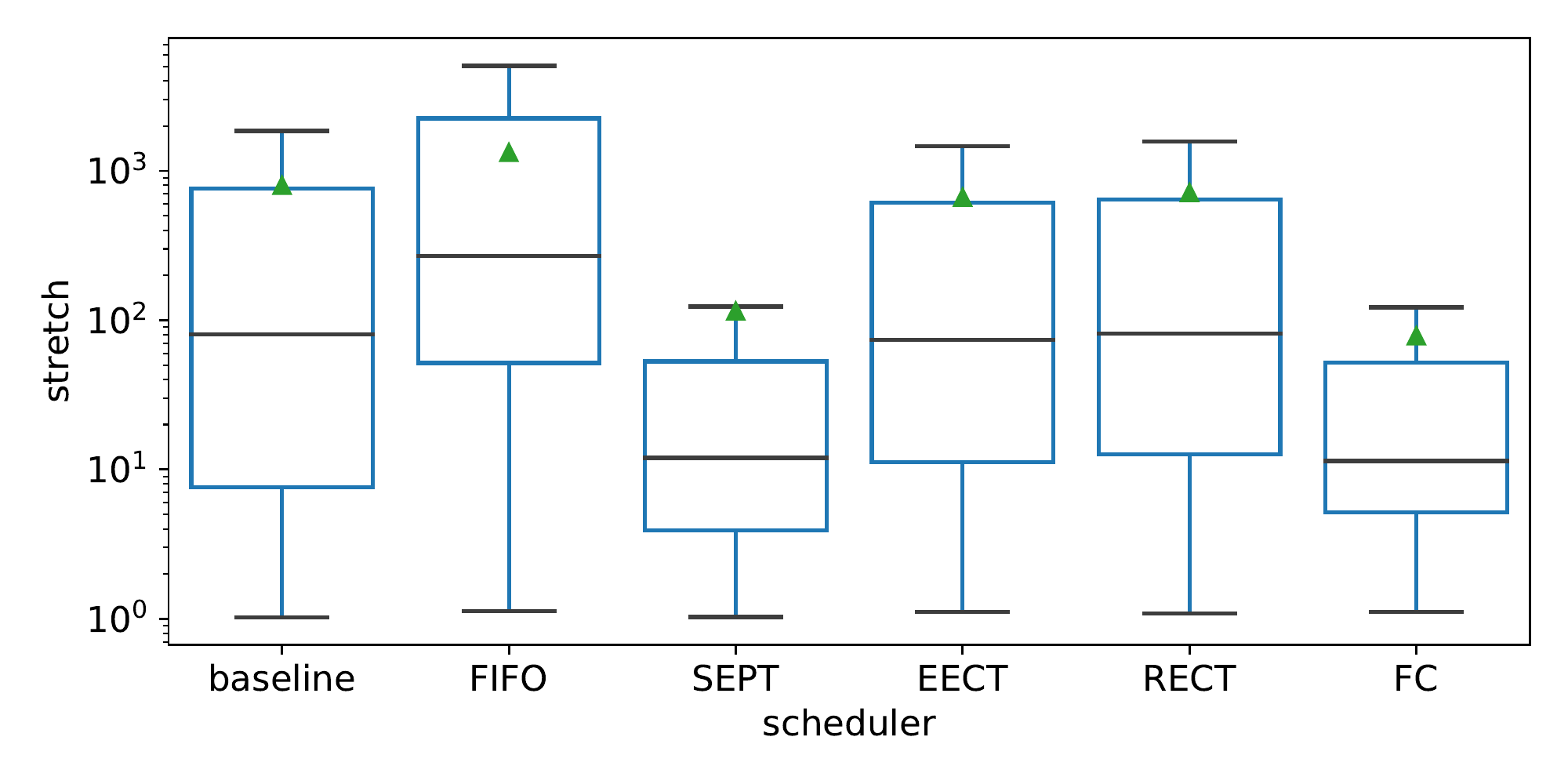}
    \chart{Experiment 3}{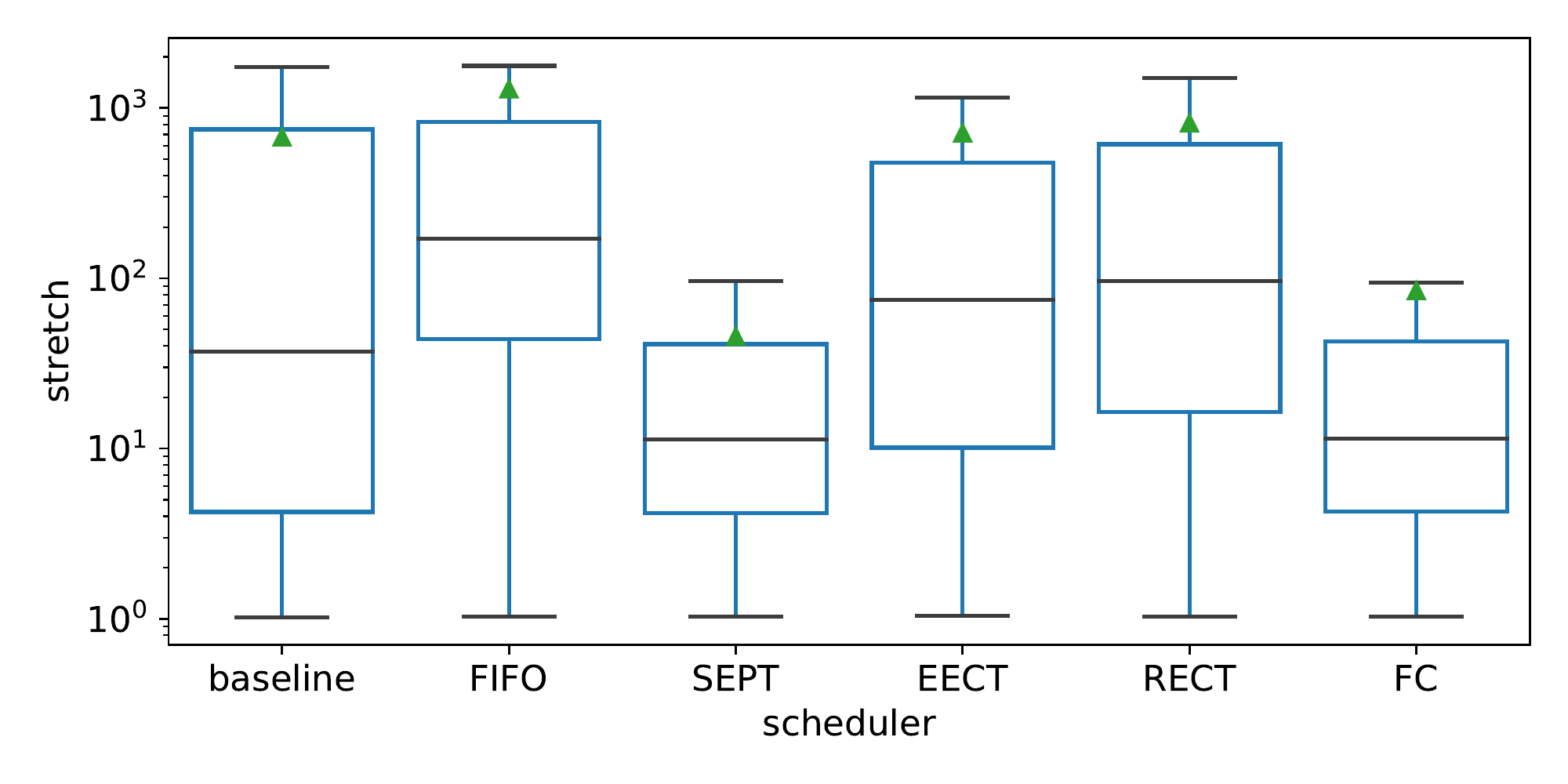}
    \\
    \chart{Experiment 4}{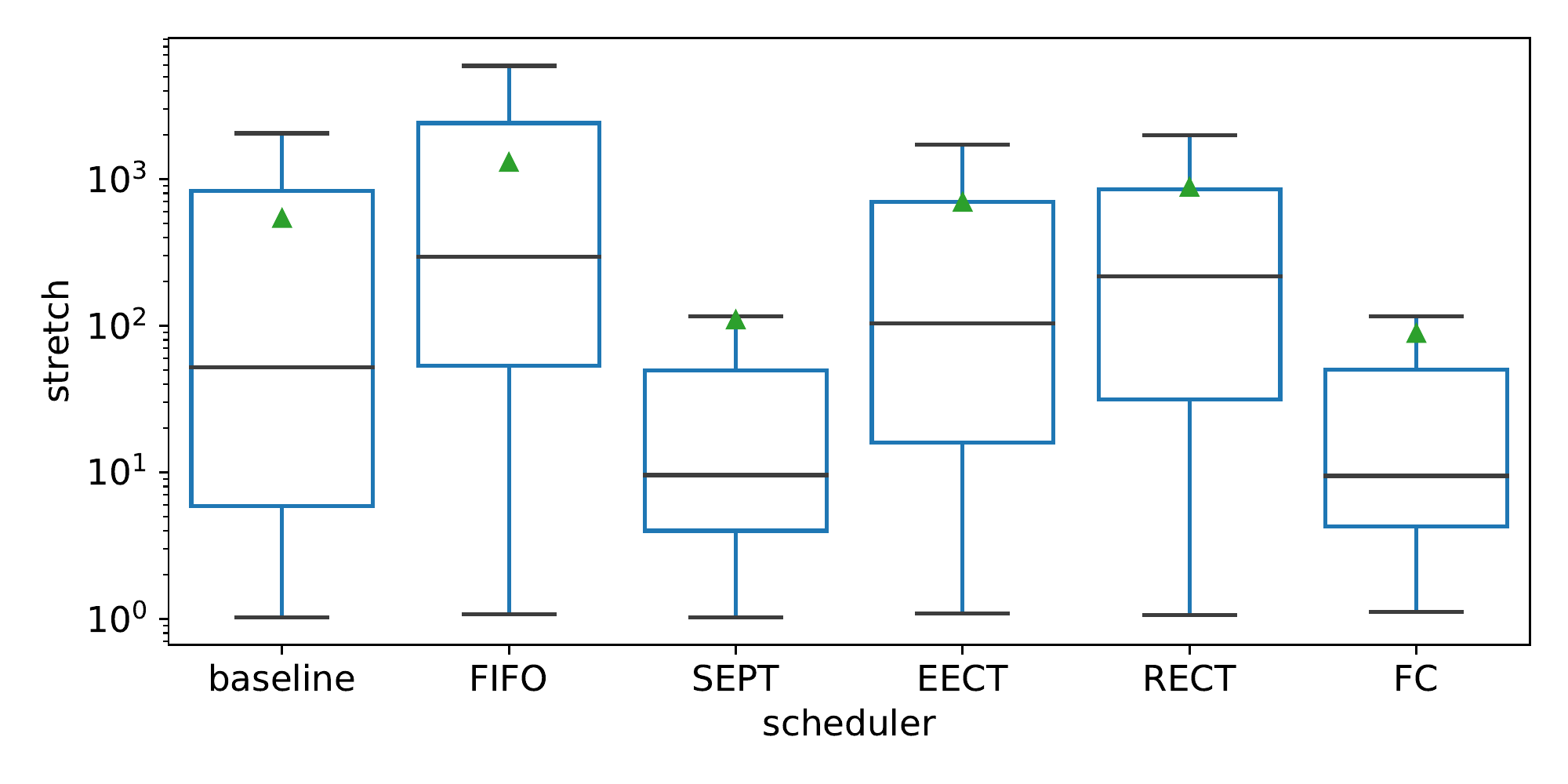}
    \chart{Experiment 5}{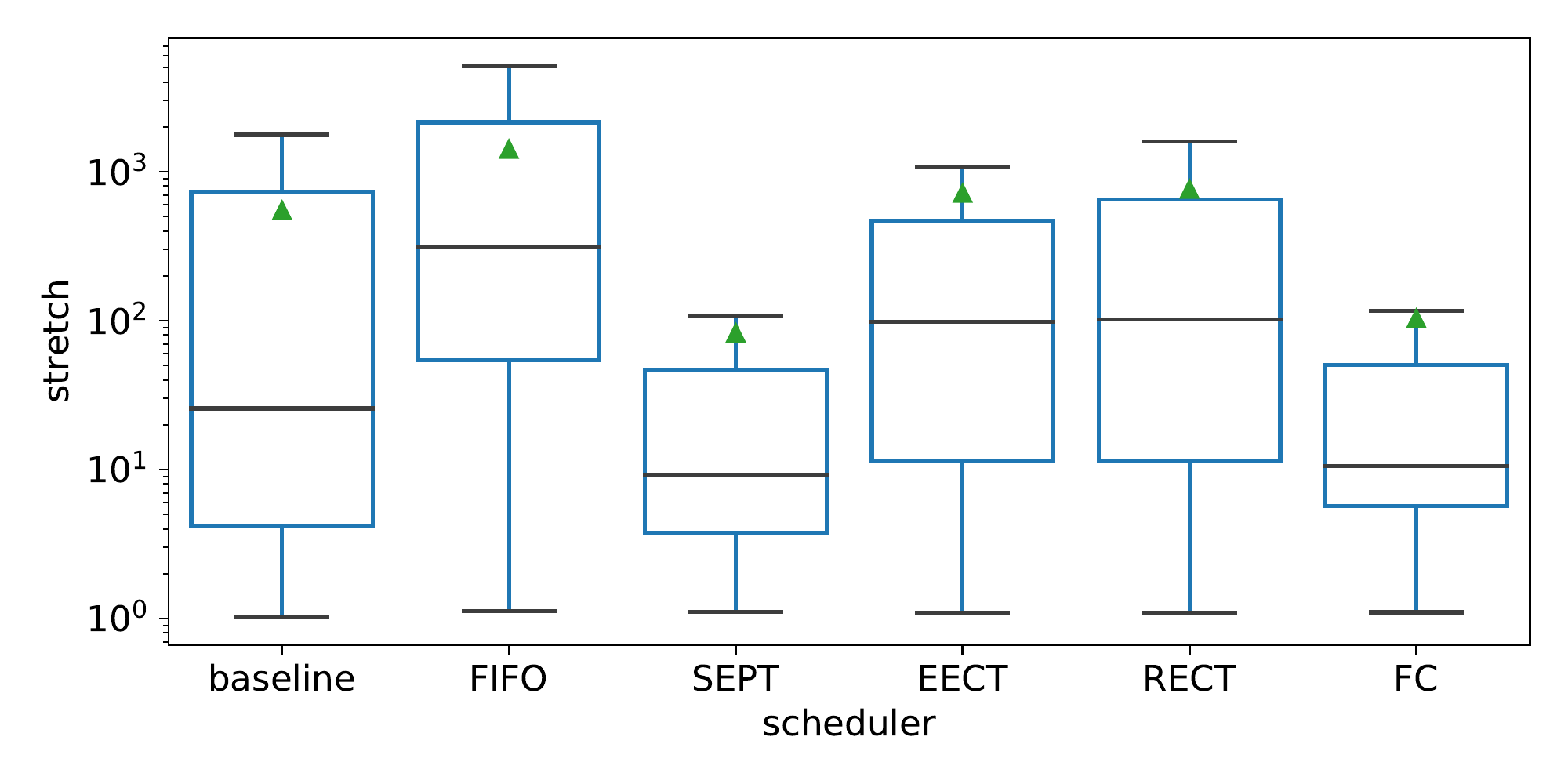}
    \\
    \caption{Stretch for different scheduling policies. Here, we present results for 5 CPUs and intensity 60. On-premise infrastructure. Average stretch presented as a green triangle.}
    \label{fig:stretch_5_60}
\end{figure*}

\begin{figure*}[h]
    \centering
    \chart{Experiment 1}{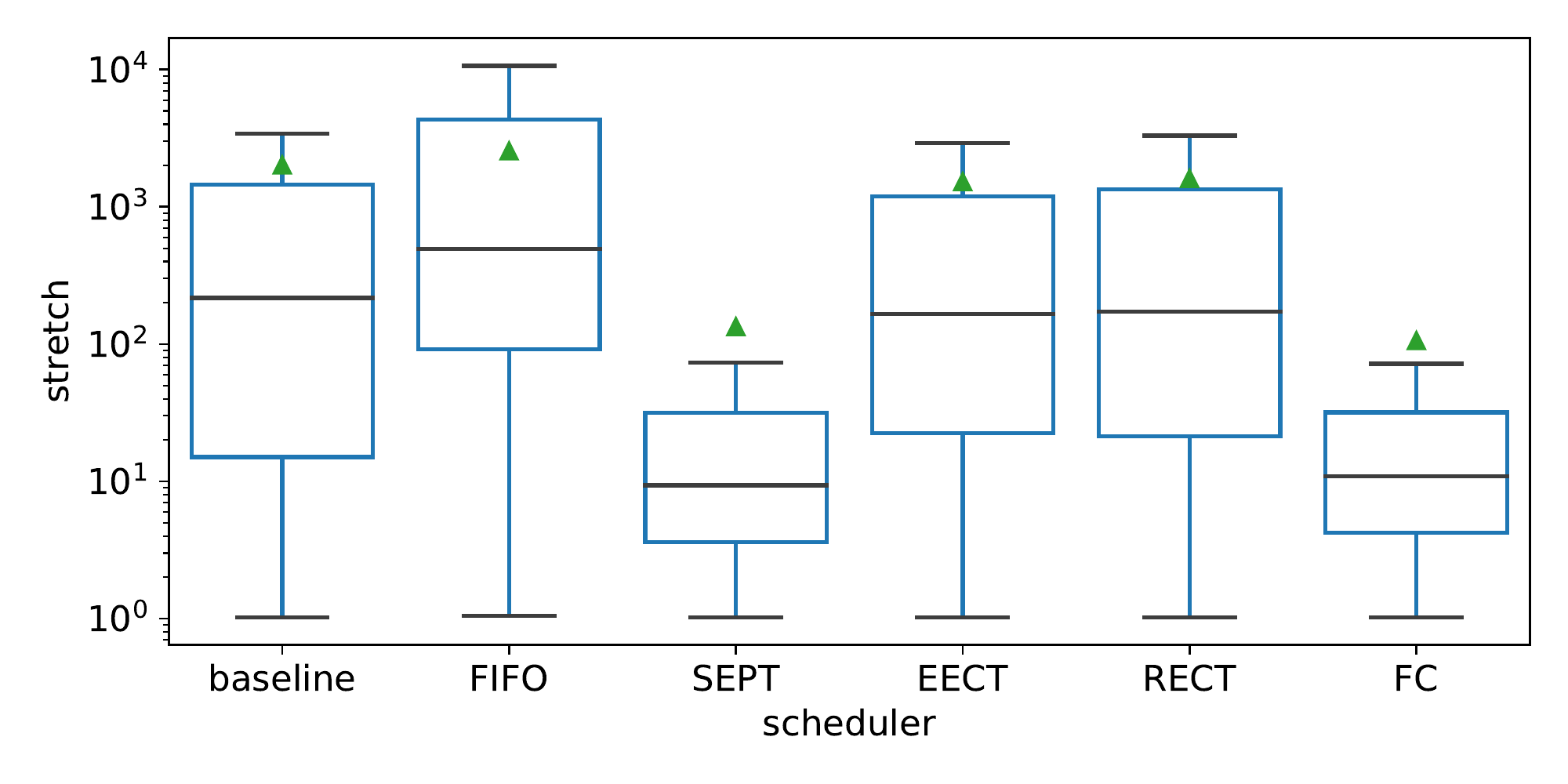}
    \chart{Experiment 2}{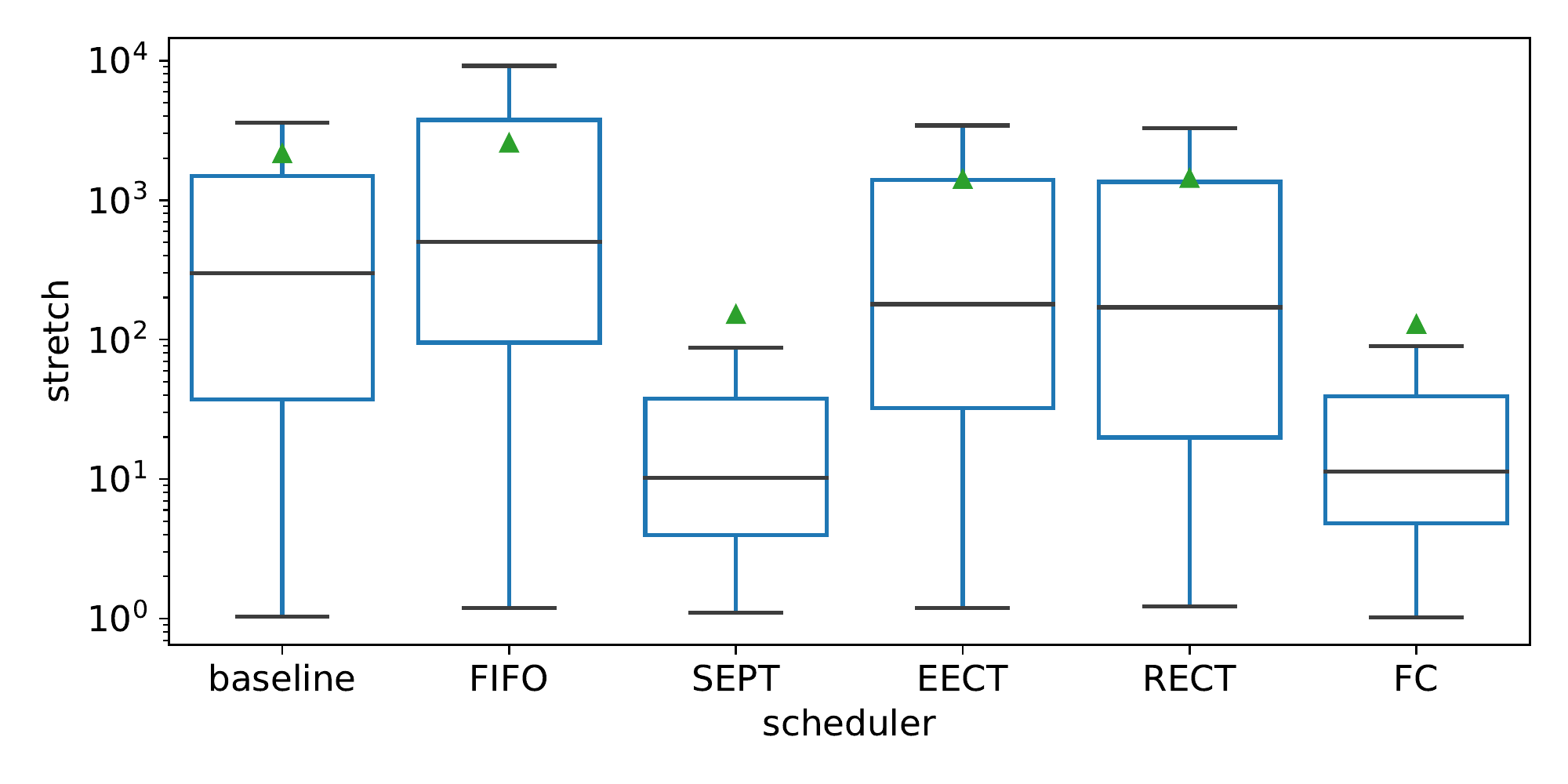}
    \chart{Experiment 3}{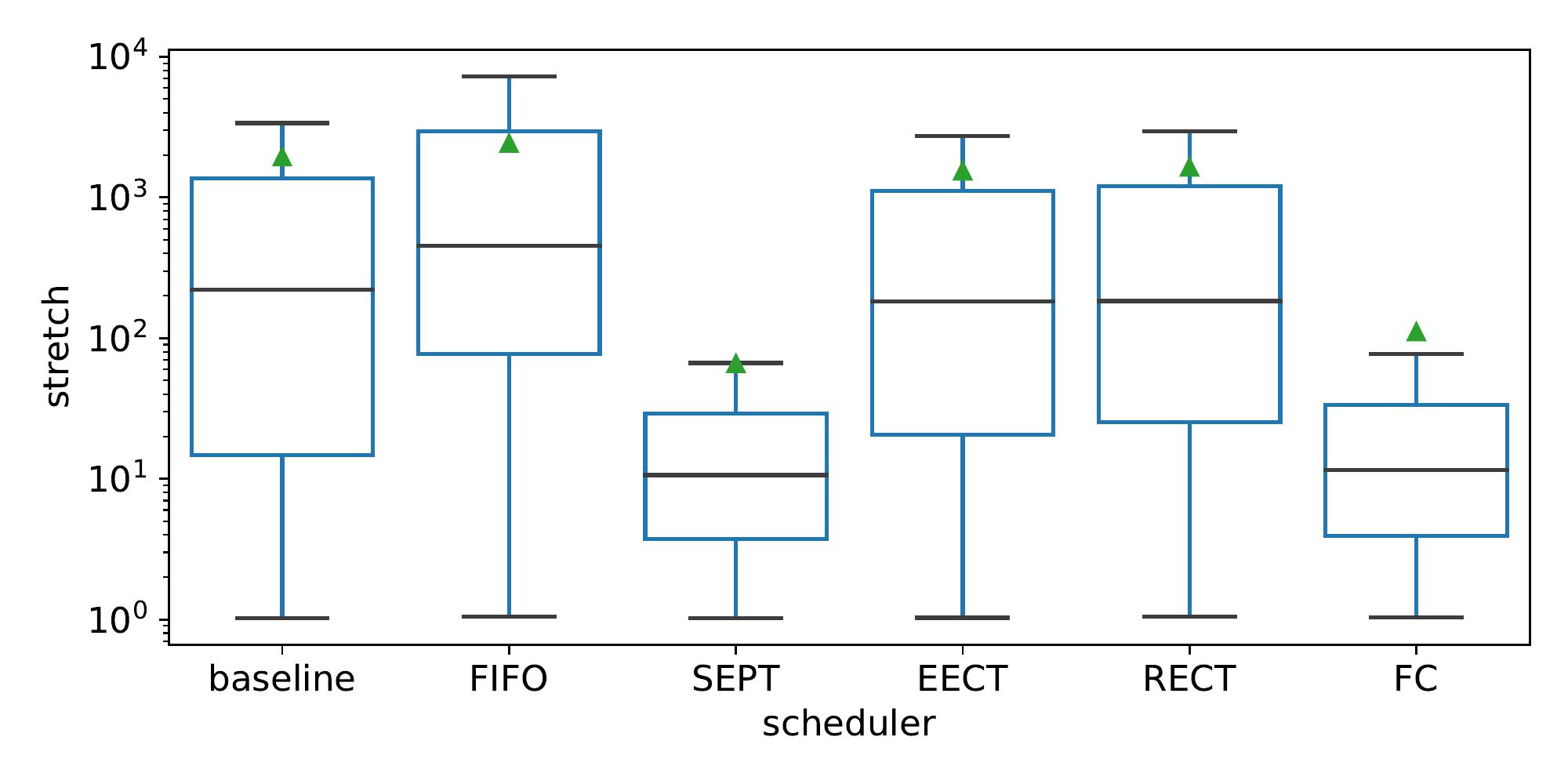}
    \\
    \chart{Experiment 4}{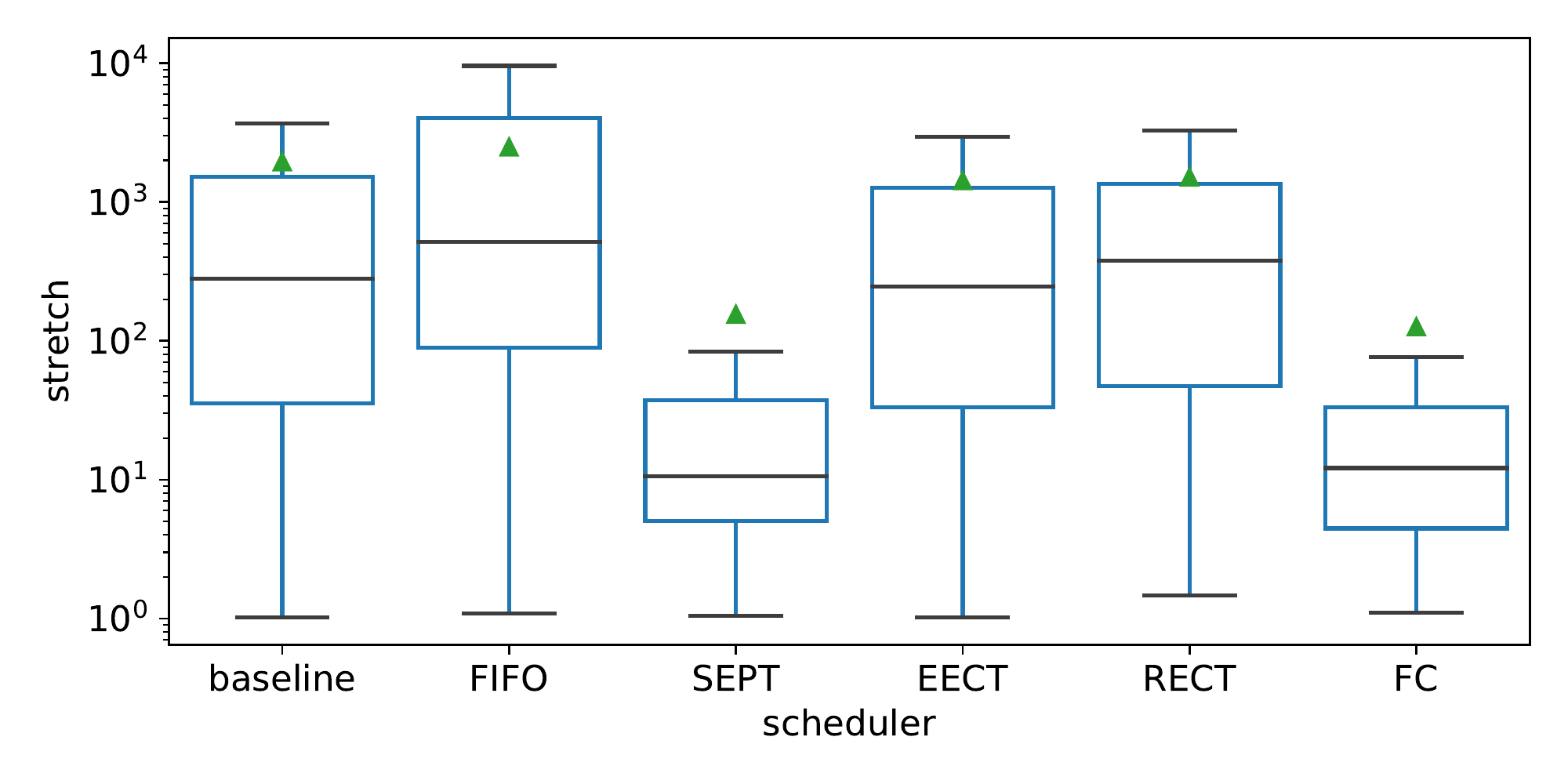}
    \chart{Experiment 5}{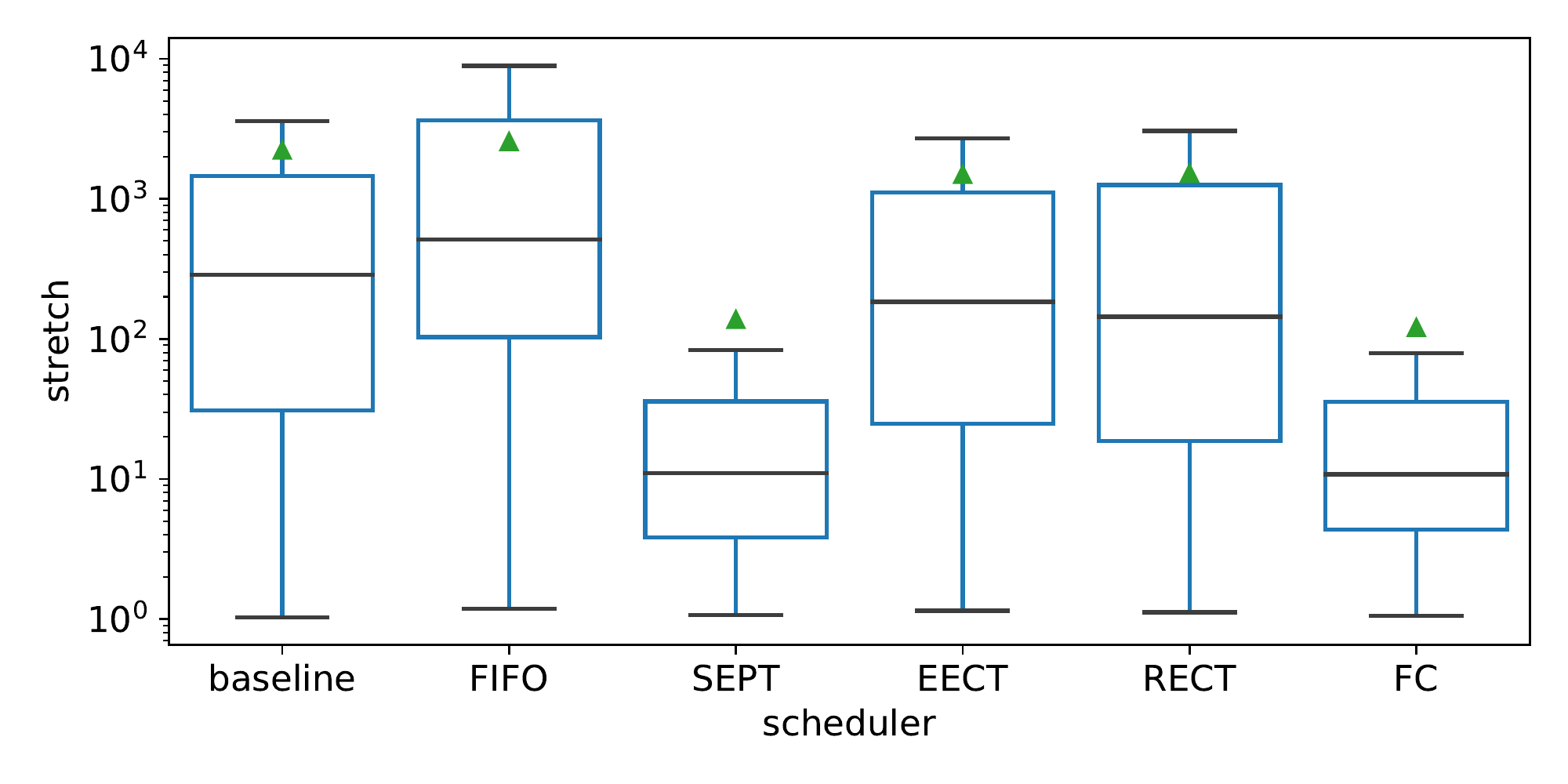}
    \\
    \caption{Stretch for different scheduling policies. Here, we present results for 5 CPUs and intensity 90. On-premise infrastructure. Average stretch presented as a green triangle.}
    \label{fig:stretch_5_90}
\end{figure*}

\begin{figure*}[h]
    \centering
    \chart{Experiment 1}{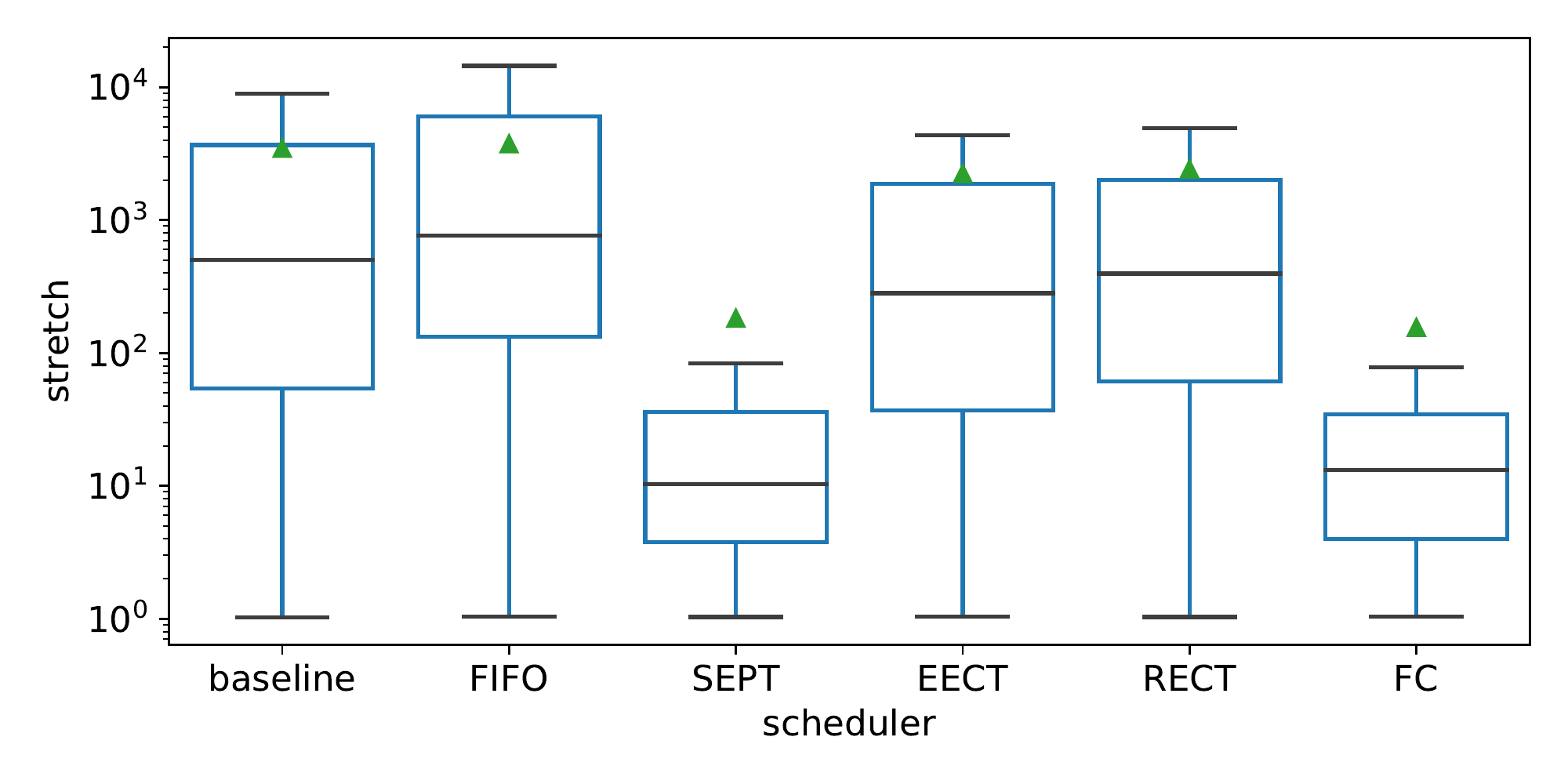}
    \chart{Experiment 2}{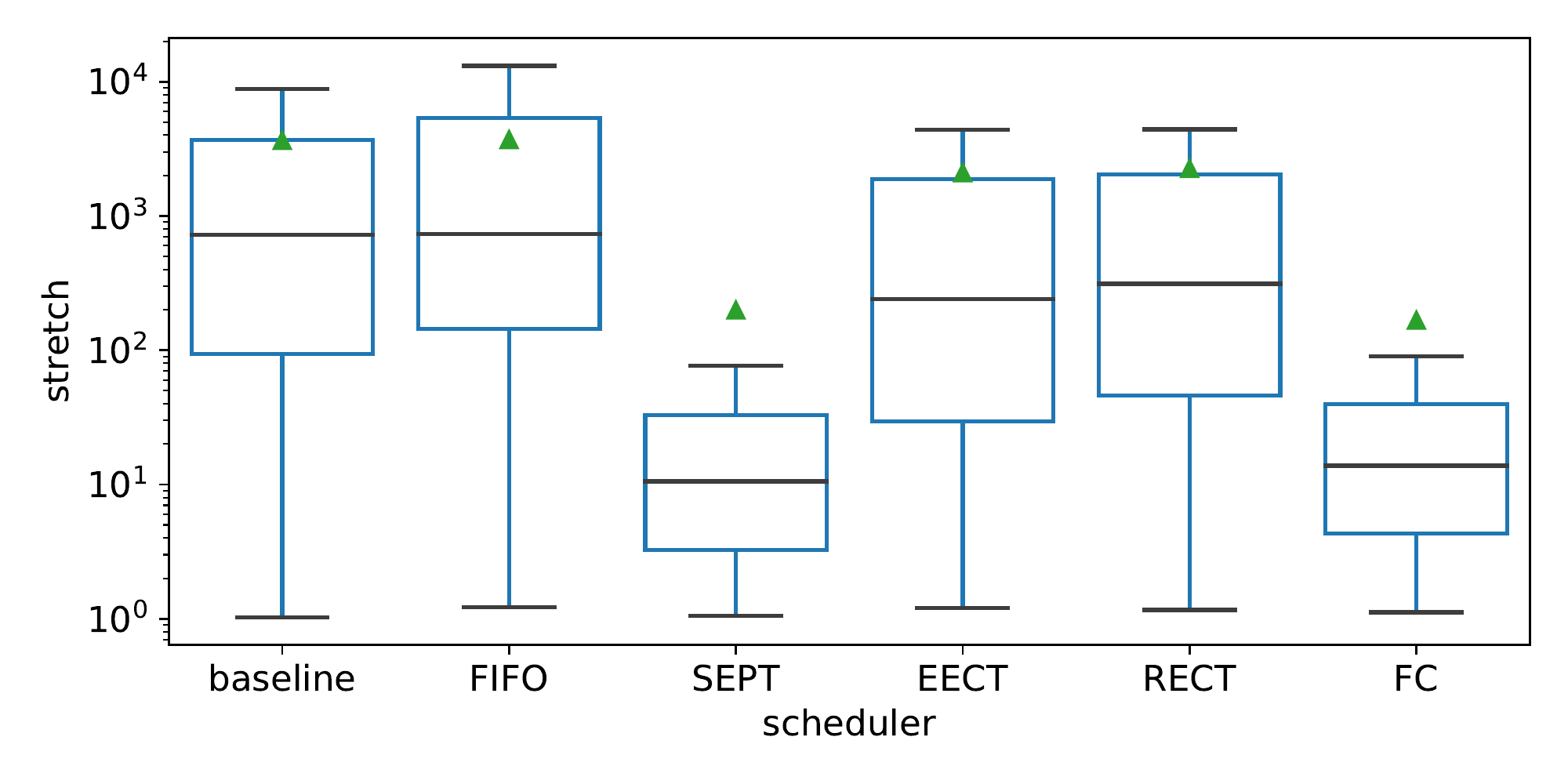}
    \chart{Experiment 3}{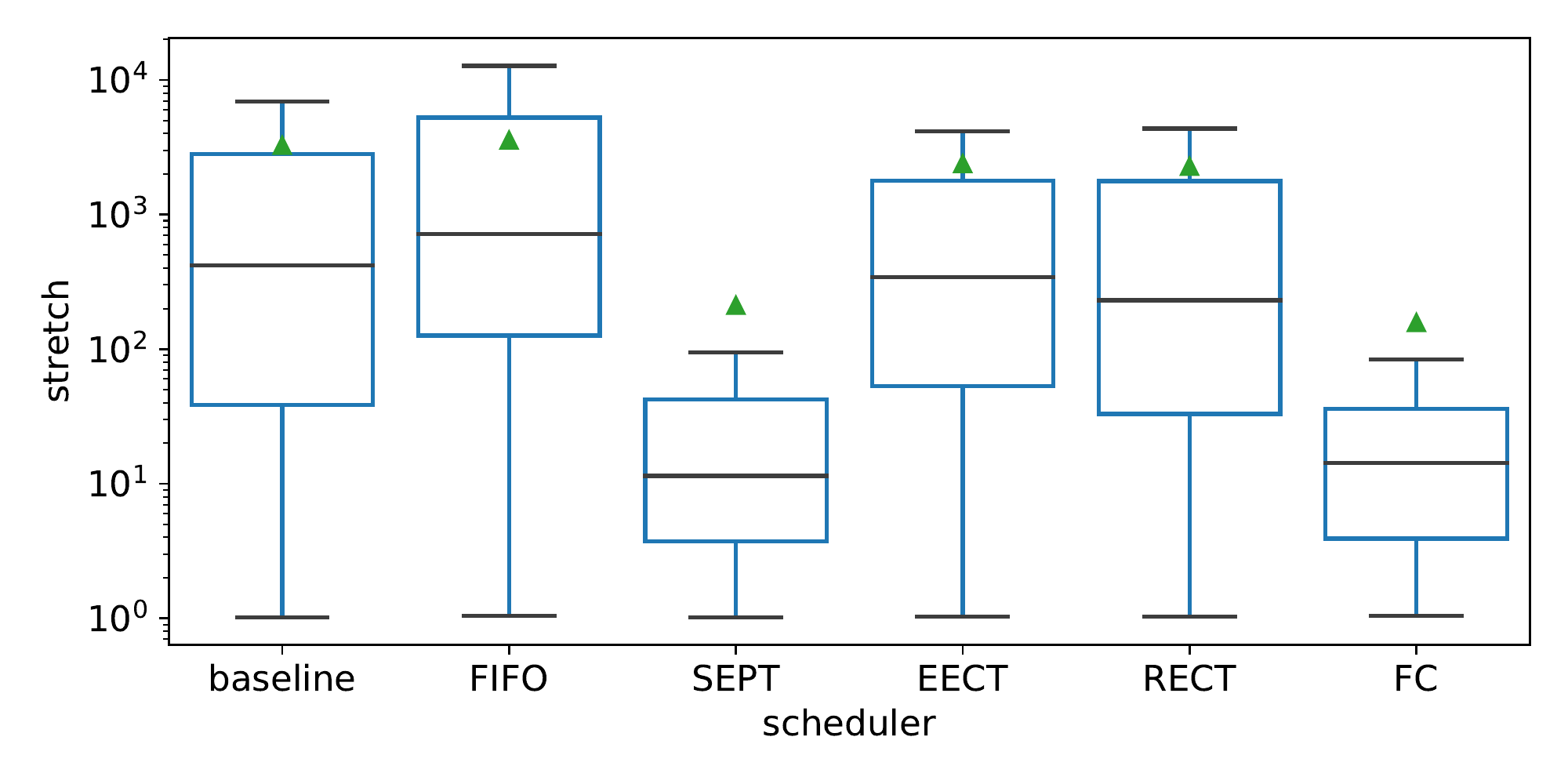}
    \\
    \chart{Experiment 4}{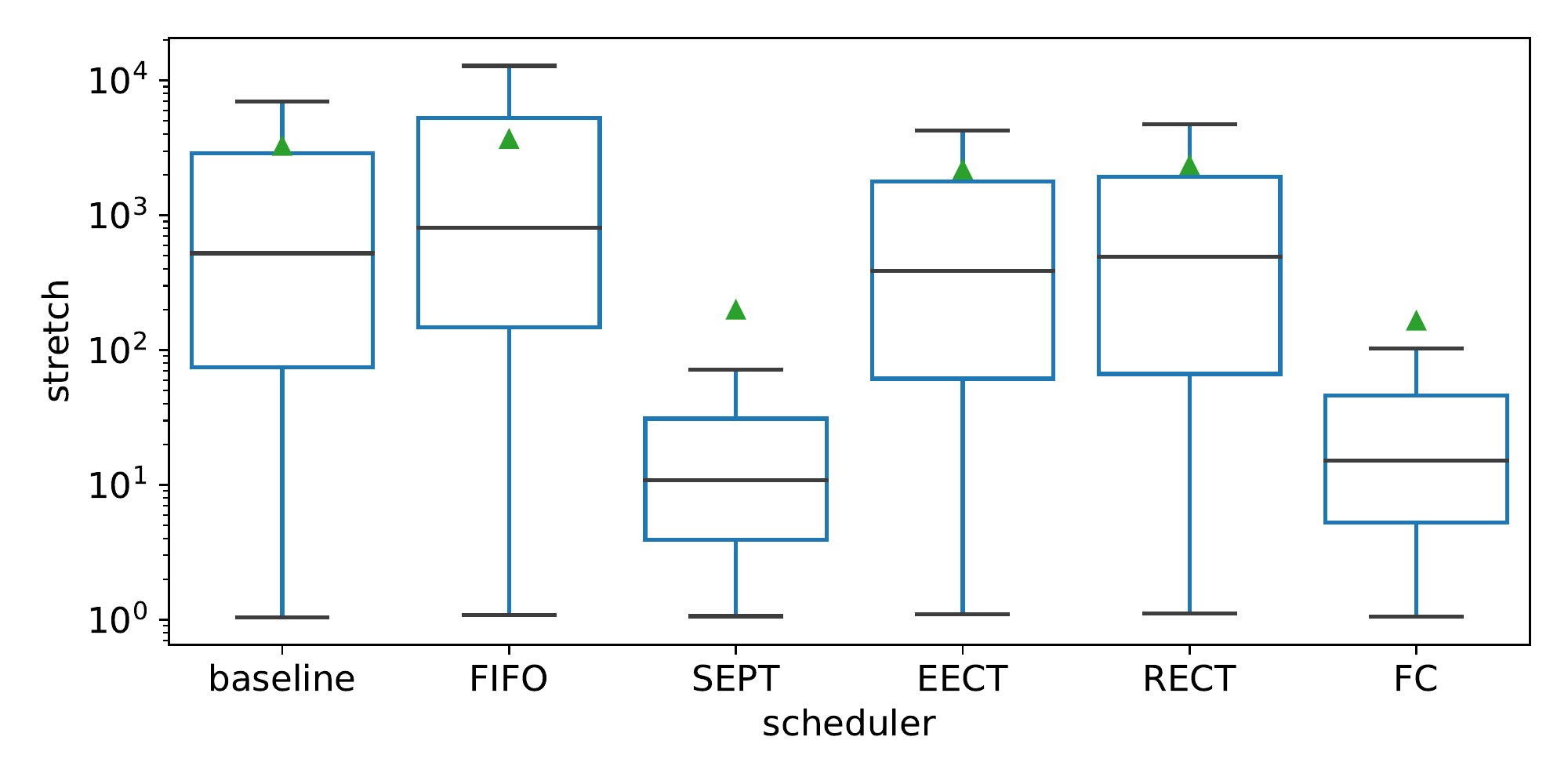}
    \chart{Experiment 5}{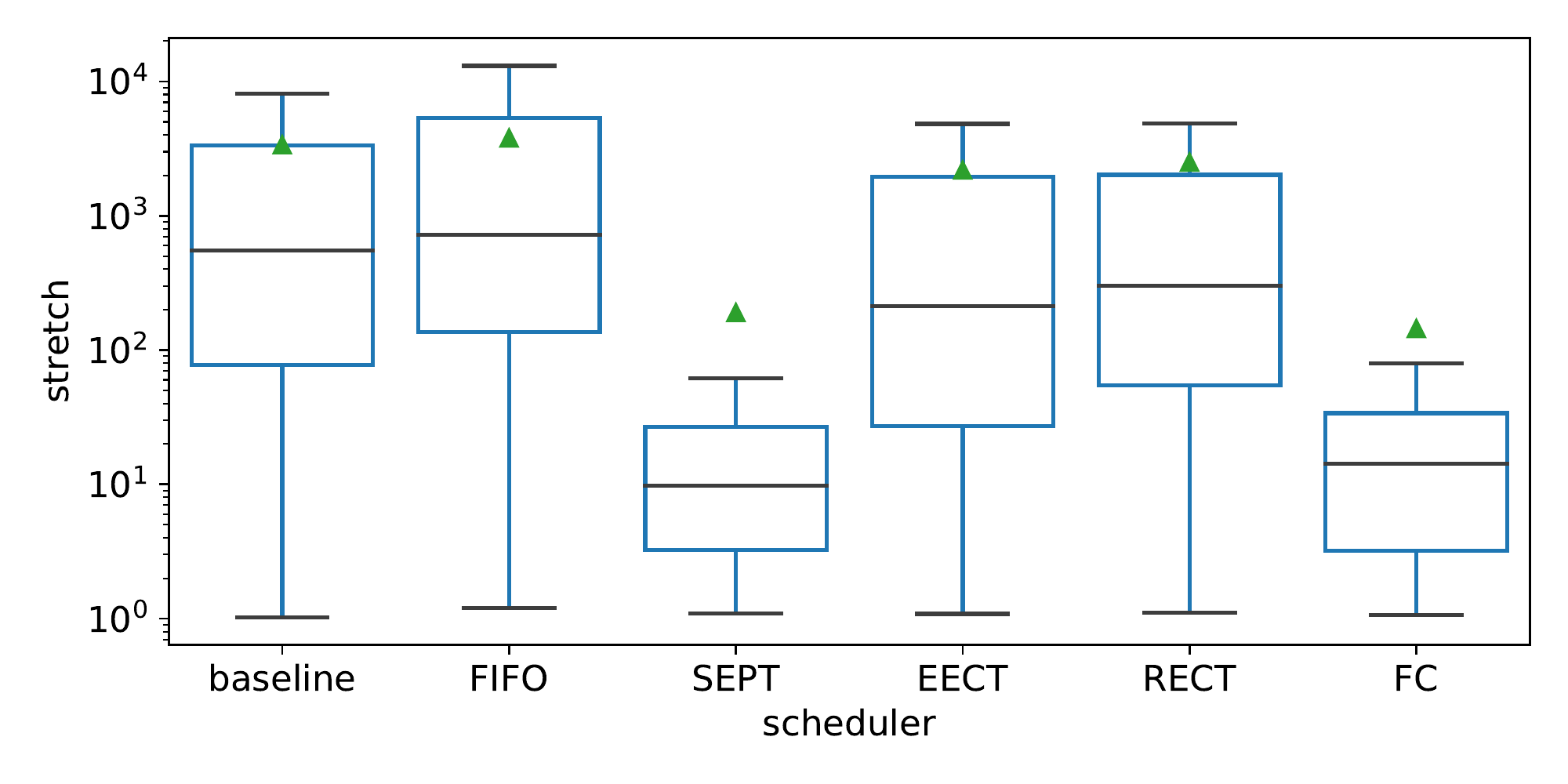}
    \\
    \caption{Stretch for different scheduling policies. Here, we present results for 5 CPUs and intensity 120. On-premise infrastructure. Average stretch presented as a green triangle.}
    \label{fig:stretch_5_120}
\end{figure*}

\begin{figure*}[h]
    \centering
    \chart{Experiment 1}{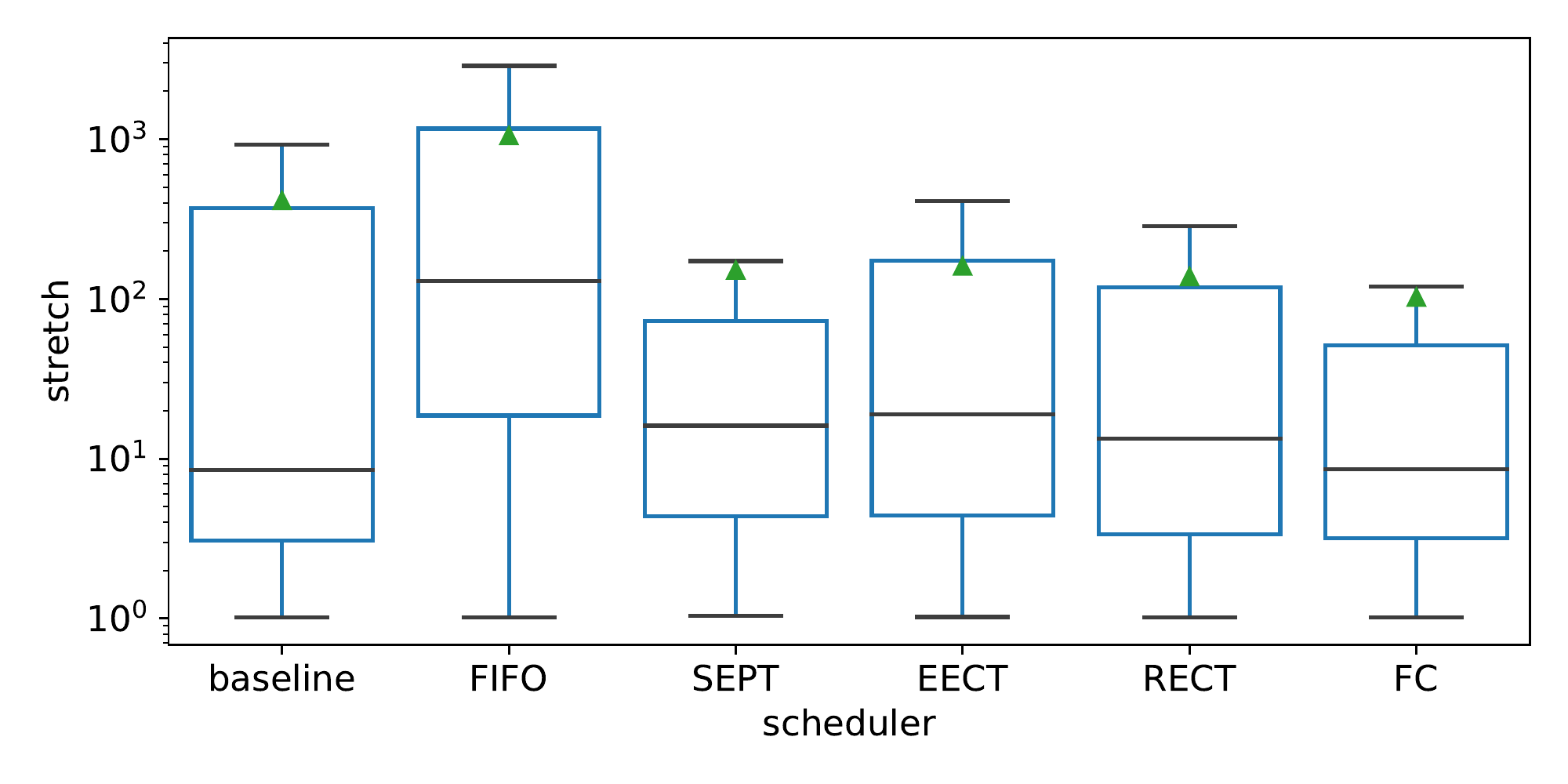}
    \chart{Experiment 2}{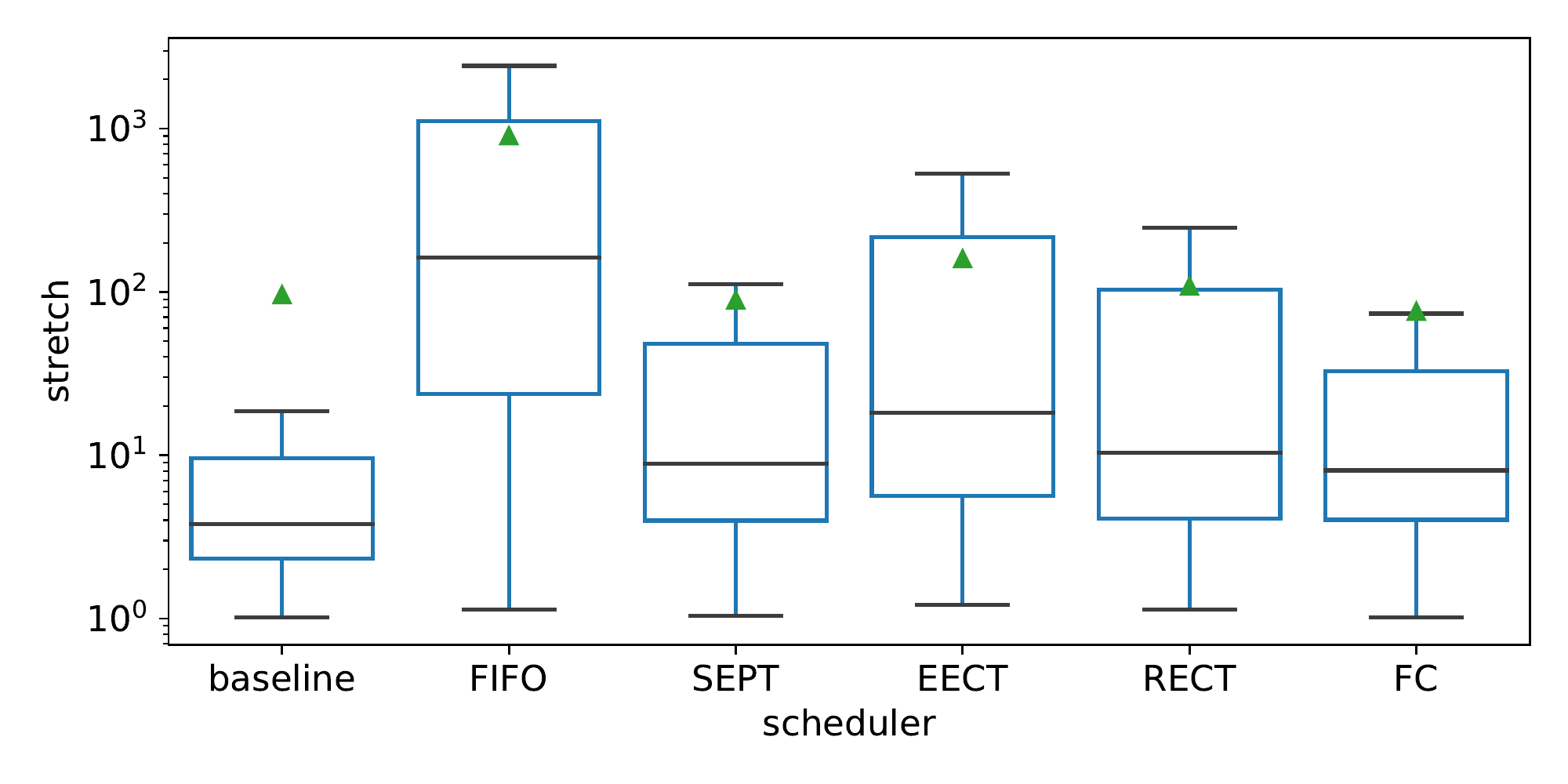}
    \chart{Experiment 3}{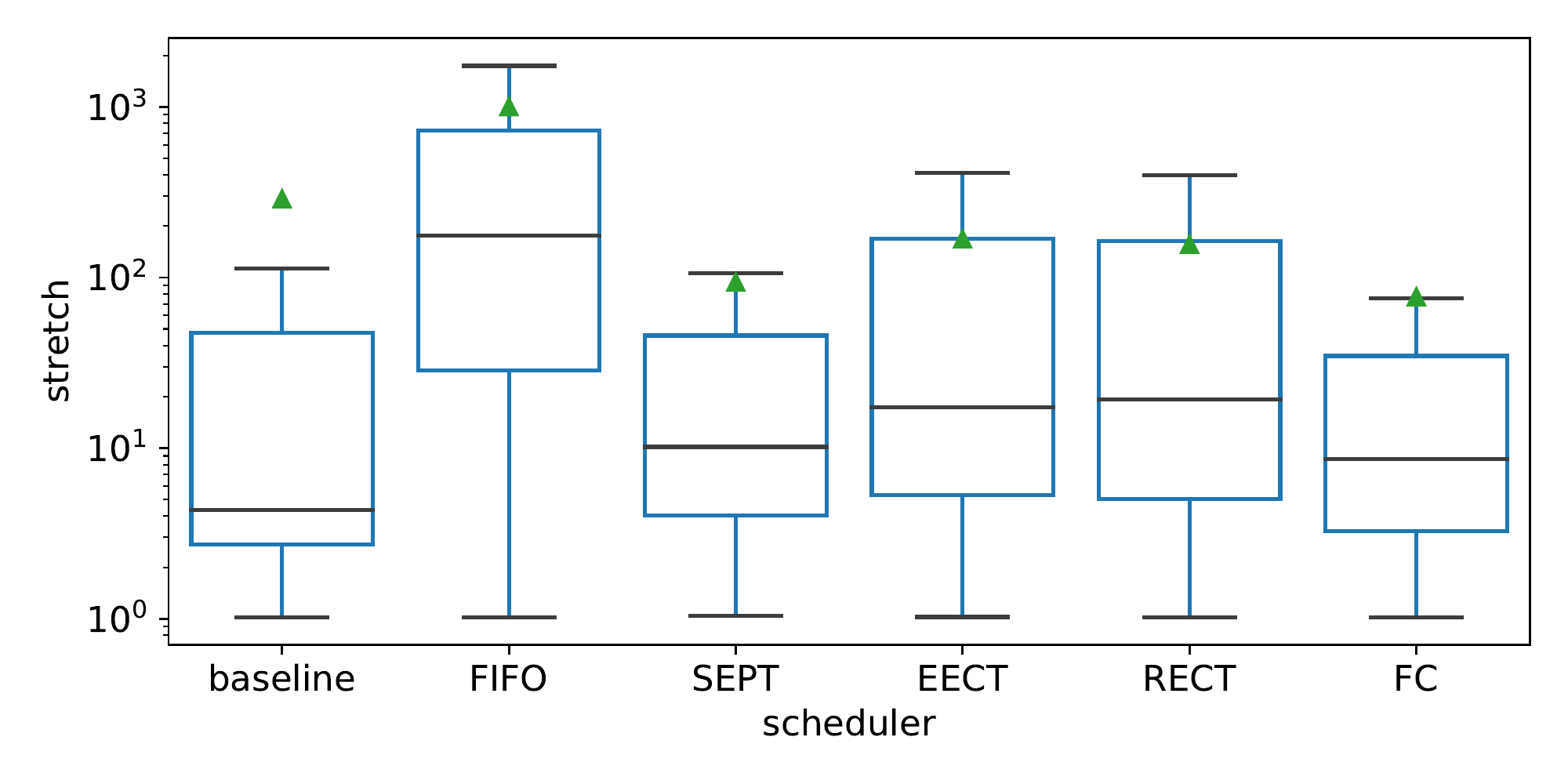}
    \\
    \chart{Experiment 4}{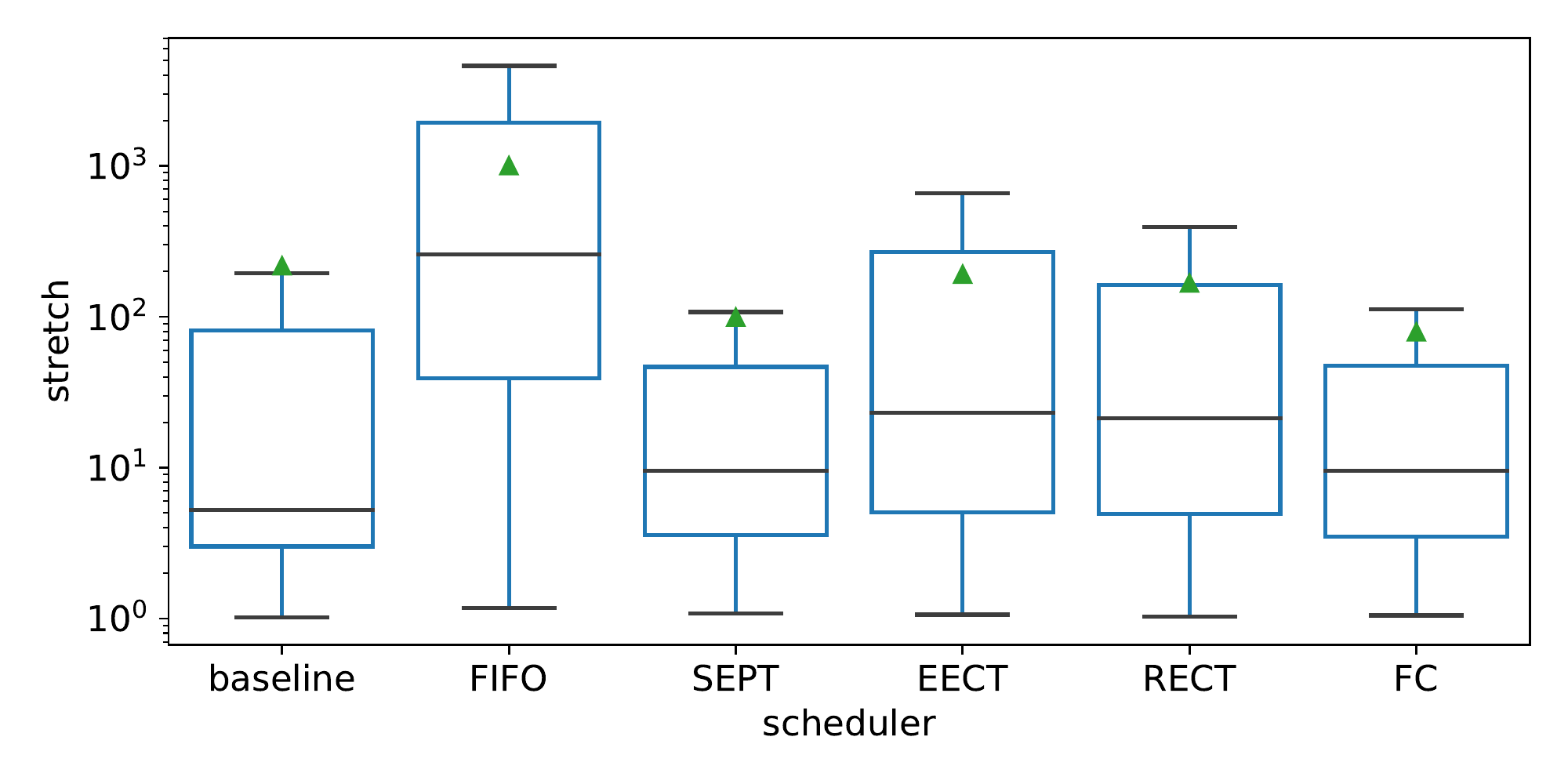}
    \chart{Experiment 5}{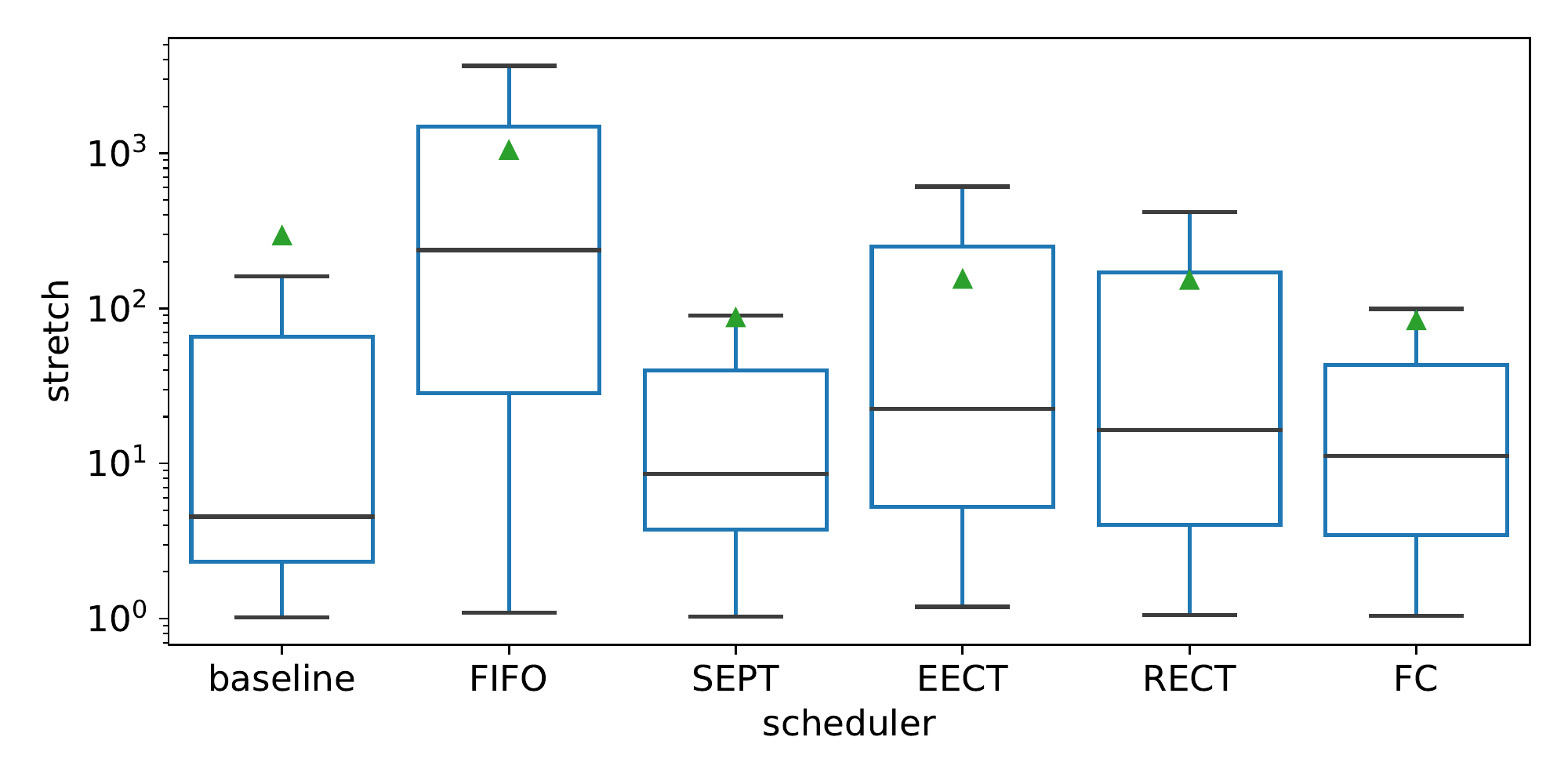}
    \\
    \caption{Stretch for different scheduling policies. Here, we present results for 10 CPUs and intensity 30. On-premise infrastructure. Average stretch presented as a green triangle.}
    \label{fig:stretch_10_30}
\end{figure*}

\begin{figure*}[h]
    \centering
    \chart{Experiment 1}{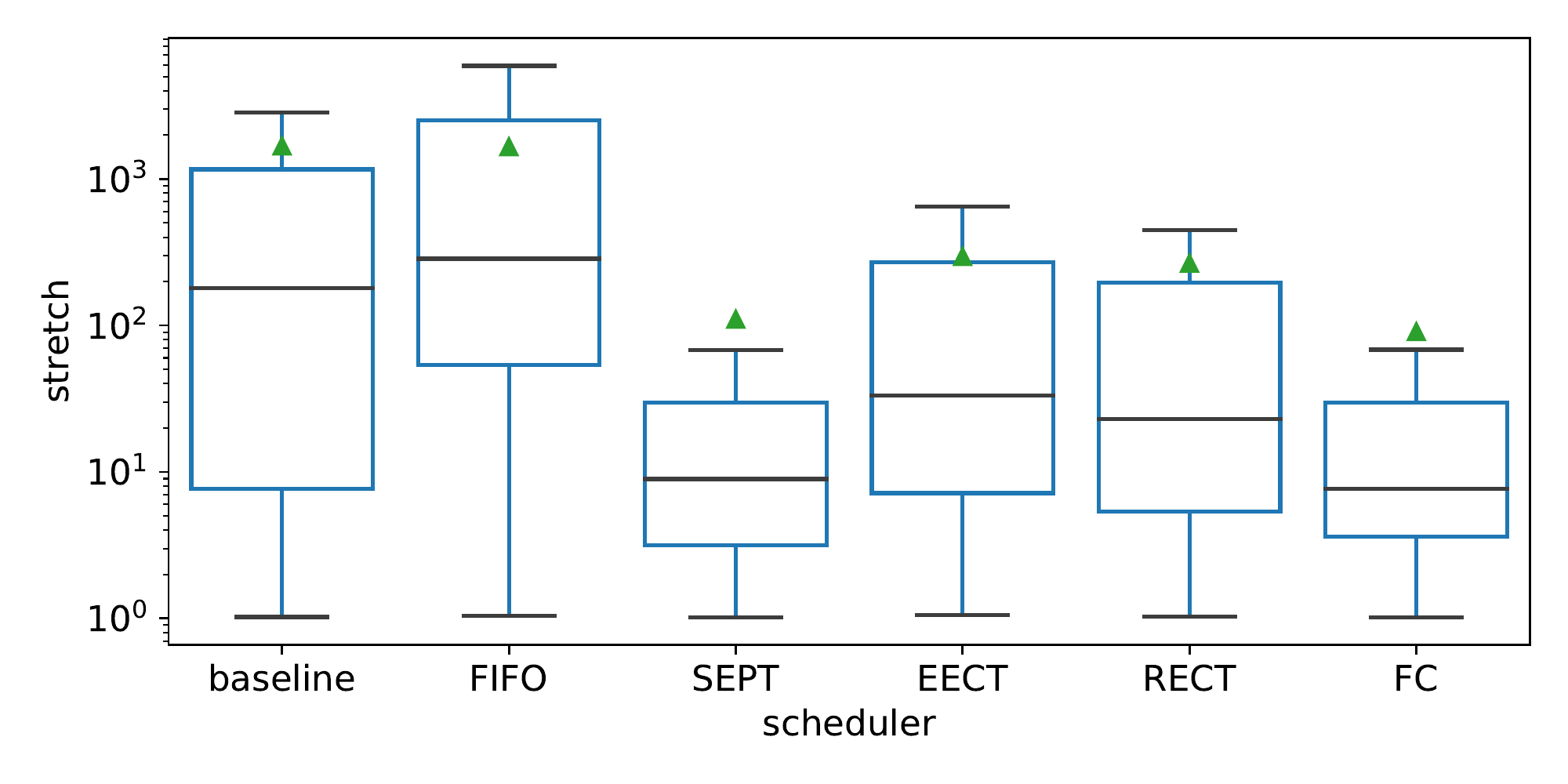}
    \chart{Experiment 2}{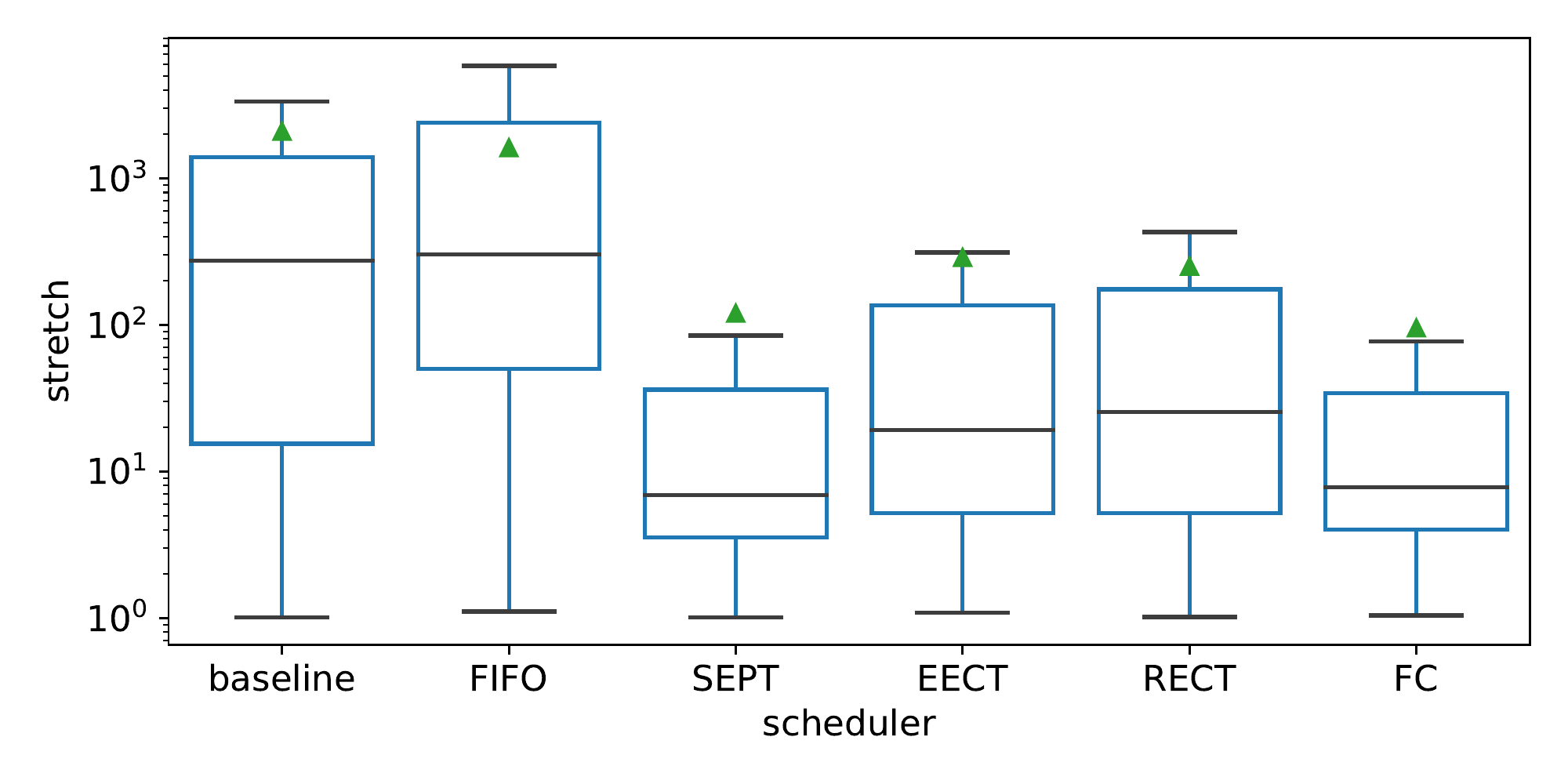}
    \chart{Experiment 3}{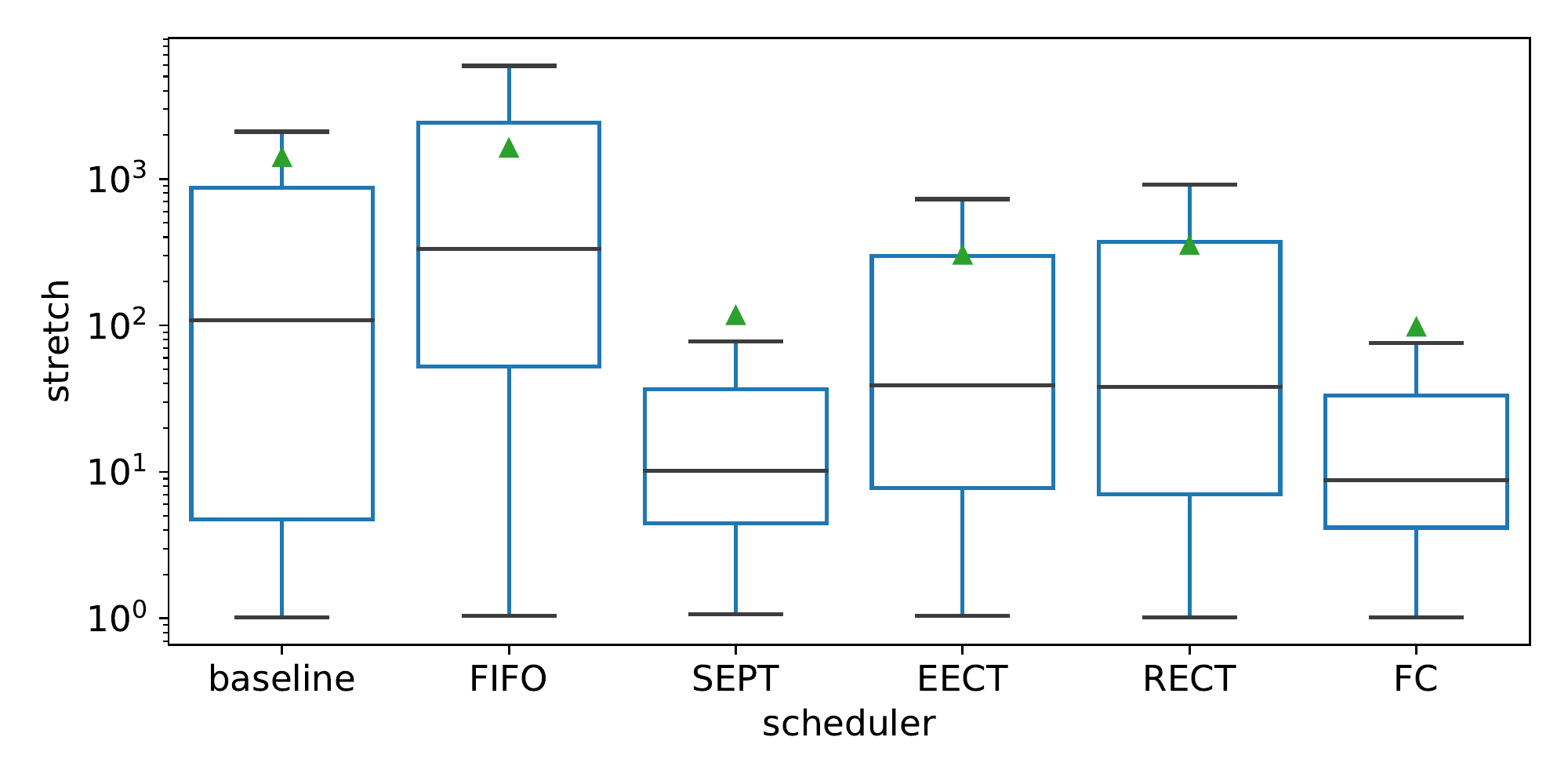}
    \\
    \chart{Experiment 4}{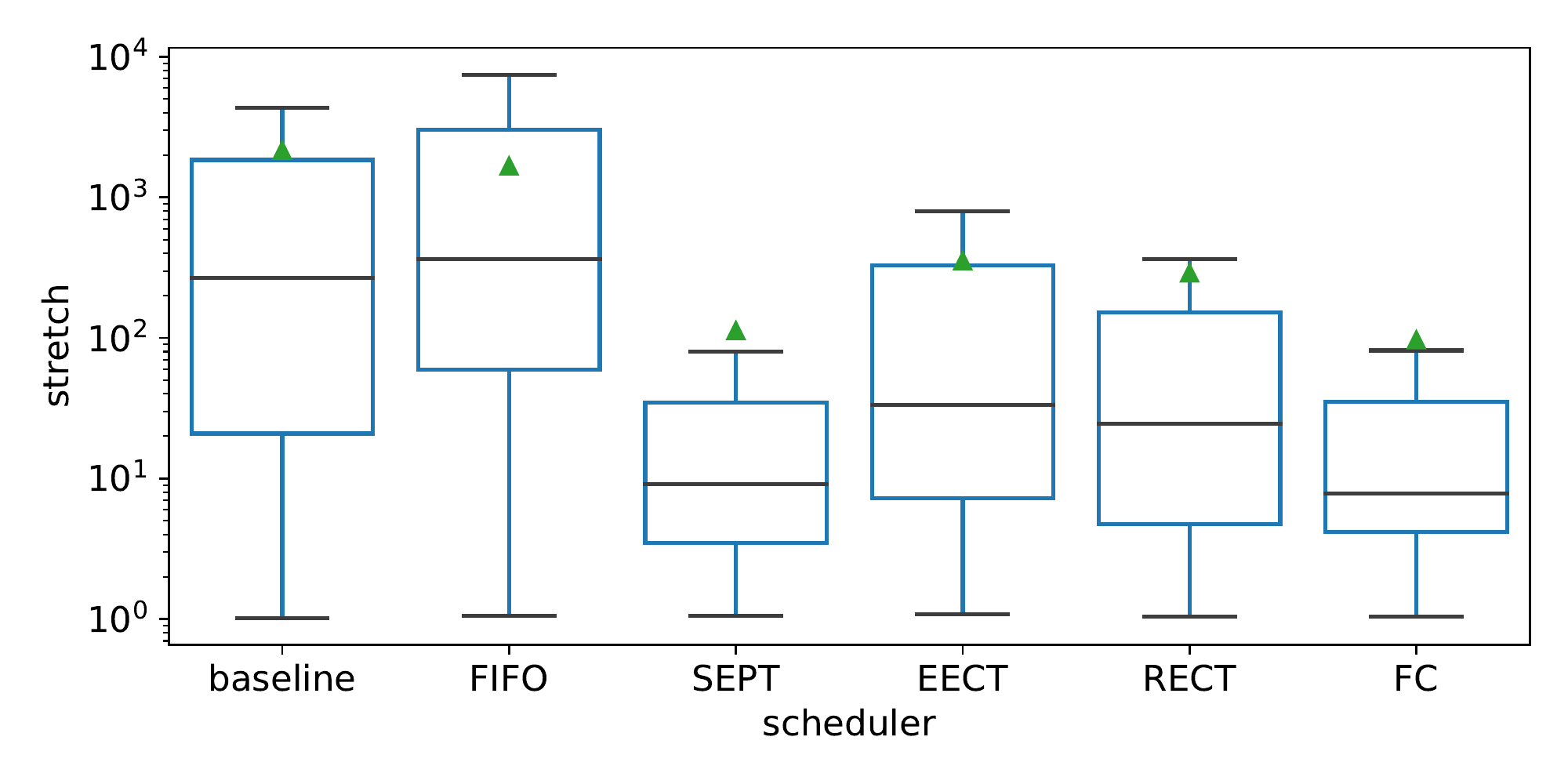}
    \chart{Experiment 5}{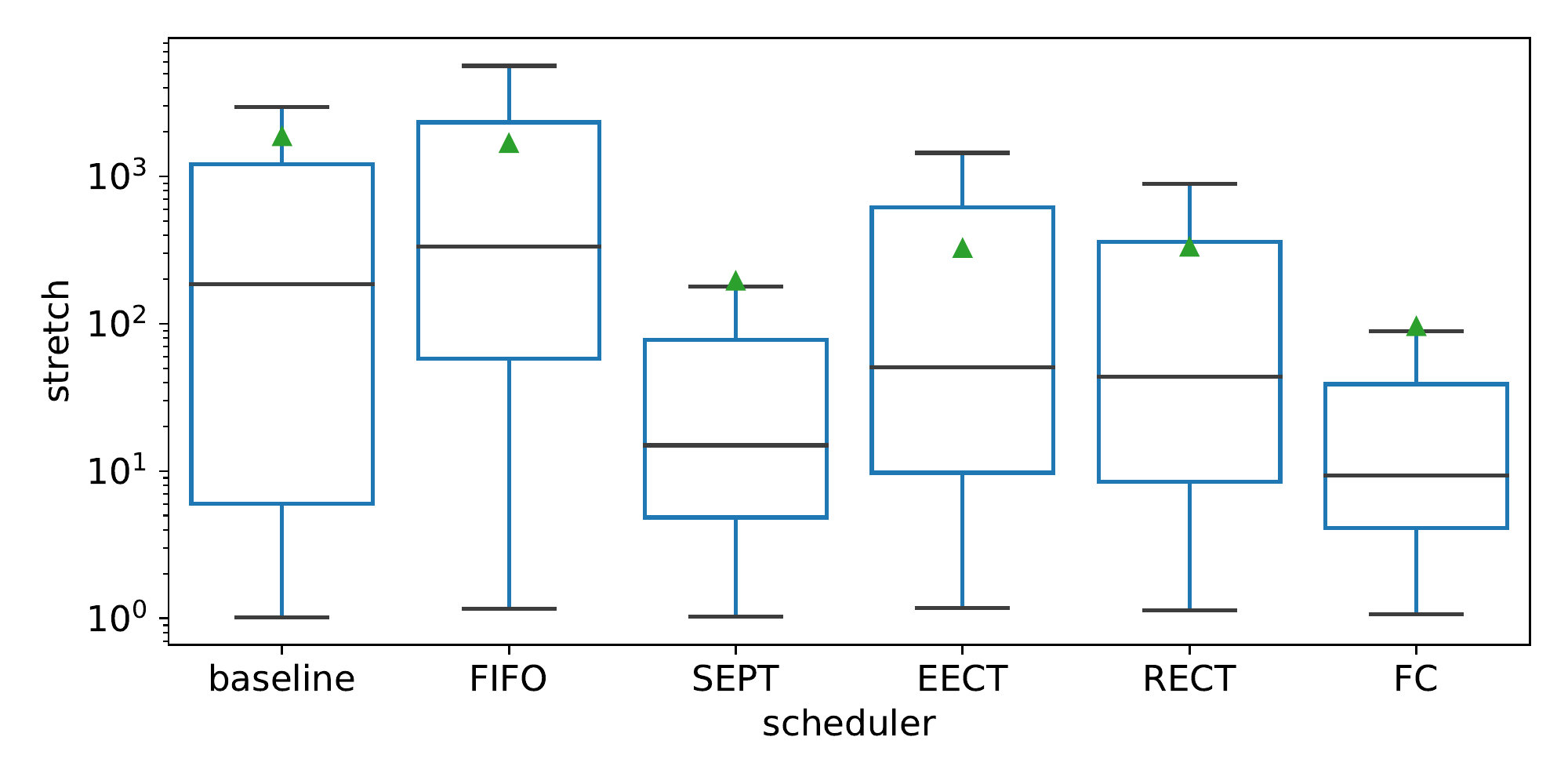}
    \\
    \caption{Stretch for different scheduling policies. Here, we present results for 10 CPUs and intensity 40. On-premise infrastructure. Average stretch presented as a green triangle.}
    \label{fig:stretch_10_40}
\end{figure*}

\begin{figure*}[h]
    \centering
    \chart{Experiment 1}{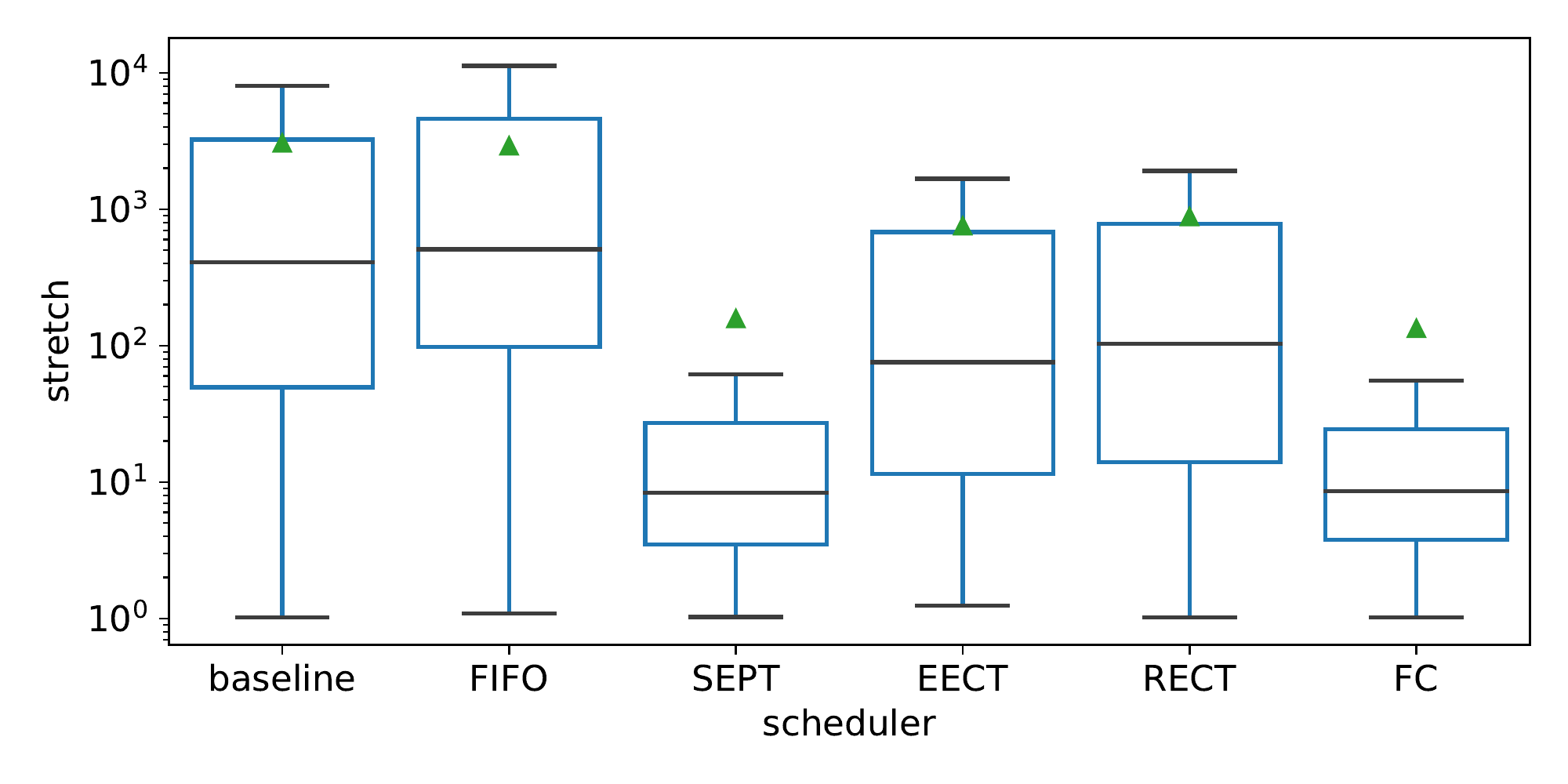}
    \chart{Experiment 2}{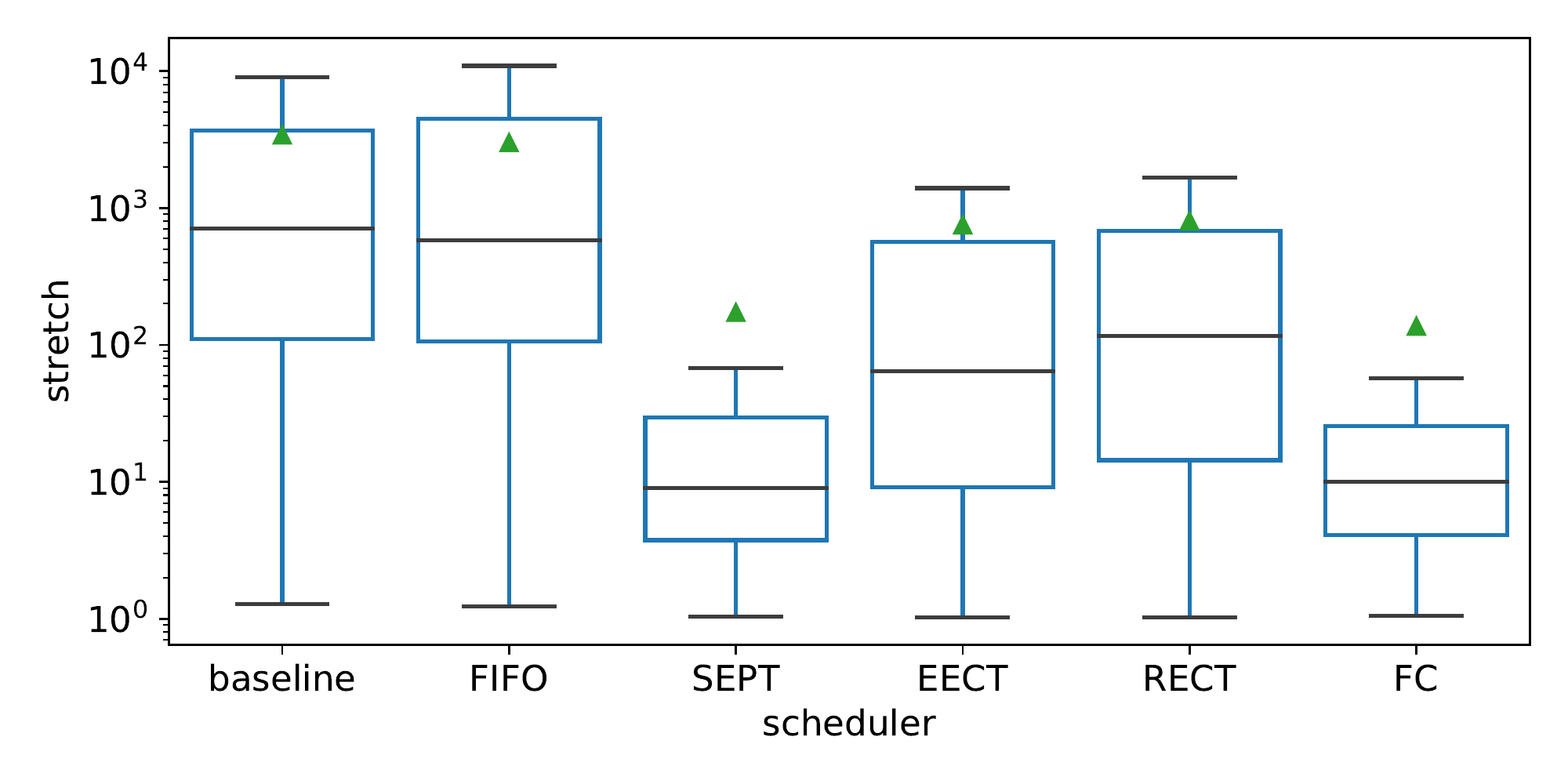}
    \chart{Experiment 3}{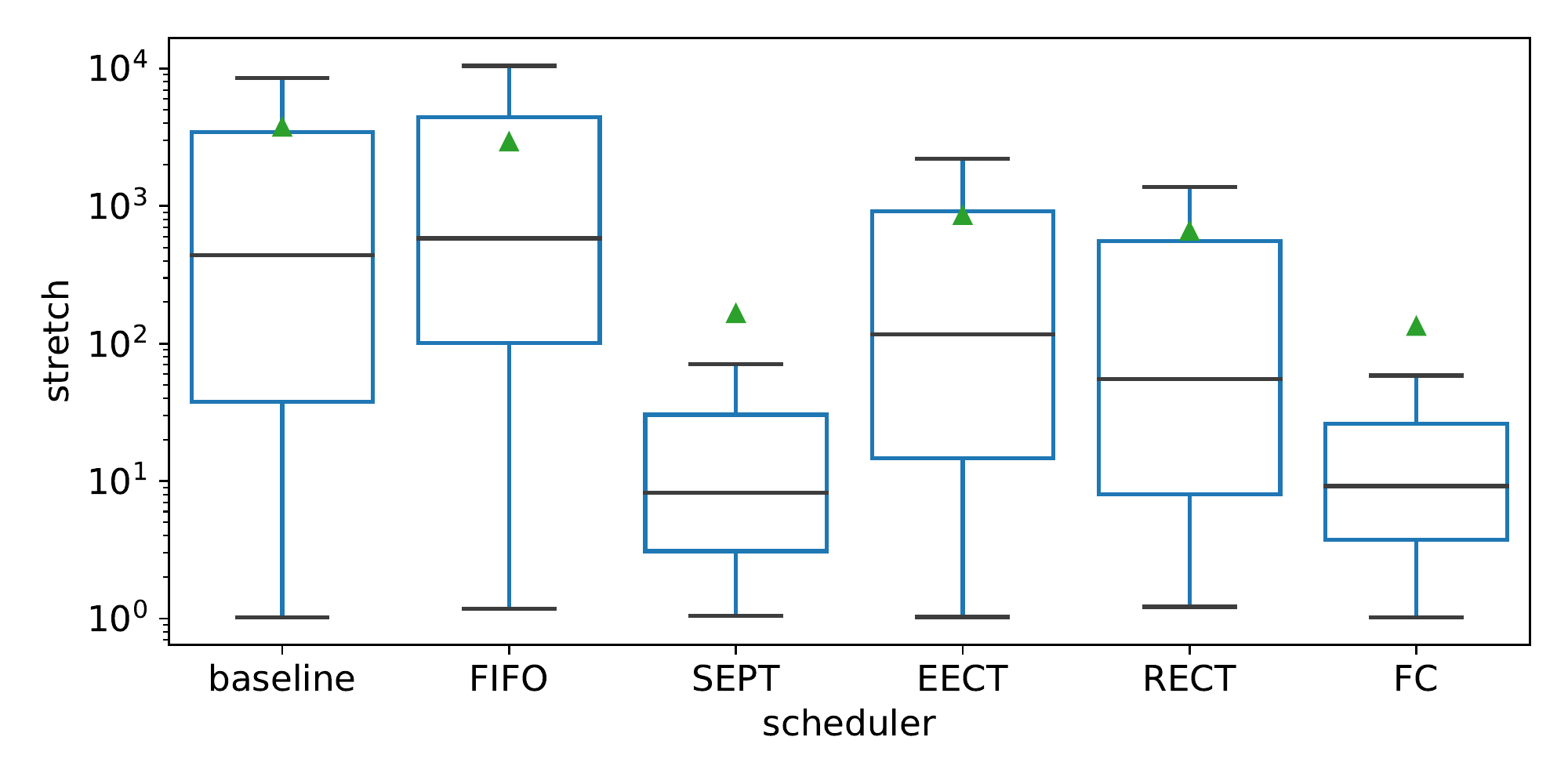}
    \\
    \chart{Experiment 4}{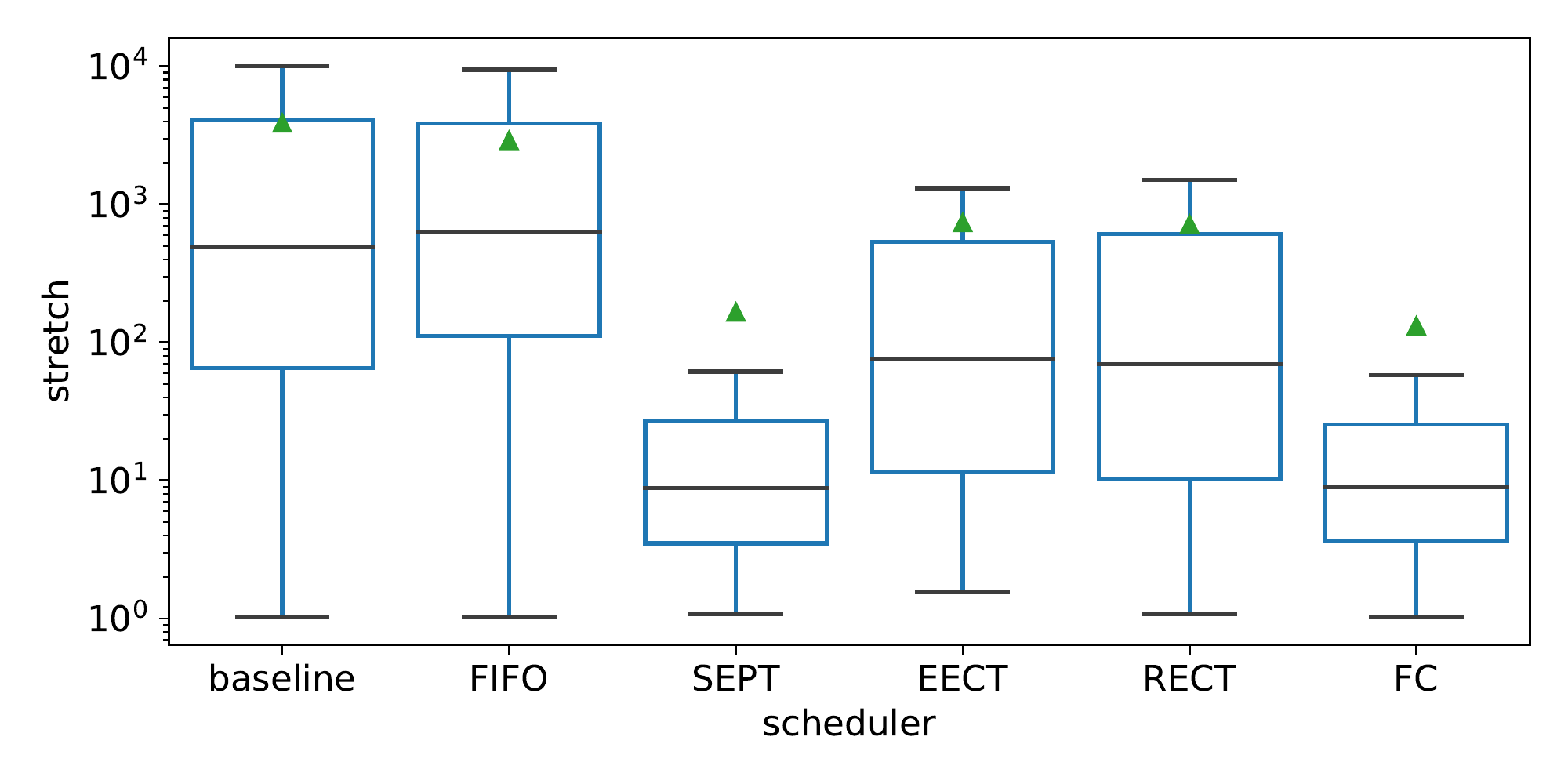}
    \chart{Experiment 5}{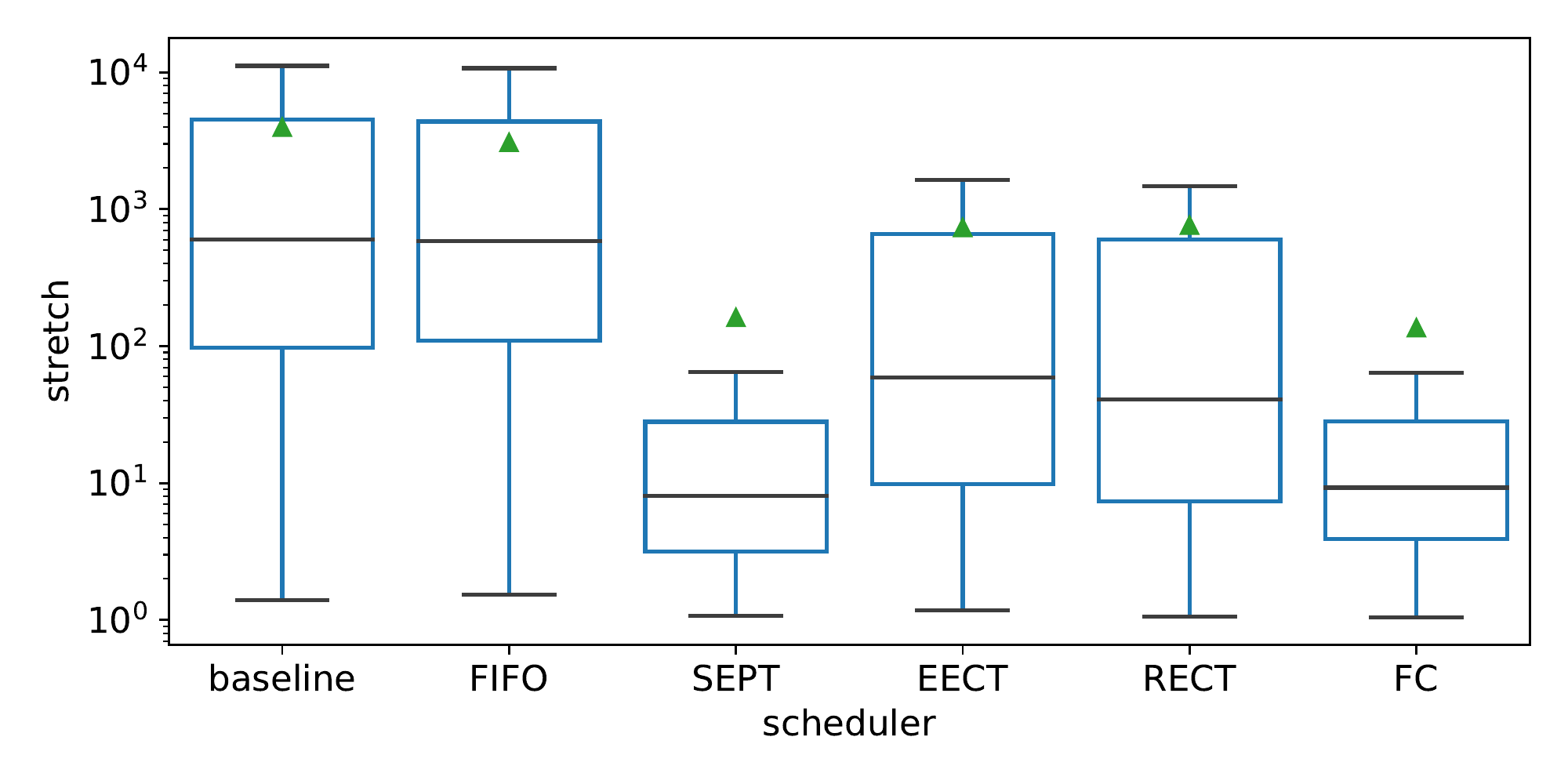}
    \\
    \caption{Stretch for different scheduling policies. Here, we present results for 10 CPUs and intensity 60. On-premise infrastructure. Average stretch presented as a green triangle.}
    \label{fig:stretch_10_60}
\end{figure*}

\begin{figure*}[h]
    \centering
    \chart{Experiment 1}{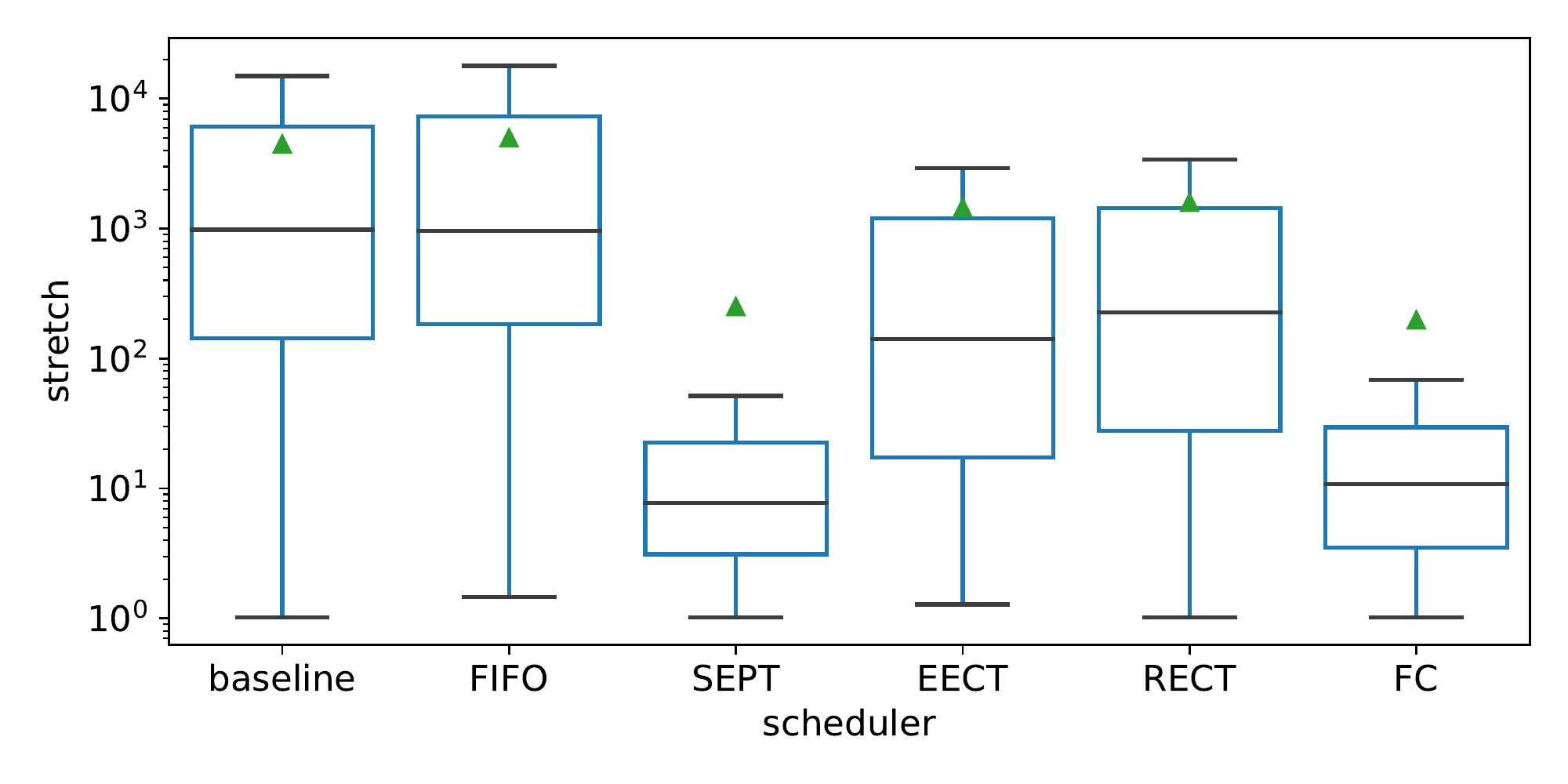}
    \chart{Experiment 2}{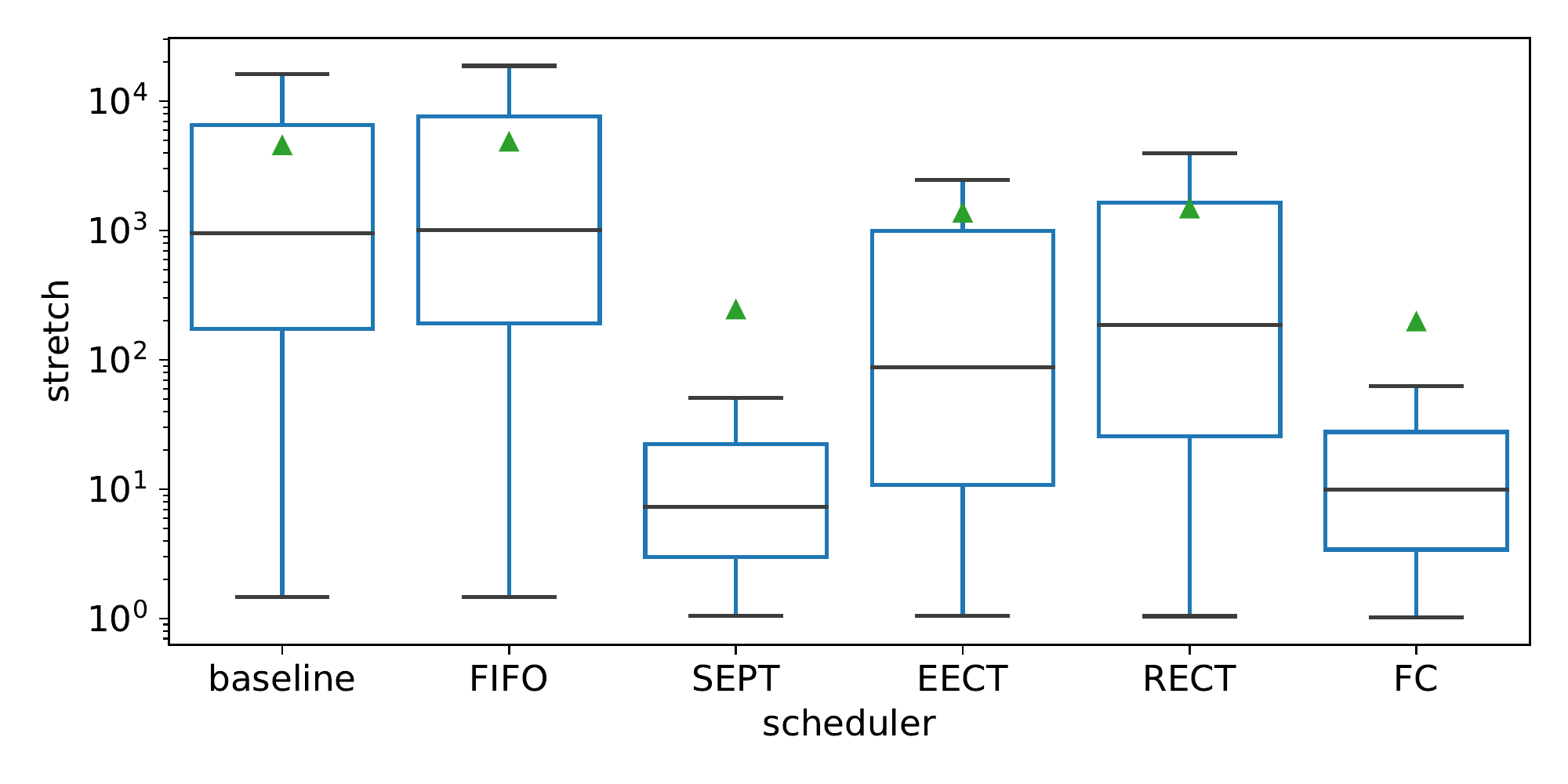}
    \chart{Experiment 3}{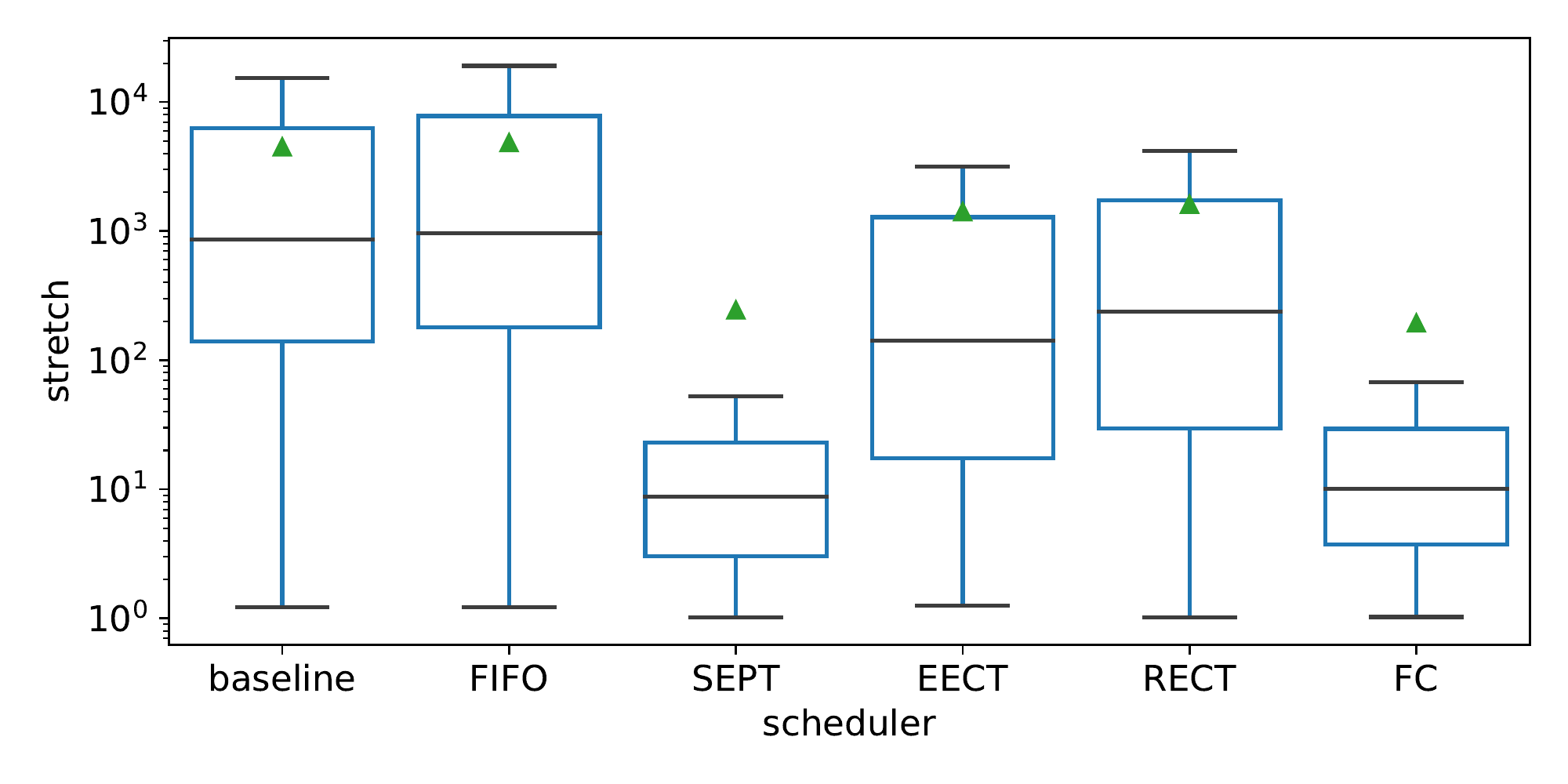}
    \\
    \chart{Experiment 4}{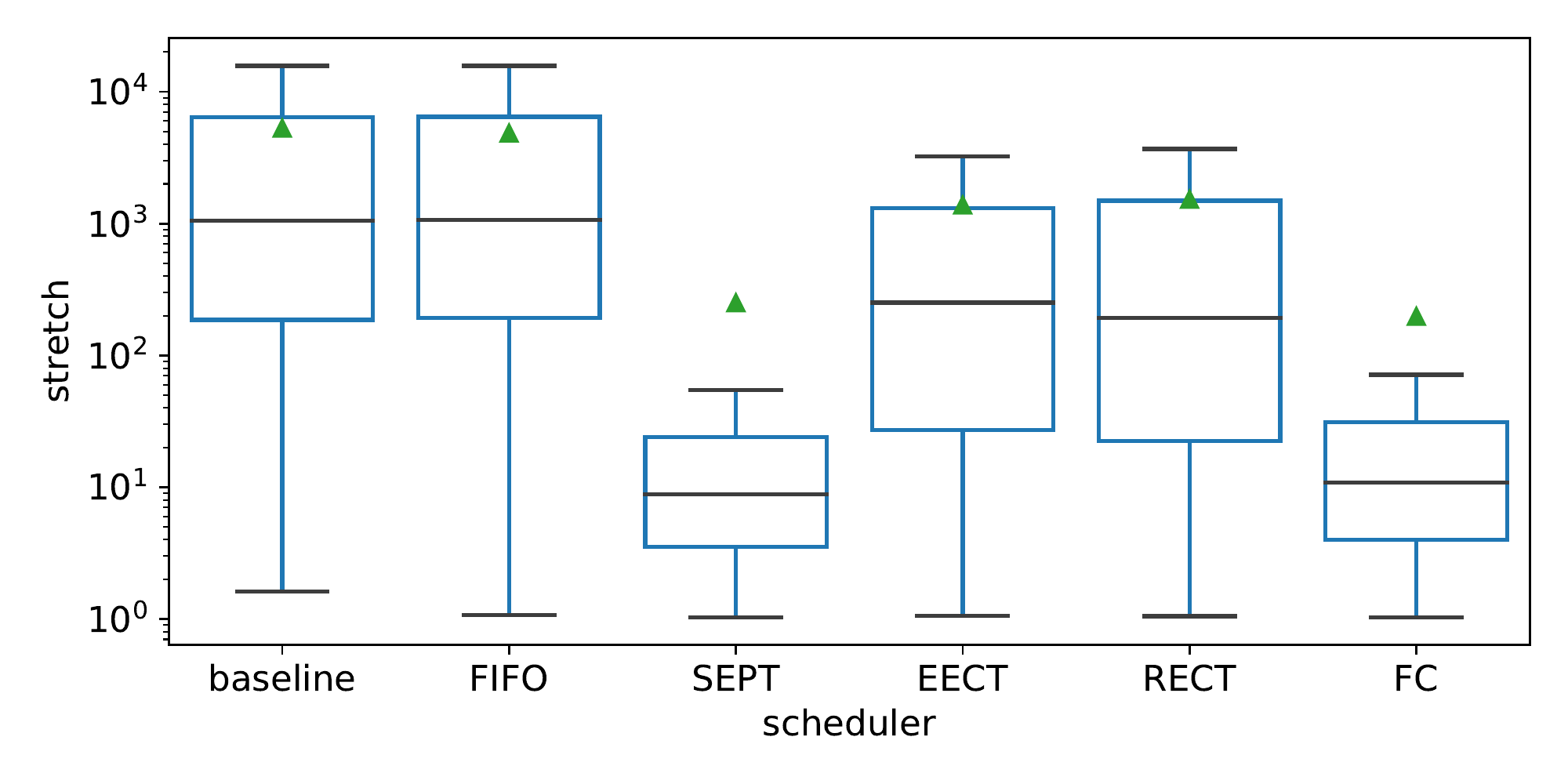}
    \chart{Experiment 5}{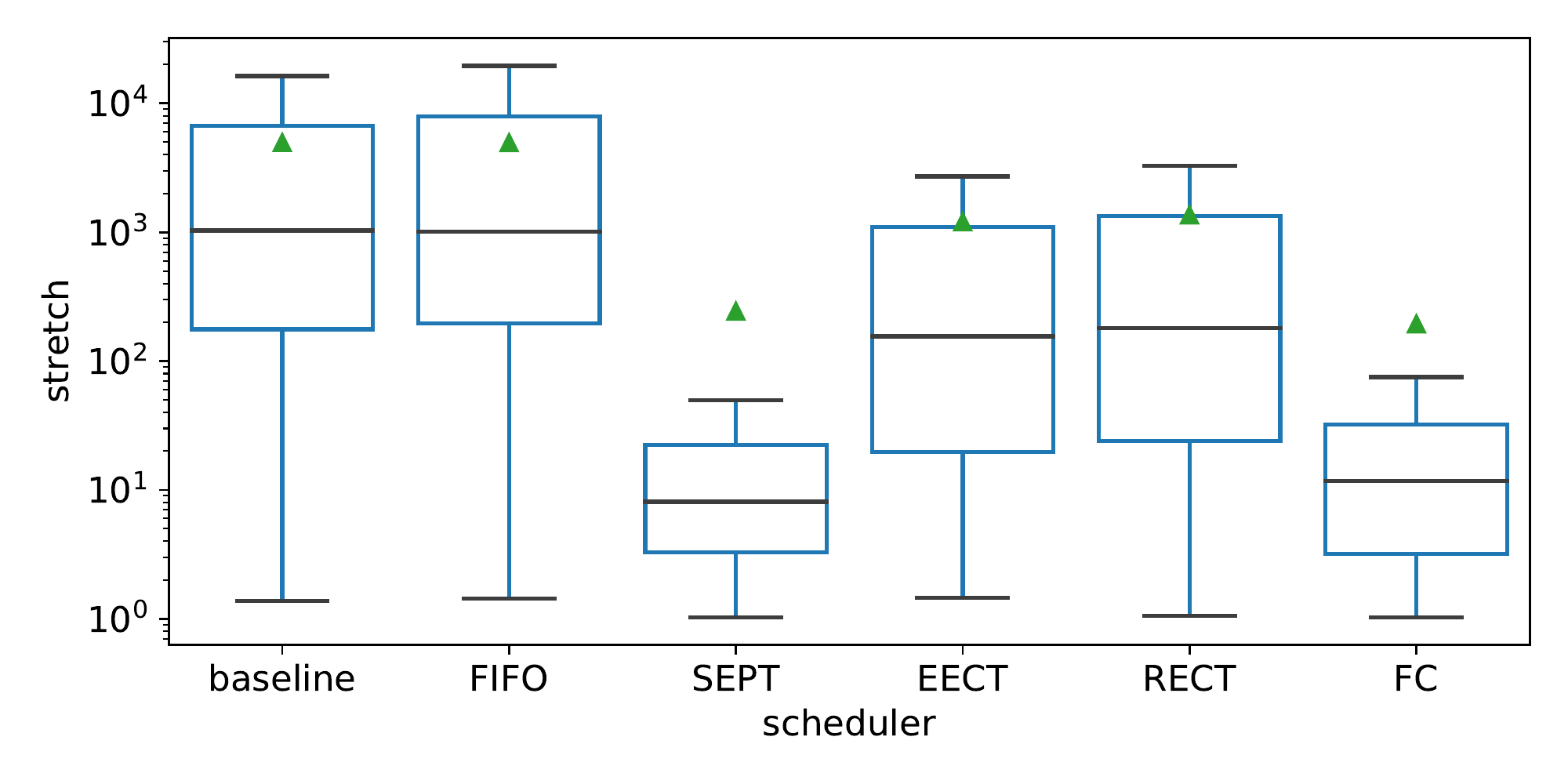}
    \\
    \caption{Stretch for different scheduling policies. Here, we present results for 10 CPUs and intensity 90. On-premise infrastructure. Average stretch presented as a green triangle.}
    \label{fig:stretch_10_90}
\end{figure*}

\begin{figure*}[h]
    \centering
    \chart{Experiment 1}{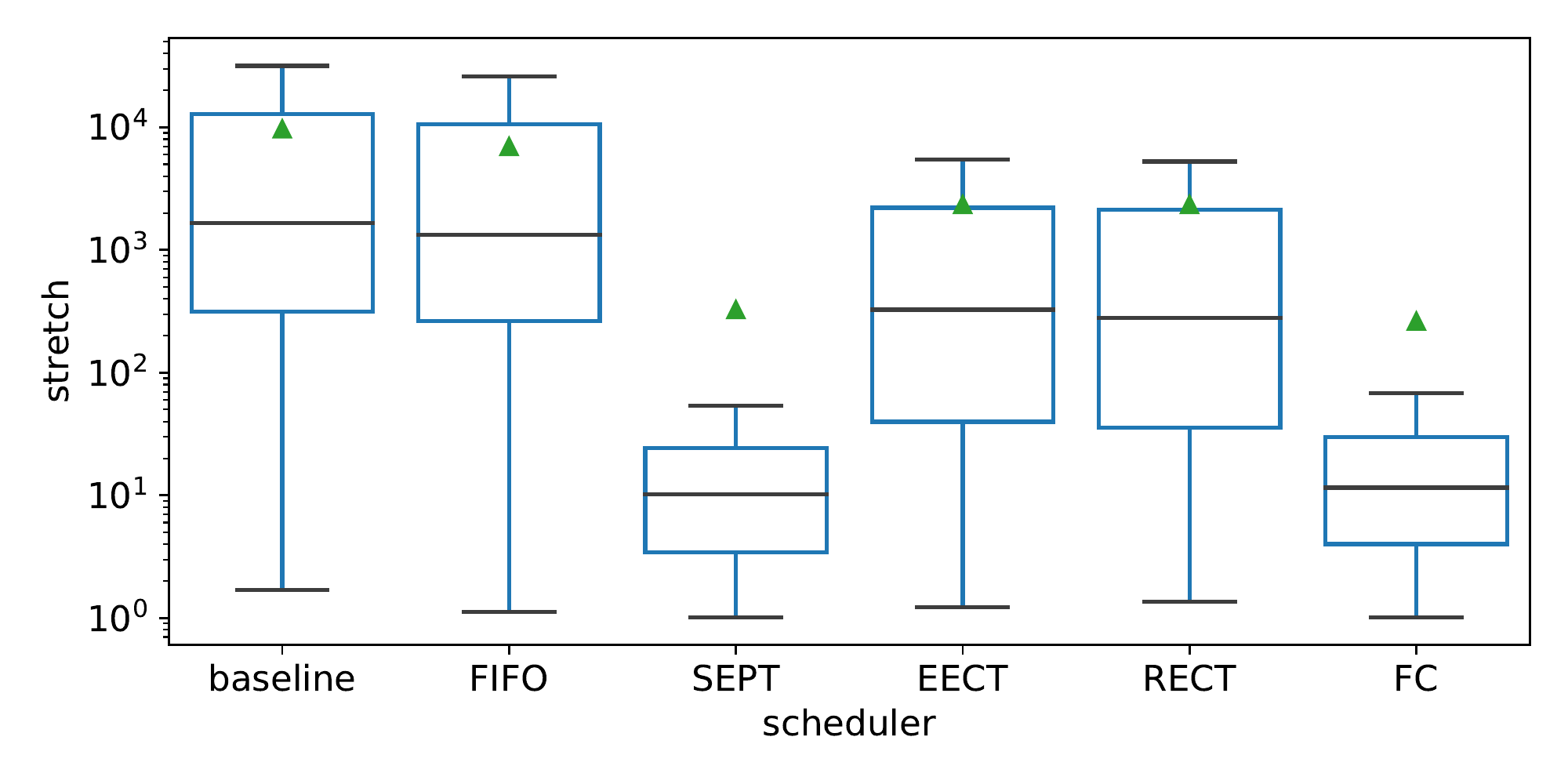}
    \chart{Experiment 2}{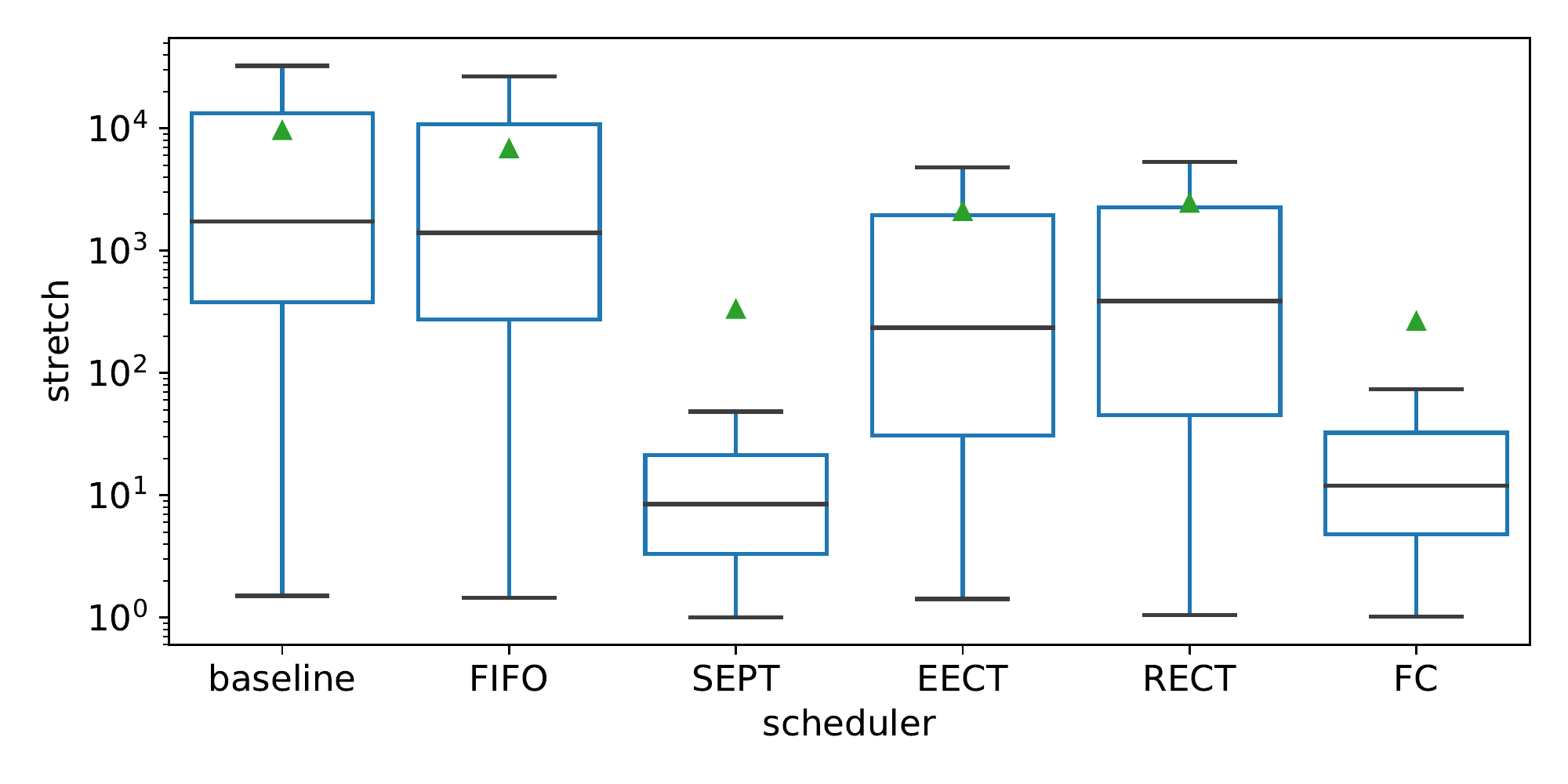}
    \chart{Experiment 3}{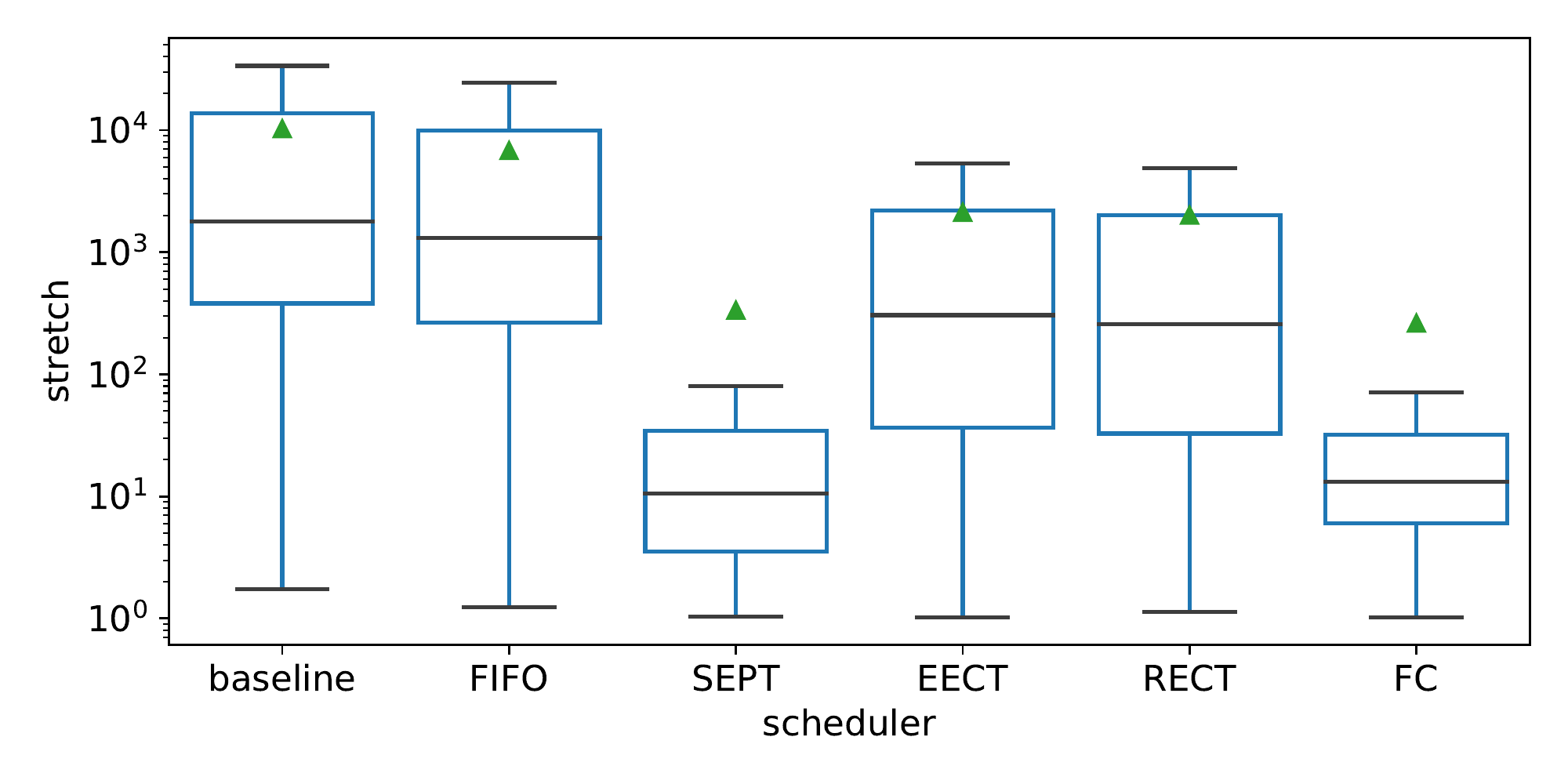}
    \\
    \chart{Experiment 4}{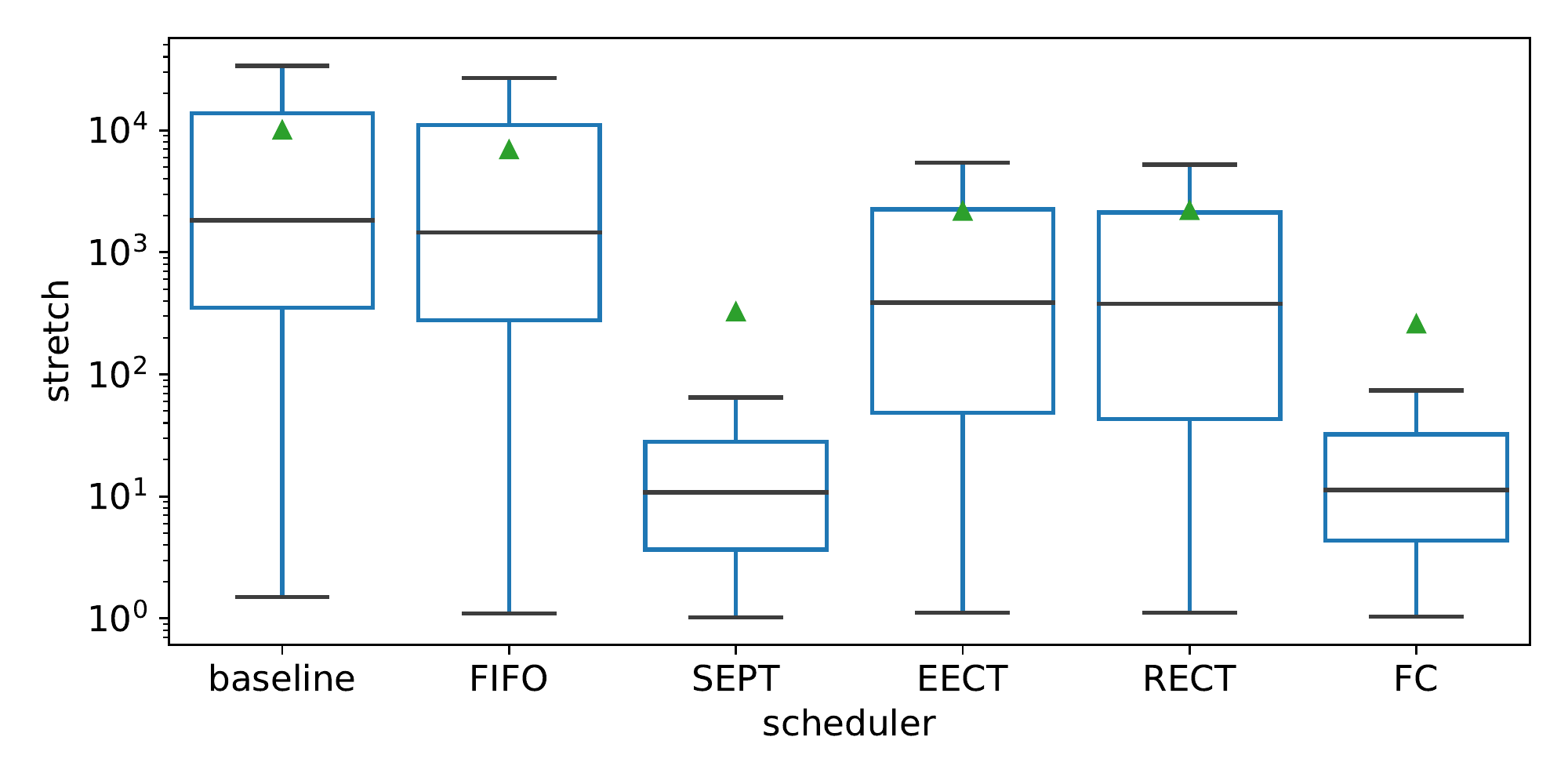}
    \chart{Experiment 5}{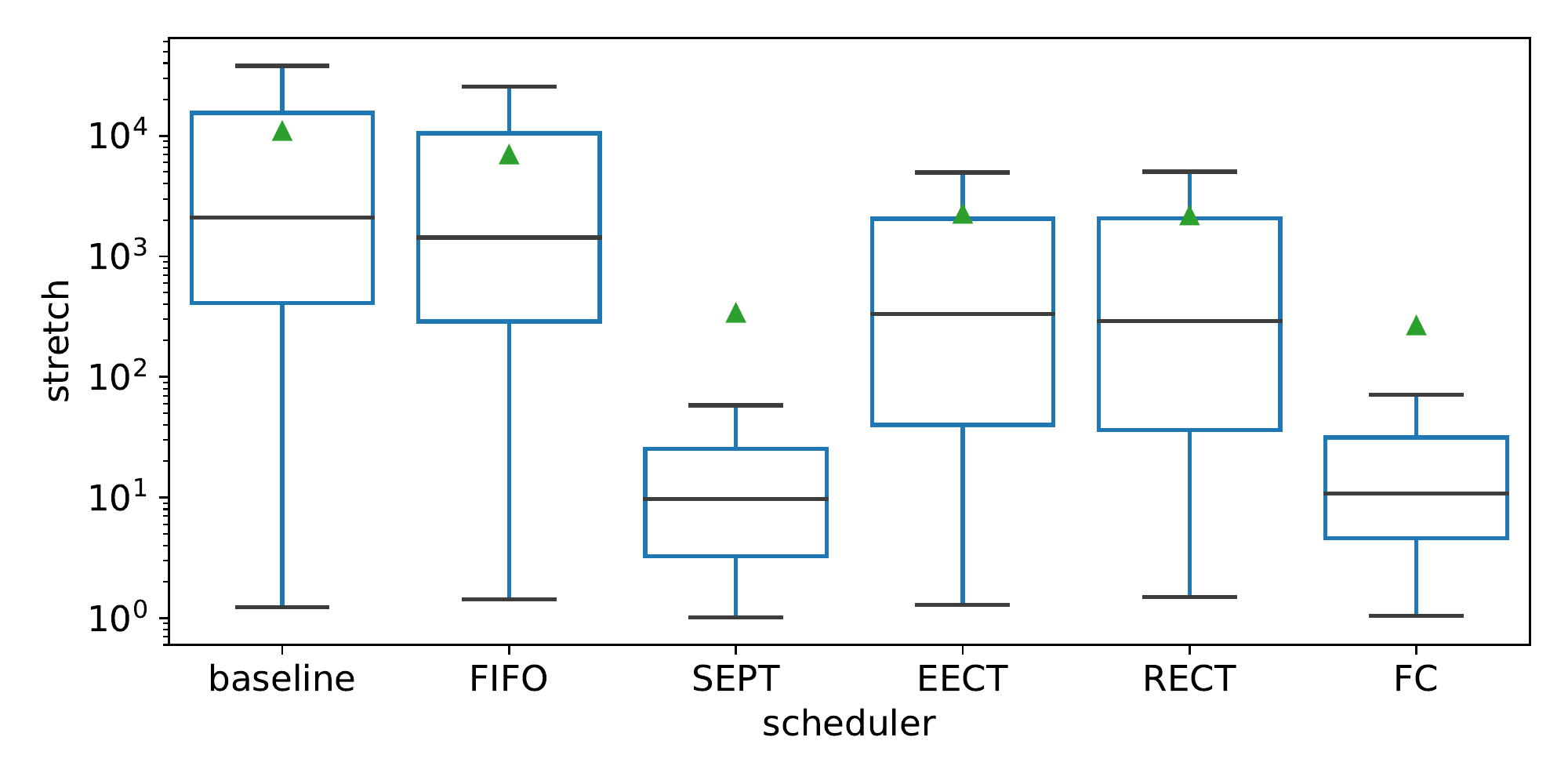}
    \\
    \caption{Stretch for different scheduling policies. Here, we present results for 10 CPUs and intensity 120. On-premise infrastructure. Average stretch presented as a green triangle.}
    \label{fig:stretch_10_120}
\end{figure*}

\begin{figure*}[h]
    \centering
    \chart{Experiment 1}{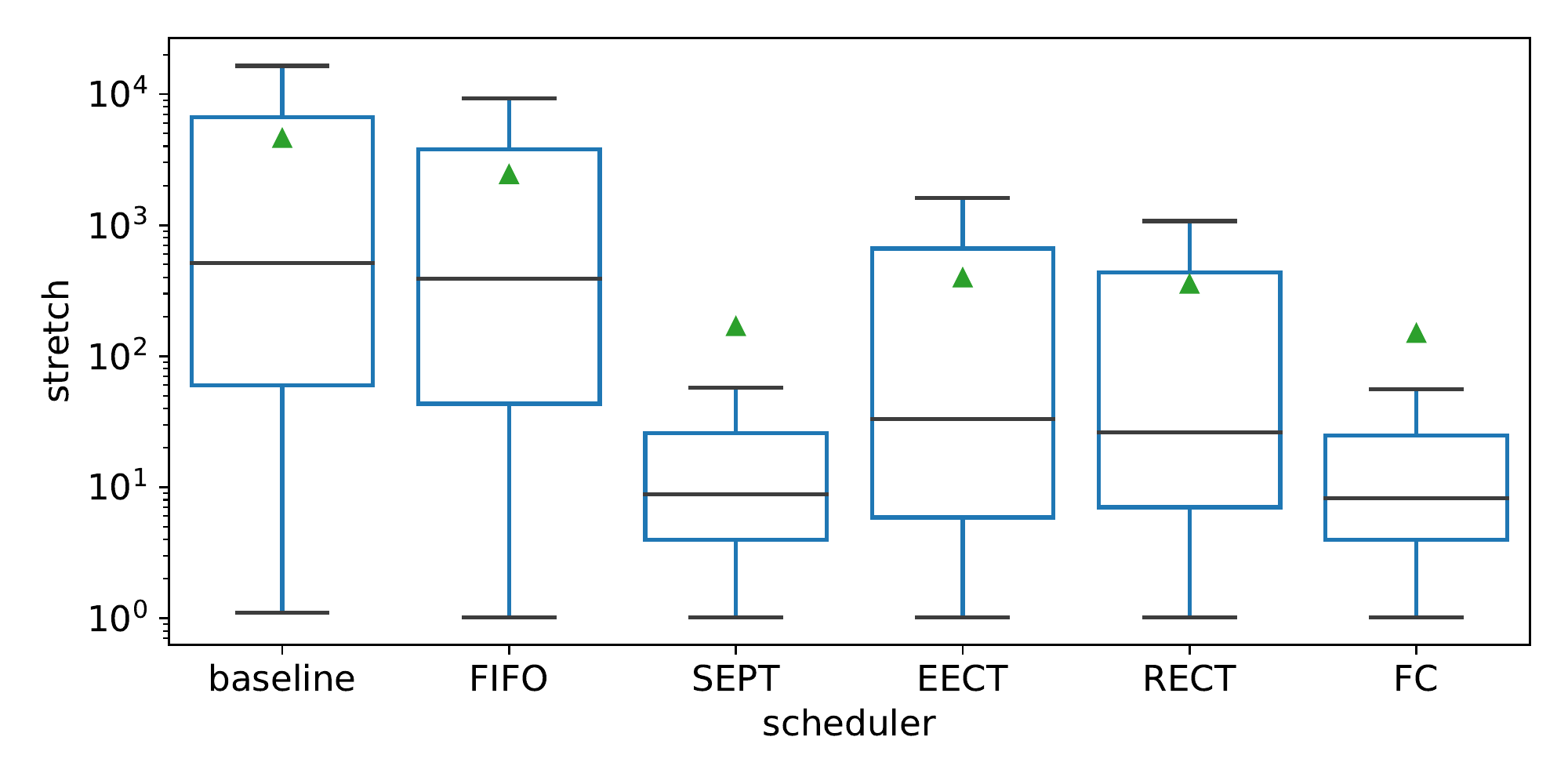}
    \chart{Experiment 2}{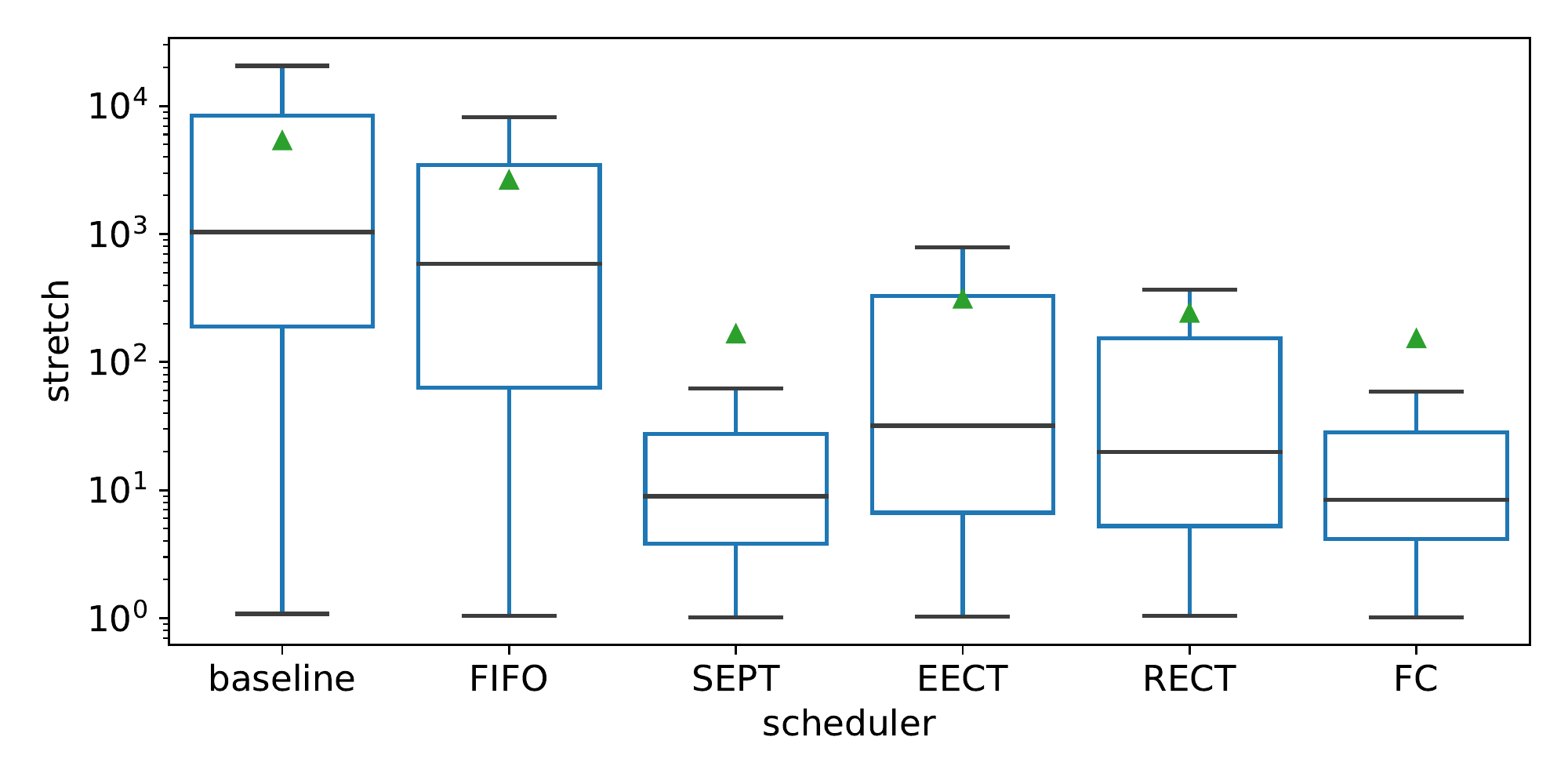}
    \chart{Experiment 3}{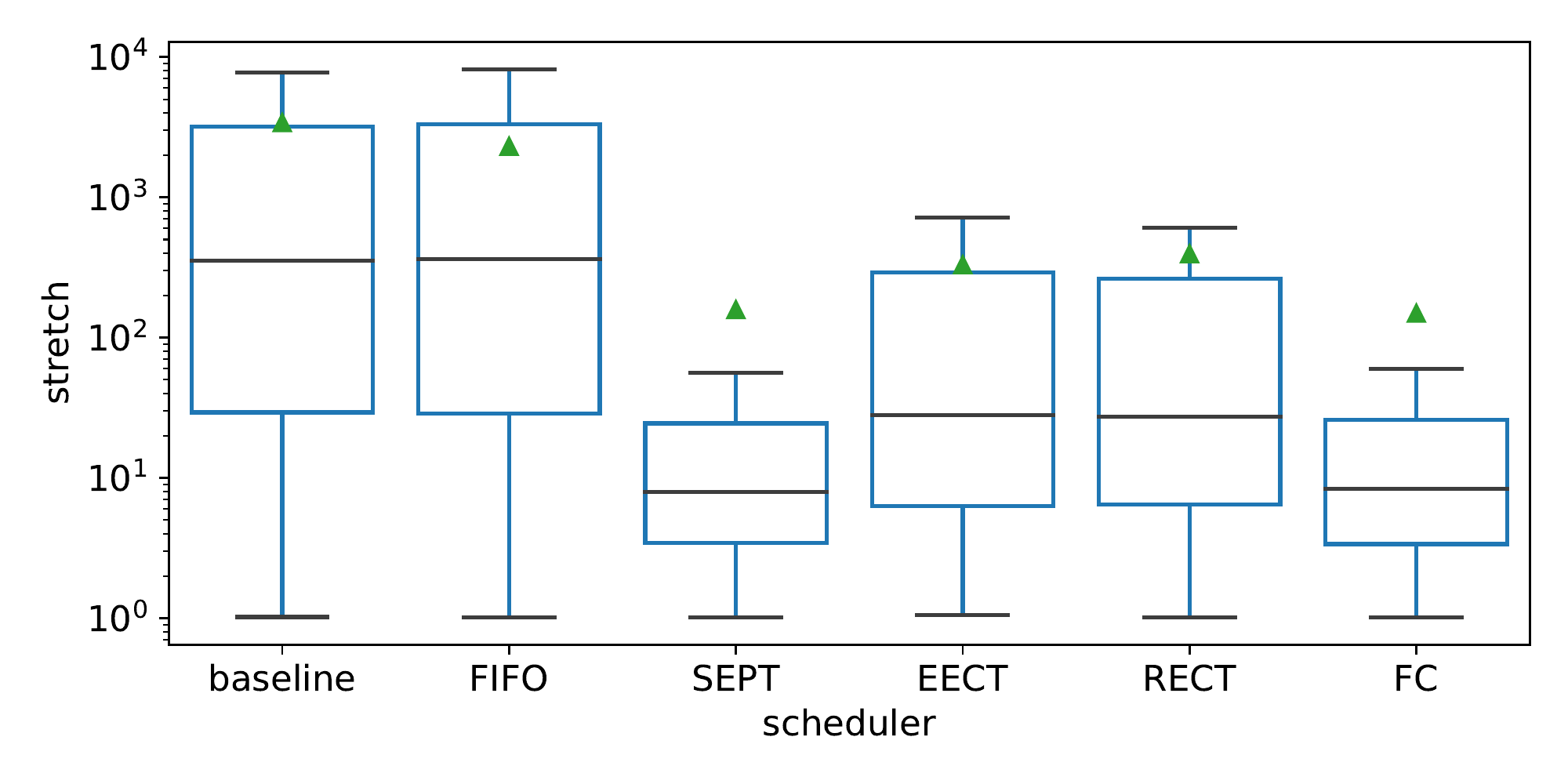}
    \\
    \chart{Experiment 4}{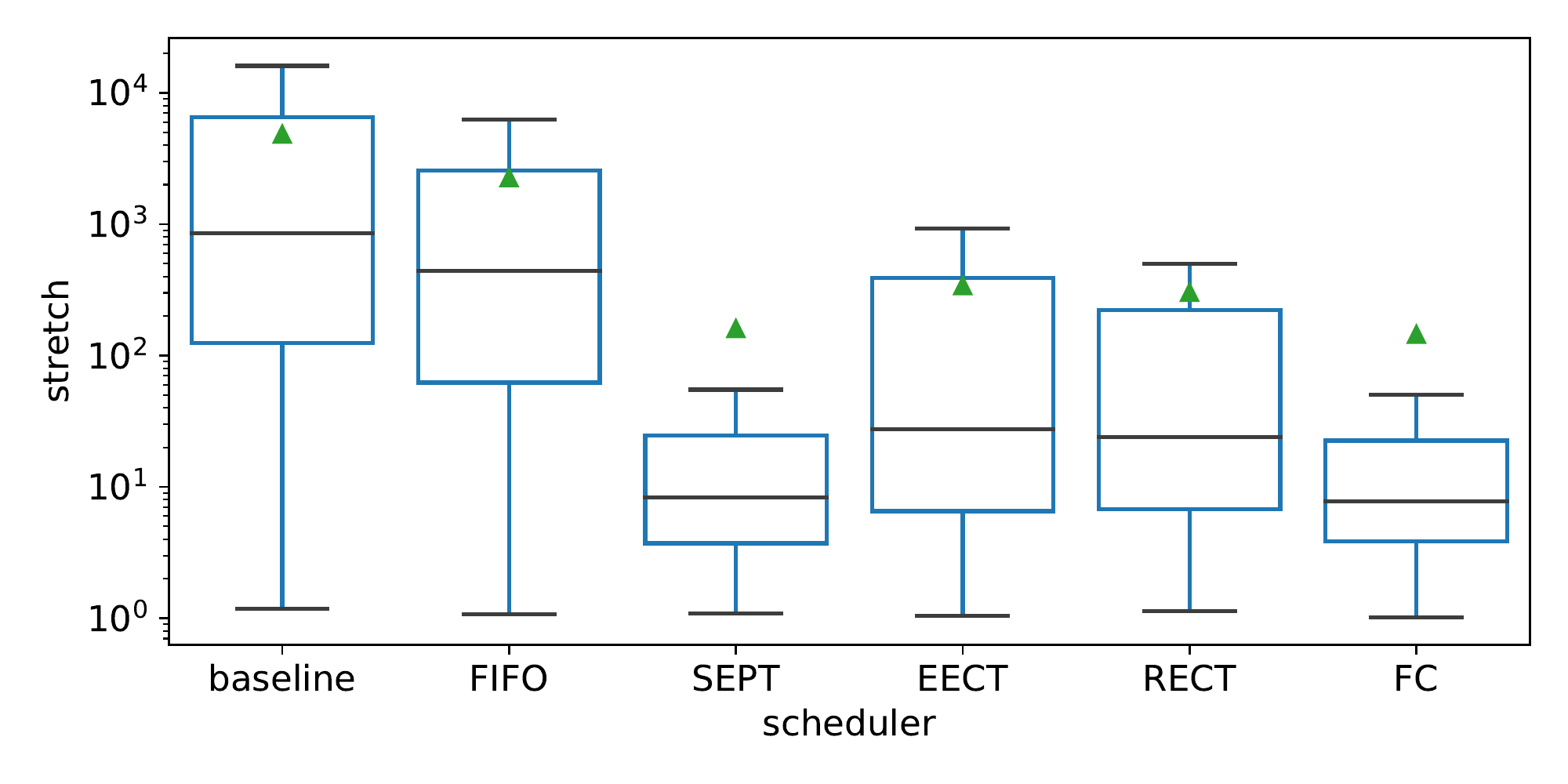}
    \chart{Experiment 5}{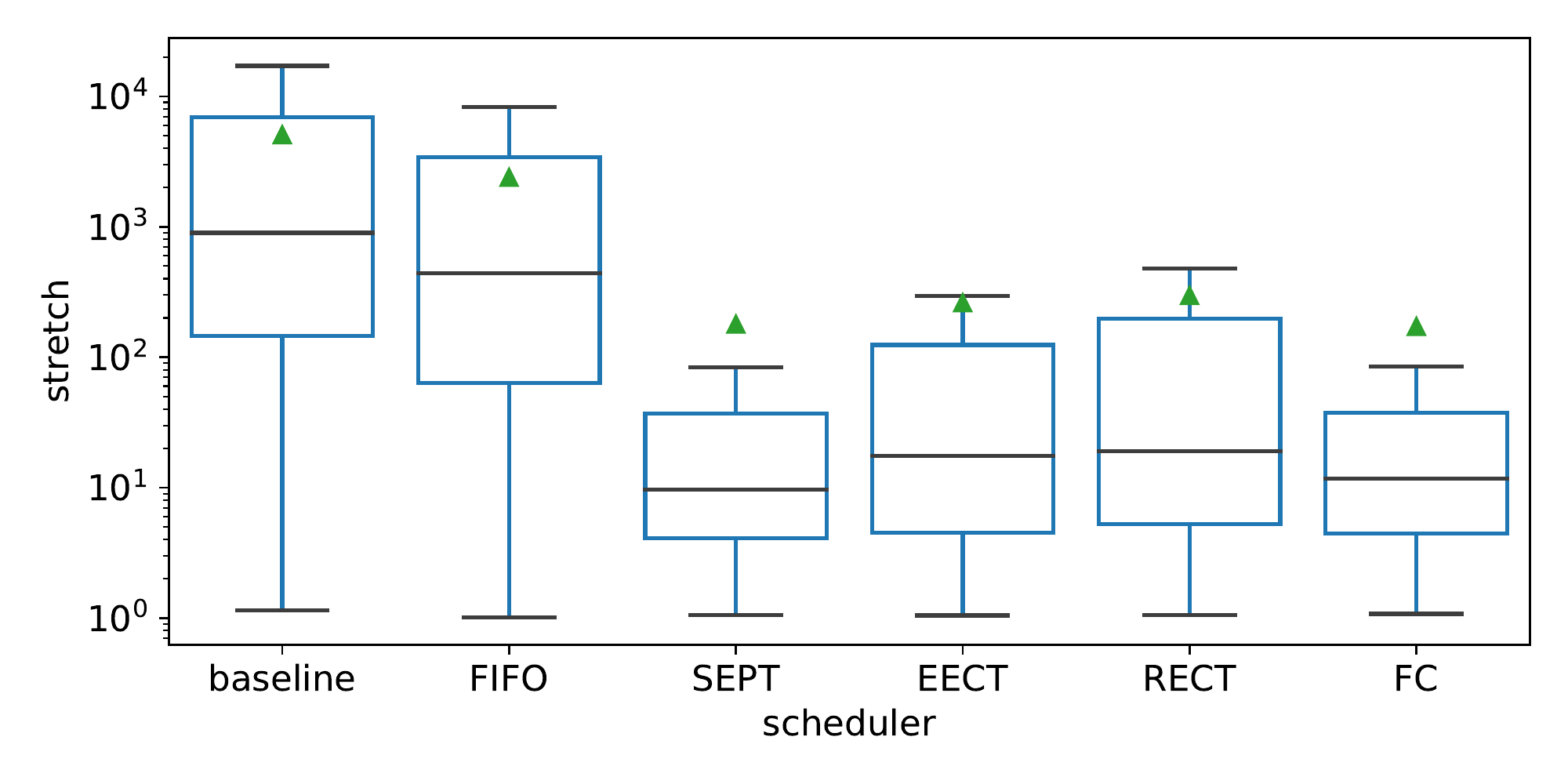}
    \\
    \caption{Stretch for different scheduling policies. Here, we present results for 20 CPUs and intensity 30. On-premise infrastructure. Average stretch presented as a green triangle.}
    \label{fig:stretch_20_30}
\end{figure*}

\begin{figure*}[h]
    \centering
    \chart{Experiment 1}{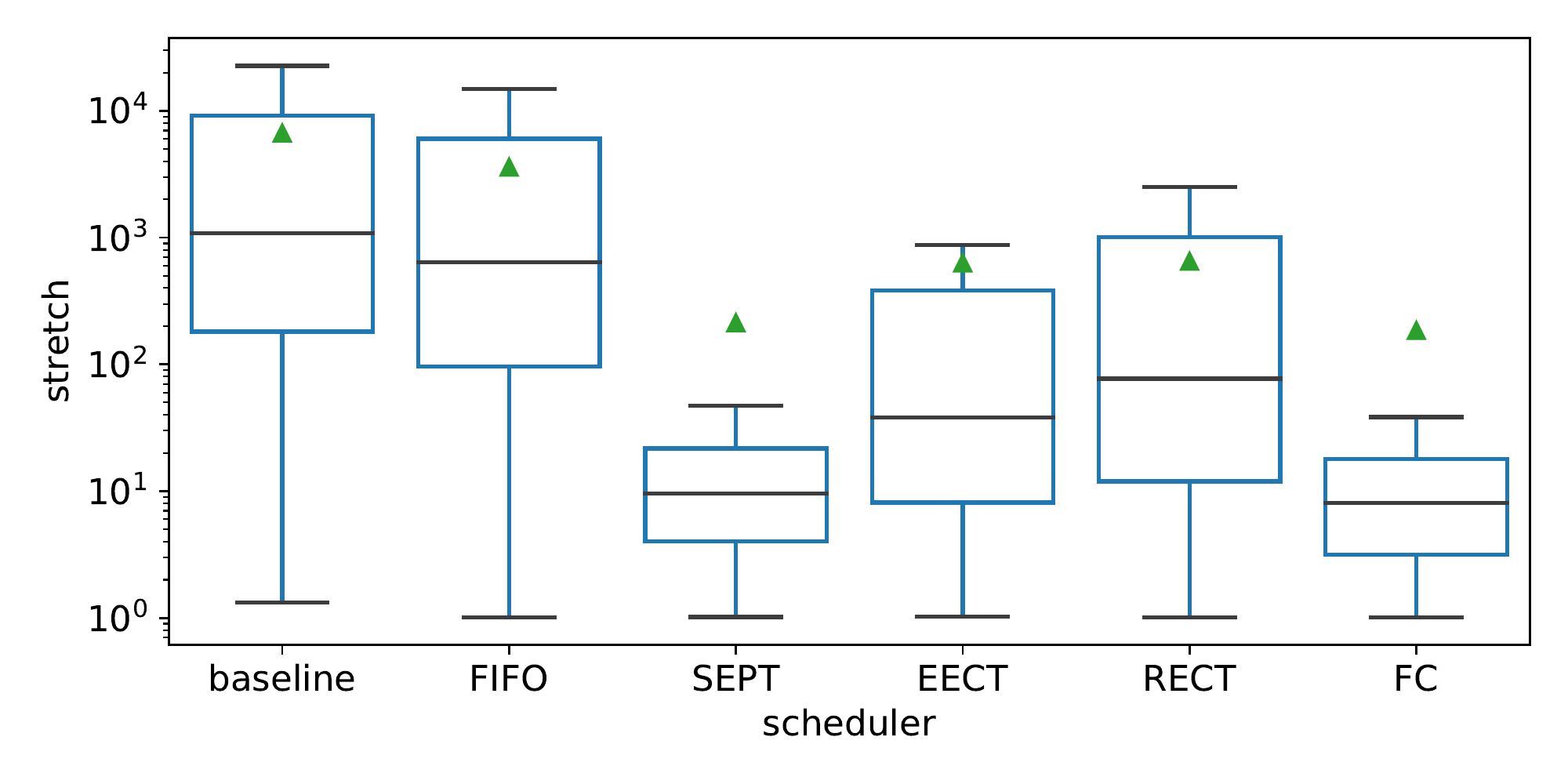}
    \chart{Experiment 2}{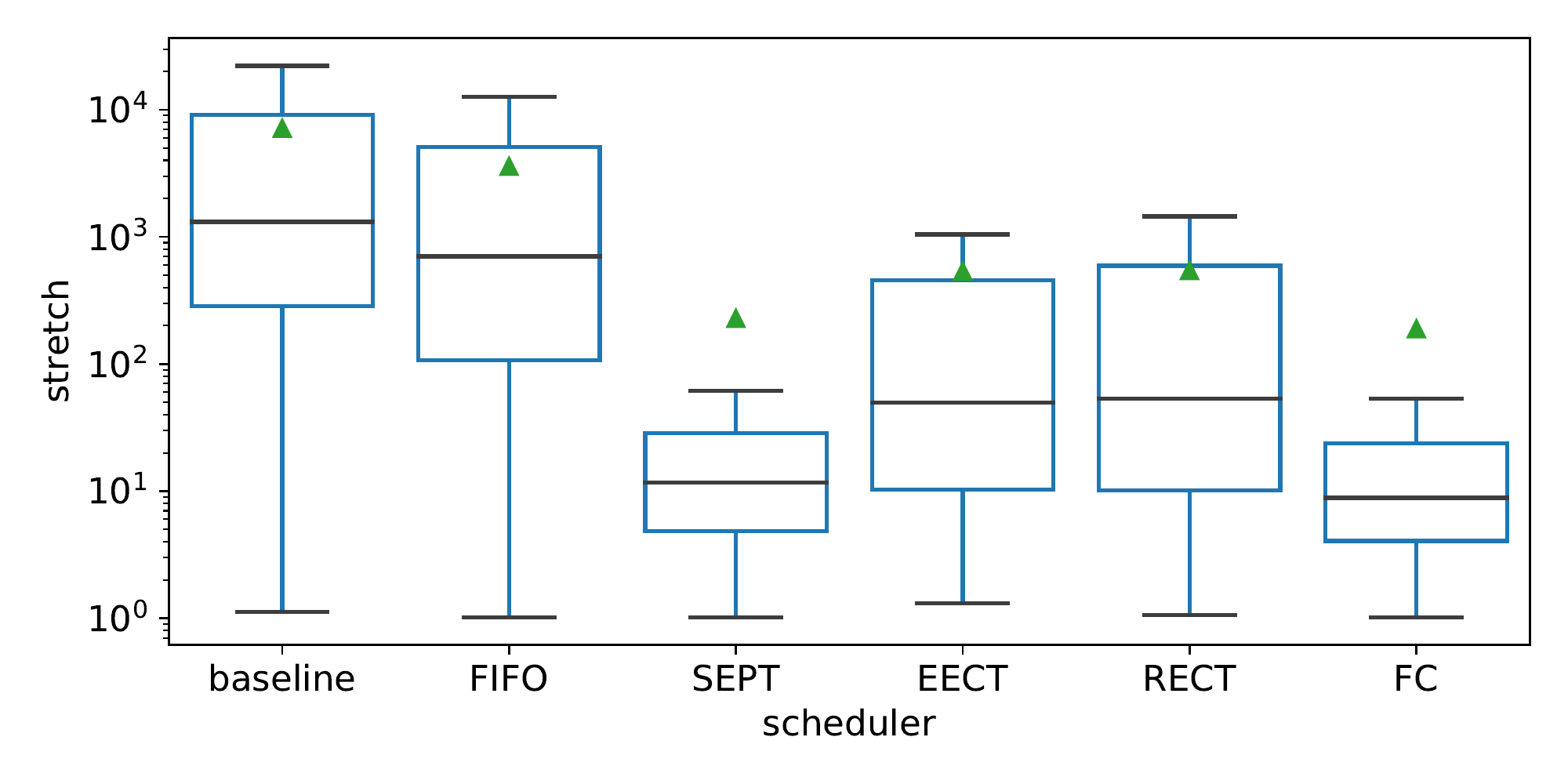}
    \chart{Experiment 3}{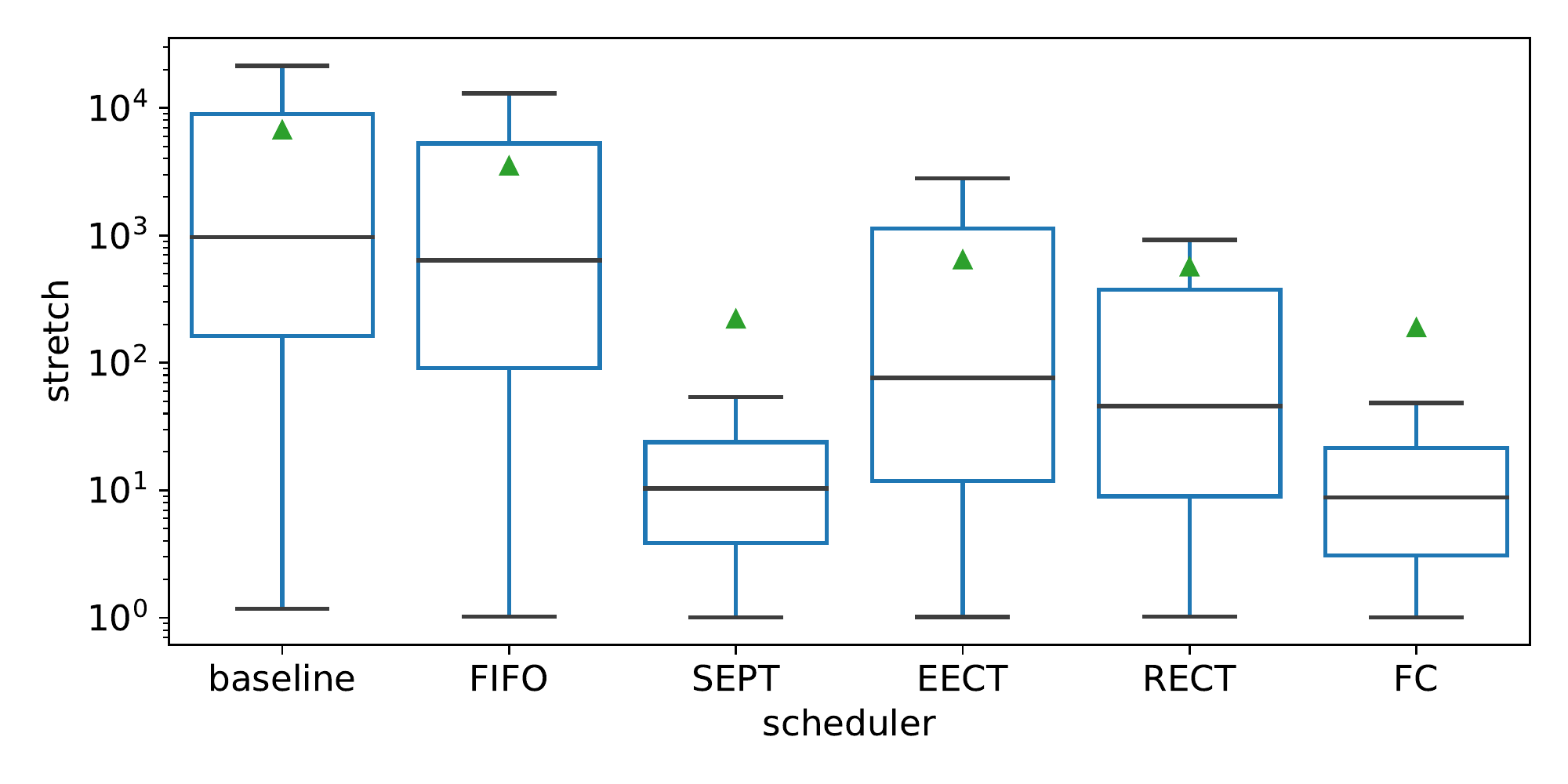}
    \\
    \chart{Experiment 4}{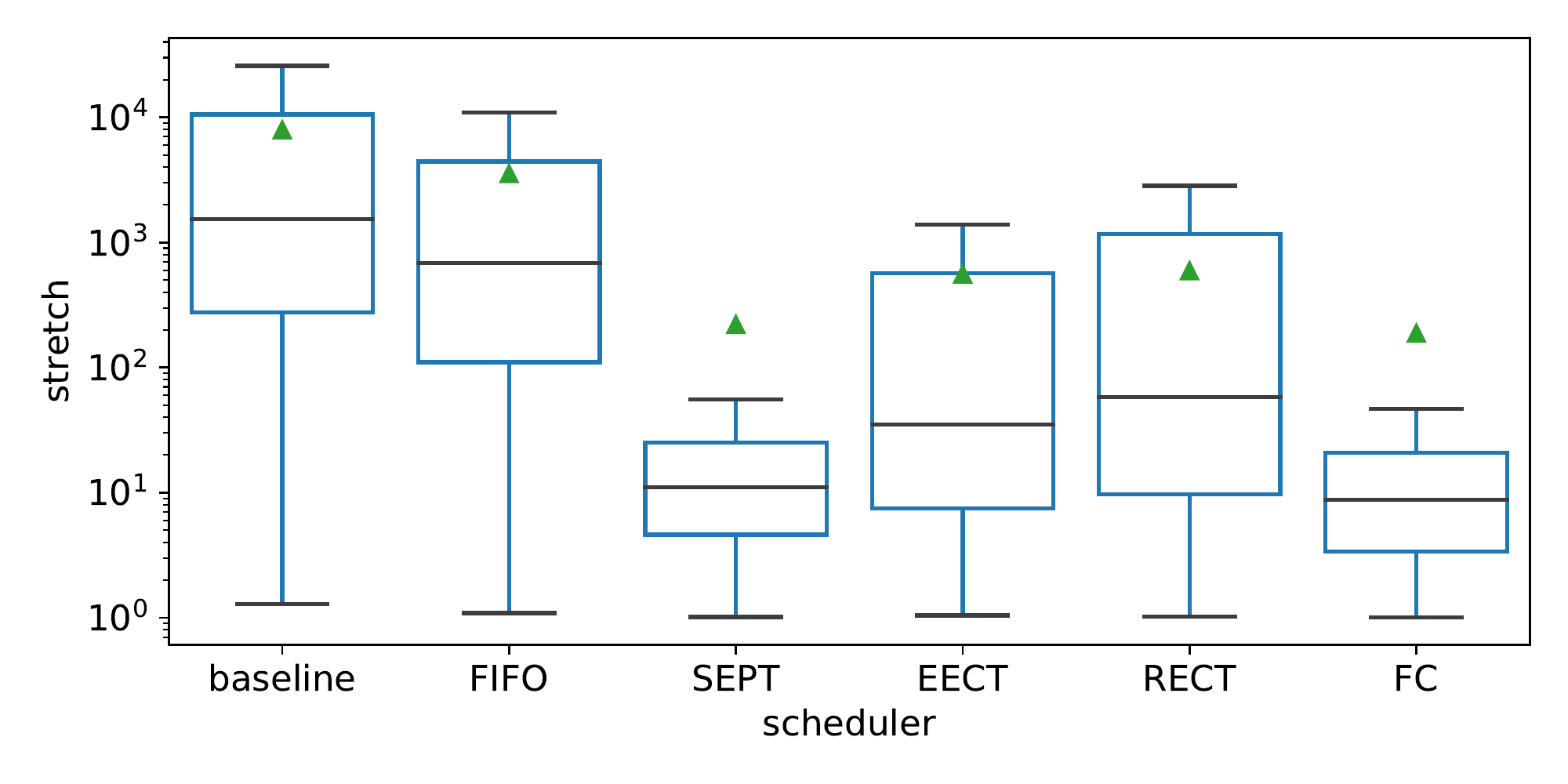}
    \chart{Experiment 5}{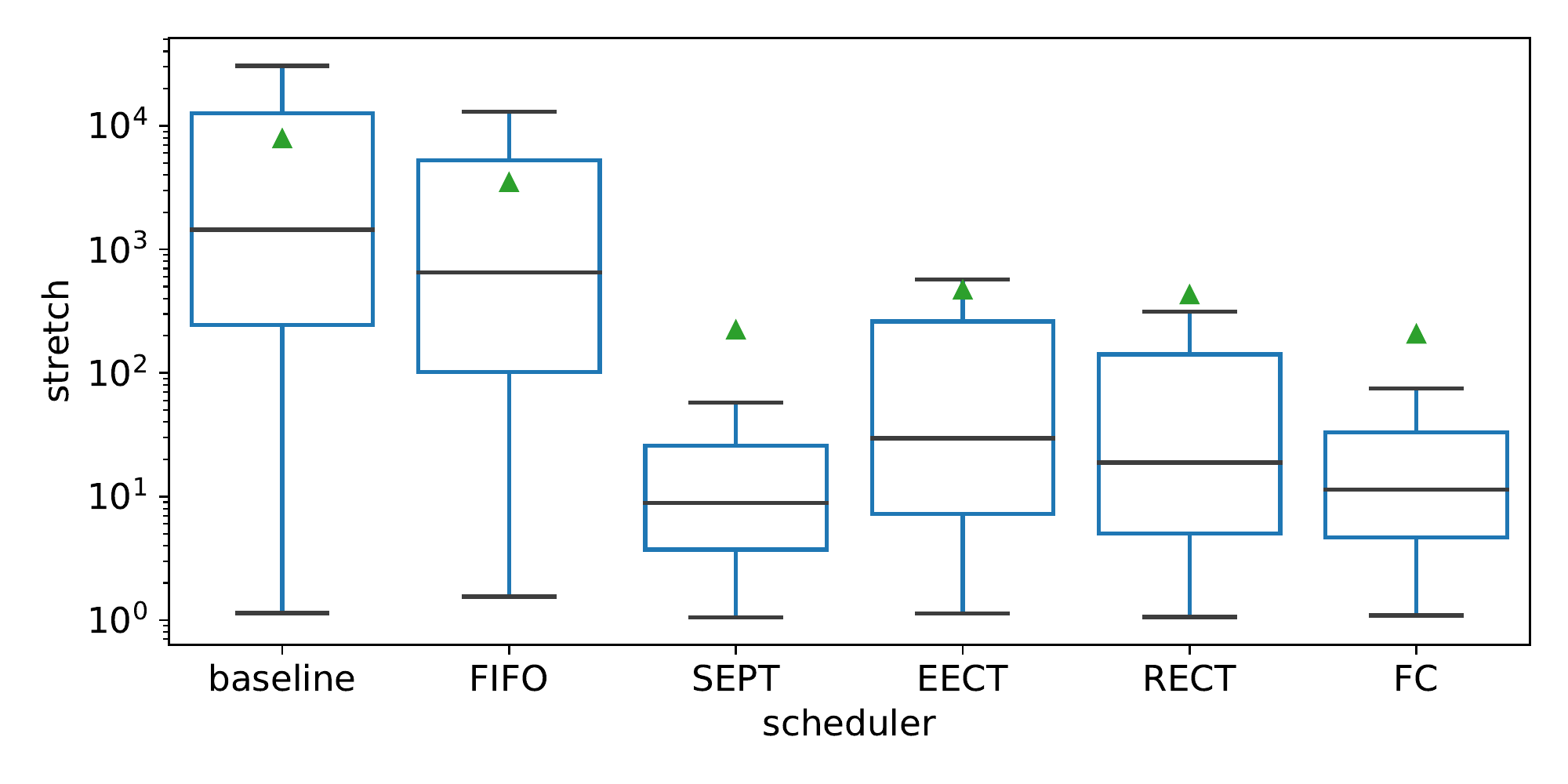}
    \\
    \caption{Stretch for different scheduling policies. Here, we present results for 20 CPUs and intensity 40. On-premise infrastructure. Average stretch presented as a green triangle.}
    \label{fig:stretch_20_40}
\end{figure*}

\begin{figure*}[h]
    \centering
    \chart{Experiment 1}{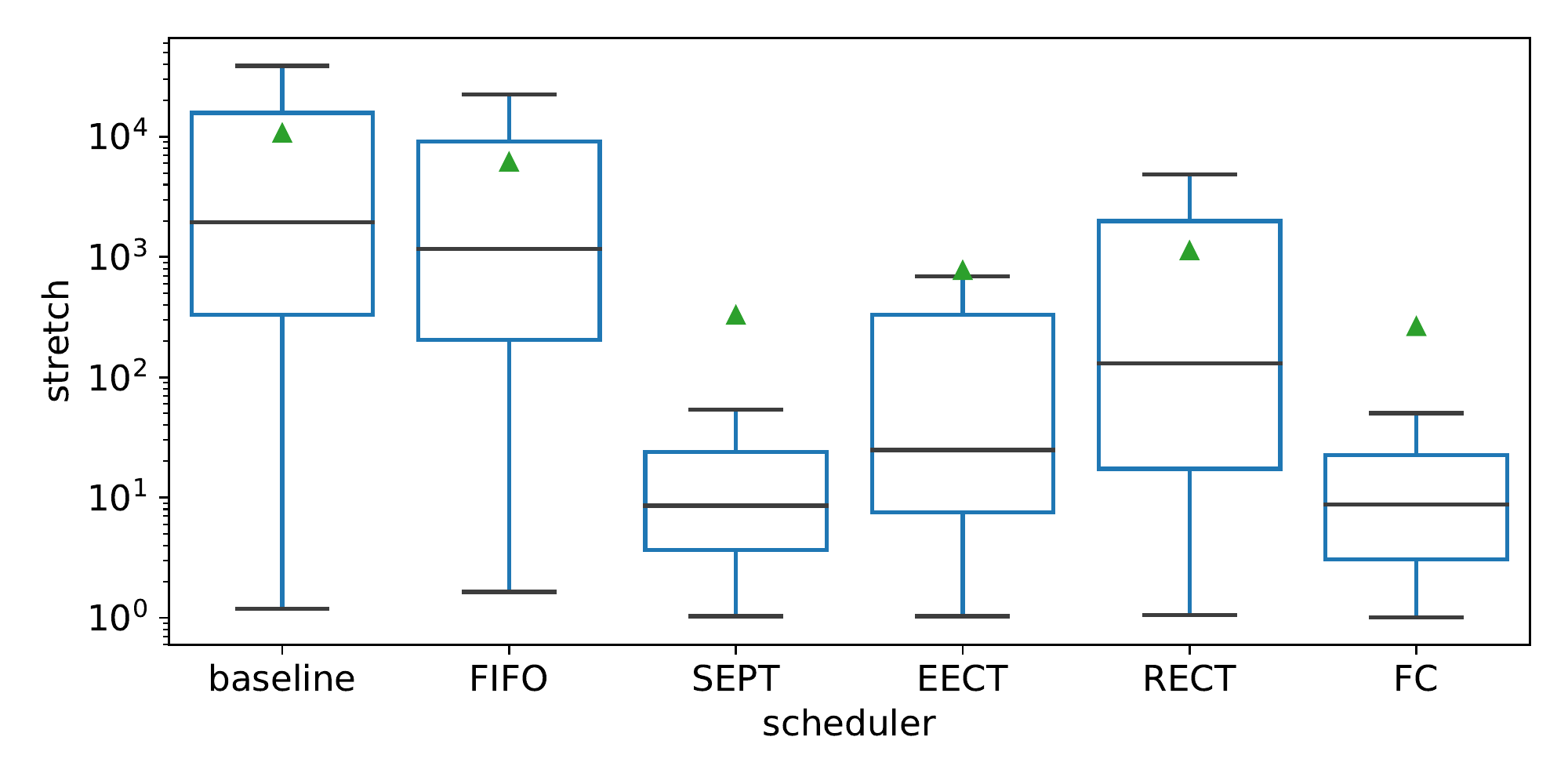}
    \chart{Experiment 2}{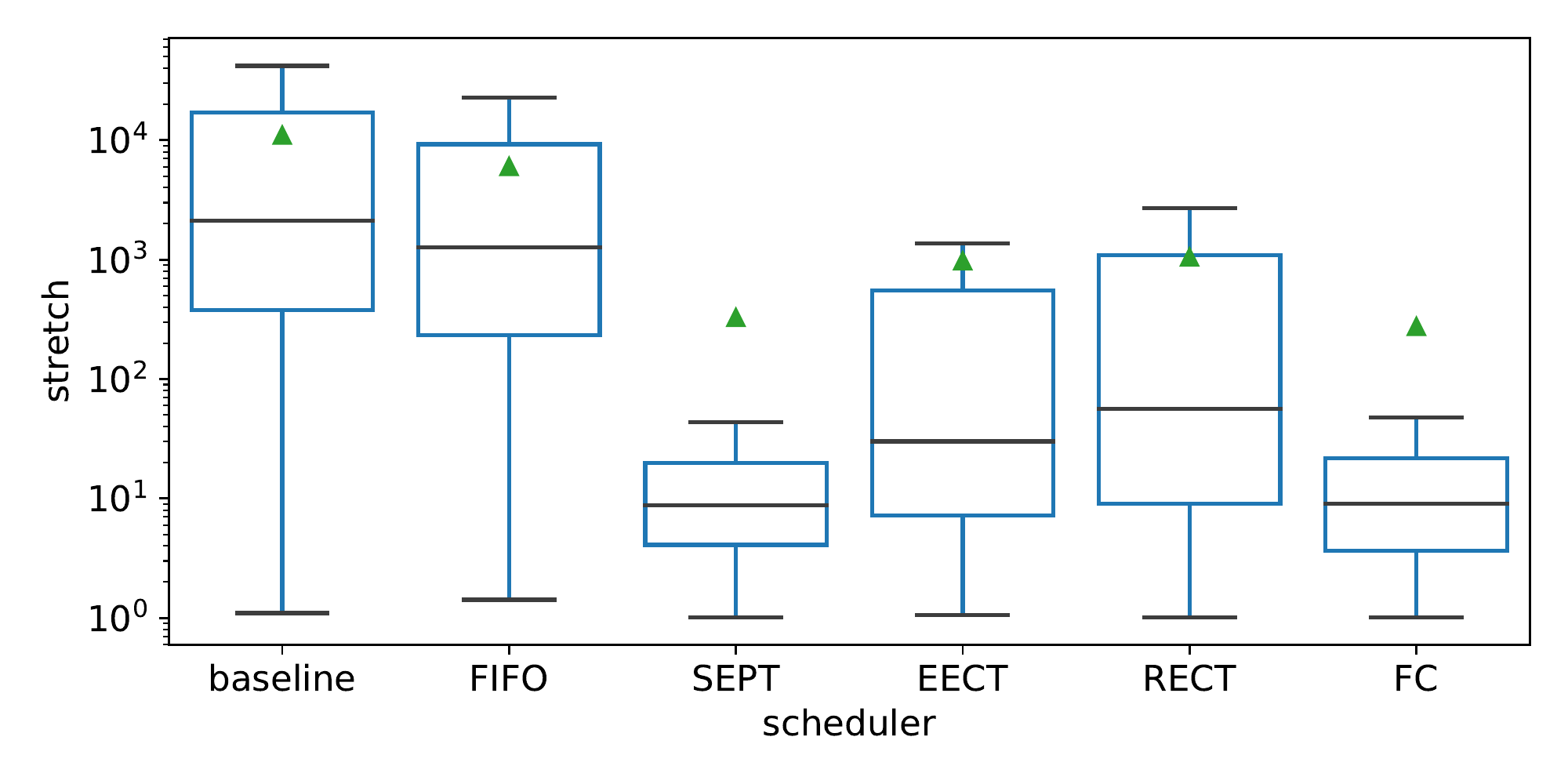}
    \chart{Experiment 3}{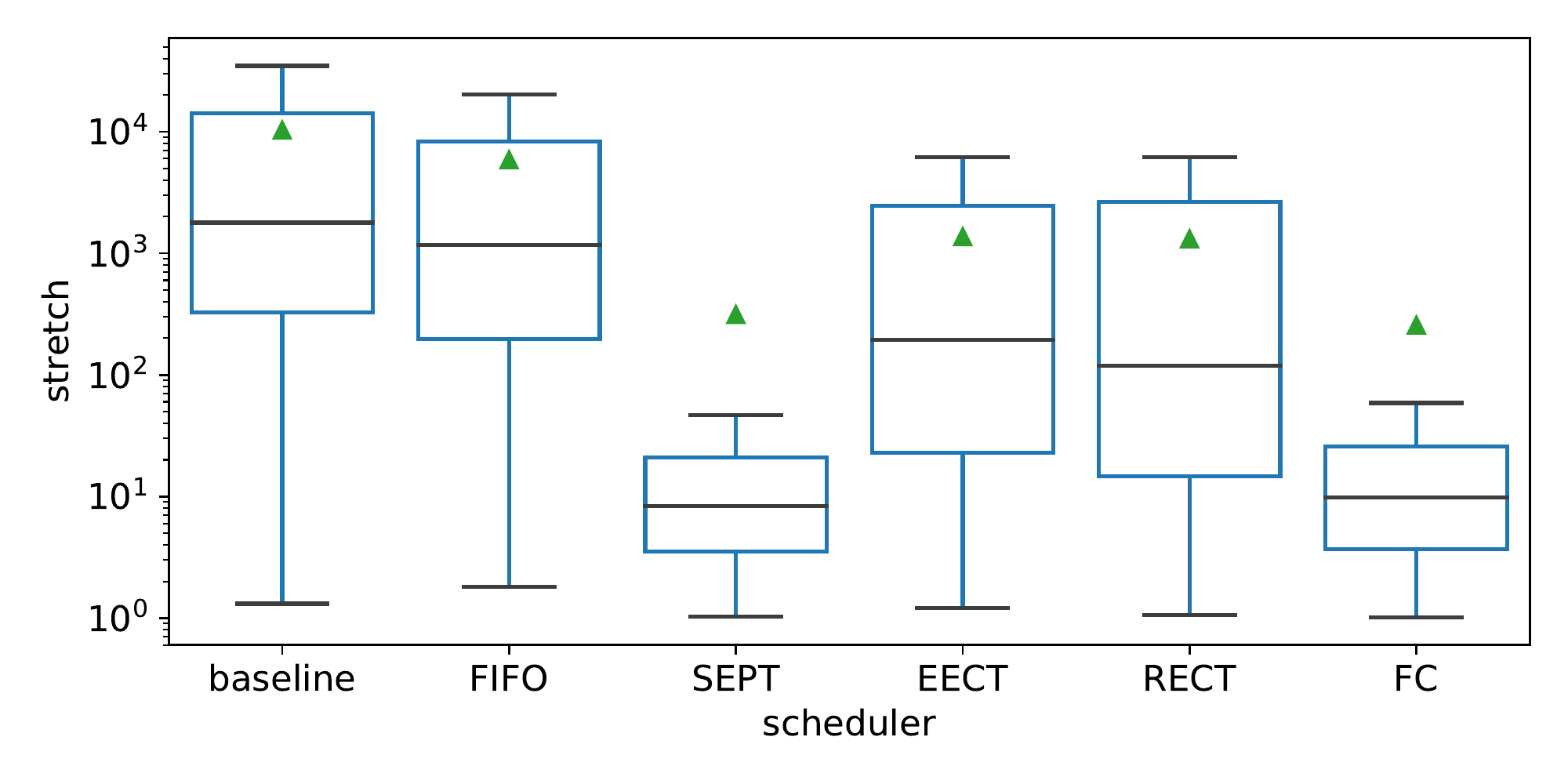}
    \\
    \chart{Experiment 4}{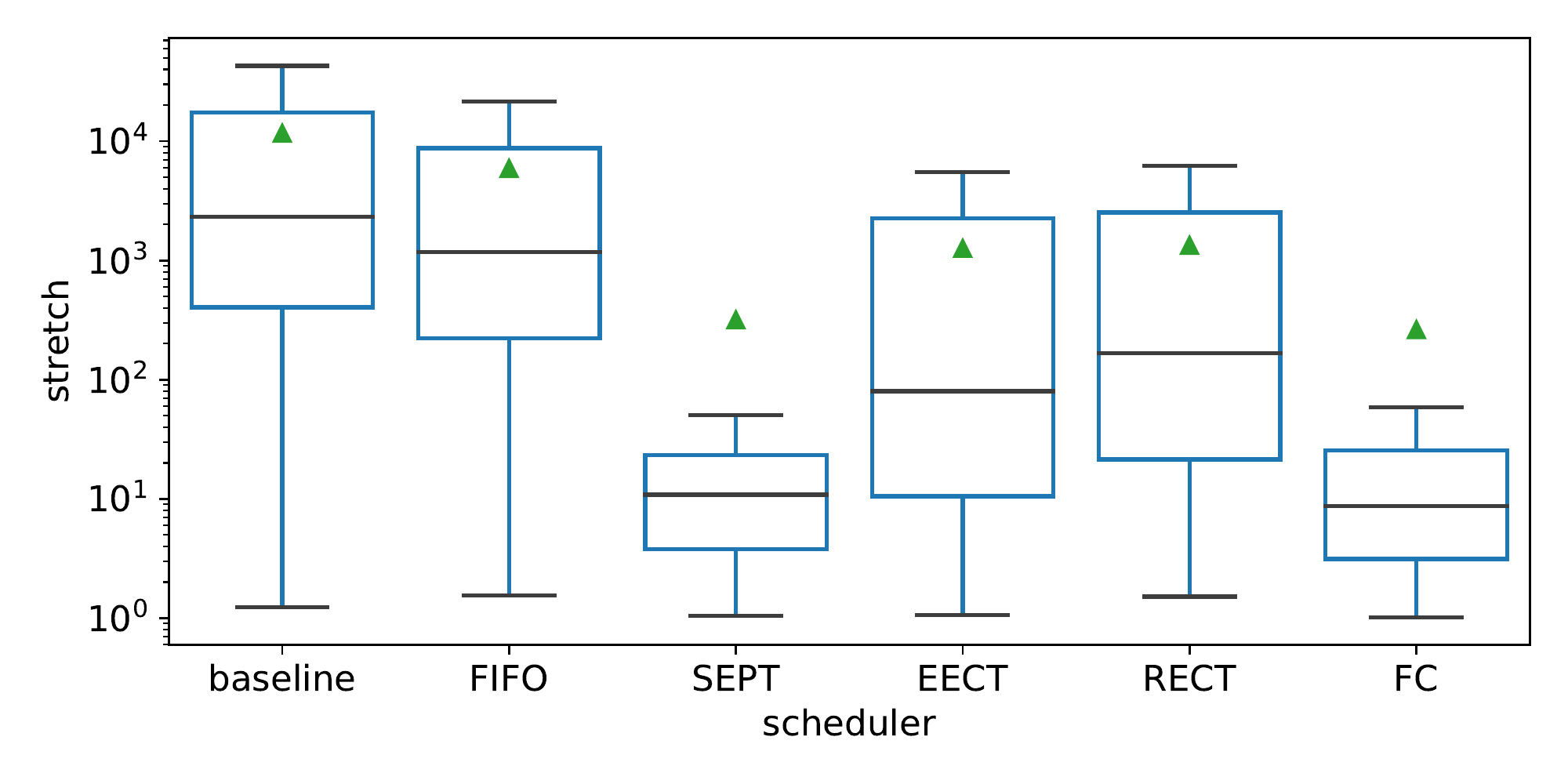}
    \chart{Experiment 5}{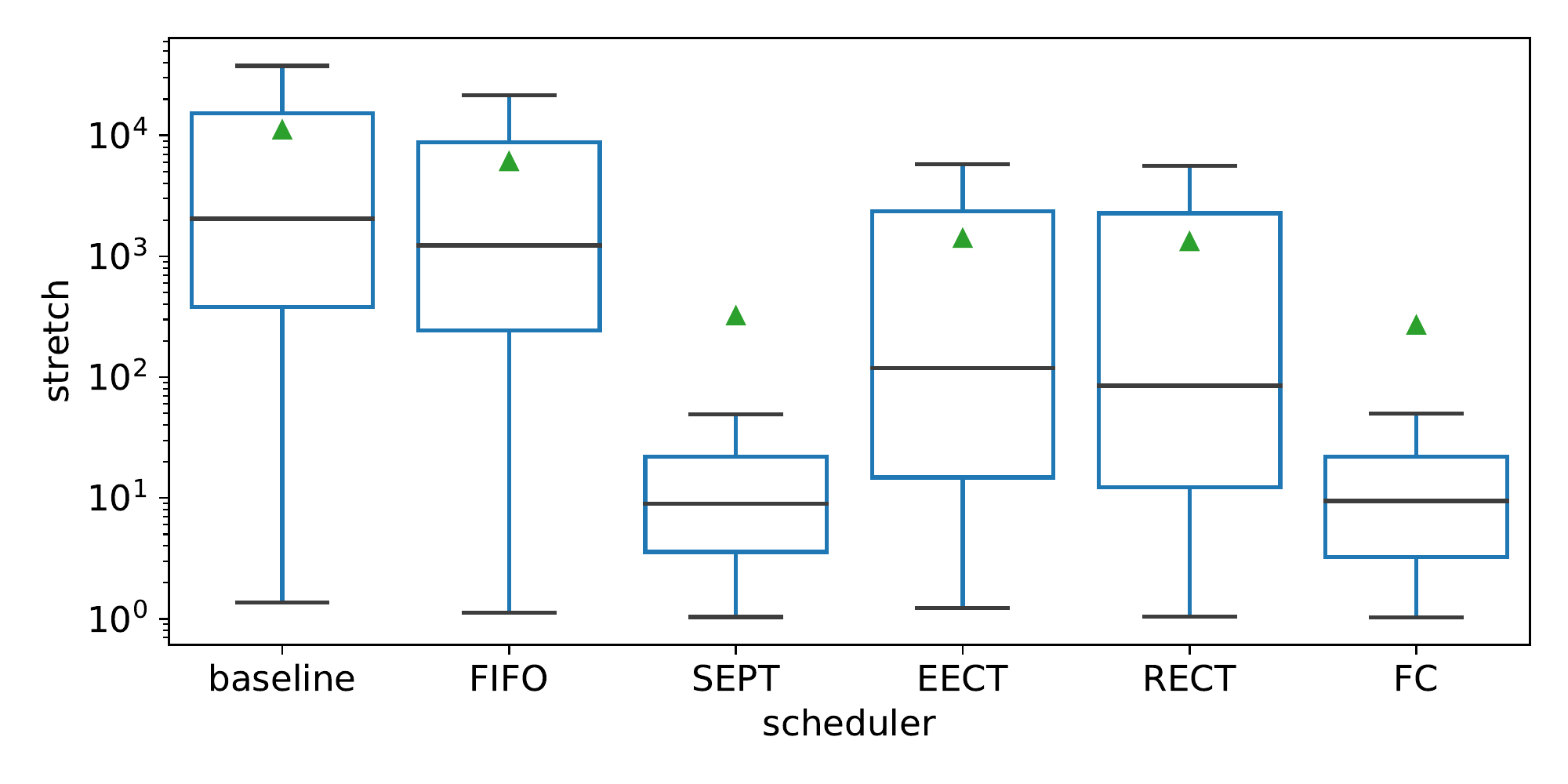}
    \\
    \caption{Stretch for different scheduling policies. Here, we present results for 20 CPUs and intensity 60. On-premise infrastructure. Average stretch presented as a green triangle.}
    \label{fig:stretch_20_60}
\end{figure*}

\begin{figure*}[h]
    \centering
    \chart{Experiment 1}{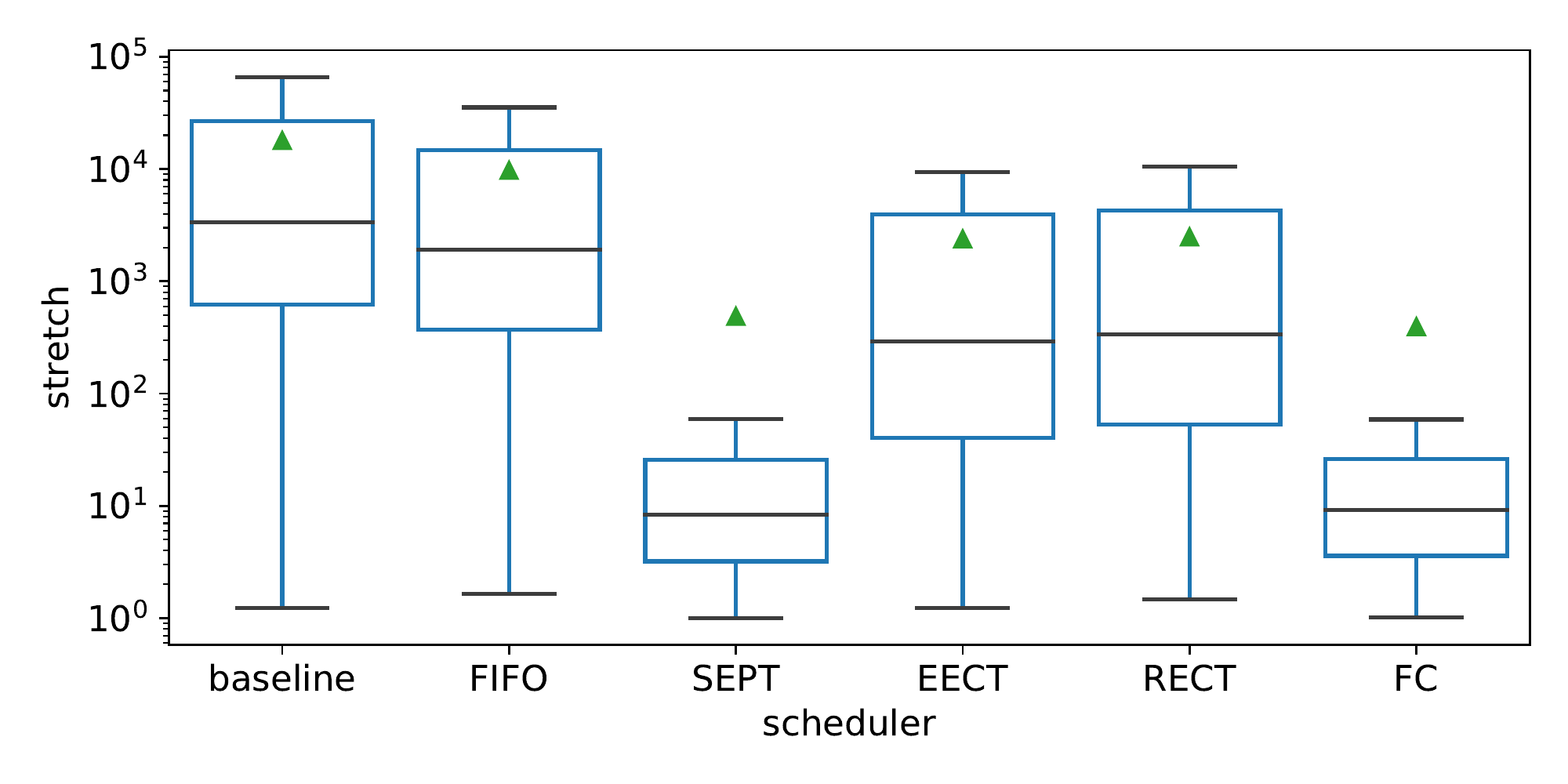}
    \chart{Experiment 2}{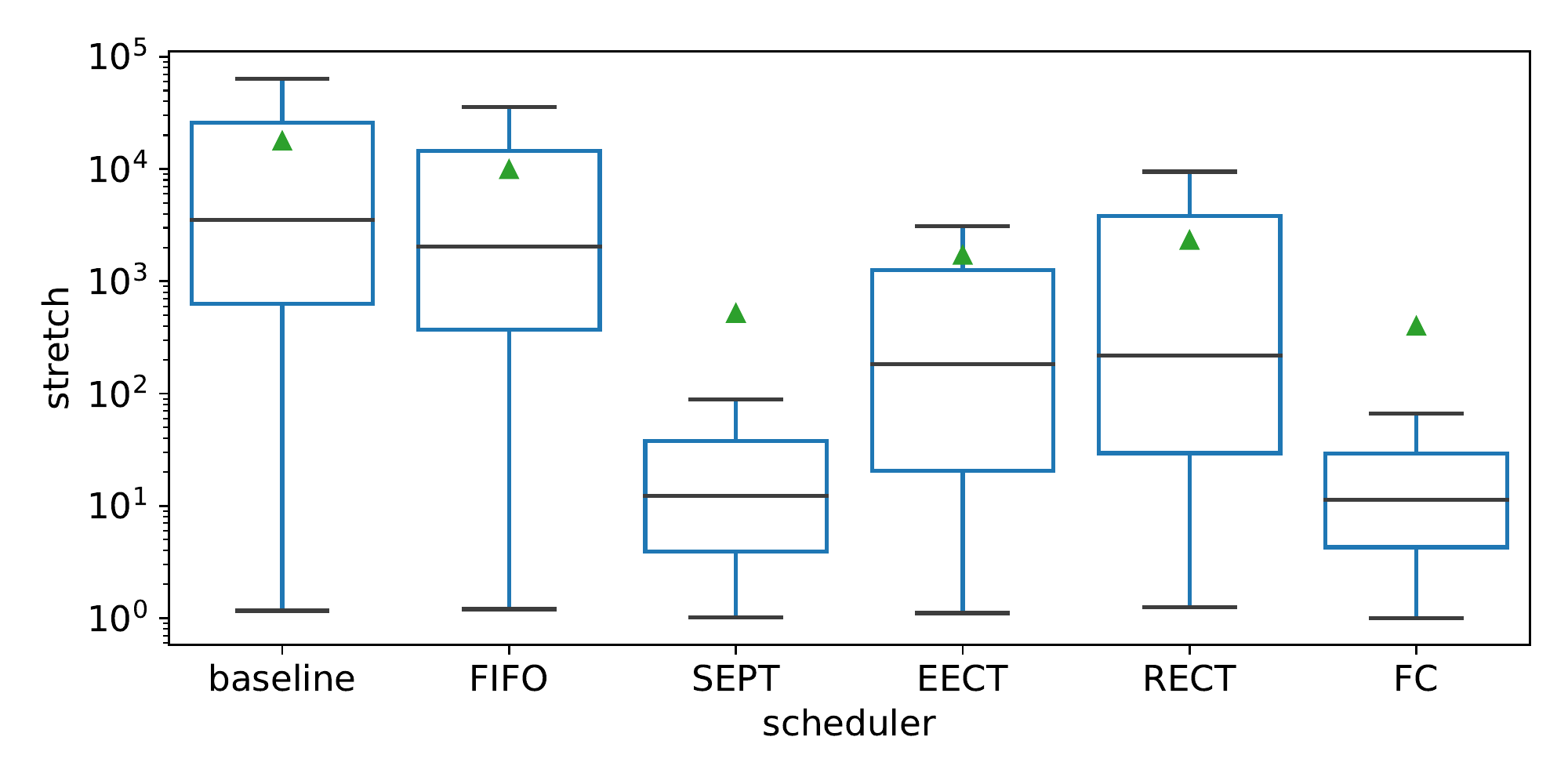}
    \chart{Experiment 3}{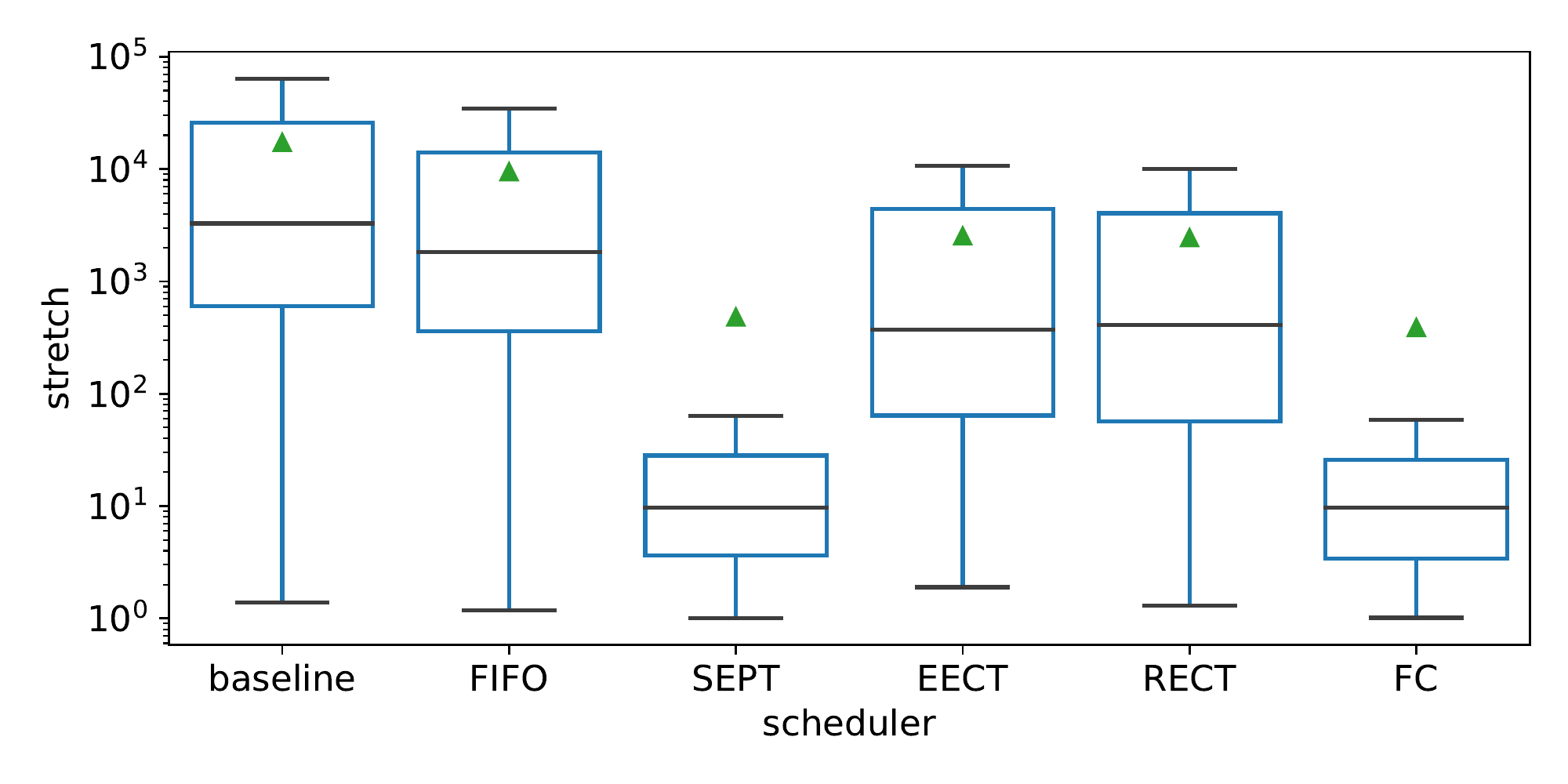}
    \\
    \chart{Experiment 4}{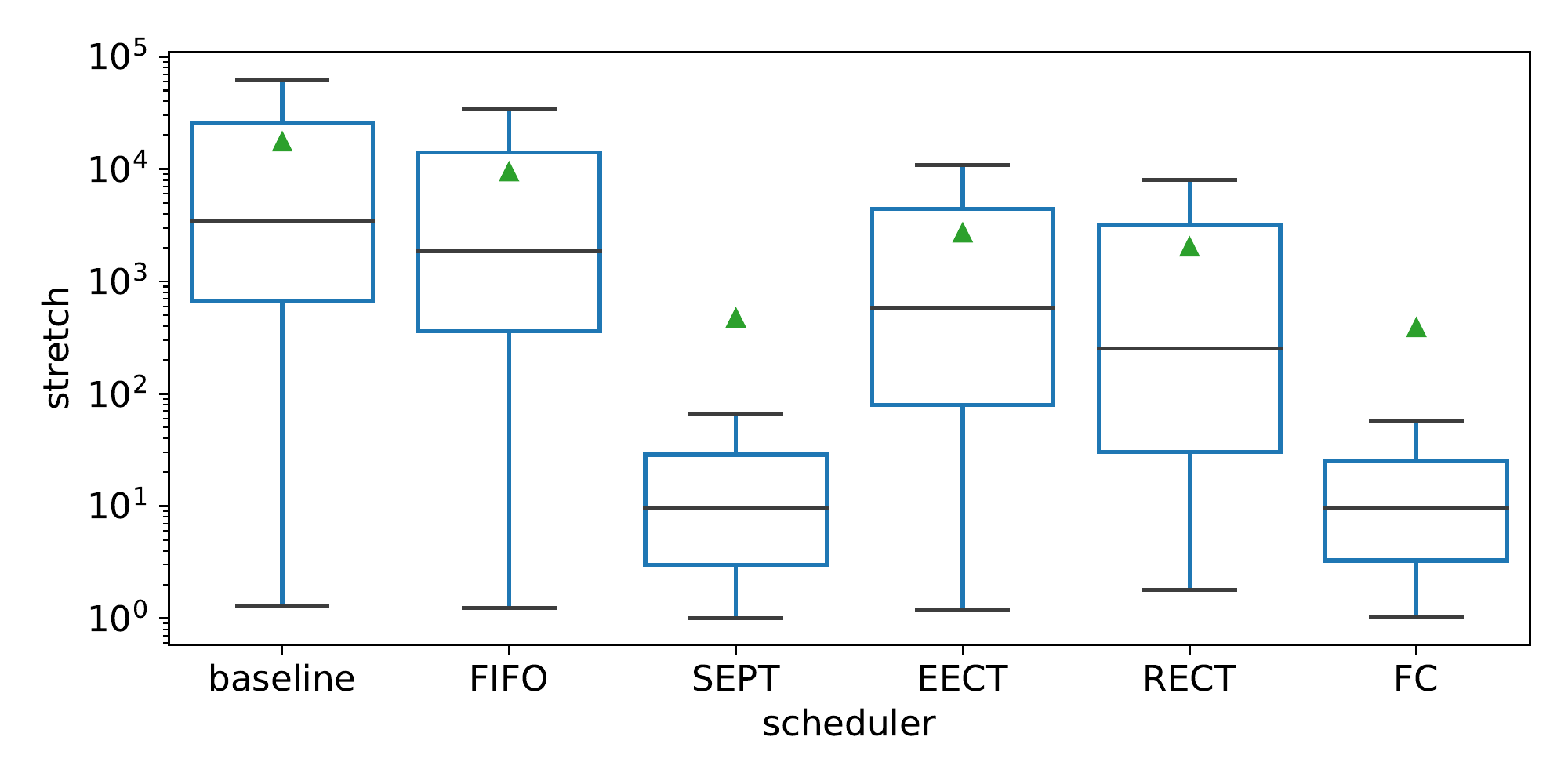}
    \chart{Experiment 5}{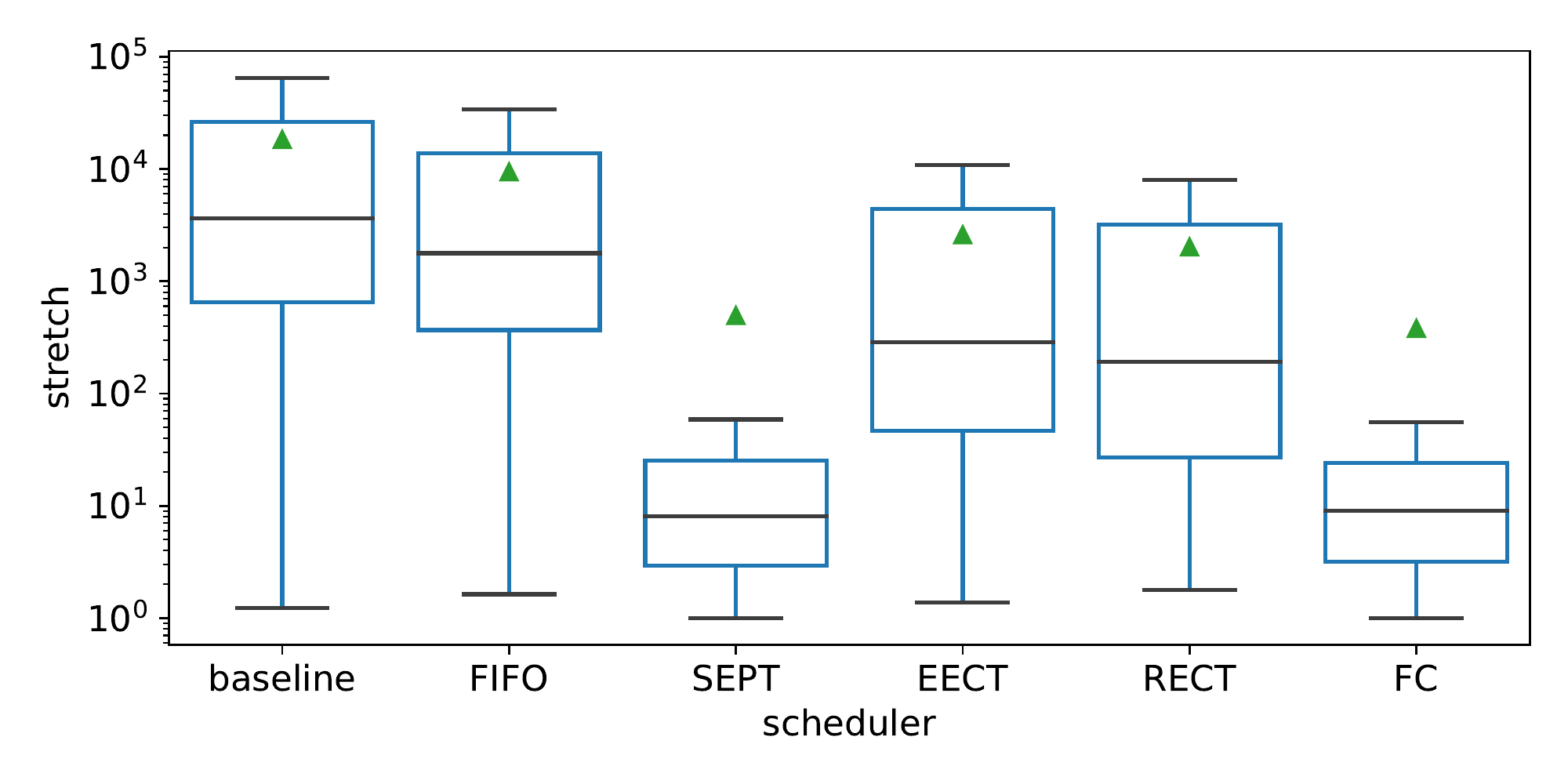}
    \\
    \caption{Stretch for different scheduling policies. Here, we present results for 20 CPUs and intensity 90. On-premise infrastructure. Average stretch presented as a green triangle.}
    \label{fig:stretch_20_90}
\end{figure*}

\begin{figure*}[h]
    \centering
    \chart{Experiment 1}{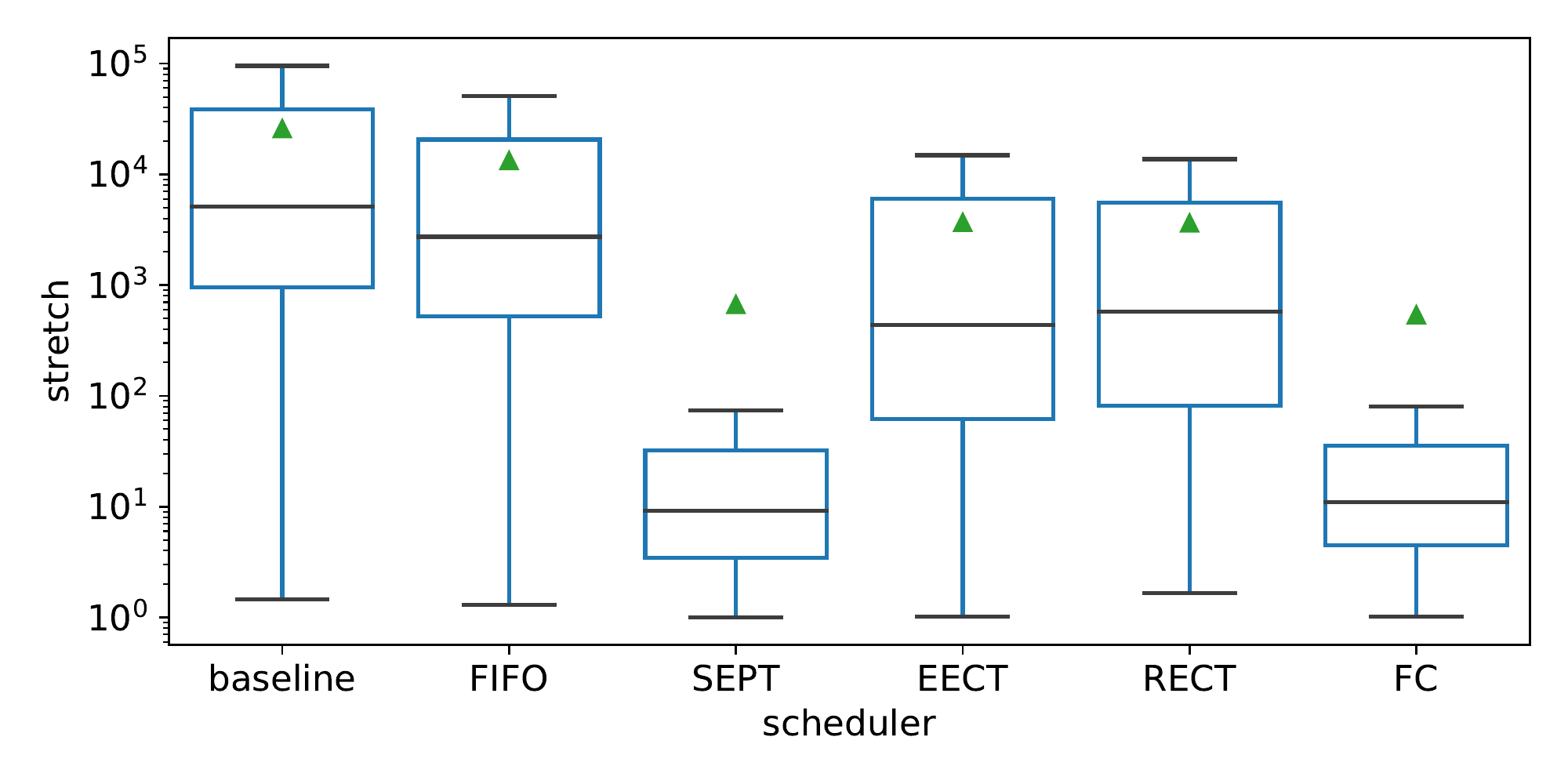}
    \chart{Experiment 2}{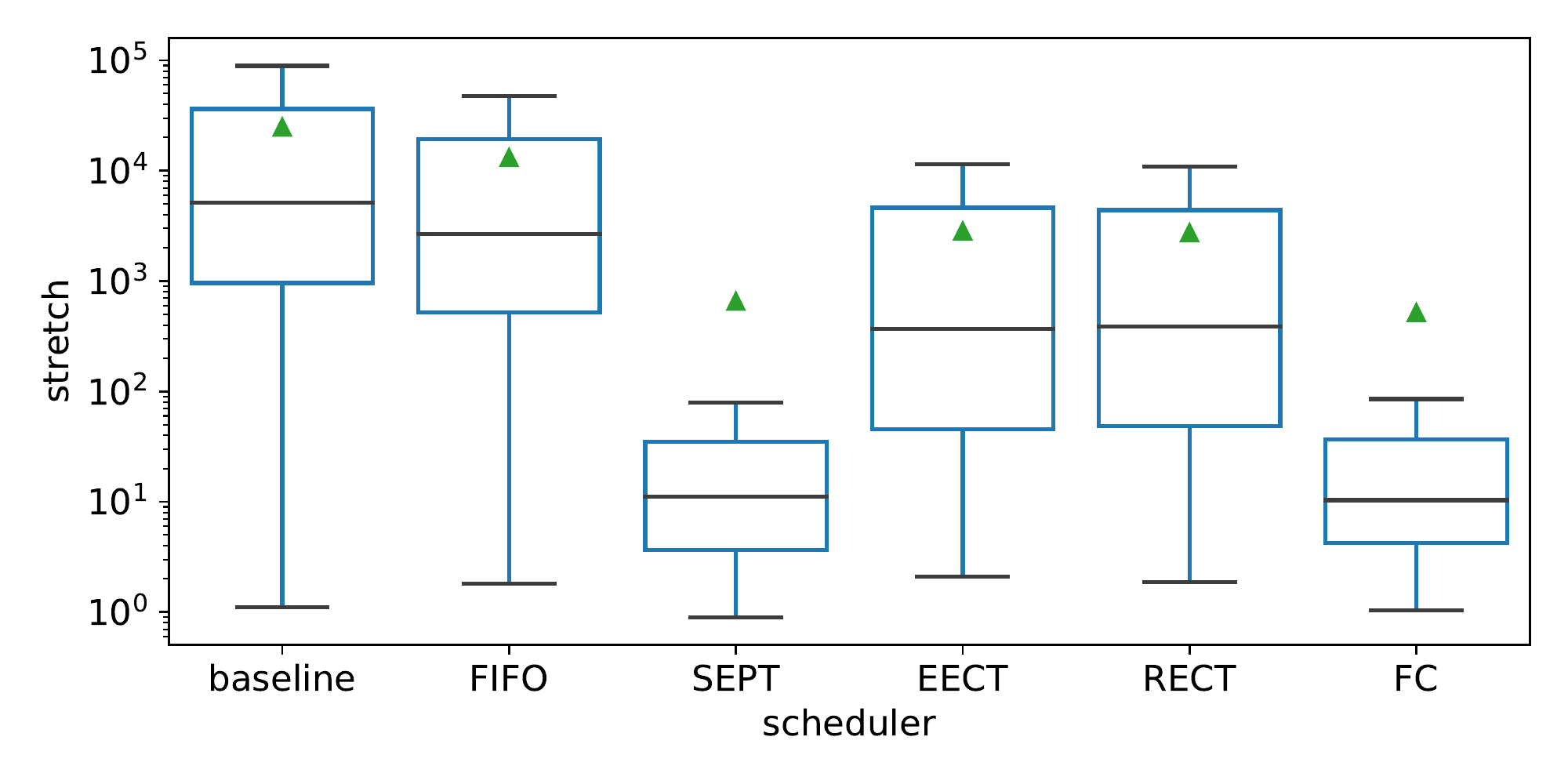}
    \chart{Experiment 3}{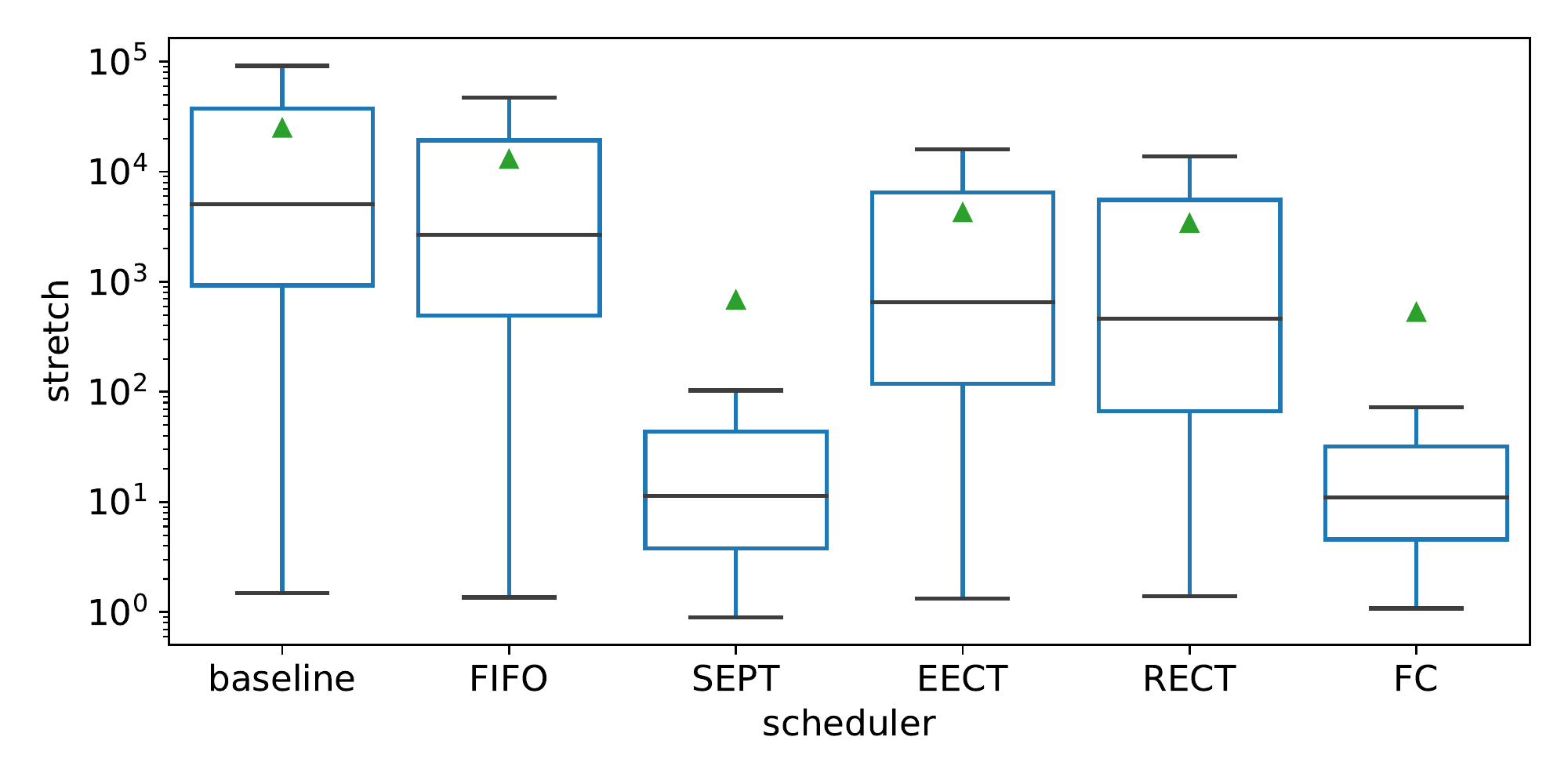}
    \\
    \chart{Experiment 4}{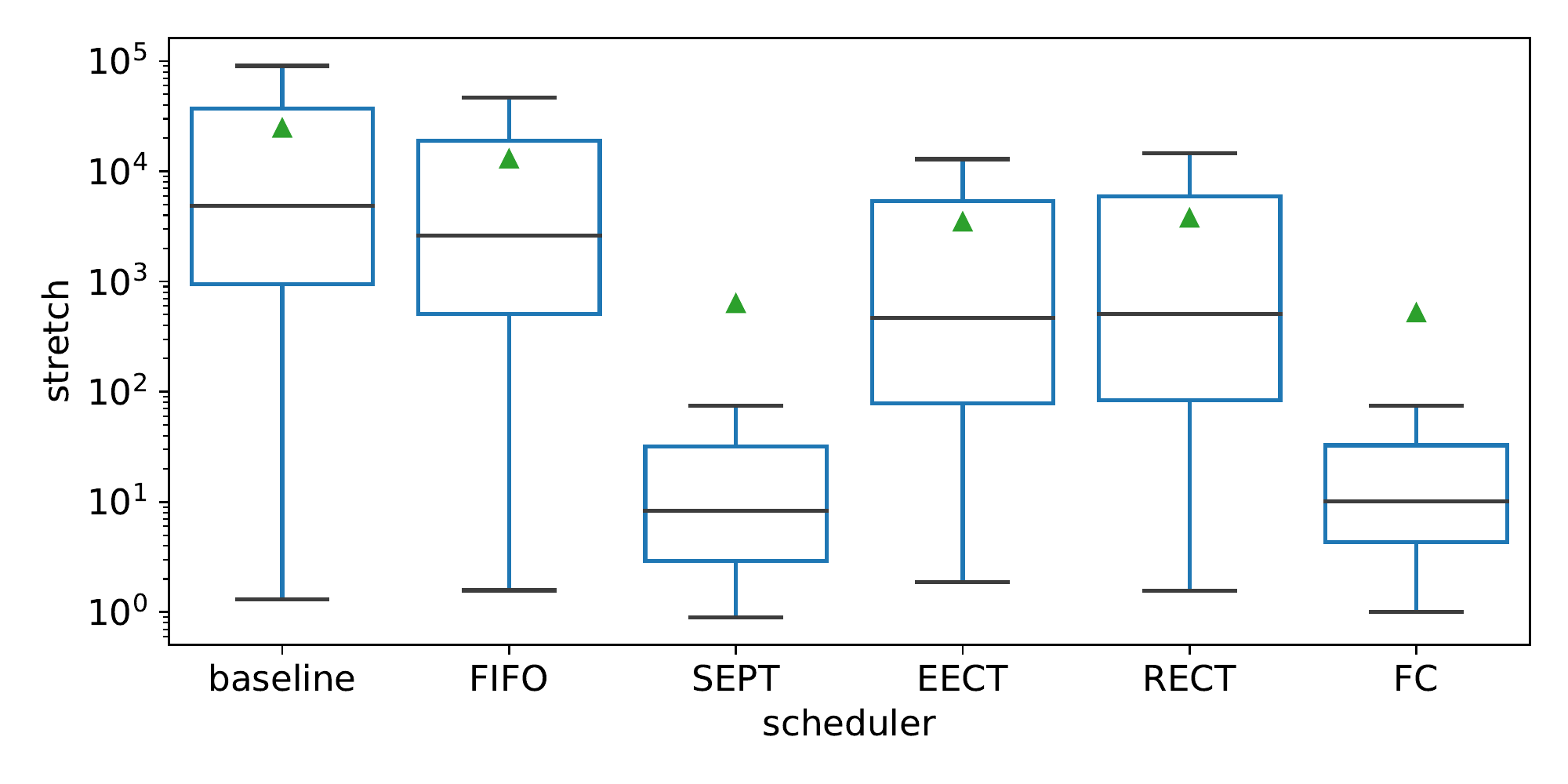}
    \chart{Experiment 5}{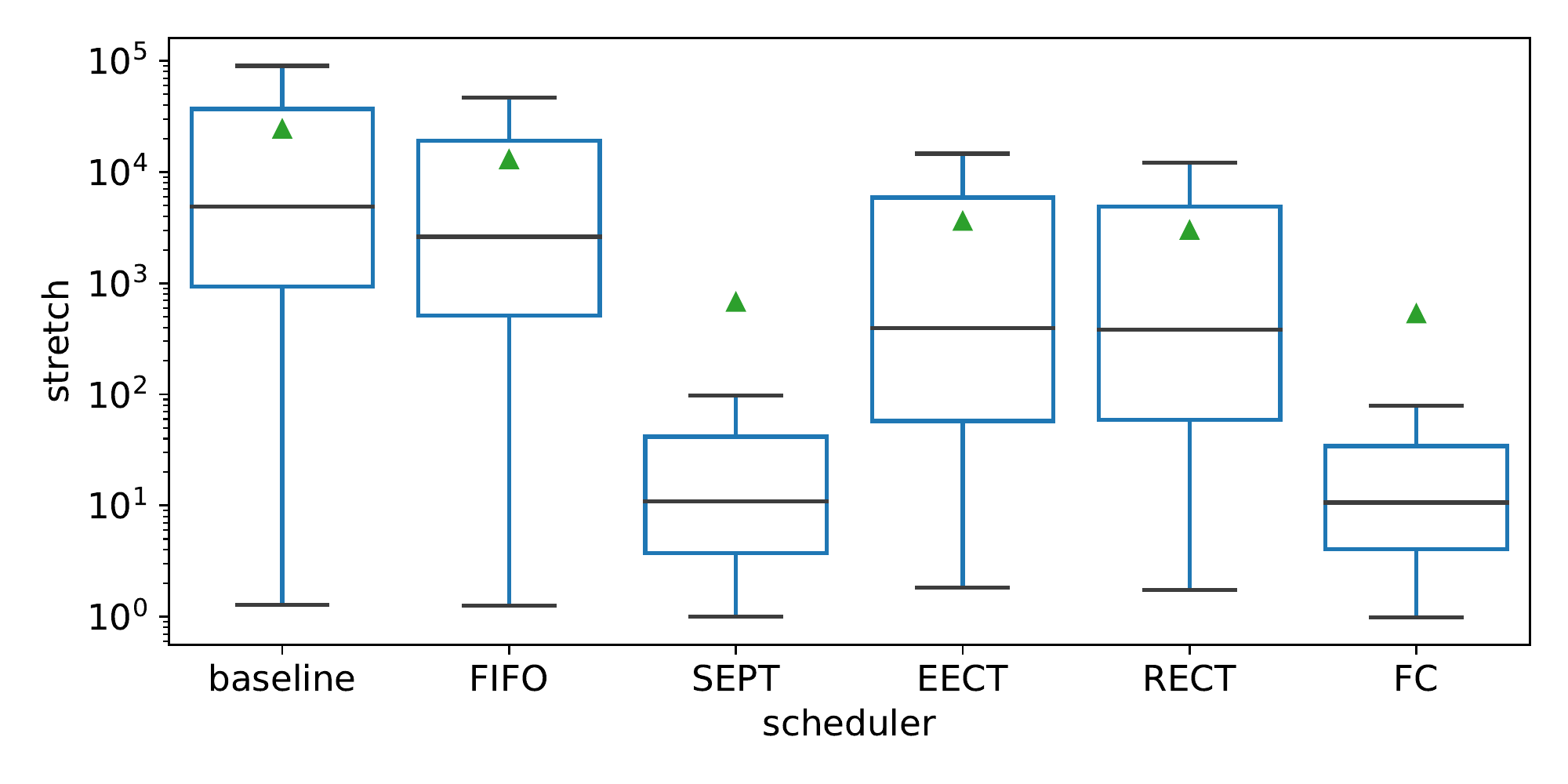}
    \\
    \caption{Stretch for different scheduling policies. Here, we present results for 20 CPUs and intensity 120. On-premise infrastructure. Average stretch presented as a green triangle.}
    \label{fig:stretch_20_120}
\end{figure*}

In Figures~\ref{fig:response_time_5_30}-\ref{fig:response_time_20_120}, we present the distributions of response times for each experiment performed under a fixed set of parameters (number of CPUs and intensity). In Figures~\ref{fig:stretch_5_30}-\ref{fig:stretch_20_120}, we present similar results for stretch.

\subsection{Multi-node experiments}

In Table~\ref{tab:num-cloud}, we present numerical results aggregated from all five experiments performed for each configuration. Then, in Table~\ref{tab:num-cloud-exp}, we present the same results in the context of each single experiment.

\begin{figure*}[h]
    \centering
    \chart{Experiment 1}{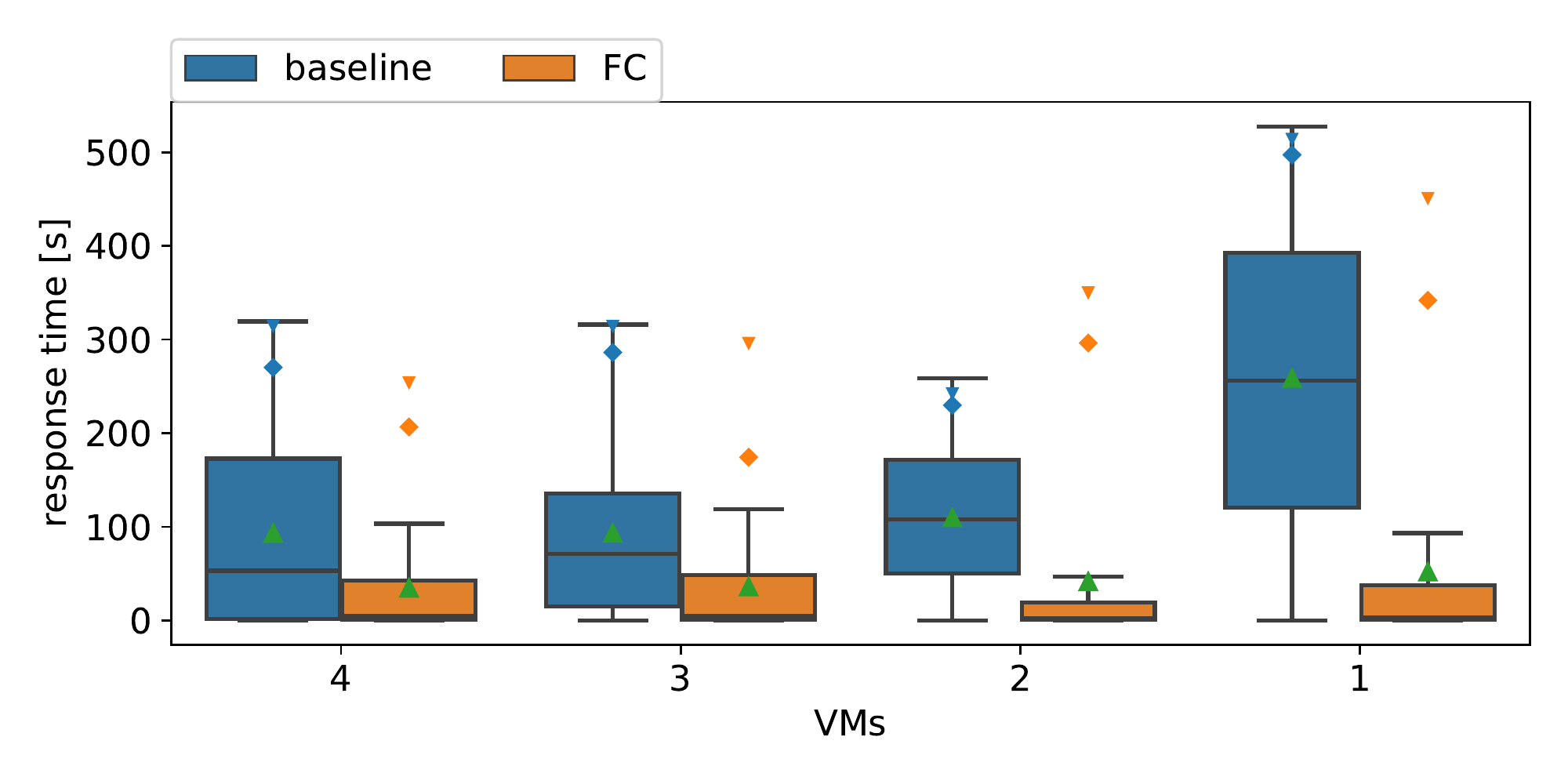}
    \chart{Experiment 2}{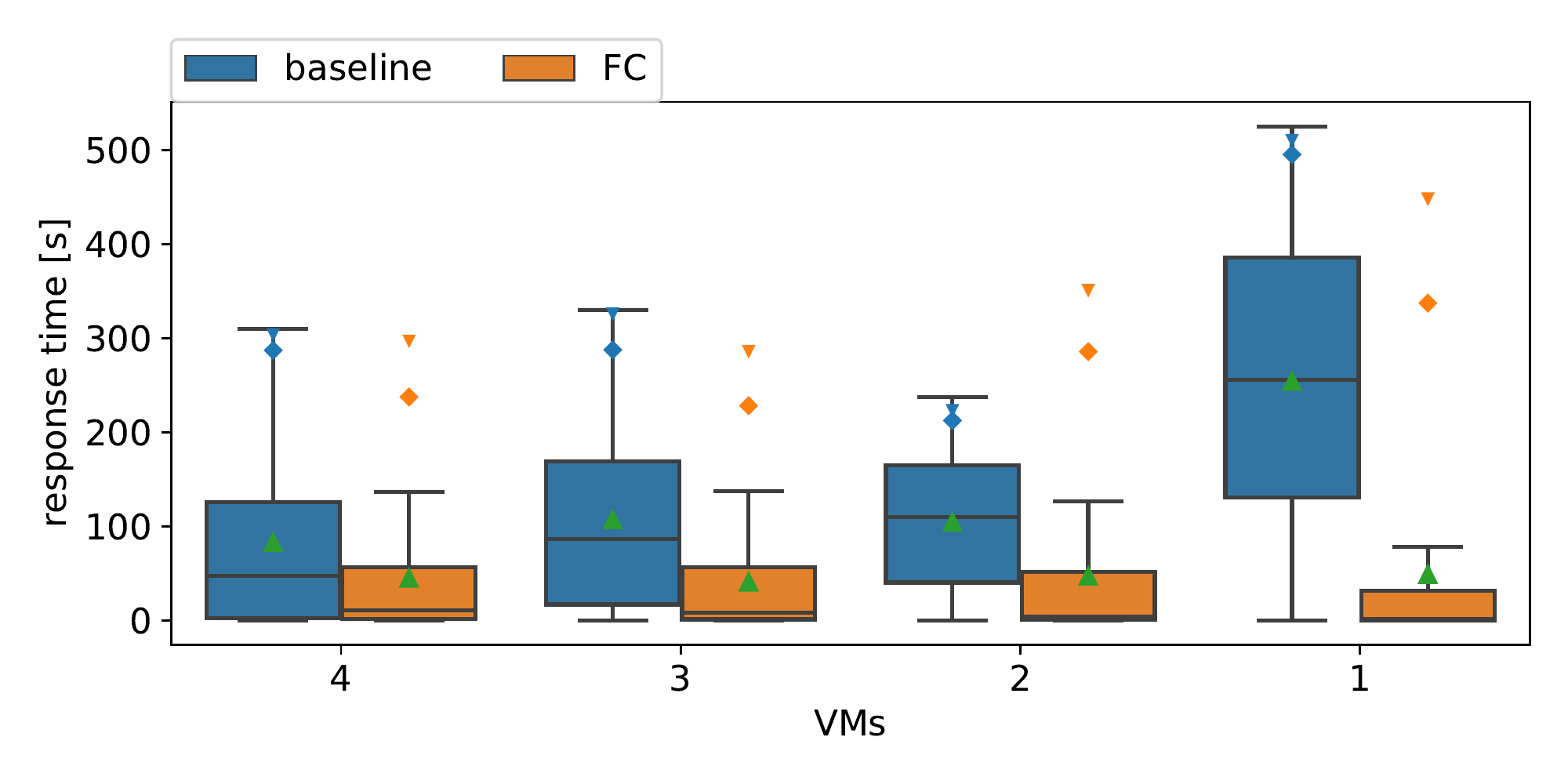}
    \chart{Experiment 3}{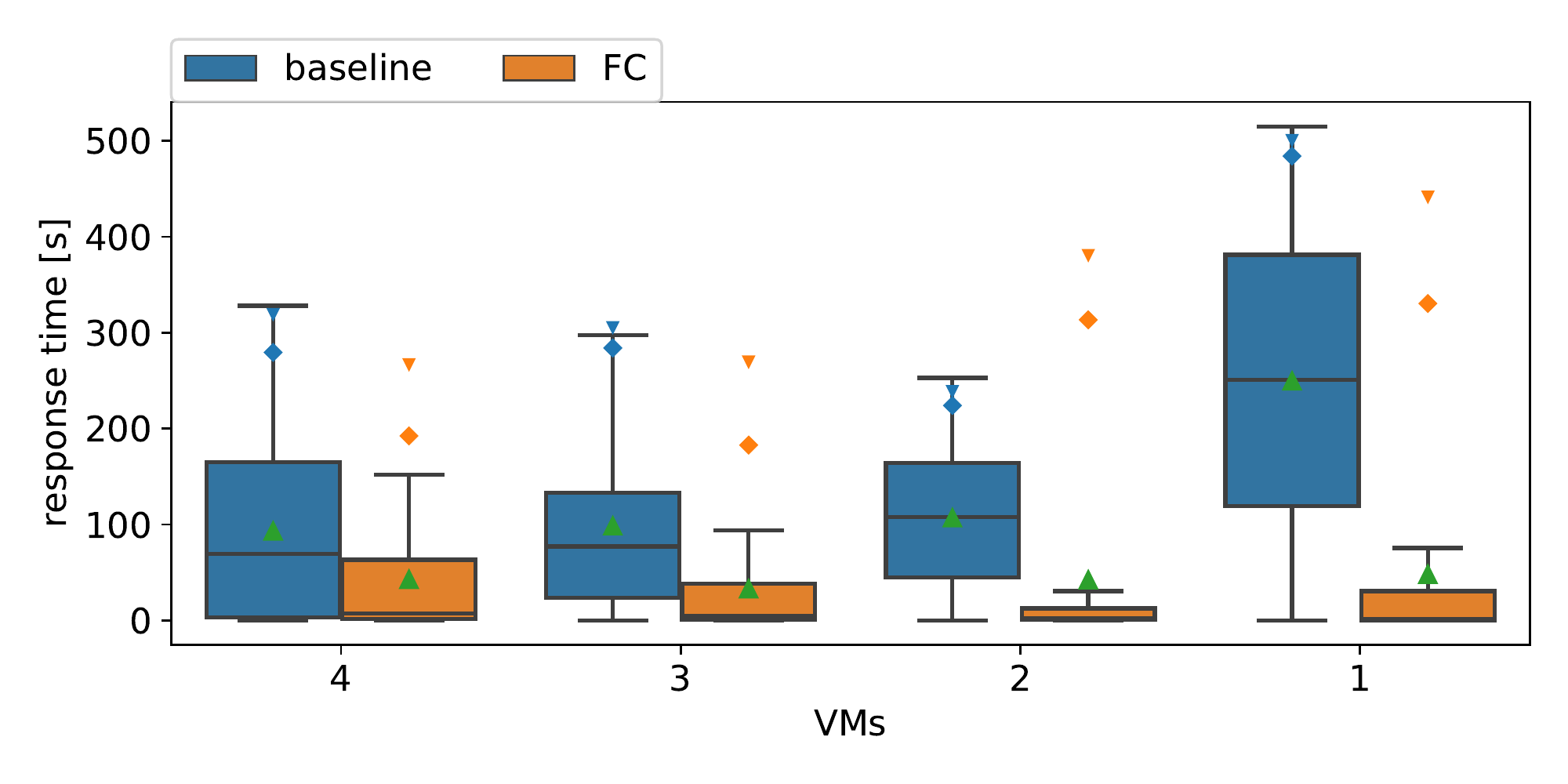}
    \\
    \chart{Experiment 4}{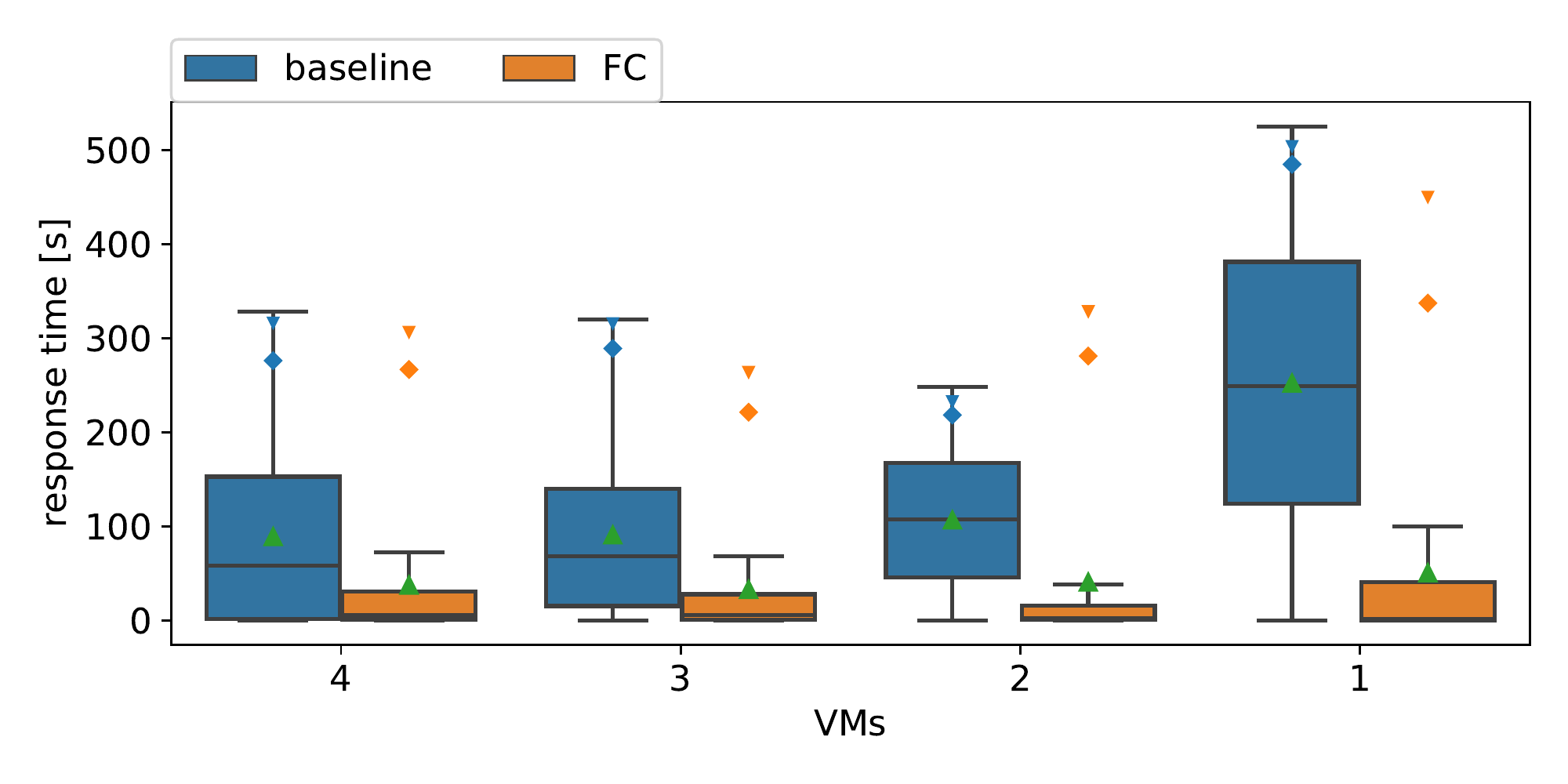}
    \chart{Experiment 5}{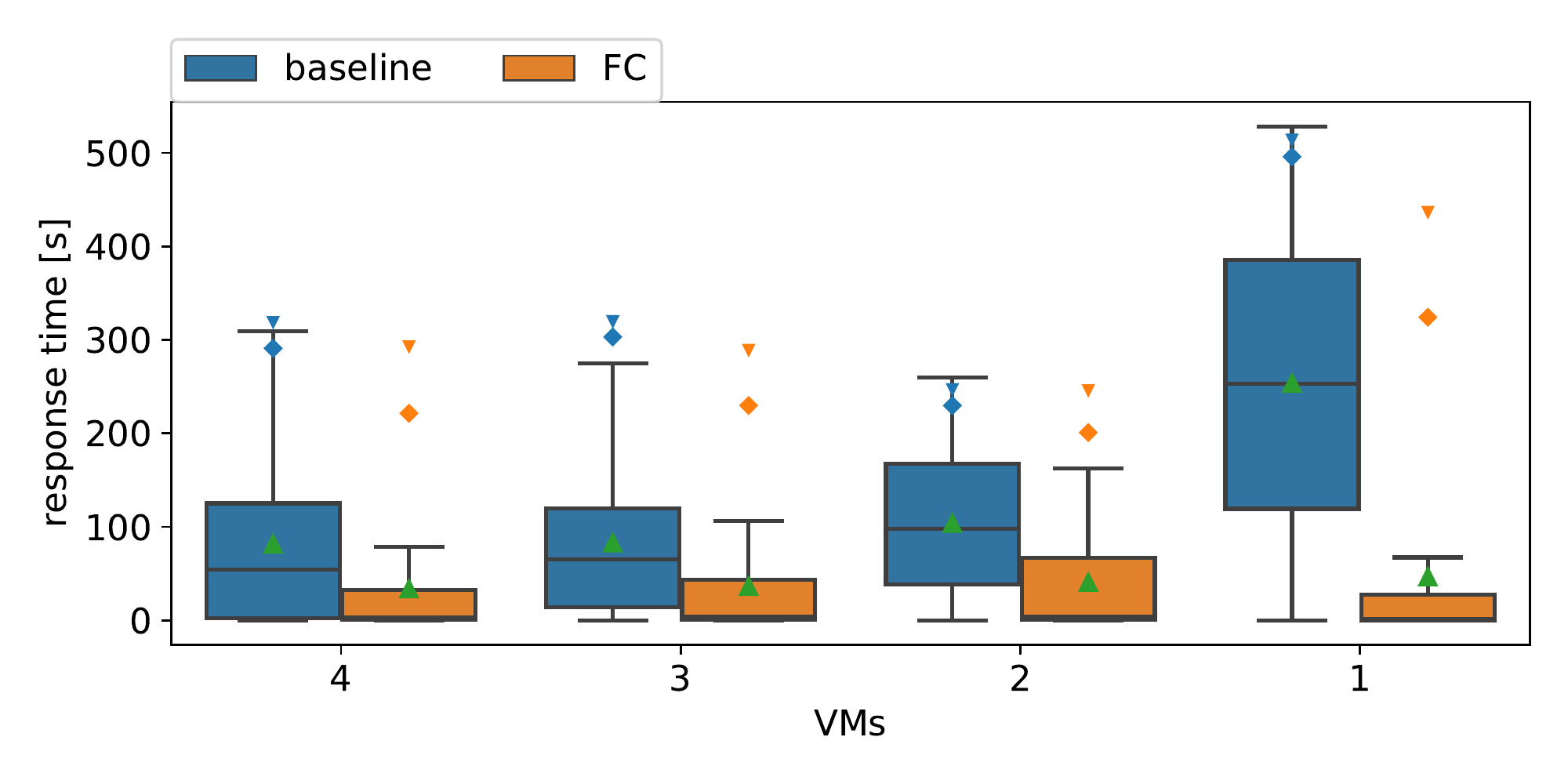}
    \\
    \caption{Response time for different scheduling policies. Here, we present results for 1-4 VMs and 10 CPUs on each VM. The load is constant (1320 requests). Cloud infrastructure. $\triangle$ is the average; $\diamond$ the 95th percentile; $\triangledown$ the 99th percentile.}
    \label{fig:response_time_cloud_10}
\end{figure*}

\begin{figure*}[h]
    \centering
    \chart{Experiment 1}{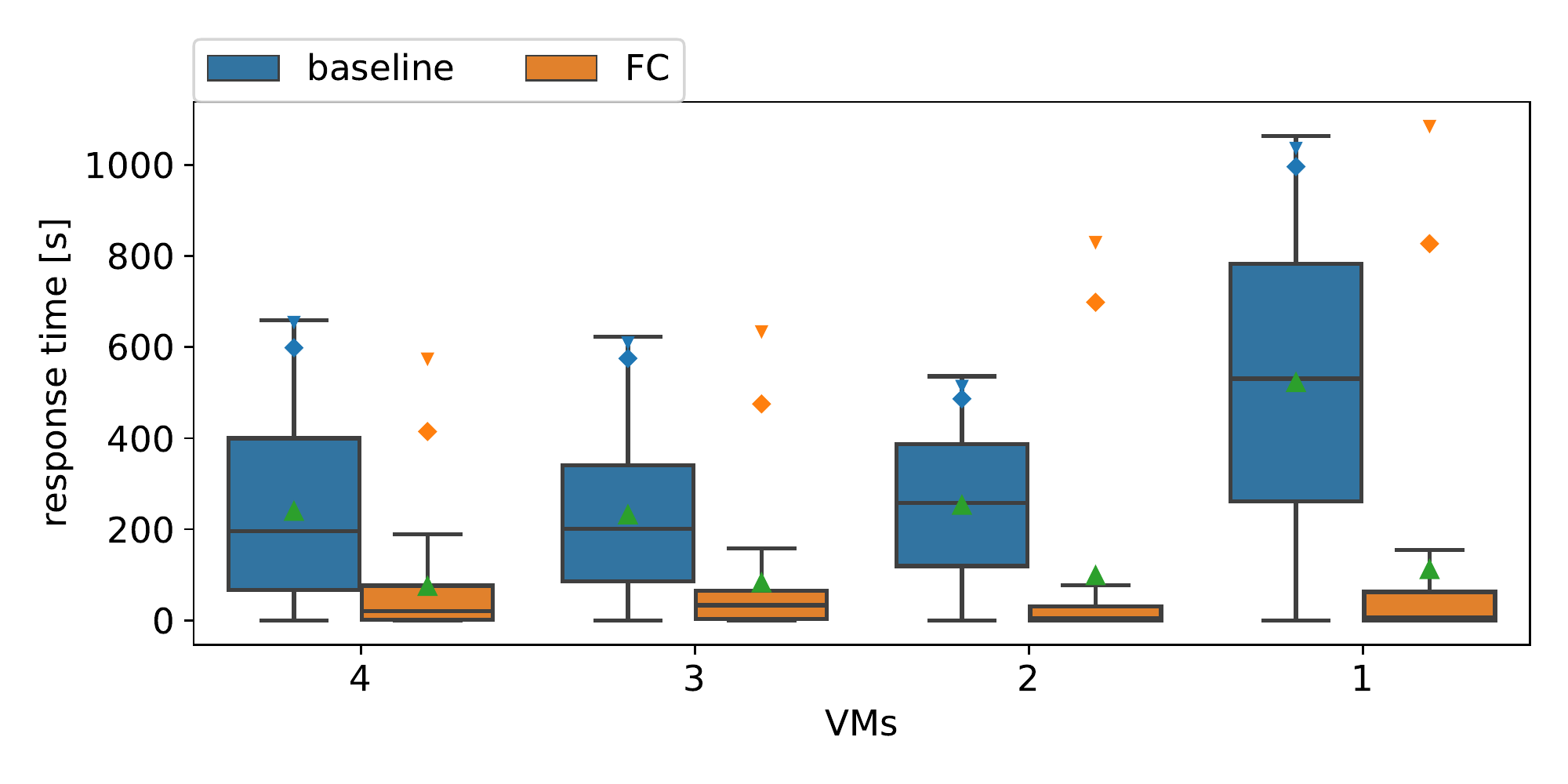}
    \chart{Experiment 2}{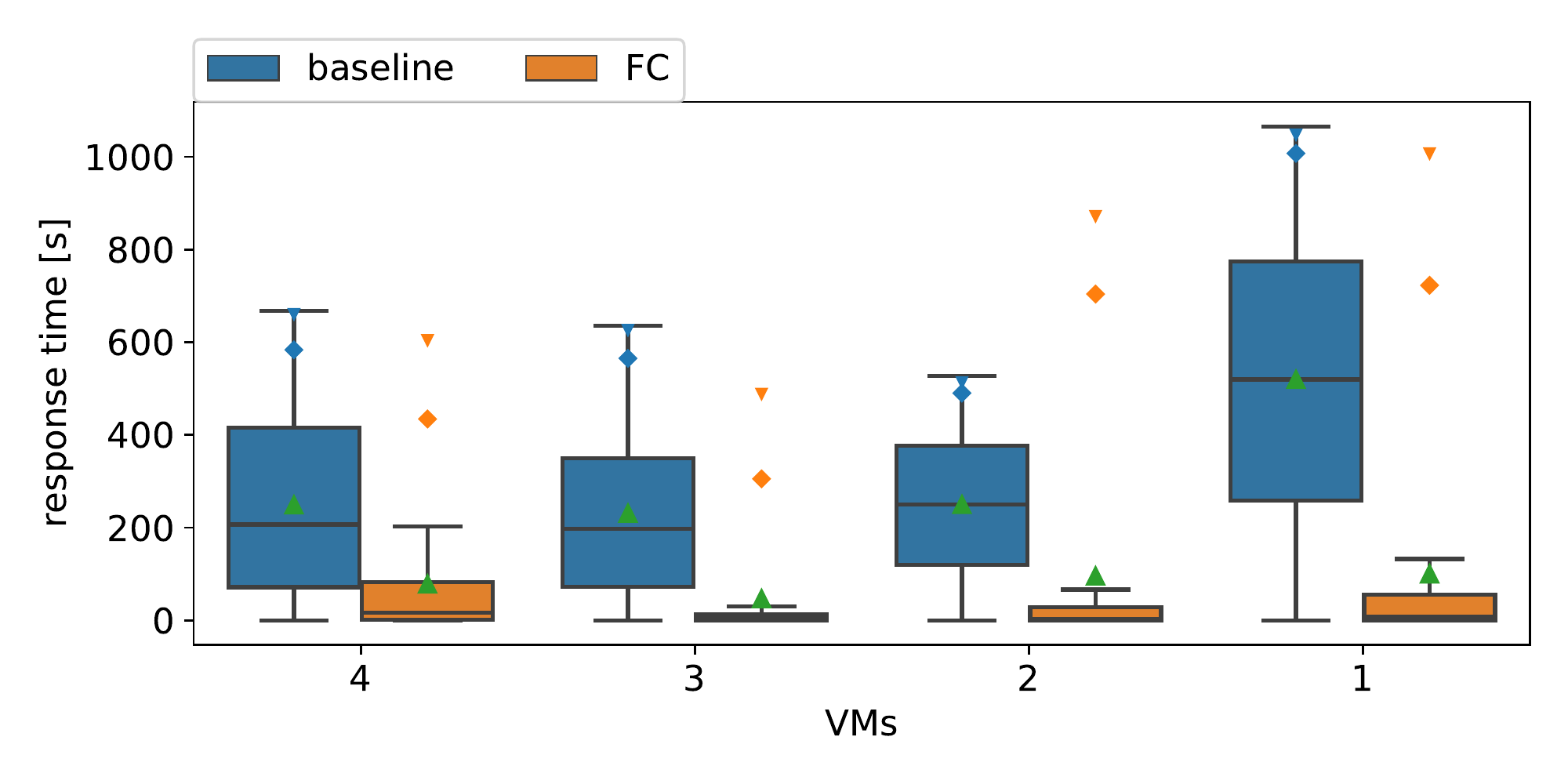}
    \chart{Experiment 3}{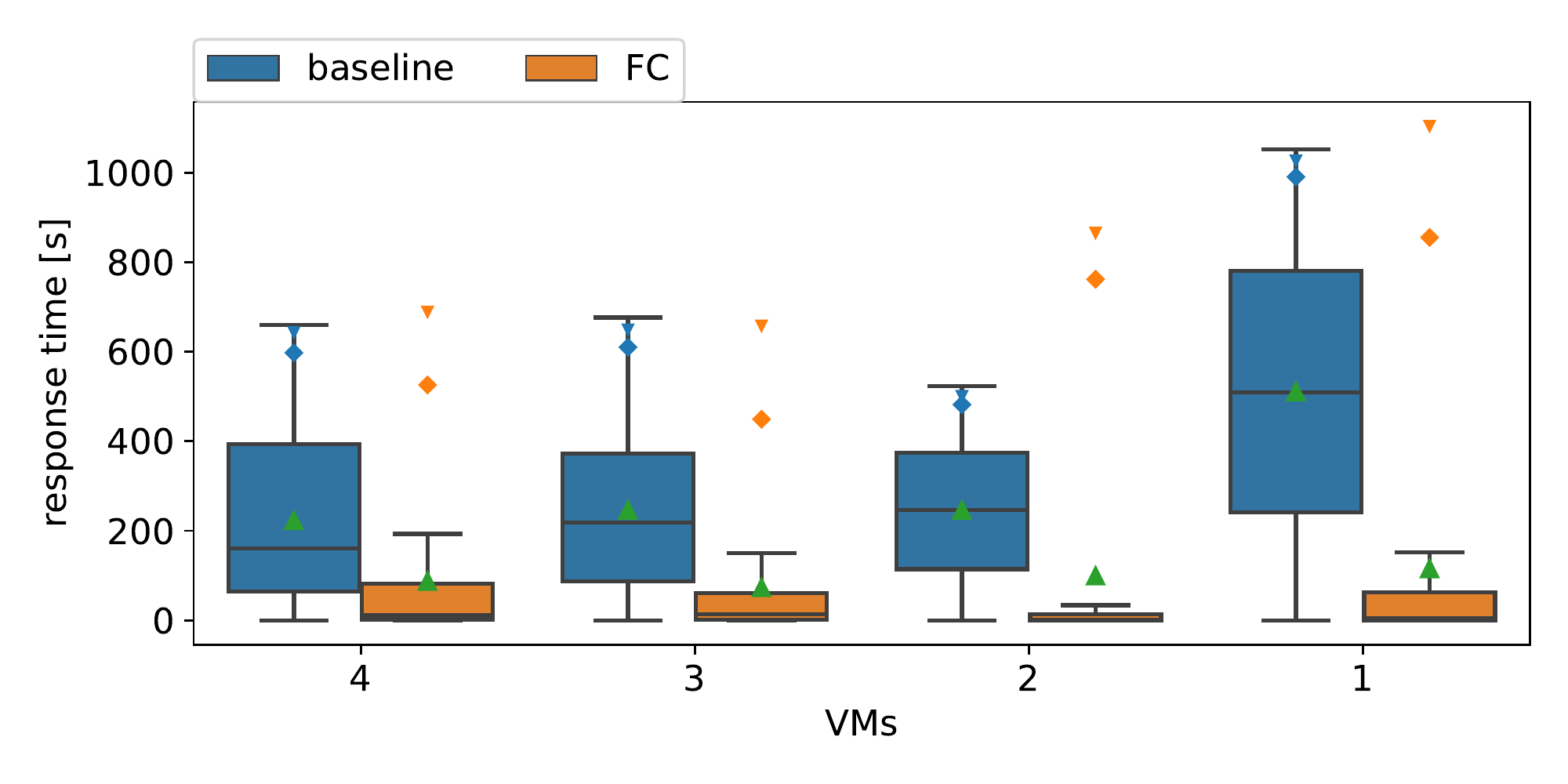}
    \\
    \chart{Experiment 4}{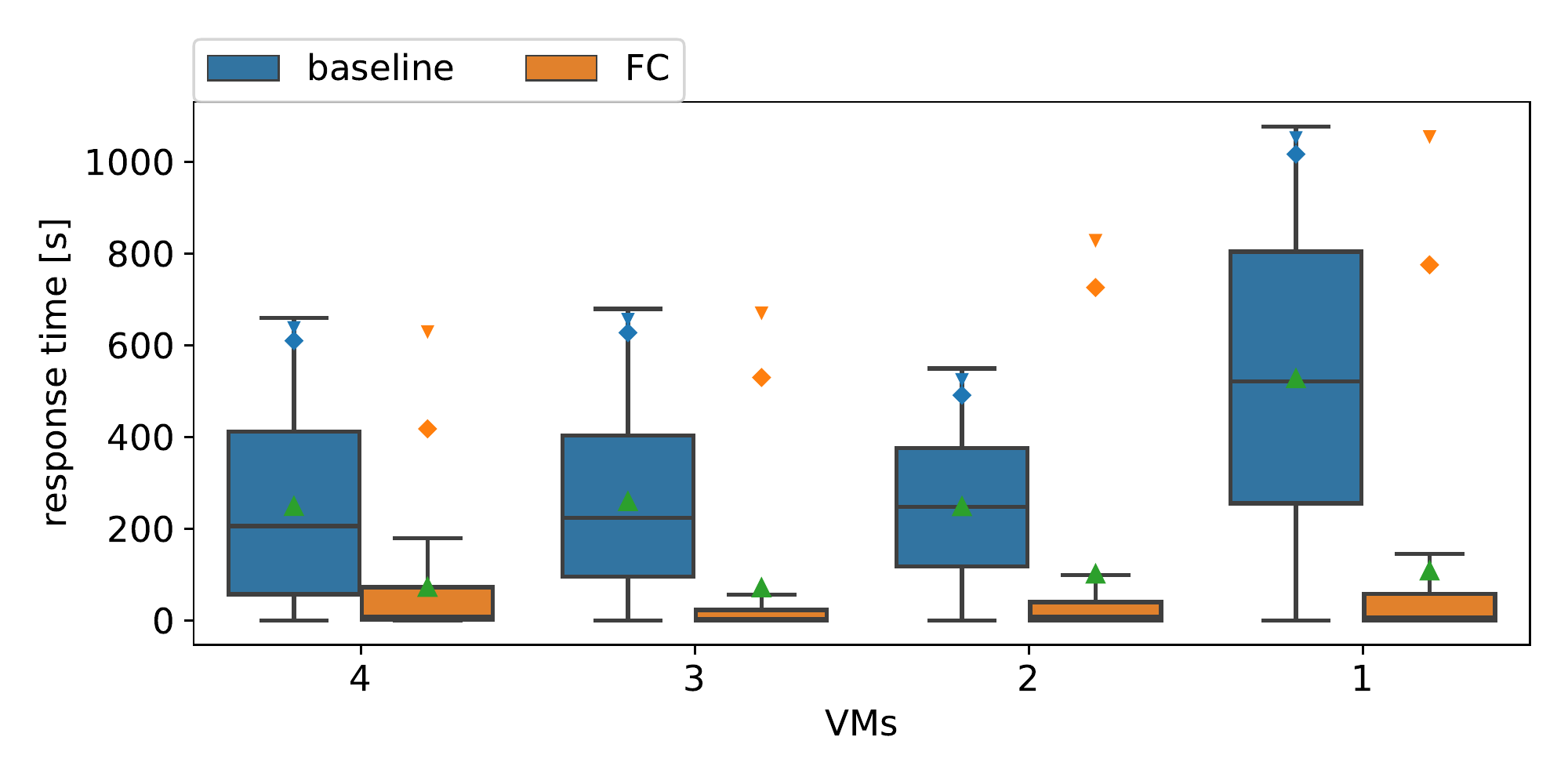}
    \chart{Experiment 5}{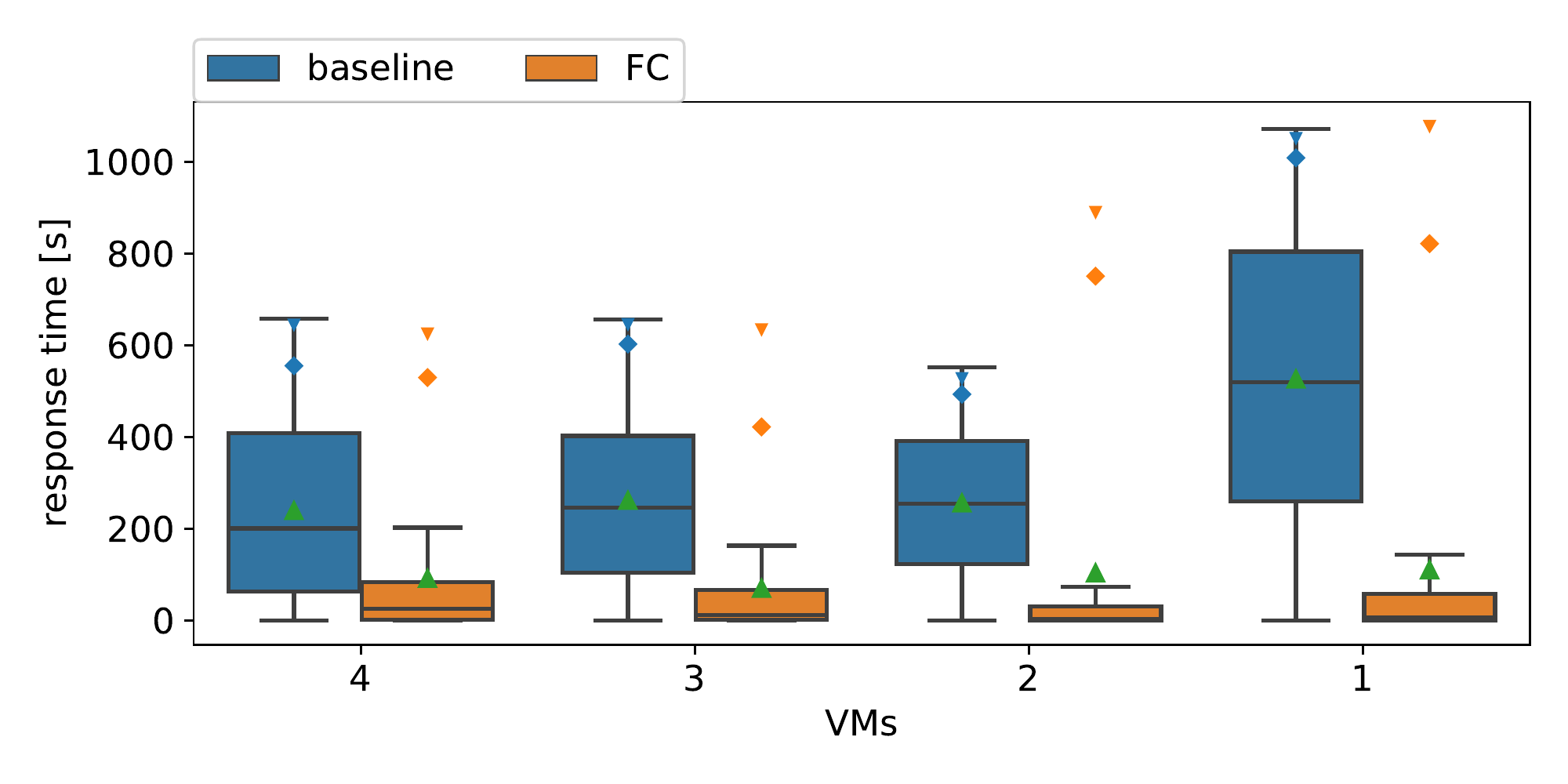}
    \\
    \caption{Response time for different scheduling policies. Here, we present results for 1-4 VMs and 18 CPUs on each VM. The load is constant (2376 requests). Cloud infrastructure. $\triangle$ is the average; $\diamond$ the 95th percentile; $\triangledown$ the 99th percentile.}
    \label{fig:response_time_cloud_18}
\end{figure*}

In Figures~\ref{fig:response_time_cloud_10} and \ref{fig:response_time_cloud_18}, we present the distributions of response times for each experiment performed under a fixed set of parameters (number of CPUs and requests) for a varying number of VMs.

\begin{table*}[h]
    \setcounter{table}{2}
    \caption{Aggregated numeric results for our experiments performed on-premises.}
    \label{tab:num-premises}
    \setlength{\tabcolsep}{4pt}
\begin{tabularx}{\textwidth}{ccl|rrrrr|rrrrr|>{\centering\arraybackslash}X}
        \toprule
        & & & \multicolumn{5}{c|}{$R(i)$ (seconds)} & \multicolumn{5}{c|}{$S(i)$} & \\
        \cmidrule(lr){4-8} \cmidrule(lr){9-13}
        CPUs & Int. & Strategy & Average & 50\%-ile & 75\%-ile & 95\%-ile & 99\%-ile & Average & 50\%-ile & 75\%-ile & 95\%-ile & 99\%-ile & $\max c(i)$\\
        \midrule
        5 & 30 & baseline & 3.79 & 0.49 & 4.11 & 18.90 & 32.14 & 18.40 & 3.83 & 5.58 & 154.34 & 265.63 & 73.53 \\
        5 & 30 & EECT & 6.43 & 3.88 & 8.00 & 25.04 & 29.57 & 99.15 & 13.62 & 126.00 & 468.43 & 787.05 & 85.57 \\
        5 & 30 & FC & 5.54 & 2.20 & 6.48 & 23.66 & 36.83 & 59.38 & 8.69 & 69.29 & 271.55 & 531.36 & 86.23 \\
        5 & 30 & FIFO & 10.79 & 10.97 & 16.34 & 22.48 & 27.57 & 267.49 & 37.72 & 317.00 & 1 310.82 & 1 541.55 & 87.56 \\
        5 & 30 & RECT & 6.74 & 3.76 & 9.27 & 25.42 & 30.84 & 110.13 & 12.27 & 134.08 & 529.12 & 918.88 & 85.89 \\
        5 & 30 & SEPT & 5.58 & 2.25 & 6.67 & 20.77 & 55.62 & 66.97 & 8.39 & 72.25 & 347.43 & 639.71 & 86.52 \\
\bottomrule
    \end{tabularx}
\end{table*}

\begin{table*}[h]
    \setcounter{table}{2}
    \caption{Aggregated numeric results for our experiments performed on-premises. (Continued)}
    \setlength{\tabcolsep}{4pt}
    \begin{tabularx}{\textwidth}{ccl|rrrrr|rrrrr|>{\centering\arraybackslash}X}
        \toprule
        & & & \multicolumn{5}{c|}{$R(i)$ (seconds)} & \multicolumn{5}{c|}{$S(i)$} & \\
        \cmidrule(lr){4-8} \cmidrule(lr){9-13}
        CPUs & Int. & Strategy & Average & 50\%-ile & 75\%-ile & 95\%-ile & 99\%-ile & Average & 50\%-ile & 75\%-ile & 95\%-ile & 99\%-ile & $\max c(i)$\\
        \midrule
        10 & 60 & baseline & 123.36 & 116.07 & 201.95 & 274.14 & 295.28 & 3 608.83 & 525.59 & 3 726.60 & 18 574.87 & 23 400.20 & 369.25 \\
        10 & 60 & EECT & 40.93 & 14.05 & 72.20 & 163.55 & 217.66 & 766.19 & 77.38 & 678.88 & 4 715.84 & 7 214.09 & 283.88 \\
        10 & 60 & FC & 22.65 & 1.07 & 5.43 & 168.50 & 213.96 & 134.24 & 9.24 & 25.96 & 1 392.55 & 1 768.29 & 280.89 \\
        10 & 60 & FIFO & 101.76 & 102.51 & 151.86 & 194.93 & 206.76 & 2 959.46 & 577.59 & 4 257.30 & 13 685.87 & 16 167.66 & 277.47 \\
        10 & 60 & RECT & 40.42 & 13.38 & 69.02 & 155.80 & 211.23 & 763.78 & 69.68 & 646.15 & 4 847.44 & 7 608.36 & 274.04 \\
        10 & 60 & SEPT & 25.14 & 1.07 & 4.55 & 179.04 & 269.92 & 164.52 & 8.50 & 28.35 & 1 479.64 & 2 230.73 & 314.87 \\
\bottomrule
    \end{tabularx}
\end{table*}

\begin{table*}[h]
    \setcounter{table}{3}
    \caption{Numeric results for our experiments performed on-premises for each of the experiments.}
    \label{tab:num-premises-exp}
    \setlength{\tabcolsep}{3.3pt}
    \begin{tabularx}{\textwidth}{cclc|rrrrr|rrrrr|>{\centering\arraybackslash}X}
        \toprule
        & & & & \multicolumn{5}{c|}{$R(i)$ (seconds)} & \multicolumn{5}{c|}{$S(i)$} & \\
        \cmidrule(lr){5-9} \cmidrule(lr){10-14}
        CPUs & Int. & Strategy & Exp. & Average & 50\%-ile & 75\%-ile & 95\%-ile & 99\%-ile & Average & 50\%-ile & 75\%-ile & 95\%-ile & 99\%-ile & $\max c(i)$\\
        \midrule
        5 & 30 & baseline & 1 & 3.30 & 0.49 & 3.68 & 14.54 & 22.28 & 15.26 & 3.58 & 5.17 & 117.16 & 184.15 & 73.01 \\
        5 & 30 & baseline & 2 & 4.55 & 0.53 & 5.19 & 26.79 & 32.80 & 23.70 & 3.89 & 5.67 & 221.43 & 271.09 & 72.25 \\
        5 & 30 & baseline & 3 & 3.46 & 0.28 & 4.37 & 15.23 & 29.69 & 14.68 & 3.67 & 5.58 & 95.35 & 245.36 & 73.53 \\
        5 & 30 & baseline & 4 & 4.98 & 0.72 & 5.63 & 29.10 & 39.11 & 26.51 & 4.50 & 6.42 & 240.49 & 323.22 & 71.46 \\
        5 & 30 & baseline & 5 & 2.65 & 0.48 & 1.92 & 11.66 & 18.78 & 11.86 & 3.37 & 5.00 & 93.74 & 155.19 & 69.40 \\
        5 & 30 & EECT & 1 & 5.52 & 3.05 & 7.26 & 21.49 & 26.81 & 81.19 & 10.51 & 81.68 & 437.25 & 612.20 & 84.58 \\
\bottomrule
    \end{tabularx}
\end{table*}

\begin{table*}[h]
    \setcounter{table}{3}
    \caption{Numeric results for our experiments performed on-premises for each of the experiments. (Continued)}
    \setlength{\tabcolsep}{3.3pt}
    \begin{tabularx}{\textwidth}{cclc|rrrrr|rrrrr|>{\centering\arraybackslash}X}
        \toprule
        & & & & \multicolumn{5}{c|}{$R(i)$ (seconds)} & \multicolumn{5}{c|}{$S(i)$} & \\
        \cmidrule(lr){5-9} \cmidrule(lr){10-14}
        CPUs & Int. & Strategy & Exp. & Average & 50\%-ile & 75\%-ile & 95\%-ile & 99\%-ile & Average & 50\%-ile & 75\%-ile & 95\%-ile & 99\%-ile & $\max c(i)$\\
        \midrule
        5 & 60 & baseline & 1 & 31.20 & 23.71 & 47.61 & 106.80 & 120.87 & 624.99 & 57.34 & 879.40 & 3 250.81 & 4 738.57 & 153.55 \\
        5 & 60 & baseline & 2 & 36.30 & 27.36 & 61.35 & 92.05 & 104.85 & 799.34 & 80.29 & 757.72 & 4 244.61 & 6 698.95 & 155.92 \\
        5 & 60 & baseline & 3 & 32.03 & 22.79 & 58.94 & 95.95 & 107.67 & 675.55 & 36.96 & 748.97 & 4 210.98 & 5 620.12 & 149.34 \\
        5 & 60 & baseline & 4 & 30.35 & 23.72 & 43.21 & 101.28 & 117.45 & 539.68 & 52.16 & 828.17 & 2 655.30 & 4 233.61 & 155.03 \\
        5 & 60 & baseline & 5 & 27.80 & 21.79 & 41.67 & 103.88 & 115.26 & 550.55 & 25.73 & 730.64 & 3 436.30 & 4 422.73 & 150.53 \\
        5 & 60 & EECT & 1 & 31.15 & 26.23 & 56.89 & 81.39 & 100.56 & 761.64 & 74.11 & 558.20 & 4 636.00 & 5 628.54 & 159.58 \\
\bottomrule
    \end{tabularx}
\end{table*}

\begin{table*}[h]
    \setcounter{table}{3}
    \caption{Numeric results for our experiments performed on-premises for each of the experiments. (Continued)}
    \setlength{\tabcolsep}{3.3pt}
    \begin{tabularx}{\textwidth}{cclc|rrrrr|rrrrr|>{\centering\arraybackslash}X}
        \toprule
        & & & & \multicolumn{5}{c|}{$R(i)$ (seconds)} & \multicolumn{5}{c|}{$S(i)$} & \\
        \cmidrule(lr){5-9} \cmidrule(lr){10-14}
        CPUs & Int. & Strategy & Exp. & Average & 50\%-ile & 75\%-ile & 95\%-ile & 99\%-ile & Average & 50\%-ile & 75\%-ile & 95\%-ile & 99\%-ile & $\max c(i)$\\
        \midrule
        5 & 120 & baseline & 1 & 122.03 & 126.81 & 190.36 & 254.81 & 268.88 & 3 462.26 & 499.79 & 3 663.00 & 17 557.67 & 21 433.94 & 336.22 \\
        5 & 120 & baseline & 2 & 127.56 & 125.78 & 198.50 & 264.33 & 277.78 & 3 646.66 & 720.64 & 3 665.63 & 17 361.50 & 22 230.94 & 345.26 \\
        5 & 120 & baseline & 3 & 117.30 & 121.54 & 189.79 & 251.63 & 266.91 & 3 261.92 & 418.81 & 2 827.27 & 17 187.56 & 20 962.98 & 333.72 \\
        5 & 120 & baseline & 4 & 117.78 & 117.25 & 187.99 & 249.21 & 260.13 & 3 252.62 & 523.47 & 2 872.29 & 17 170.21 & 20 718.30 & 324.33 \\
        5 & 120 & baseline & 5 & 117.87 & 116.66 & 189.00 & 249.18 & 255.61 & 3 374.07 & 549.97 & 3 334.76 & 17 053.94 & 20 837.54 & 324.51 \\
        5 & 120 & EECT & 1 & 87.50 & 87.06 & 151.12 & 203.14 & 248.01 & 2 234.71 & 280.96 & 1 869.49 & 11 953.09 & 15 407.09 & 312.97 \\
\bottomrule
    \end{tabularx}
\end{table*}

\begin{table*}[h]
    \setcounter{table}{3}
    \caption{Numeric results for our experiments performed on-premises for each of the experiments. (Continued)}
    \setlength{\tabcolsep}{3.3pt}
    \begin{tabularx}{\textwidth}{cclc|rrrrr|rrrrr|>{\centering\arraybackslash}X}
        \toprule
        & & & & \multicolumn{5}{c|}{$R(i)$ (seconds)} & \multicolumn{5}{c|}{$S(i)$} & \\
        \cmidrule(lr){5-9} \cmidrule(lr){10-14}
        CPUs & Int. & Strategy & Exp. & Average & 50\%-ile & 75\%-ile & 95\%-ile & 99\%-ile & Average & 50\%-ile & 75\%-ile & 95\%-ile & 99\%-ile & $\max c(i)$\\
        \midrule
        10 & 40 & baseline & 1 & 59.88 & 57.94 & 101.90 & 147.43 & 155.88 & 1 676.88 & 180.47 & 1 161.32 & 9 180.49 & 12 268.83 & 219.78 \\
        10 & 40 & baseline & 2 & 71.13 & 71.74 & 120.46 & 162.31 & 163.55 & 2 088.48 & 273.68 & 1 391.91 & 10 665.65 & 13 571.08 & 231.75 \\
        10 & 40 & baseline & 3 & 51.40 & 39.37 & 90.18 & 138.24 & 143.77 & 1 397.22 & 108.02 & 866.32 & 8 384.85 & 10 880.68 & 205.13 \\
        10 & 40 & baseline & 4 & 75.80 & 62.93 & 130.36 & 181.03 & 184.49 & 2 158.32 & 266.77 & 1 848.00 & 12 045.26 & 15 078.37 & 251.03 \\
        10 & 40 & baseline & 5 & 63.95 & 61.67 & 109.26 & 149.12 & 155.58 & 1 864.75 & 186.38 & 1 208.51 & 9 458.74 & 12 262.58 & 215.87 \\
        10 & 40 & EECT & 1 & 21.77 & 5.33 & 29.05 & 114.52 & 137.21 & 294.80 & 33.11 & 268.81 & 1 262.54 & 2 762.69 & 199.08 \\
\bottomrule
    \end{tabularx}
\end{table*}

\begin{table*}[h]
    \setcounter{table}{3}
    \caption{Numeric results for our experiments performed on-premises for each of the experiments. (Continued)}
    \setlength{\tabcolsep}{3.3pt}
    \begin{tabularx}{\textwidth}{cclc|rrrrr|rrrrr|>{\centering\arraybackslash}X}
        \toprule
        & & & & \multicolumn{5}{c|}{$R(i)$ (seconds)} & \multicolumn{5}{c|}{$S(i)$} & \\
        \cmidrule(lr){5-9} \cmidrule(lr){10-14}
        CPUs & Int. & Strategy & Exp. & Average & 50\%-ile & 75\%-ile & 95\%-ile & 99\%-ile & Average & 50\%-ile & 75\%-ile & 95\%-ile & 99\%-ile & $\max c(i)$\\
        \midrule
        10 & 90 & baseline & 1 & 155.48 & 154.60 & 241.52 & 302.53 & 330.13 & 4 460.32 & 979.73 & 6 089.49 & 20 976.80 & 25 013.86 & 389.01 \\
        10 & 90 & baseline & 2 & 157.45 & 159.78 & 239.82 & 299.68 & 327.21 & 4 537.49 & 952.11 & 6 571.50 & 21 181.51 & 24 765.16 & 395.02 \\
        10 & 90 & baseline & 3 & 155.03 & 155.34 & 236.77 & 312.36 & 324.80 & 4 482.98 & 857.62 & 6 280.72 & 21 708.50 & 25 846.05 & 391.85 \\
        10 & 90 & baseline & 4 & 178.74 & 172.02 & 273.39 & 363.44 & 370.88 & 5 300.02 & 1 054.70 & 6 409.69 & 26 440.68 & 30 450.22 & 442.46 \\
        10 & 90 & baseline & 5 & 170.33 & 161.02 & 262.91 & 352.69 & 364.21 & 4 959.93 & 1 031.42 & 6 671.69 & 22 749.39 & 29 040.72 & 431.29 \\
        10 & 90 & EECT & 1 & 70.55 & 28.83 & 131.15 & 244.07 & 338.36 & 1 444.40 & 140.87 & 1 194.19 & 8 789.56 & 13 688.38 & 409.96 \\
\bottomrule
    \end{tabularx}
\end{table*}

\begin{table*}[h]
    \setcounter{table}{3}
    \caption{Numeric results for our experiments performed on-premises for each of the experiments. (Continued)}
    \setlength{\tabcolsep}{3.3pt}
    \begin{tabularx}{\textwidth}{cclc|rrrrr|rrrrr|>{\centering\arraybackslash}X}
        \toprule
        & & & & \multicolumn{5}{c|}{$R(i)$ (seconds)} & \multicolumn{5}{c|}{$S(i)$} & \\
        \cmidrule(lr){5-9} \cmidrule(lr){10-14}
        CPUs & Int. & Strategy & Exp. & Average & 50\%-ile & 75\%-ile & 95\%-ile & 99\%-ile & Average & 50\%-ile & 75\%-ile & 95\%-ile & 99\%-ile & $\max c(i)$\\
        \midrule
        20 & 30 & baseline & 1 & 153.53 & 149.16 & 240.27 & 317.83 & 327.07 & 4 594.96 & 511.28 & 6 670.27 & 22 287.83 & 26 578.84 & 394.92 \\
        20 & 30 & baseline & 2 & 181.52 & 183.58 & 273.00 & 345.14 & 360.10 & 5 364.05 & 1 036.06 & 8 429.92 & 25 081.90 & 28 892.60 & 421.43 \\
        20 & 30 & baseline & 3 & 117.25 & 109.35 & 197.52 & 261.82 & 278.40 & 3 407.55 & 353.51 & 3 191.72 & 17 836.66 & 21 461.51 & 345.78 \\
        20 & 30 & baseline & 4 & 164.33 & 163.66 & 256.31 & 331.12 & 342.63 & 4 836.34 & 849.64 & 6 553.56 & 23 354.22 & 27 814.65 & 405.09 \\
        20 & 30 & baseline & 5 & 169.02 & 167.67 & 259.77 & 340.77 & 348.62 & 5 077.63 & 899.17 & 6 952.13 & 23 676.51 & 28 591.09 & 417.72 \\
        20 & 30 & EECT & 1 & 28.95 & 15.02 & 23.74 & 171.77 & 244.46 & 396.10 & 33.13 & 662.41 & 1 902.10 & 2 025.64 & 306.61 \\
\bottomrule
    \end{tabularx}
\end{table*}

\begin{table*}[h]
    \setcounter{table}{3}
    \caption{Numeric results for our experiments performed on-premises for each of the experiments. (Continued)}
    \setlength{\tabcolsep}{3.3pt}
    \begin{tabularx}{\textwidth}{cclc|rrrrr|rrrrr|>{\centering\arraybackslash}X}
        \toprule
        & & & & \multicolumn{5}{c|}{$R(i)$ (seconds)} & \multicolumn{5}{c|}{$S(i)$} & \\
        \cmidrule(lr){5-9} \cmidrule(lr){10-14}
        CPUs & Int. & Strategy & Exp. & Average & 50\%-ile & 75\%-ile & 95\%-ile & 99\%-ile & Average & 50\%-ile & 75\%-ile & 95\%-ile & 99\%-ile & $\max c(i)$\\
        \midrule
        20 & 60 & baseline & 1 & 359.16 & 356.25 & 548.21 & 701.22 & 732.80 & 10 713.60 & 1 952.68 & 15 799.82 & 49 687.07 & 59 286.36 & 803.90 \\
        20 & 60 & baseline & 2 & 372.35 & 381.47 & 577.16 & 734.52 & 758.69 & 11 005.84 & 2 129.04 & 17 028.13 & 50 673.56 & 60 859.33 & 831.31 \\
        20 & 60 & baseline & 3 & 348.31 & 347.68 & 546.77 & 696.55 & 731.74 & 10 310.75 & 1 788.12 & 14 142.40 & 48 705.72 & 58 087.92 & 804.17 \\
        20 & 60 & baseline & 4 & 392.49 & 392.74 & 595.12 & 755.12 & 789.06 & 11 662.68 & 2 318.82 & 17 450.02 & 54 921.48 & 63 084.97 & 862.45 \\
        20 & 60 & baseline & 5 & 374.33 & 375.75 & 580.82 & 748.76 & 772.44 & 11 129.08 & 2 047.78 & 15 342.95 & 51 800.26 & 63 131.00 & 843.06 \\
        20 & 60 & EECT & 1 & 60.01 & 6.37 & 61.69 & 376.07 & 525.96 & 771.89 & 24.86 & 329.62 & 4 354.79 & 5 604.77 & 590.55 \\
\bottomrule
    \end{tabularx}
\end{table*}

\begin{table*}[h]
    \setcounter{table}{3}
    \caption{Numeric results for our experiments performed on-premises for each of the experiments. (Continued)}
    \setlength{\tabcolsep}{3.3pt}
    \begin{tabularx}{\textwidth}{cclc|rrrrr|rrrrr|>{\centering\arraybackslash}X}
        \toprule
        & & & & \multicolumn{5}{c|}{$R(i)$ (seconds)} & \multicolumn{5}{c|}{$S(i)$} & \\
        \cmidrule(lr){5-9} \cmidrule(lr){10-14}
        CPUs & Int. & Strategy & Exp. & Average & 50\%-ile & 75\%-ile & 95\%-ile & 99\%-ile & Average & 50\%-ile & 75\%-ile & 95\%-ile & 99\%-ile & $\max c(i)$\\
        \midrule
        20 & 120 & baseline & 1 & 861.95 & 855.41 & 1 304.81 & 1 664.29 & 1 734.61 & 25 814.07 & 5 133.92 & 38 745.78 & 118 916.86 & 141 152.01 & 1 815.17 \\
        20 & 120 & baseline & 2 & 830.83 & 830.35 & 1 260.50 & 1 589.97 & 1 648.10 & 24 813.51 & 5 151.70 & 36 334.94 & 115 938.52 & 133 576.22 & 1 717.03 \\
        20 & 120 & baseline & 3 & 832.58 & 836.40 & 1 263.11 & 1 574.69 & 1 651.94 & 24 797.73 & 5 071.07 & 37 247.99 & 115 558.62 & 131 745.55 & 1 733.27 \\
        20 & 120 & baseline & 4 & 824.36 & 820.36 & 1 254.98 & 1 599.85 & 1 660.99 & 24 653.89 & 4 891.06 & 36 888.60 & 112 978.09 & 136 446.87 & 1 734.12 \\
        20 & 120 & baseline & 5 & 817.69 & 813.87 & 1 238.41 & 1 571.08 & 1 636.58 & 24 348.56 & 4 906.58 & 36 925.35 & 109 845.26 & 131 351.80 & 1 715.70 \\
        20 & 120 & EECT & 1 & 177.78 & 126.37 & 222.38 & 728.28 & 1 020.37 & 3 678.36 & 435.71 & 6 002.87 & 17 613.34 & 22 086.78 & 1 146.48 \\
\bottomrule
    \end{tabularx}
\end{table*}

\begin{table*}[h]
    \setcounter{table}{4}
    \caption{Aggregated numeric results for our experiments performed on multiple nodes.}
    \label{tab:num-cloud}
    \setlength{\tabcolsep}{4pt}
    \begin{tabularx}{\textwidth}{ccccl|rrrrr|>{\centering\arraybackslash}X}
        \toprule
        & & & & & \multicolumn{5}{c|}{$R(i)$ (seconds)} & \\
        \cmidrule(lr){6-10}
        Nodes & CPUs (per node) & Total CPUs & Int. & Strategy & Average & 50\%-ile & 75\%-ile & 95\%-ile & 99\%-ile & $\max c(i)$\\
        \midrule
        1 & 10 & 10 & 120 & baseline & 253.74 & 253.68 & 385.12 & 490.51 & 511.45 & 586.21 \\
        1 & 10 & 10 & 120 & FC & 49.15 & 1.68 & 33.12 & 337.01 & 446.11 & 548.03 \\
        2 & 10 & 20 & 60 & baseline & 106.39 & 106.54 & 167.49 & 220.35 & 240.22 & 317.15 \\
        2 & 10 & 20 & 60 & FC & 42.40 & 2.46 & 30.15 & 270.63 & 346.99 & 467.53 \\
        3 & 10 & 30 & 40 & baseline & 94.50 & 73.19 & 137.27 & 287.51 & 315.08 & 381.75 \\
        3 & 10 & 30 & 40 & FC & 35.73 & 5.03 & 41.94 & 203.59 & 281.38 & 364.24 \\
        4 & 10 & 40 & 30 & baseline & 87.96 & 54.84 & 147.22 & 283.36 & 315.95 & 376.84 \\
        4 & 10 & 40 & 30 & FC & 38.65 & 5.68 & 45.93 & 217.24 & 292.32 & 373.19 \\
        1 & 18 & 18 & 120 & baseline & 521.15 & 519.76 & 789.13 & 1 003.64 & 1 045.16 & 1 136.16 \\
        1 & 18 & 18 & 120 & FC & 108.96 & 6.00 & 59.48 & 803.26 & 1 063.21 & 1 232.69 \\
        2 & 18 & 36 & 60 & baseline & 250.52 & 251.49 & 381.16 & 487.78 & 518.81 & 609.21 \\
        2 & 18 & 36 & 60 & FC & 99.55 & 2.93 & 28.97 & 728.20 & 859.13 & 1 009.59 \\
        3 & 18 & 54 & 40 & baseline & 245.87 & 215.44 & 377.28 & 597.07 & 643.72 & 737.64 \\
        3 & 18 & 54 & 40 & FC & 68.62 & 6.00 & 54.02 & 443.97 & 638.19 & 756.19 \\
        4 & 18 & 72 & 30 & baseline & 239.86 & 193.97 & 406.44 & 599.57 & 649.21 & 723.27 \\
        4 & 18 & 72 & 30 & FC & 80.72 & 15.02 & 80.27 & 461.29 & 627.30 & 831.40 \\
        \bottomrule
    \end{tabularx}
\end{table*}

\begin{table*}[h]
    \setcounter{table}{5}
    \caption{Aggregated numeric results for our experiments performed on multiple nodes for each of the experiments.}
    \label{tab:num-cloud-exp}
    \setlength{\tabcolsep}{4pt}
    \begin{tabularx}{\textwidth}{cccclc|rrrrr|>{\centering\arraybackslash}X}
        \toprule
        & & & & & & \multicolumn{5}{c|}{$R(i)$ (seconds)} & \\
        \cmidrule(lr){7-11}
        Nodes & CPUs (per node) & Total CPUs & Int. & Strategy & Exp. & Average & 50\%-ile & 75\%-ile & 95\%-ile & 99\%-ile & $\max c(i)$\\
        \midrule
        1 & 10 & 10 & 120 & baseline & 1 & 258.70 & 256.09 & 392.30 & 497.28 & 513.62 & 585.90 \\
        1 & 10 & 10 & 120 & baseline & 2 & 254.33 & 255.72 & 385.77 & 495.21 & 509.98 & 583.72 \\
        1 & 10 & 10 & 120 & baseline & 3 & 249.56 & 251.05 & 381.26 & 484.18 & 500.15 & 573.55 \\
        1 & 10 & 10 & 120 & baseline & 4 & 252.33 & 249.29 & 381.32 & 485.15 & 503.65 & 584.74 \\
        1 & 10 & 10 & 120 & baseline & 5 & 253.77 & 253.18 & 385.35 & 495.83 & 513.37 & 586.21 \\
        1 & 10 & 10 & 120 & FC & 1 & 51.78 & 3.00 & 37.66 & 341.80 & 450.44 & 548.03 \\
\bottomrule
    \end{tabularx}
\end{table*}

\begin{table*}[h]
    \setcounter{table}{5}
    \caption{Aggregated numeric results for our experiments performed on multiple nodes for each of the experiments. (Continued)}
    \setlength{\tabcolsep}{4pt}
    \begin{tabularx}{\textwidth}{cccclc|rrrrr|>{\centering\arraybackslash}X}
        \toprule
        & & & & & & \multicolumn{5}{c|}{$R(i)$ (seconds)} & \\
        \cmidrule(lr){7-11}
        Nodes & CPUs (per node) & Total CPUs & Int. & Strategy & Exp. & Average & 50\%-ile & 75\%-ile & 95\%-ile & 99\%-ile & $\max c(i)$\\
        \midrule
        2 & 18 & 36 & 60 & baseline & 1 & 252.61 & 257.73 & 386.53 & 486.38 & 513.20 & 593.55 \\
        2 & 18 & 36 & 60 & baseline & 2 & 249.82 & 250.08 & 376.92 & 490.16 & 512.13 & 585.58 \\
        2 & 18 & 36 & 60 & baseline & 3 & 246.20 & 245.94 & 374.04 & 481.82 & 499.38 & 581.16 \\
        2 & 18 & 36 & 60 & baseline & 4 & 248.05 & 247.31 & 375.07 & 490.75 & 524.18 & 608.34 \\
        2 & 18 & 36 & 60 & baseline & 5 & 255.90 & 254.93 & 391.11 & 492.95 & 526.50 & 609.21 \\
        2 & 18 & 36 & 60 & FC & 1 & 98.04 & 3.84 & 31.30 & 698.30 & 829.07 & 959.14 \\
\bottomrule
    \end{tabularx}
\end{table*}

\end{document}